\documentclass[longauth]{aa}
\usepackage{graphicx}
\usepackage{txfonts}
\usepackage{color}
\usepackage{subcaption}
\usepackage[dvipsnames]{xcolor}
\usepackage{float}
\usepackage{gensymb}
\usepackage{siunitx} 
\usepackage{hyperref}
\hypersetup{
    colorlinks,
    linkcolor={blue},
    citecolor={blue},
    urlcolor={blue}
}

\graphicspath{{./figures/}}
\begin{document}

\title{Strong lensing models of eight CLASH clusters from extensive spectroscopy: accurate total mass reconstructions in the cores}

\titlerunning{Strong lens modelling of eight CLASH clusters from extensive spectroscopy}
\authorrunning{G.~B.~Caminha et al.} 

\author{G.~B.~Caminha       \inst{\ref{Kapteyn}}                        
                            \thanks{e-mail address: \href{mailto:caminha[at]astro.rug.nl}{caminha[at]astro.rug.nl}. The full redshift catalogues from the MUSE observations (Table \ref{tab:full_z_cat}) is available at the CDS via anonymous ftp cdsarc.u-strasbg.fr (130.79.128.5) or via http://cdsarc.u-strasbg.fr/viz-bin/cat/J/A+A/632/A36.},
        P.~Rosati           \inst{\ref{unife},\,\ref{inafbologna}}      \and
        C.~Grillo           \inst{\ref{unimilano},\,\ref{dark}}         \and
        G.~Rosani           \inst{\ref{Kapteyn}}                        \and
        K.~I.~Caputi        \inst{\ref{Kapteyn},\,\ref{dawn}}           \and
        M.~Meneghetti       \inst{\ref{inafbologna},\,\ref{infnbologna}}\and
        A.~Mercurio         \inst{\ref{inafcapo}}                       \and
        I.~Balestra         \inst{\ref{obs_munich},\,\ref{inaftrieste}} \and
        P.~Bergamini        \inst{\ref{inafpd}}                         \and
        A.~Biviano          \inst{\ref{inaftrieste}}                    \and
        M.~Nonino           \inst{\ref{inaftrieste}}                    \and
        K.~Umetsu           \inst{\ref{sinica}}                         \and
        E.~Vanzella         \inst{\ref{inafbologna}}                    \and
		M.~Annunziatella    \inst{\ref{tufts}}                          \and
        T.~Broadhurst       \inst{\ref{bilbao},\,\ref{sebastian},\,\ref{ikerbasque}} \and
        C.~Delgado-Correal  \inst{\ref{unife}}                          \and
        R.~Demarco          \inst{\ref{concepcion}}                     \and
        A.~M.~Koekemoer     \inst{\ref{stsi}}                           \and
        M.~Lombardi         \inst{\ref{unimilano}}                      \and
        C.~Maier            \inst{\ref{vienna}}                         \and
        M.~Verdugo          \inst{\ref{vienna}}                         \and
        A.~Zitrin           \inst{\ref{BenGurion}}                      
        }
\institute{
Kapteyn Astronomical Institute, University of Groningen, Postbus 800, 9700 AV Groningen, The Netherlands \label{Kapteyn}\and
Dipartimento di Fisica e Scienze della Terra, Universit\`a degli Studi di Ferrara, Via Saragat 1, I-44122 Ferrara, Italy\label{unife}\and
INAF - Osservatorio Astronomico di Bologna, via Gobetti 93/3, 40129 Bologna, Italy\label{inafbologna}\and
Dipartimento di Fisica, Universit\`a  degli Studi di Milano, via Celoria 16, I-20133 Milano, Italy\label{unimilano} \and
Dark Cosmology Centre, Niels Bohr Institute, University of Copenhagen, Juliane Maries Vej 30, DK-2100 Copenhagen, Denmark\label{dark}\and
The Cosmic Dawn Center, Niels Bohr Institute, University of Copenhagen, Juliane Maries Vej 30, DK-2100 Copenhagen {\O}, Denmark\label{dawn}\and
INFN - Sezione di Bologna, viale Berti Pichat 6/2, 40127 Bologna, Italy\label{infnbologna} \and
INAF - Osservatorio Astronomico di Capodimonte, Via Moiariello 16, I-80131 Napoli, Italy\label{inafcapo} \and
University Observatory Munich, Scheinerstrasse 1, 81679 Munich, Germany\label{obs_munich} \and
Dipartimento di Fisica e Astronomia “G. Galilei”, Università degli Studi di Padova, Vicolo dell’Osservatorio 3, I-35122, Italy \label{inafpd} \and
INAF - Osservatorio Astronomico di Trieste, via G. B. Tiepolo 11, I-34143, Trieste, Italy\label{inaftrieste} \and
Academia Sinica Institute of Astronomy and Astrophysics (ASIAA), No. 1, Section 4, Roosevelt Road, Taipei 10617, Taiwan \label{sinica} \and
Physics Department, Tufts University, 574 Boston Ave., Medford, 02155, MA \label{tufts} \and
Department of Theoretical Physics, University of the Basque Country UPV/EHU, Bilbao, Spain \label{bilbao} \and
Donostia International Physics Center (DIPC), 20018 Donostia-San Sebastian (Gipuzkoa), Spain \label{sebastian} \and
IKERBASQUE, Basque Foundation for Science, Bilbao, Spain \label{ikerbasque} \and
Departamento de Astronom\'ia, Facultad de Ciencias F\'isicas y Matem\'aticas, Universidad de Concepci\'on, Concepci\'on, Chile \label{concepcion} \and
Space Telescope Science Institute, 3700 San Martin Dr., Baltimore MD 21218, USA \label{stsi} \and
University of Vienna, Department of Astrophysics, Tuerkenschanzstrasse 17, 1180 Vienna, Austria \label{vienna} \and
Physics Department, Ben-Gurion University of the Negev, P.O. Box 653, Be'er-Sheva 84105, Israel \label{BenGurion}
}

\abstract
{We carried out a detailed strong lensing analysis of a sub-sample of eight galaxy clusters of the Cluster Lensing And Supernova survey with Hubble (CLASH) in the redshift range of $ z_{\rm cluster} = [0.23 - 0.59]$ using extensive spectroscopic information, primarily from the Multi Unit Spectroscopic Explorer (MUSE) archival data and complemented with CLASH-VLT redshift measurements.
The observed positions of the multiple images of strongly lensed background sources were used to constrain parametric models describing the cluster total mass distributions.
Different models were tested in each cluster depending on the complexity of its mass distribution and on the number of detected multiple images.
Four clusters show more than five spectroscopically confirmed multiple image families.
In this sample, we did not make use of families that are only photometrically identified in order to reduce model degeneracies between the values of the total mass of a cluster source redshifts, in addition to systematics due to the potential misidentifications of multiple images.
For the remaining four clusters, we used additional families without any spectroscopic confirmation to increase the number of strong lensing constraints up to the number of free parameters in our parametric models.
We present spectroscopic confirmation of 27 multiply lensed sources, with no previous spectroscopic measurements, spanning over the redshift range of $z_{\rm src}=[0.7-6.1]$.
Moreover, we confirm an average of $48$ galaxy members in the core of each cluster thanks to the high efficiency and large field of view of MUSE.
We used this information to derive precise strong lensing models, projected total mass distributions, and magnification maps.
We show that, despite having different properties (i.e. number of mass components, total mass, redshift, etc), the projected total mass and mass density profiles of all clusters have very similar shapes when rescaled by independent measurements of $M_{200c}$ and $R_{200c}$.
Specifically, we measured the mean value of the projected total mass of our cluster sample within 10 (20)\% of $R_{200c}$ to be 0.13 (0.32) of $M_{200c}$, with a remarkably small scatter of 5 (6)\%.
Furthermore, the large number of high-z sources and the precise magnification maps derived in this work for four clusters add up to the sample of high-quality gravitational telescopes to be used to study the faint and distant Universe.
}


\maketitle

\section{Introduction}
\label{sec:introduction}

\begin{figure*}
  \centering
  \includegraphics[width = 0.675\columnwidth]{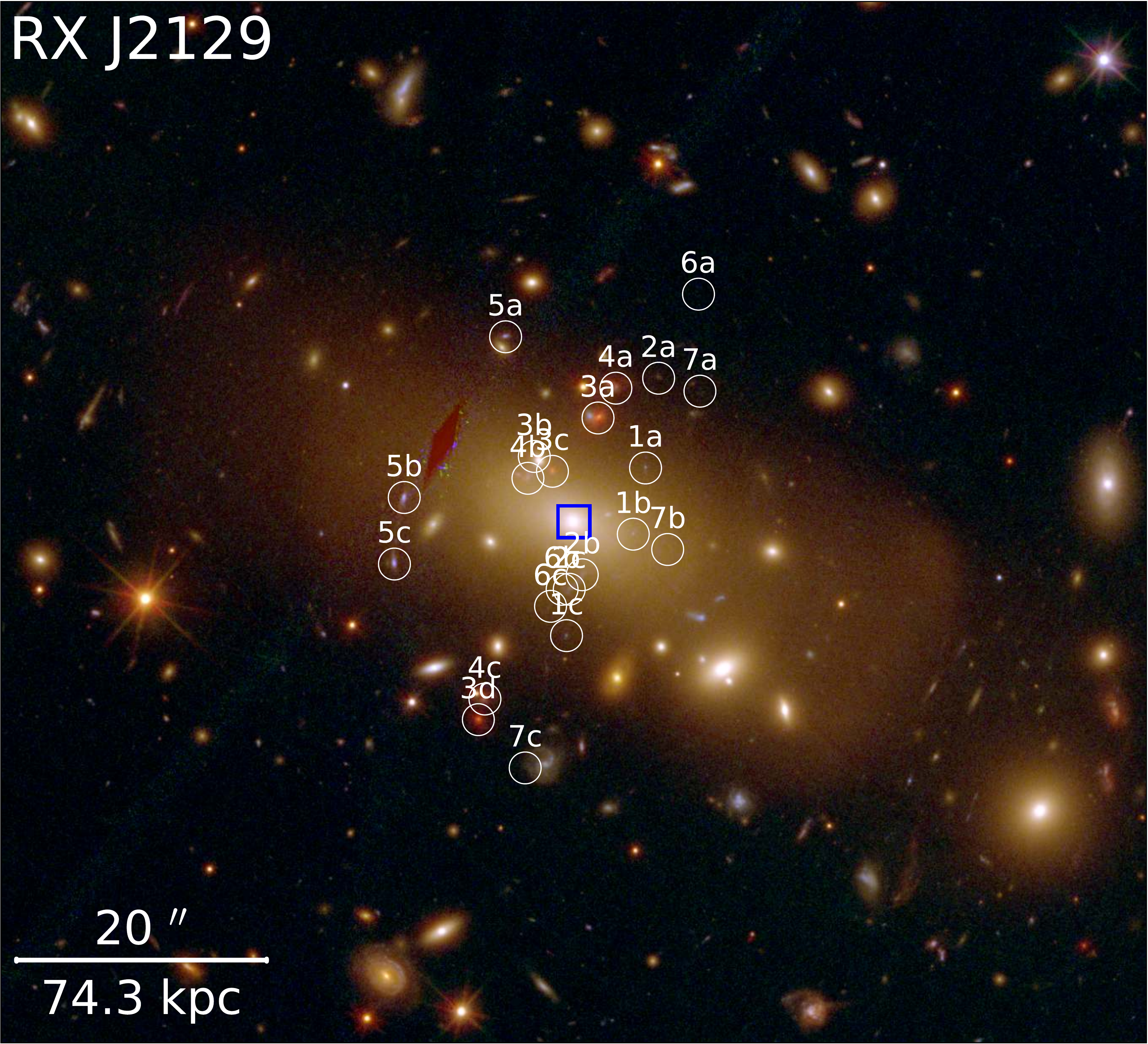}
  \includegraphics[width = 0.675\columnwidth]{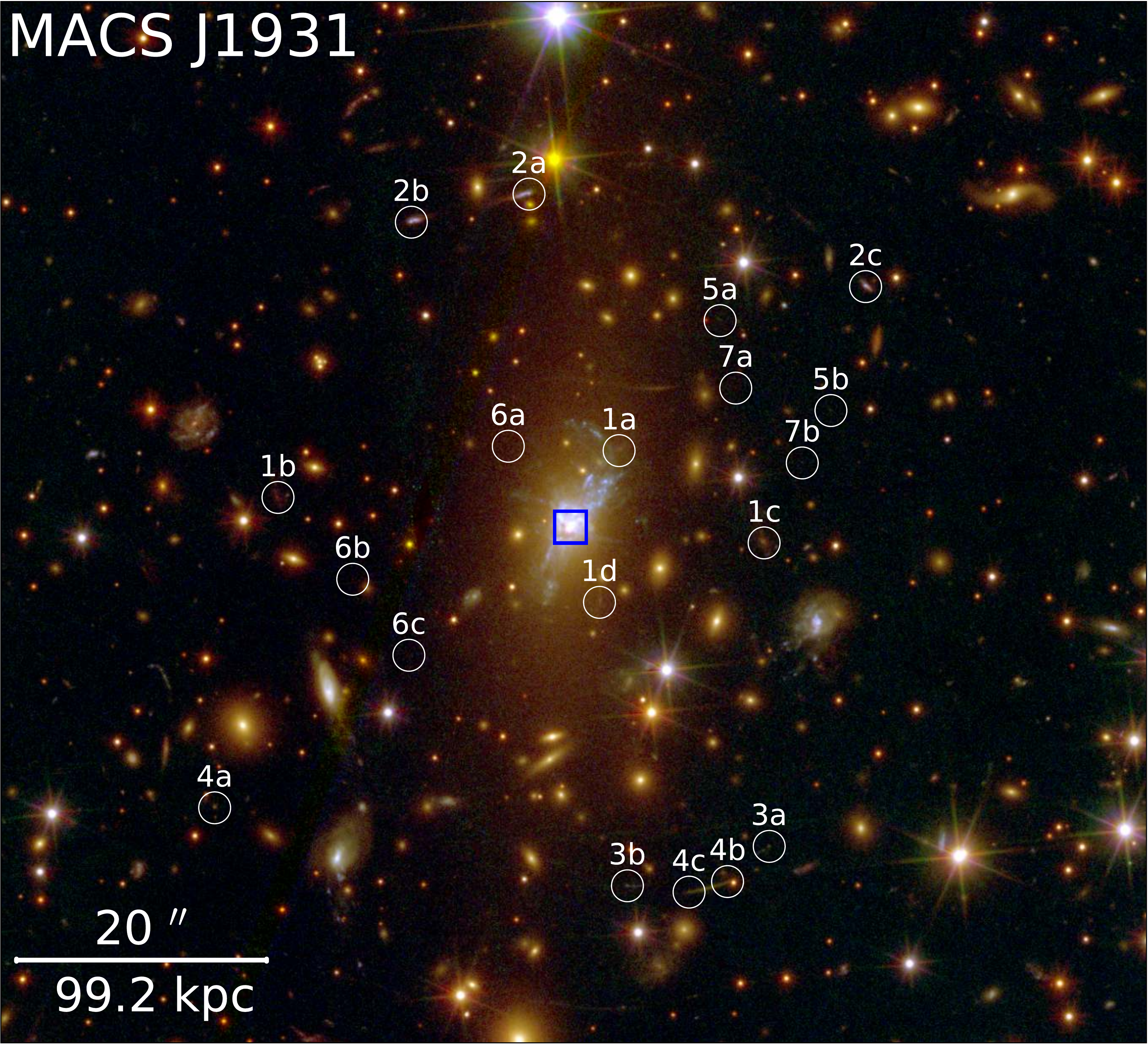}  
  \includegraphics[width = 0.675\columnwidth]{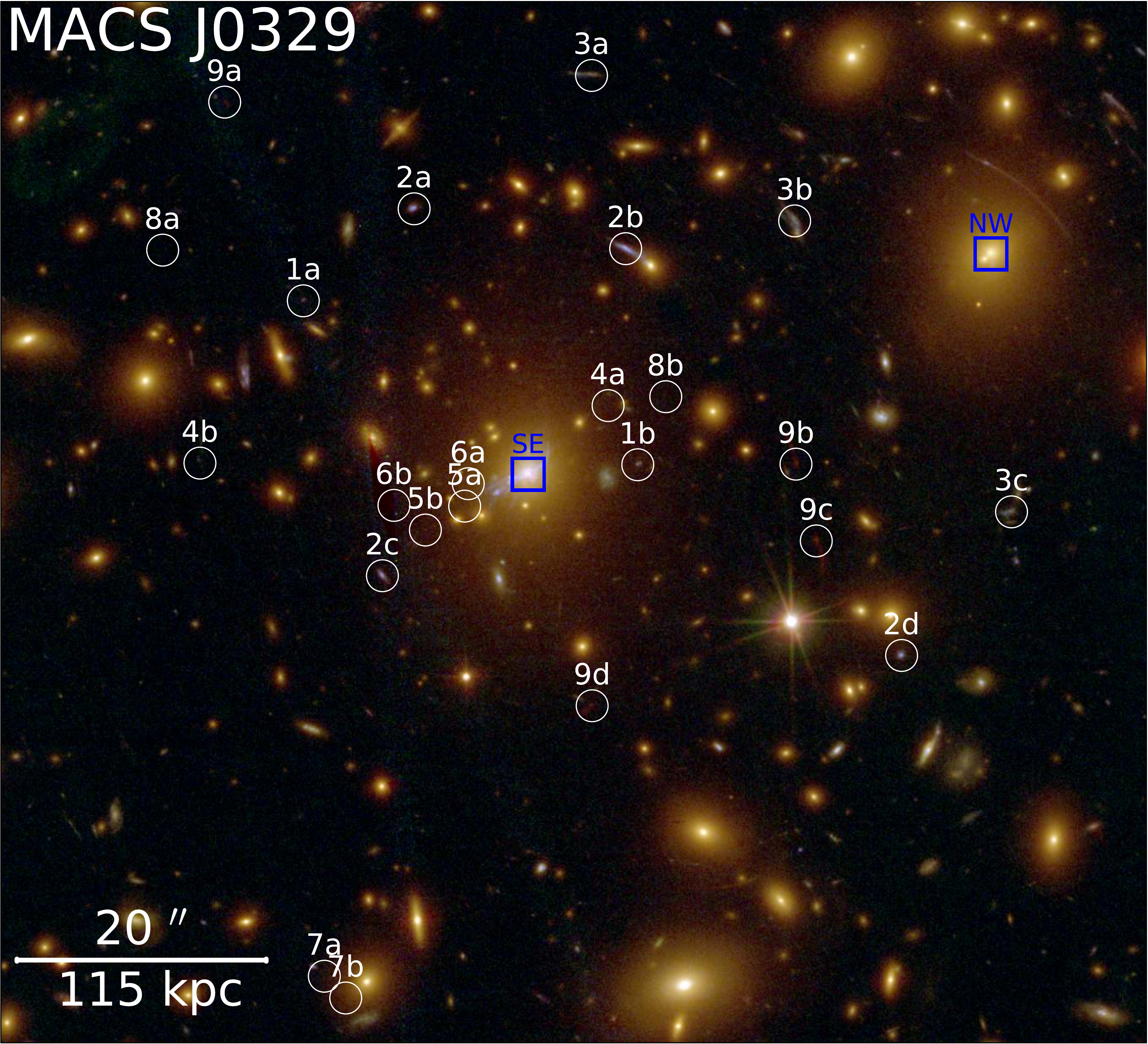}

  \includegraphics[width = 0.675\columnwidth]{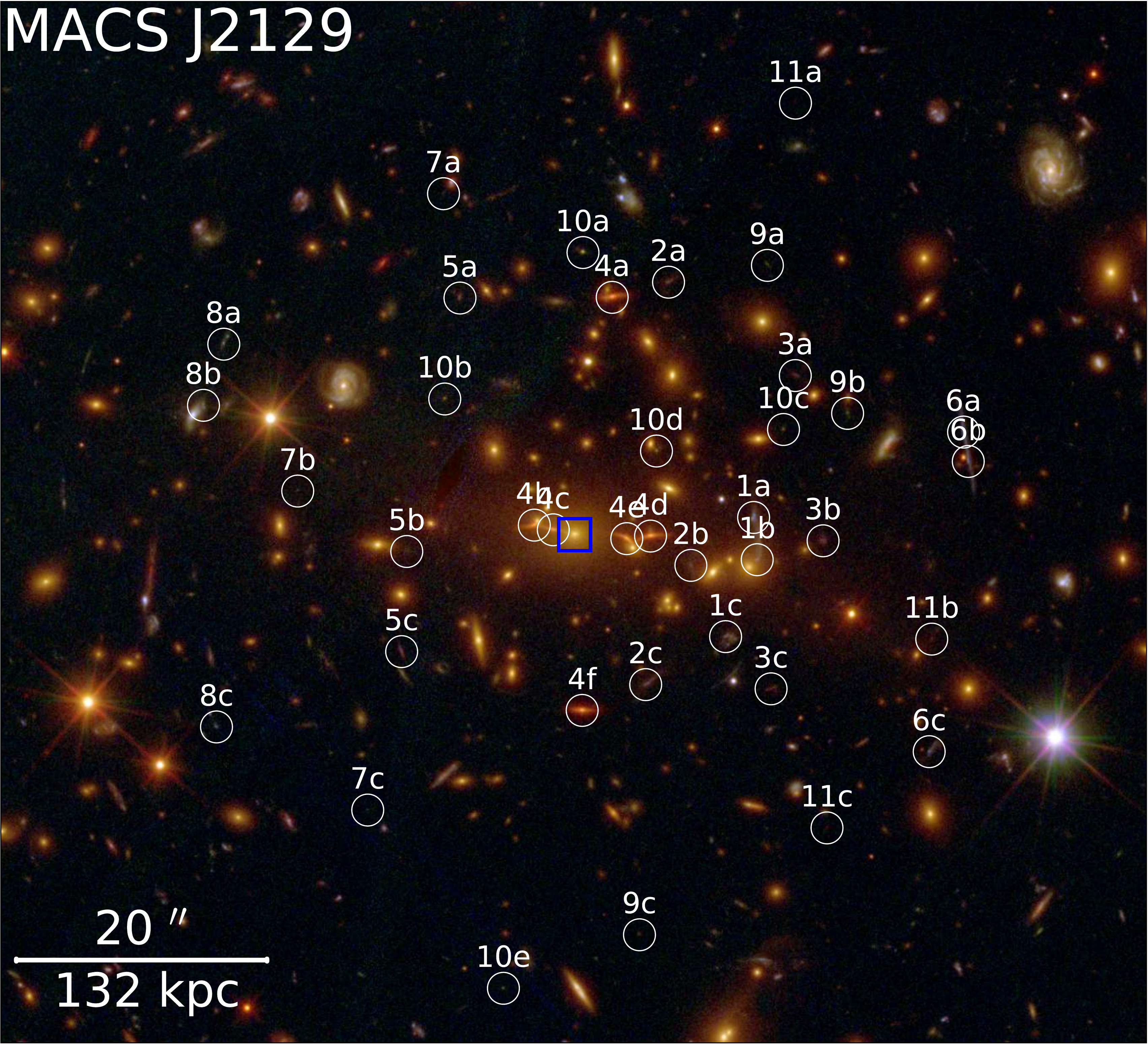}
  \hspace{0.675\columnwidth}
  \includegraphics[width = 0.675\columnwidth]{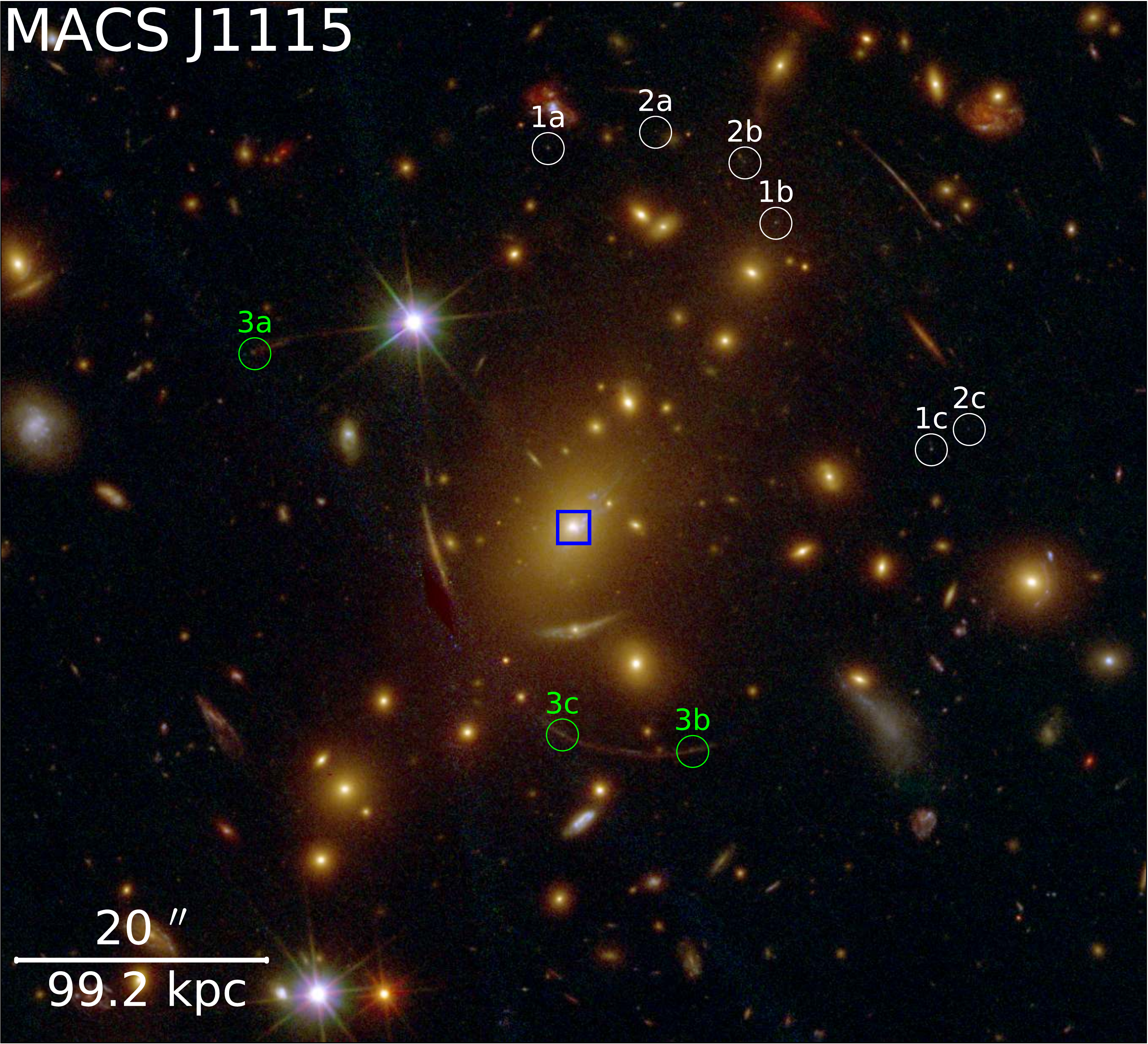}
  
  \includegraphics[width = 0.675\columnwidth]{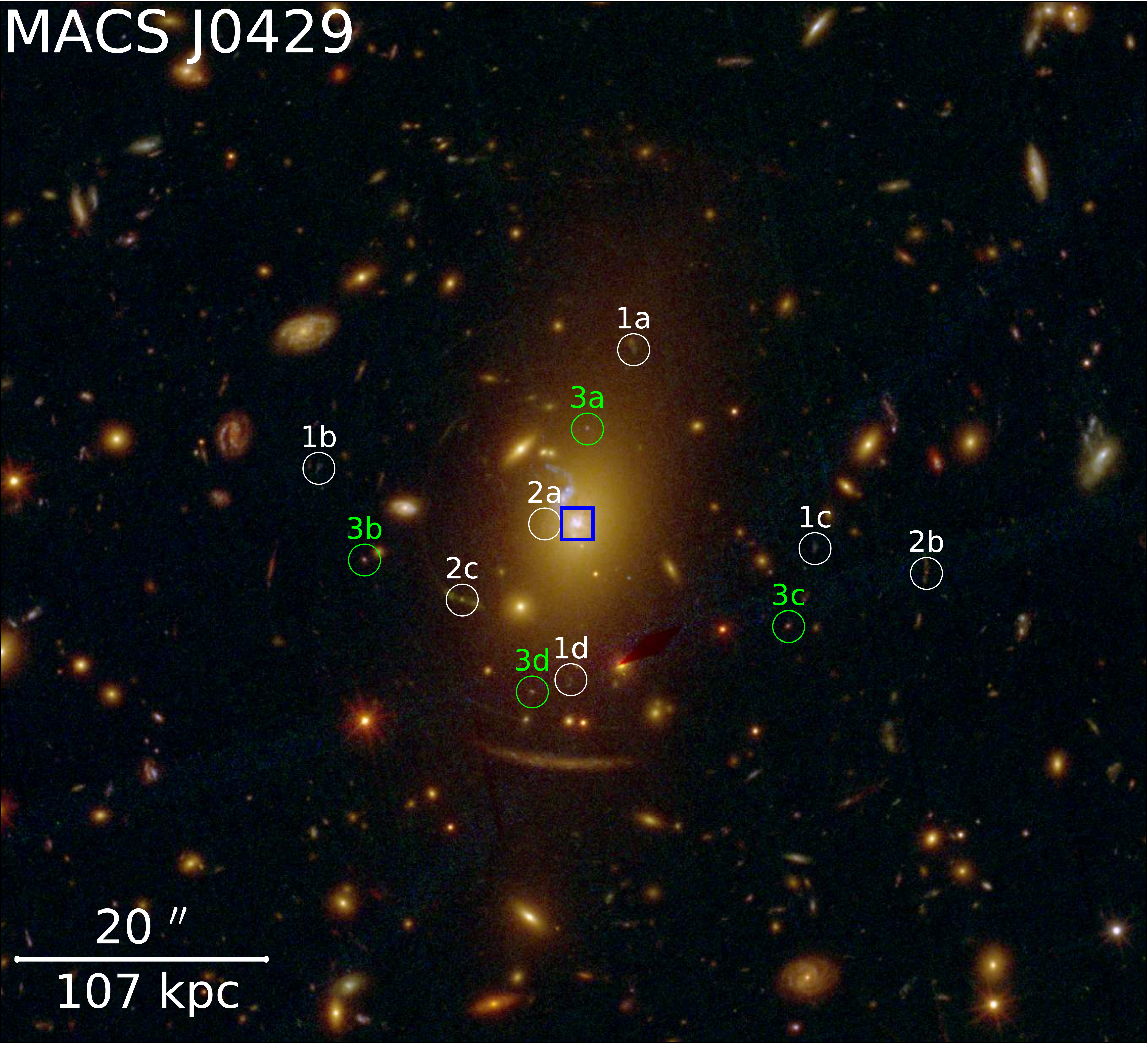}
  \includegraphics[width = 0.675\columnwidth]{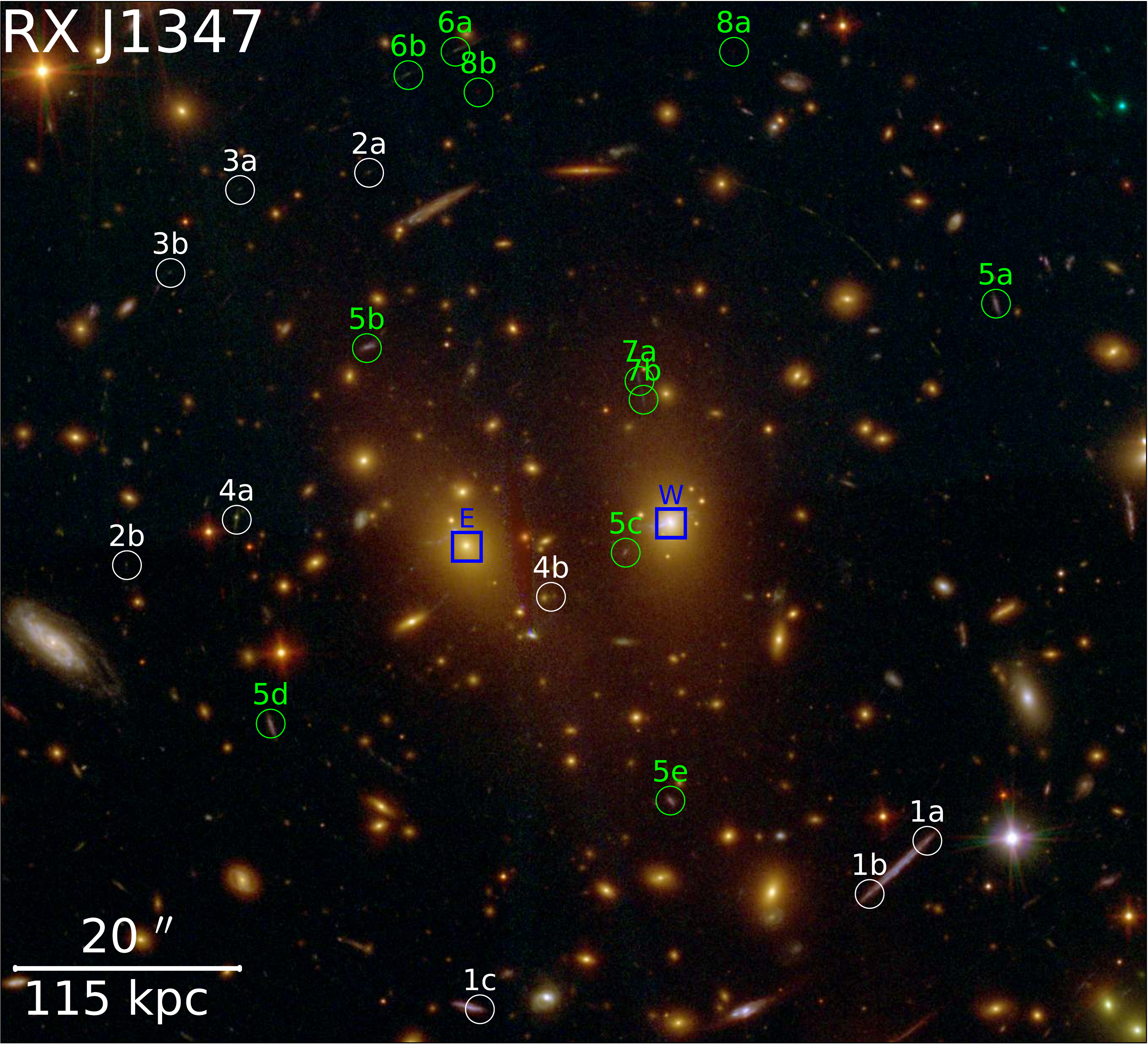}
  \includegraphics[width = 0.675\columnwidth]{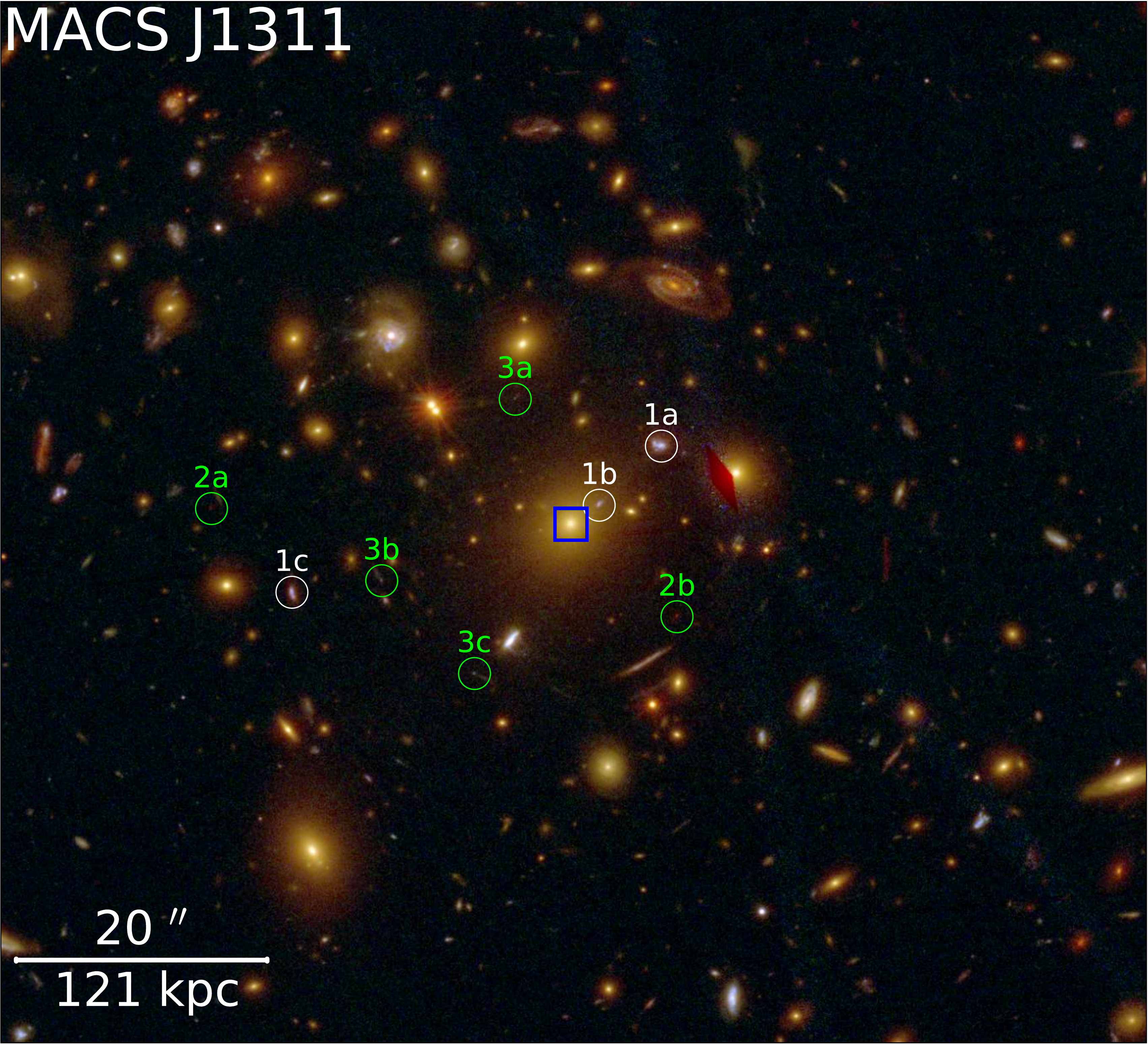}
  \caption{Colour composite images of the eight clusters in our sample created from the combination of the HST/ACS and WFC3 imaging. Circles show the multiple image positions used as model constraints in our strong lens modelling. We identify in white the families with spectroscopic redshift measurements and in green those for which the redshift value is a free parameter in our models (considered only in our silver sample, see Section \ref{sec:strong_lensing_models} for more details). Blue boxes indicate the BCG positions.}
  \label{fig:multiple_images}
\end{figure*}

The importance of galaxy cluster strong lensing in cosmological studies has increased significantly in recent years thanks to high quality data from extensive observational programmes, using photometry and spectroscopy.
Strong lenses can be used for different purposes, such as to study the details of the total mass distribution in galaxy clusters, to identify and characterise intrinsically faint but highly magnified sources at high-redshifts, and to probe the background geometry of the Universe \citep[for a review see][]{2011A&ARv..19...47K}.
For instance, gravitational lensing studies \citep[e.g.][]{2004ApJ...604...88S,2013ApJ...765...25N,2013ApJ...765...24N} have indicated that the inner slope of the dark-matter mass density profile of massive clusters is flatter than the canonical Navarro-Frenk-White profile \citep[NFW,][]{1996ApJ...462..563N, 1997ApJ...490..493N, 1998ApJ...499L...5M, 2012MNRAS.425.2169G}.
On the other hand, hydrodynamical simulations have shown contrasting results, whether baryonic processes produce a shallow dark matter profile \citep{2012MNRAS.422.3081M} or whether they are in agreement with the NFW model \citep{2015MNRAS.452..343S}.
To clarify observationally this discrepancy found in simulations, it is necessary to perform accurate strong lensing analyses using a large sample of spectroscopically confirmed multiple image families, that is, strong lensing constraints, in order to provide reliable measurements of the inner total mass density.

Moreover, the gravitational lensing magnification effect produced by clusters across a relatively large area of the sky has been used to select candidates of the most distant galaxies in the Universe \citep[$z\gtrsim9$,][]{2013ApJ...762...32C, 2014ApJ...795..126B, 2014ApJ...793L..12Z, 2018ApJ...864L..22S}.
Therefore, lensing fields are good targets to push the detection limits of current instrumentation towards the faint and far population of galaxies \citep{2015ApJ...800...18A, 2018MNRAS.479.5184A, 2017ApJ...843..129B}.
Studies of background faint galaxies at lower redshifts $z \approx 3-6$ \citep[][]{2016A&A...595A.100C, 2016MNRAS.456.4191P, 2017ApJ...849...82H, 2017MNRAS.467.3306S, 2017MNRAS.465.3803V, 2017MNRAS.467.4304V, 2019MNRAS.483.3618V} have also greatly benefited from the enhanced spatial resolution and amplified flux produced by the gravitational lensing effect.
This all leads to some important insights on the evolution and characterization of galaxies, indicating that this faint population might be important for the reionisation of the Universe \citep{2014MNRAS.443L..20Y, 2015ApJ...811..140B, 2015ApJ...802L..19R}.
In pursuing this large variety of applications, several observational efforts, such as the Cluster Lensing And Supernova survey with Hubble \citep[CLASH,][]{2012ApJS..199...25P}, the Hubble Frontier Fields \citep[HFF,][]{2017ApJ...837...97L} and, more recently, the REionisation Lensing Cluster Survey \citep[RELICS,][]{2017arXiv171008930S, 2019arXiv190302002C}, have spent a total of $\approx 1300$ orbits of the Hubble Space Telescope (HST).
This then provides homogeneous photometry in optical and near infra-red filters with limiting magnitudes in the range of $\rm mag_{F160W} = 27.5 - 28.7$ ($5\sigma$ detection for point sources within $0\arcsec.8$ radius) for 68 lens galaxy clusters.
However, it is crucial to have extensive spectroscopic information about multiple image families and cluster members to build precise and high-resolution
mass and magnification maps in order to probe the total mass distribution of lens clusters and the physical properties of lensed background sources in detail \citep[see e.g.][]{2015ApJ...800...38G, 2016AA...587A..80C, 2016ApJ...832...82J}.
Therefore, different spectroscopic campaigns have targeted sub-samples of these clusters confirming high-redshift candidates, multiple images and cluster members.
Using HST, the Grism Lens-Amplified Survey from Space \citep[GLASS,][]{2015ApJ...812..114T} has obtained relatively low resolution grism spectroscopy in the inner core ($\approx 5 \; {\rm arcmin}^{2}$) of ten clusters.
The CLASH-VLT ESO Large programme carried out a panoramic spectroscopic campaign of 13 southern CLASH clusters using the VIsible Multi-Object Spectrograph (VIMOS), covering $\approx 0.1\;{\rm deg}^{2}$ in each cluster \citep[Rosati et al. in prep,][]{2015ApJ...800...38G, 2016ApJS..224...33B, 2017MNRAS.466.4094M}.

The Multi Unit Spectroscopic Explorer \citep[MUSE,][]{2014Msngr.157...13B} has driven a revolution in strong lens studies of galaxy clusters.
Its efficiency, field of view of $1 \; {\rm arcmin}^{2}$, and capability of detecting faint emission lines out to $z\approx 6.5$ without any source pre-selection are being exploited to expand significantly the spectroscopic confirmation of multiply lensed sources and cluster members that are then used to constrain lens models \citep[see e.g.][]{2016ApJ...822...78G, 2017AA...600A..90C, 2017AA...607A..93C, 2017MNRAS.469.3946L, 2018MNRAS.473..663M, 2019MNRAS.483.3082J}. 
The spectroscopic confirmation of many multiple images is essential to remove some degeneracies between the source redshifts and the cluster total mass.
Moreover, this information avoids wrong identifications of counterimages that can significantly limit the accuracy and precision with which the cluster mass and magnification maps are reconstructed.

In this work, we selected a sub-sample of eight CLASH clusters with MUSE observations in order to carry out detailed strong lensing analyses and improve their total mass distribution measurements in the cores.
The cluster extended names are listed in Table \ref{tab:summary_muse} and hereafter we will use their abbreviated names, namely RX~J2129, MACS~J1931, MACS~J0329, MACS~J2129, MACS~J1115, MACS~J0429, RX~J1347 and MACS~J1311.
We also incorporated a relatively small set of CLASH-VLT spectroscopic redshifts to cover the regions external to the MUSE field of view (Rosati et al. in prep.). 
Our cluster sample spans the redshift and virial mass ranges of $z_{\rm cluster} = 0.234 - 0.587$ and $M_{200c}^{WL} = (4.6 - 35.4)\times 10^{14} {\rm M_{\odot}}$, respectively \citep[these last values were obtained from weak lensing measurements,][]{2015ApJ...806....4M, 2018ApJ...860..104U}.
With the exception of MACS~J2129, these clusters were selected within the CLASH survey to be dynamically relaxed, based on Chandra X-ray observations \citep{2012ApJS..199...25P}.
The combination of these eight clusters with previous works on MACS~J0416, MACS~J1206 and Abell~1063 \citep{2016AA...587A..80C, 2017AA...600A..90C, 2017AA...607A..93C} constitutes the sample of all CLASH clusters with available MUSE spectroscopy to date.

Previous studies have carried out strong lensing analyses on this cluster sample.
To mention some recent works, \citet{2017MNRAS.466.4094M} and \citet{2018ApJ...866...48U} studied the specific clusters MACS~J2129 and RX~J1347, respectively.
Moreover, the full CLASH 25 cluster sample has been strong (and weak-)-lens analysed by \citet{2015ApJ...801...44Z}.
Our work, using the MUSE spectroscopic data, adds a significant number of new confirmed multiple images used to constrain the lens models. 

This paper is organised as follows.
In Section \ref{sec:data}, we present the data used in this work and describe the MUSE observations and redshift measurements.
The methodology used in our strong lens modelling is explained in Section \ref{sec:strong_lensing_models}, and in Section \ref{sec:results} we discuss the results on the total mass distribution of the clusters.
Finally, in Section \ref{sec:conclusions}, we summarise our conclusions.

Throughout this work, we adopt the a flat $\Lambda$CDM cosmological model, with $\Omega_m=0.3$ and $H_{0} = {\rm 70\,km\,s^{-1}\,Mpc^{-1}}$.
The images are oriented with north at top and east to the left, and the angles are measured from the west and oriented counterclockwise.

\begin{table}[!]
\centering
\tiny
\caption{MUSE observation summary.}
\begin{tabular}{l c c c c c c l} \hline \hline
Cluster field of view & $ z_{cluster}$ & $t_{exp}$[hrs] & $ N_{exp.}^{a}$ & seeing$^b$ ($\arcsec$) \\
\hline
RX~J2129.7$+$0005-East   &  0.234 & 2.46 & 6 & $0.61\pm0.08$ \\
RX~J2129.7$+$0005-West   &  0.234 & 3.29 & 8 & $0.53\pm0.17$ \\
MACS~J1931.8$-$2635      &  0.352 & 2.46 & 6 & $0.96\pm0.05$ \\
MACS~J0329.7$-$0211      &  0.450 & 2.47 & 6 & $0.83\pm0.12$ \\
MACS~J2129.4$-$0741-East &  0.587 & 2.46 & 6 & $0.81\pm0.17$ \\
MACS~J2129.4$-$0741-West &  0.587 & 3.25 & 8 & $0.83\pm0.10$ \\
\hline
MACS~J1115.9$+$0129      &  0.352 & 2.47 & 6 & $1.14\pm0.21$ \\
MACS~J0429.6$-$0253-North&  0.399 & 2.45 & 6 & $1.07\pm0.50$ \\
MACS~J0429.6$-$0253-South&  0.399 & 1.63 & 4 & $0.98\pm0.14$ \\
RX~J1347.5$-$1145        &  0.451 & 0.81 & 2 & $0.77\pm0.04$ \\
MACS~J1311.0$-$0310      &  0.494 & 0.81 & 2 & $0.54\pm0.02$ \\
\hline
\end{tabular}
\label{tab:summary_muse}
\tablefoot{The final MUSE field of view for each cluster is shown in Figures \ref{fig:members} and \ref{fig:members_2}.
\tablefoottext{a}{Number of exposures in each pointing used to create the final data-cube.}
\tablefoottext{b}{Median value of the seeing, measured from the DIMM station, and its standard deviation.}
}
\end{table}

\section{Data}
\label{sec:data}

\subsection{HST-CLASH}

The cluster sample presented in this work has been observed by the CLASH survey, using the ACS and WFC3 cameras onboard HST, in 16 filters from the UV through the NIR \citep{2012ApJS..199...25P}.
The data has been reduced using the {\tt Mosaicdrizzle} pipeline \citep{2011ApJS..197...36K} and the final co-added science imaging is made publicly available\footnote{\url{https://archive.stsci.edu/prepds/clash/}} with two different spatial resolutions of 30mas and 65mas per pixel.
In Figure \ref{fig:multiple_images}, we show the colour composite images of the eight clusters studied in this work, produced from the combination of optical and near IR filters by using the {\tt Trilogy} code \citep{2012ApJ...757...22C}.

\subsection{Spectroscopic data}
\label{sec:spectroscopic_data}
Up to now, a total of 11 CLASH clusters have been observed by MUSE in different programmes.
In previous works, we have used deep observations on three targets, MACS~J1206, Abell~1063 and MACS~J0416 \citep{2016AA...587A..80C, 2017AA...600A..90C, 2017AA...607A..93C}, where the last two clusters are also part of the HFF initiative.
In this work, we made use of archival MUSE data from the ESO programme IDs 095.A-0525, 096.A-0105, 097.A-0909 and 098.A-0590 (P.I. J.-P.~Kneib) on the remaining eight clusters.
The observations were carried out during the period between 2015-June and 2017-January, with observation blocks (OBs) consisting of two exposures of $\approx 1465$~seconds.
The fields of view of RX~J2129, MACS~J2129 and MACS~J0429 are composed of two pointings with overlapping areas of $\approx 0.29$~arcmin$^2$, $\approx 0.12$~arcmin$^2$ and zero~arcmin$^2$, respectively.
The remaining five clusters were observed with one single pointing, of which MACS~J1115 and RX~J1347 observations are off-centred by $\approx 30\arcsec$ from the corresponding brightest cluster galaxies (BCGs).
Although having similar exposure times, the dither and rotation pattern of each OB is slightly different in each cluster.

\begin{figure*}
  \centering
  \includegraphics[width = 1.015\columnwidth]{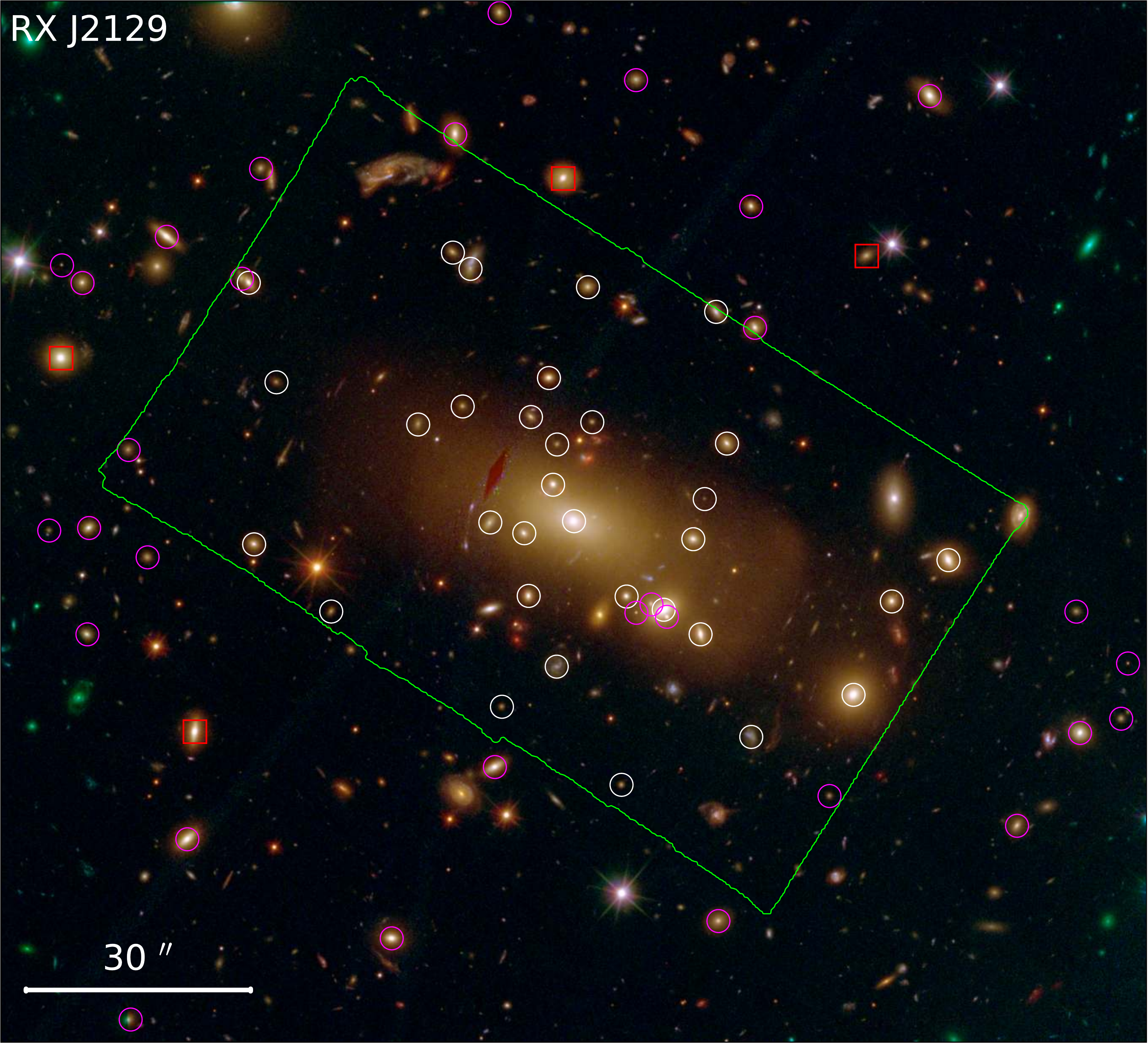}
  \includegraphics[width = 1.015\columnwidth]{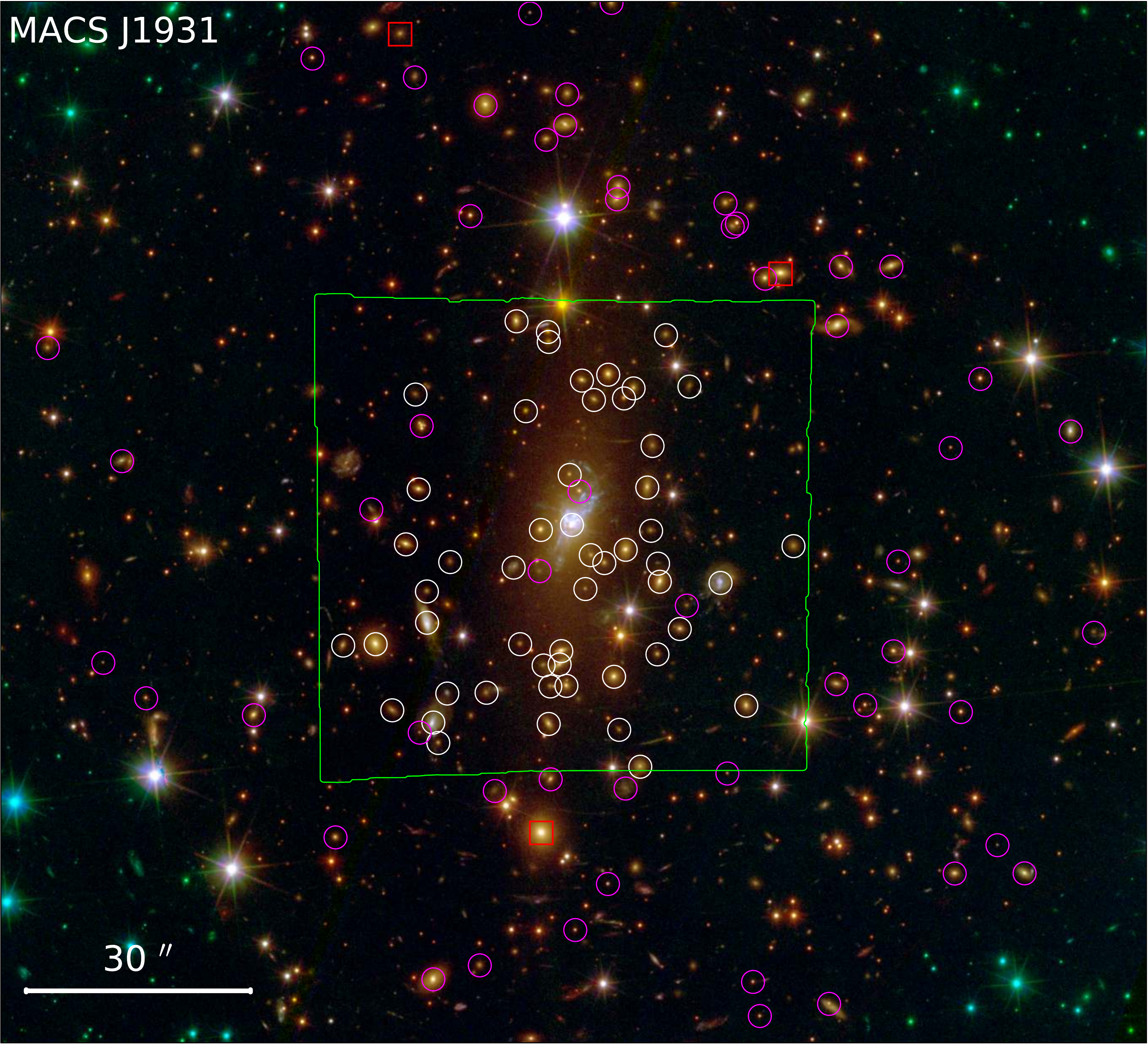}

  \includegraphics[width = 1.015\columnwidth]{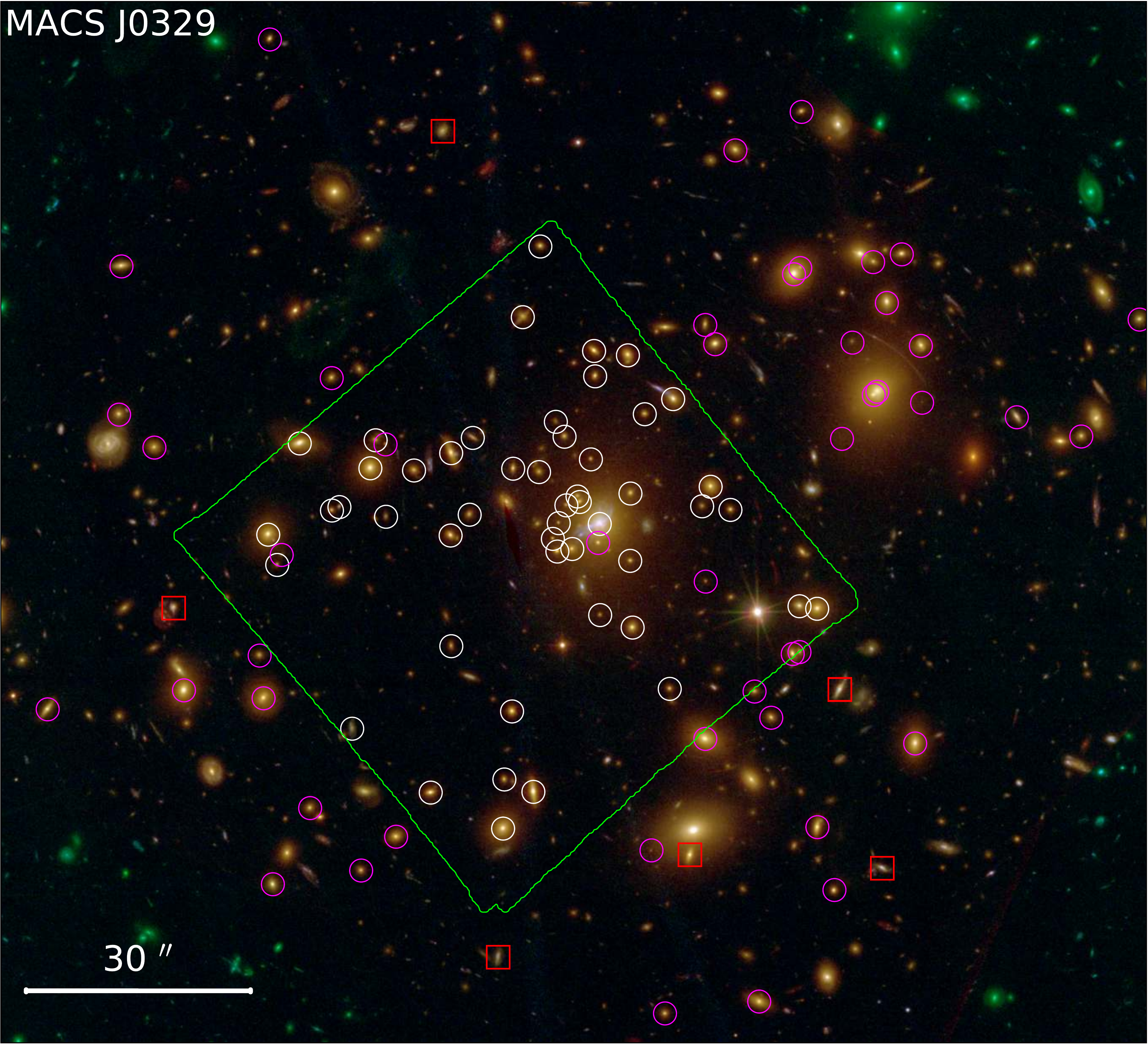}
  \includegraphics[width = 1.015\columnwidth]{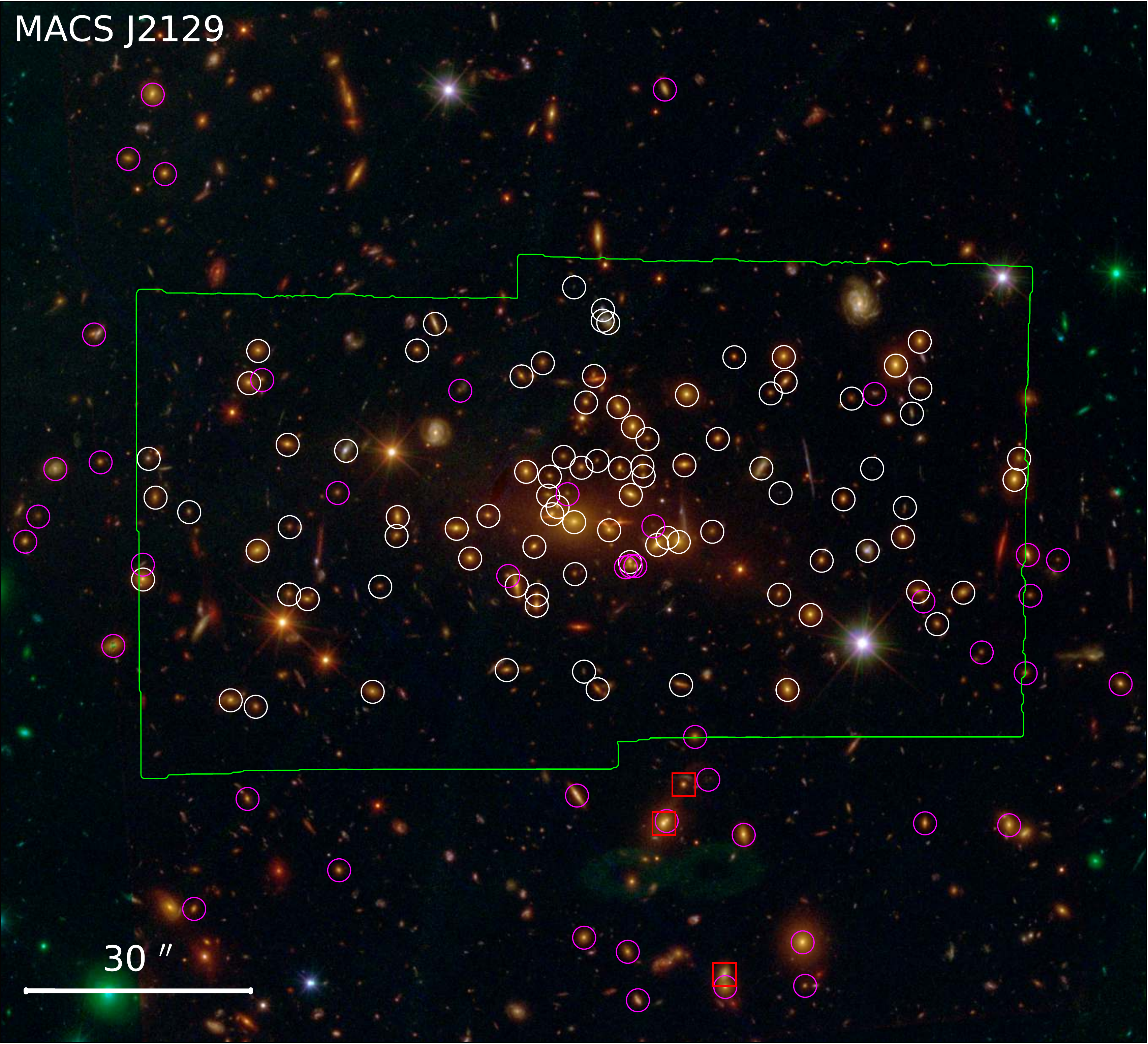}

  \caption{Cluster member selection. White and magenta circles show the position of the cluster members with MUSE spectroscopic confirmation and selected from photometry, respectively. Red boxes show the cluster members spectroscopically confirmed by CLASH-VLT and thus included in the final member catalogue, but  not selected by our criteria. The green regions show the area covered by the MUSE observations.}
  \label{fig:members}
\end{figure*}

\begin{figure*}
  \centering
  \includegraphics[width = 1.015\columnwidth]{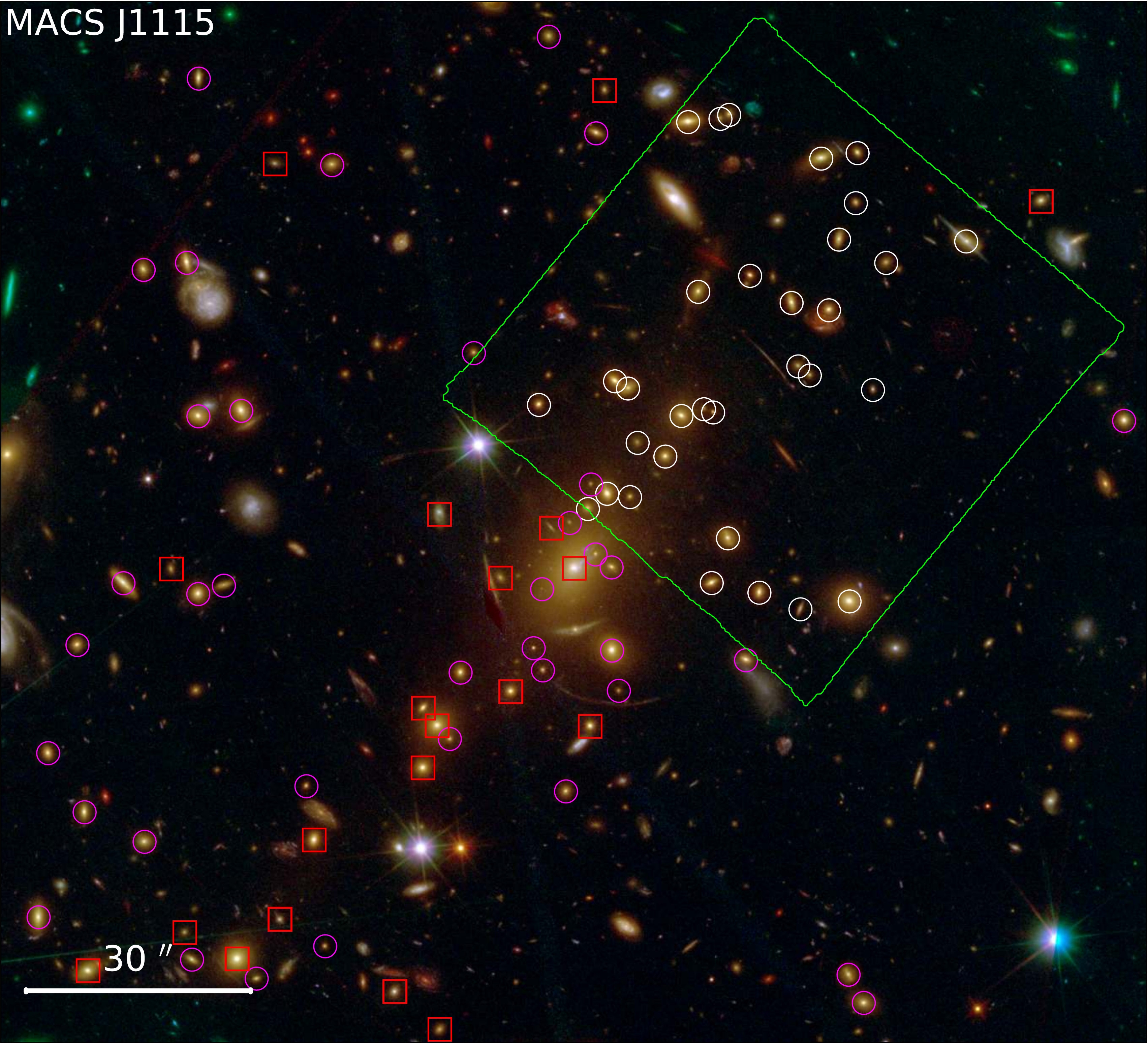}
  \includegraphics[width = 1.015\columnwidth]{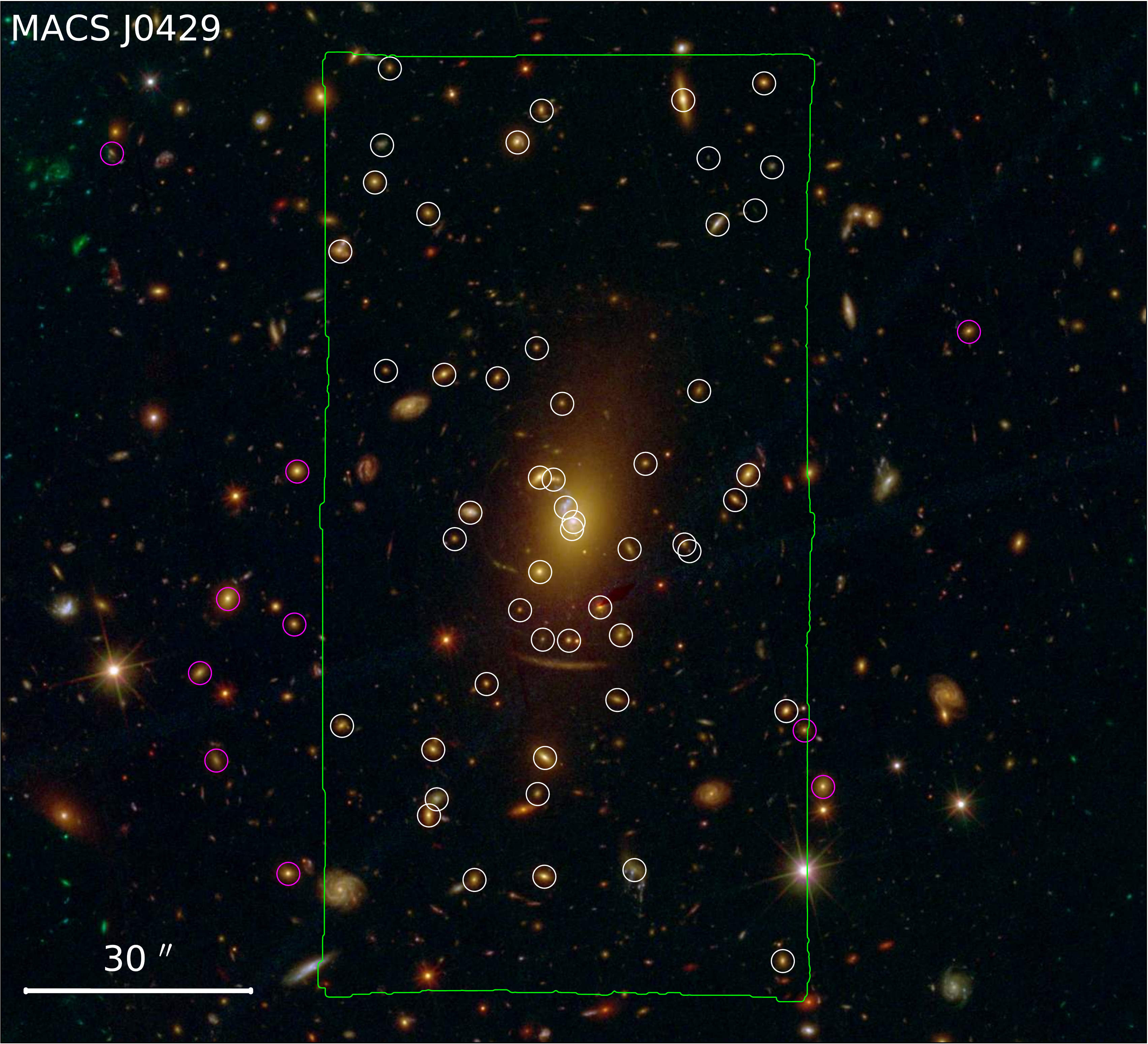}

  \includegraphics[width = 1.015\columnwidth]{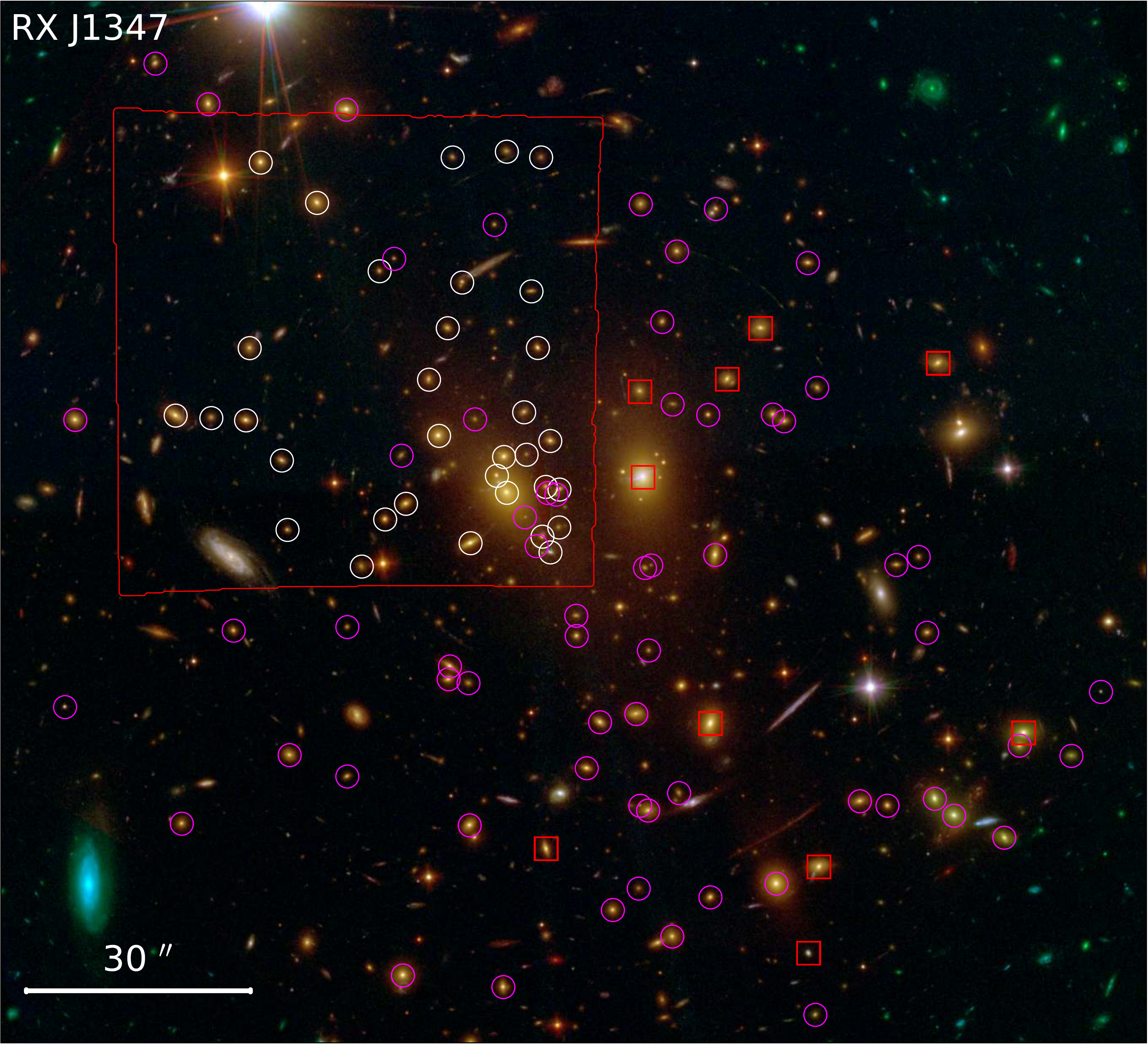}
  \includegraphics[width = 1.015\columnwidth]{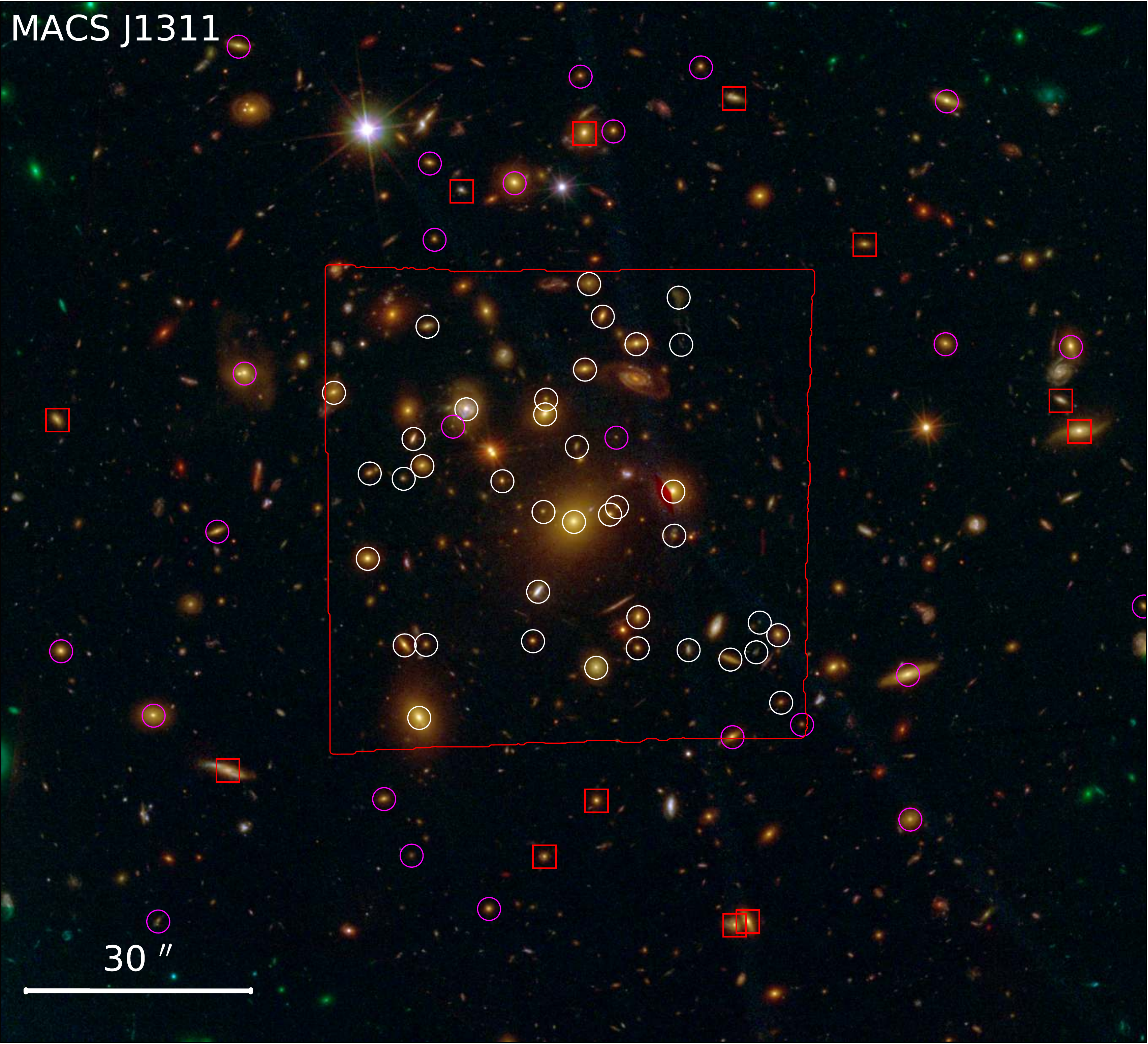}

  \caption{Same as in Figure \ref{fig:members} but for the silver sample of galaxy clusters. Red regions indicate MUSE exposures with less than one hour on target. We note that MACS~J0429 is not a CLASH-VLT cluster.}
  \label{fig:members_2}
\end{figure*}

Final exposure times on target of each pointing vary from shallow 0.8 hours to 3.2 hours and the overlapping regions in RX~J2129 and MACS~J2129 reach $\approx 5.6$ hours depth.
The summary of MUSE observations is presented in Table \ref{tab:summary_muse} and the final fields of view in Figures \ref{fig:members} and \ref{fig:members_2}.
For all exposures, a small dither pattern and $90^{\circ}$ rotations with respect to the original position angle (PA) were applied.
We note that for the clusters MACS~J2129 and MACS~J1931, although they include more than two exposures, only two different rotation angles were used, the original PA and $\rm PA+90^{\circ}$.
For the other clusters with more than two exposures, all four $90^{\circ}$ rotations were applied.
Because of that, in some final MUSE data-cubes the instrumental features were not optimally removed, in particular in the regions between single instrument IFUs.

We followed the standard MUSE reduction procedure, similarly to our previous analyses of other clusters \citep{2015A&A...574A..11K, 2016ApJ...822...78G, 2017A&A...599A..28K, 2017AA...600A..90C, 2017AA...607A..93C}.
We used the MUSE pipeline version 2.4.1 \citep{2014ASPC..485..451W} to process all raw exposures.
Each object exposure is corrected using {\tt BIAS} and {\tt FLAT} calibrations of the corresponding night, as well as {\tt ILLUMINATION} exposures.
Moreover, wavelength and flux calibration are applied in order to create the {\tt PIXELTABLE}s and data-cubes of each exposure.
We inspected the wavelength collapsed (white) images and the data-cubes of each exposure, finding no large variations between the observational conditions of most pointings.
Two object exposures on RX~J2129 presented technical problems and were not used in the final stacks (they are not included in the numbers presented in Table \ref{tab:summary_muse}).
For the same cluster, one exposure shows a bright satellite track across the field of view, which we mask out by removing the affected slices in order not to contaminate the final stack.
After that, we combined the {\tt PIXELTABLE}s into a final data-cube with the final depth and covering the entire field of view of each cluster (see Figures \ref{fig:members} and \ref{fig:members_2}).
Since the standard reduction pipeline does not have an optimal sky subtraction, we applied the {\tt ZAP} tool \citep[version 2.1,][]{2016MNRAS.458.3210S} to remove remaining sky residuals.
To do that, we defined the sky regions using a combination of the MUSE white images and HST data.
The MUSE data-cubes extend from $4750\,\AA$ to $\rm 9350\,\AA$ in wavelength, with an almost constant dispersion of $\approx 1.25\,\AA$, and spatial sampling of $0\arcsec.2$.
The final seeing varies from $\approx 0\arcsec.5$ to $1\arcsec.1$ in the final stacked data-cubes (see Table \ref{tab:summary_muse}).
The final fields of view of all observations are shown in Figures \ref{fig:members} and \ref{fig:members_2}.

Since MUSE does not cover the entire HST field of view, we also used a small subset of redshifts from the ESO large programme CLASH-VLT (ID 186.A-0798, P.I.: Rosati) that has observed seven of the eight clusters analysed in this work, using VIMOS high-multiplexing capabilities \citep[][]{2003SPIE.4841.1670L}.
MACS~J0429 is the only cluster in our sample not included in that programme.
We refer to the spectroscopic campaign of MACS~J0416 \citep{2016ApJS..224...33B} as an example of CLASH-VLT observations, which typically extend to approximately two virial radii.
A full description of all observations and data processing of the 13 CLASH-VLT clusters will be presented in Rosati et al. in prep..

While VIMOS provides an efficient coverage of the entire cluster volume, over a field of view of $\approx 25\arcmin$ across, the central $\approx 1-2\arcmin$ is not adequately sampled with the standard multi-slit strategy due to the rapid increase of targets in the cluster cores (multiple images and cluster galaxies).
To this respect, MUSE integral field spectroscopy represents an ideal complement to the VIMOS observations in the cluster cores.
Finally, we also incorporated redshifts from the GLASS \citep{2015ApJ...812..114T} survey that has observed RX~J1347 and MACS~J2129.

\subsection{MUSE redshift measurements}

Following our previous works, we identify and measure source redshifts in two different manners.
Firstly, we use HST detections as a prior to extract spectra centred on these positions and with an aperture radius of $0\arcsec.8$ from the data-cubes.
This aperture value is similar to the observational seeing of the observations and found to be a good compromise to reduce the contamination from nearby sources and maximise the signal to noise.
In the case of some extended sources and objects with large lensing distortion and with low surface brightness, for which we could not estimate a secure redshift with the previously mentioned apertures, we use a customised extraction that takes into account the extent of the emission observed in the MUSE data.
Secondly, a continuum subtracted data-cube is created by subtracting a median kernel (with 151 spectral pixel width) in the wavelength axis from each spaxel.
The result is a data-cube where continuum emission is removed and emission lines are more easily identified.
We visually inspect these continuum subtracted data-cubes in order to identify emission lines from sources with no HST continuum detection and to disentangle emission from nearby sources in projection.

\begin{table*}[!]
\centering
\small
\caption{Full redshift catalogue.}
\begin{tabular}{c c c c c c c l} \hline \hline
ID &  RA & Dec & $z_{\rm MUSE}$ & QF & mult.\\
(A)&  (B)& (C) & (D)            & (E)& (F)\\
\hline
R2129-J$212939.77$$+000458.74$ & 322.4157062 & 0.0829825& 0.1288 & 3 & 1 \\
R2129-J$212940.68$$+000509.74$ & 322.4195150 & 0.0860391& 0.1344 & 3 & 1 \\
R2129-J$212938.71$$+000510.68$ & 322.4112998 & 0.0863003& 0.1348 & 3 & 1 \\
R2129-J$212937.16$$+000524.20$ & 322.4048393 & 0.0900552& 0.1374 & 3 & 1 \\
R2129-J$212940.84$$+000555.17$ & 322.4201773 & 0.0986576& 0.1378 & 3 & 1 \\
\vdots & \vdots & \vdots & \vdots & \vdots & \vdots \\
\hline
\end{tabular}
\label{tab:full_z_cat}
\tablefoot{Five entries of the full redshift catalogue that is available in the electronic version of the paper. The columns are: (A) the ID build from the cluster name and object RA and Dec; (B) and (C) are the observed right ascension and declination in degrees; (D) and (E) are the spectroscopic redshift and its quality flag; (F) is the number of entries of the same object in this catalogue used to indicate multiply lensed sources.
}
\end{table*}
To measure redshifts, we used the one-dimension (spatially averaged) and pseudo two-dimension (in two perpendicular directions) spectra.
Similarly to our previous works \citep[see e. g., ][]{2017AA...600A..90C,2017AA...607A..93C}, we analysed the data in order to identify spectral features to measure redshifts, such as emission or absorptions lines.
In the cases of sources with continuum emission but no evident features, we used a template matching to help with the identification of faint spectral features.
To each redshift, we attributed a quality flag that is related to the reliability of the measurement.
The quality flag system has four different categories: $\rm QF=1$, the measurement is not reliable; $\rm QF=2$, the redshift value is based on a faint, but clearly detected, feature and is likely to be correct; $\rm QF=3$, secure measurement from more than one absorption and/or emission lines; $\rm QF=9$, redshift based on a single narrow emission.
Cases of objects having a single emission line with clear features that allow us to identify its nature (for instance the Lyman-$\alpha$ profile or \ion{O}{II} doublet) are considered to have secure ($\rm QF=3$) redshift measurements.
The full sample of all fields presented in this work contains $\approx 900$ secure (i.e., $\rm QF>1$) redshift measurements, of which 114 are stars, 62 are foreground galaxies, 390 are cluster members and 395 are background galaxies.
Regarding high redshift sources with $z>2.9$, where the Lyman-$\alpha$ line falls within the MUSE wavelength range, the number of unique detections (i.e. accounting for the lensing multiplicity) is 116.
The first entries of the full catalogue are presented in Table \ref{tab:full_z_cat} and the entire catalogue is available in the electronic version of the paper.

\section{Strong lensing models}
\label{sec:strong_lensing_models}
To perform the strong lensing modelling, we adopted the same methodology applied in our previous studies of the galaxy clusters Abell~1063, MACS~J0416 and MACS~J1206.
In this work we use the software {\tt lenstool} \citep{1996ApJ...471..643K, 2007NJPh....9..447J}.
We adopted parametric models to describe the different components of the total mass distribution of each cluster.
Moreover, the multiple image positions of the lensed sources were used as constraints to estimate the best-fitting values of the model parameters.
In the next subsections, we provide a short description of our methodology and refer to \citet{2016AA...587A..80C, 2017AA...600A..90C, 2017AA...607A..93C} for more details.

\subsection{Overall description}
\label{sec:overall_description}

For the smooth mass components (i.e., dark matter, intracluster light and hot gas), we used a pseudo-isothermal elliptical mass distribution \citep[PIEMD,][]{1993ApJ...417..450K}, that has been shown to describe well such component in previous studies \citep[e.g.][]{2015ApJ...800...38G, 2017AA...607A..93C}.
A PIEMD profile is characterised by a total of six parameters: the central velocity dispersion ($\sigma_{\rm v}$) and core radius ($r_{\rm core}$); the orientation angle ($\theta$) and ellipticity ($\varepsilon$); and the position of the centre ($x,y$).
The ellipticity is given by $\varepsilon \equiv 1 - b/a$, where $a$ and $b$ are the major and minor axis, respectively.
The orientation angle $\theta$ is defined to be zero in the east-west direction and increases counterclockwise.

In order to test possible systematic effects related to the adopted cluster mass parametrization, we also used a generalised NFW profile \citep[gNFW,][]{1996MNRAS.278..488Z,2000ApJ...529L..69J,2001ApJ...555..504W} for the cluster smooth mass component.
This model has the value of the central 3-dimensional slope $\gamma_{\rm gNFW}$, besides those of the concentration $c$ and scale radius $r_{\rm s}$, as free parameters.
In this model, the total number of free parameters is seven (including the position of the centre and the values of ellipticity and orientation angle).
A value of $\gamma_{\rm gNFW}$ equal to one translates into the standard NFW model.
We remark that this model has a pseudo-elliptical implementation in the {\tt lenstool} software, in which the ellipticity is introduced in the projected lens potential and not in the projected mass density profile.
This approximation enables a faster solution of the ray-tracing equation.
However, for high values of ellipticity ($\varepsilon \gtrsim 0.5$) it is known to create nonphysical dumbbell shaped projected mass density distributions \citep{2002A&A...390..821G,2012A&A...544A..83D}.
We remark here that we used this profile only to compare with our reference PIEMD parametrization.

\begin{figure*}
  \centering
  \includegraphics[width = 0.504\columnwidth]{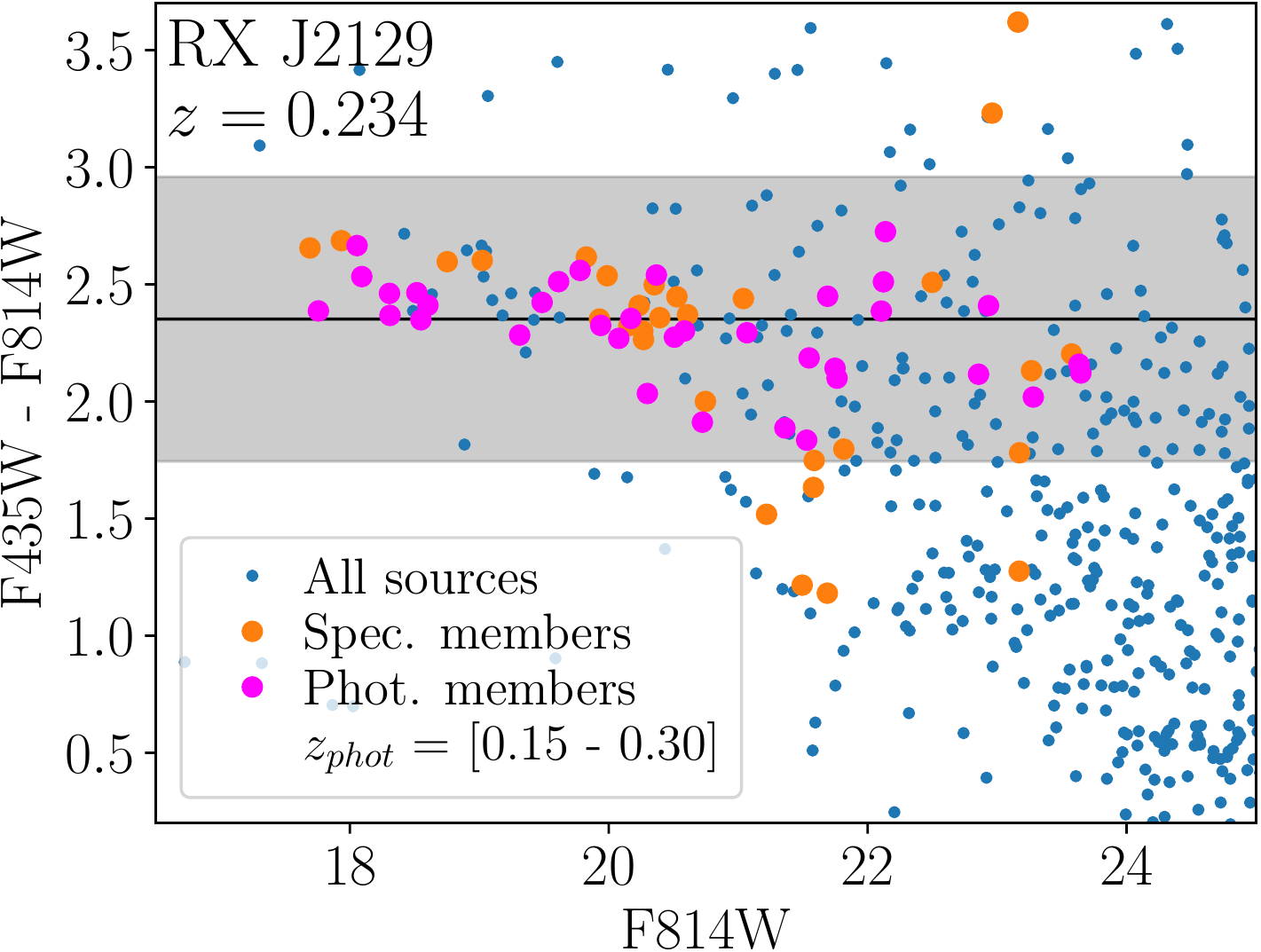}
  \includegraphics[width = 0.504\columnwidth]{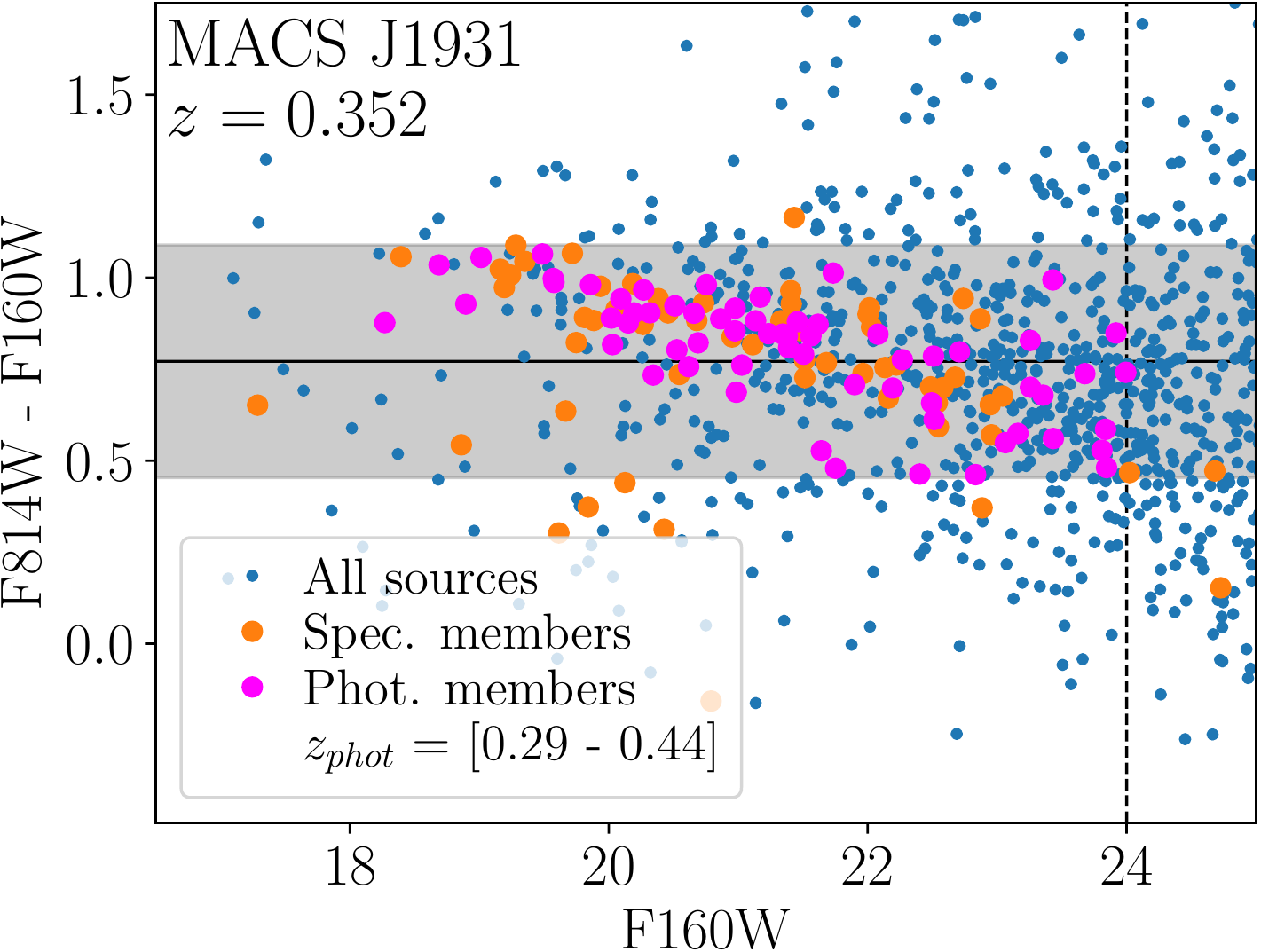}
  \includegraphics[width = 0.504\columnwidth]{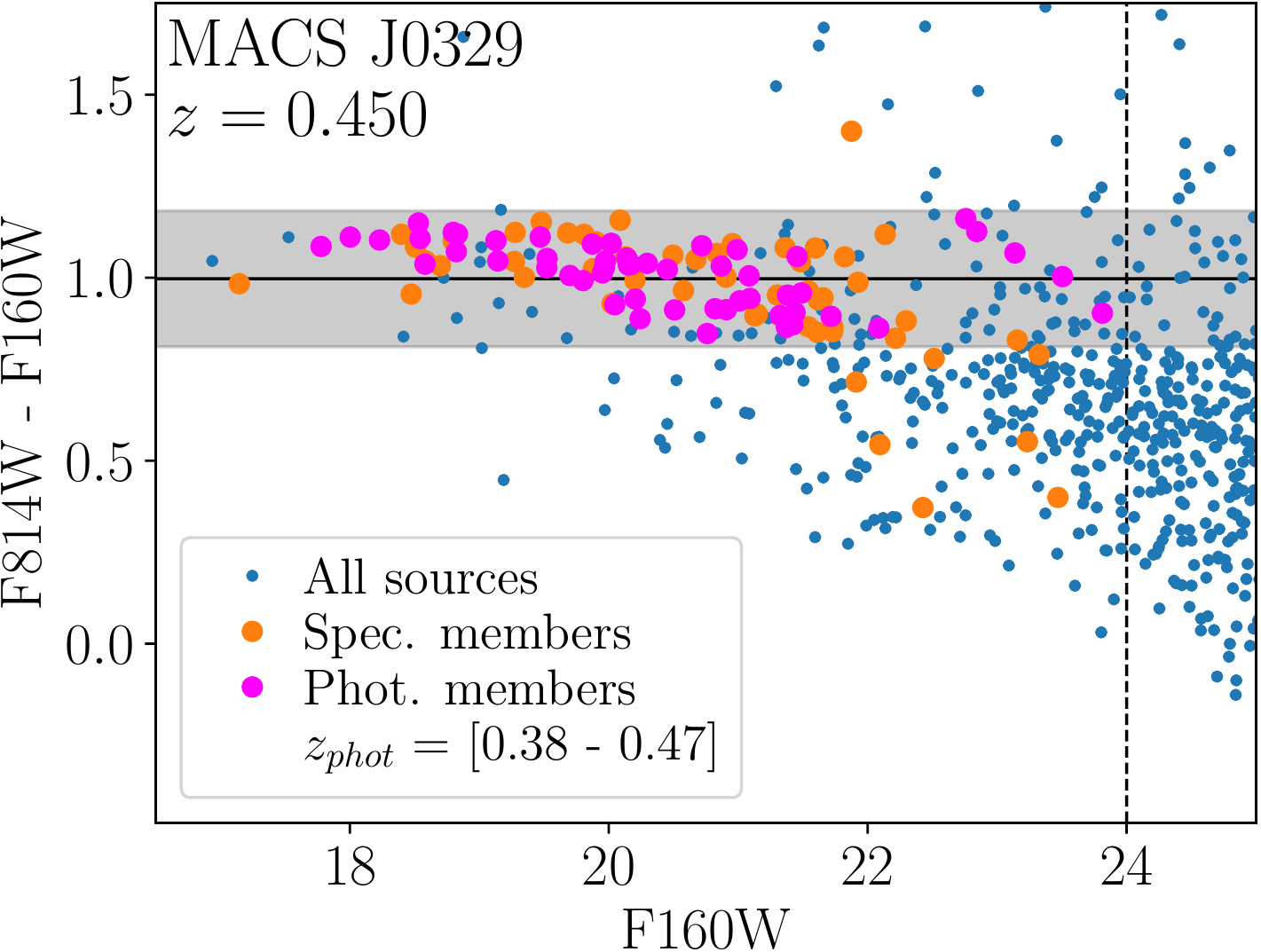}  
  \includegraphics[width = 0.504\columnwidth]{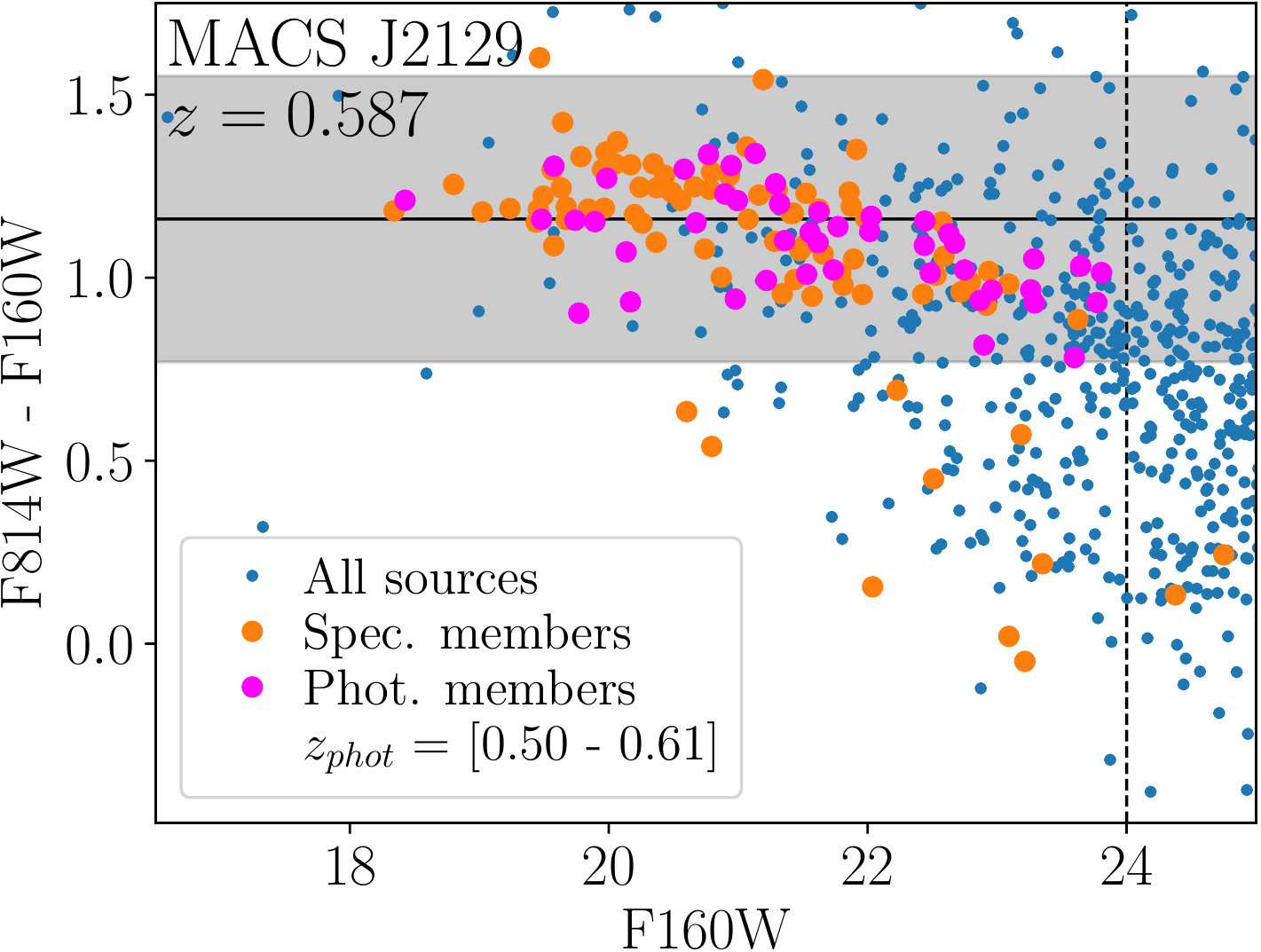}

  \vspace{0.25cm}

  \includegraphics[width = 0.504\columnwidth]{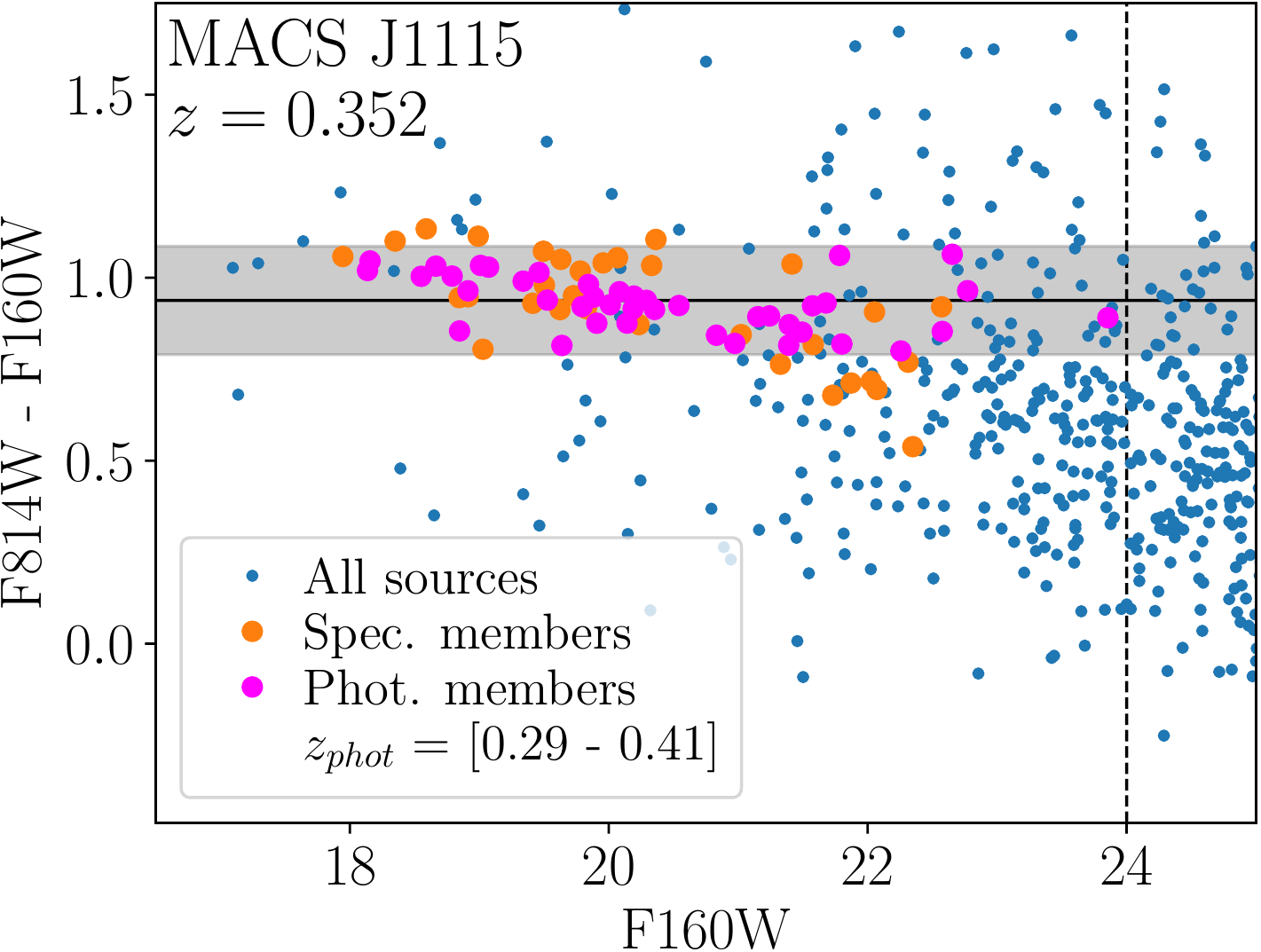}
  \includegraphics[width = 0.504\columnwidth]{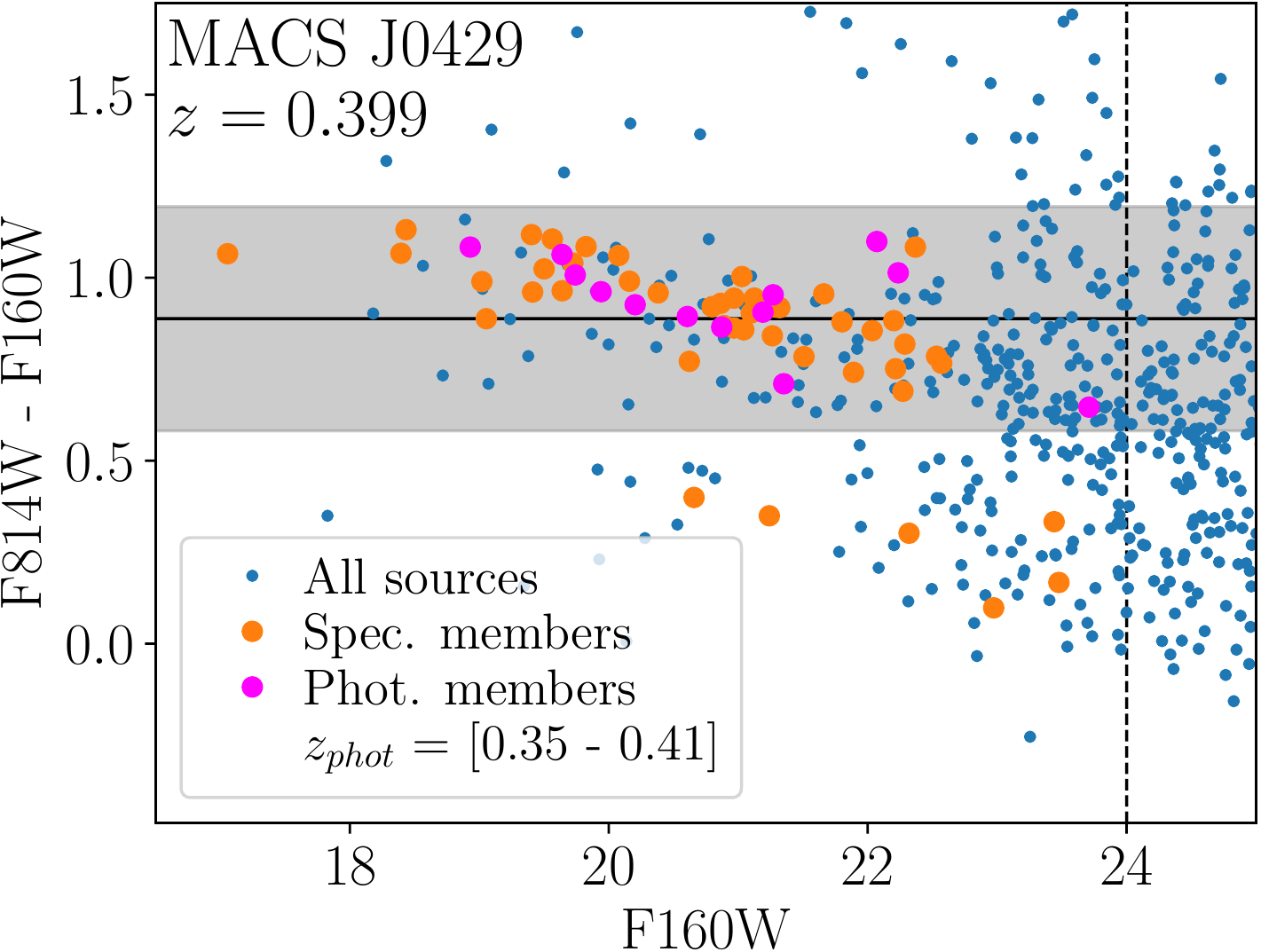}
  \includegraphics[width = 0.504\columnwidth]{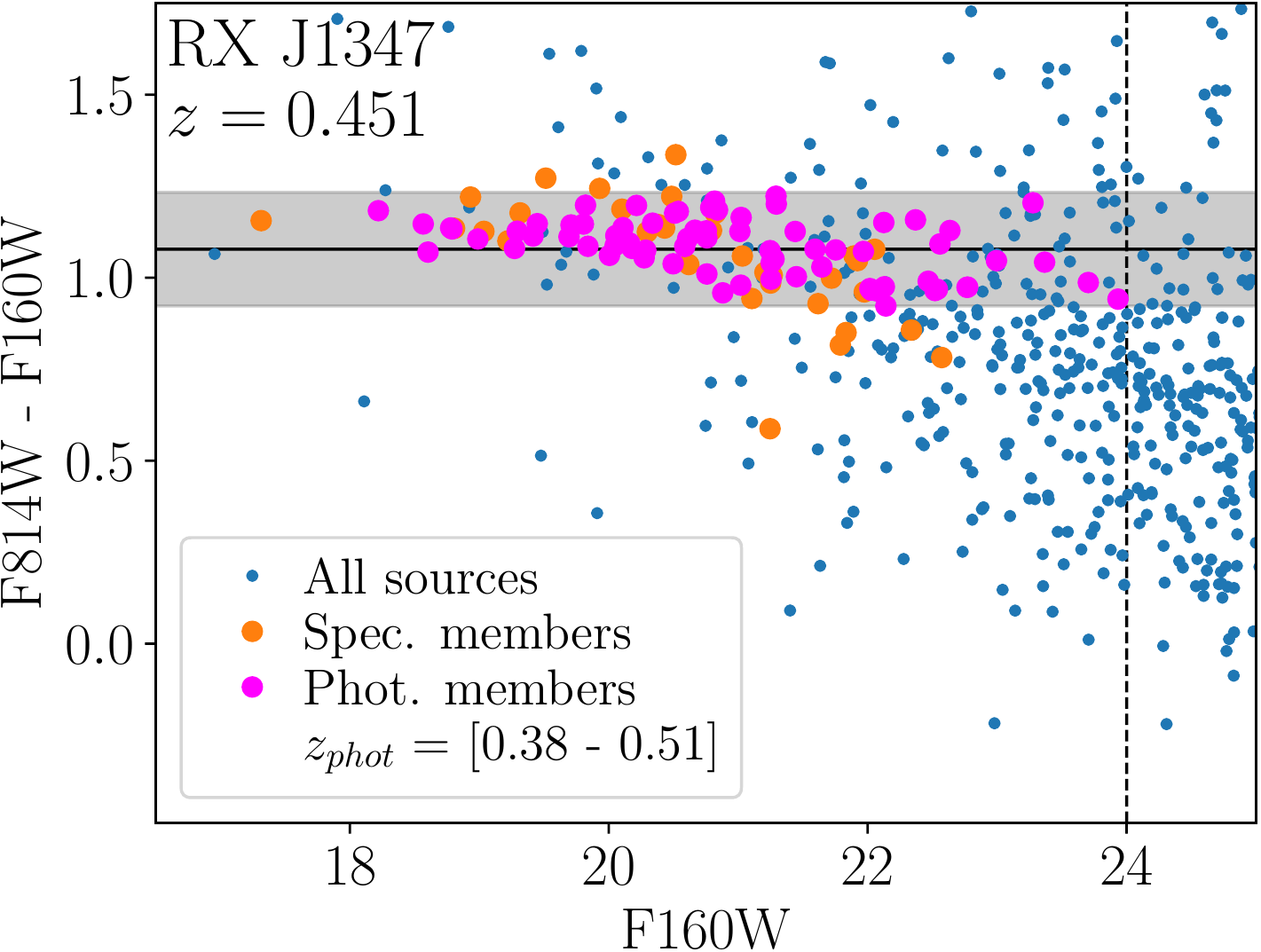}
  \includegraphics[width = 0.504\columnwidth]{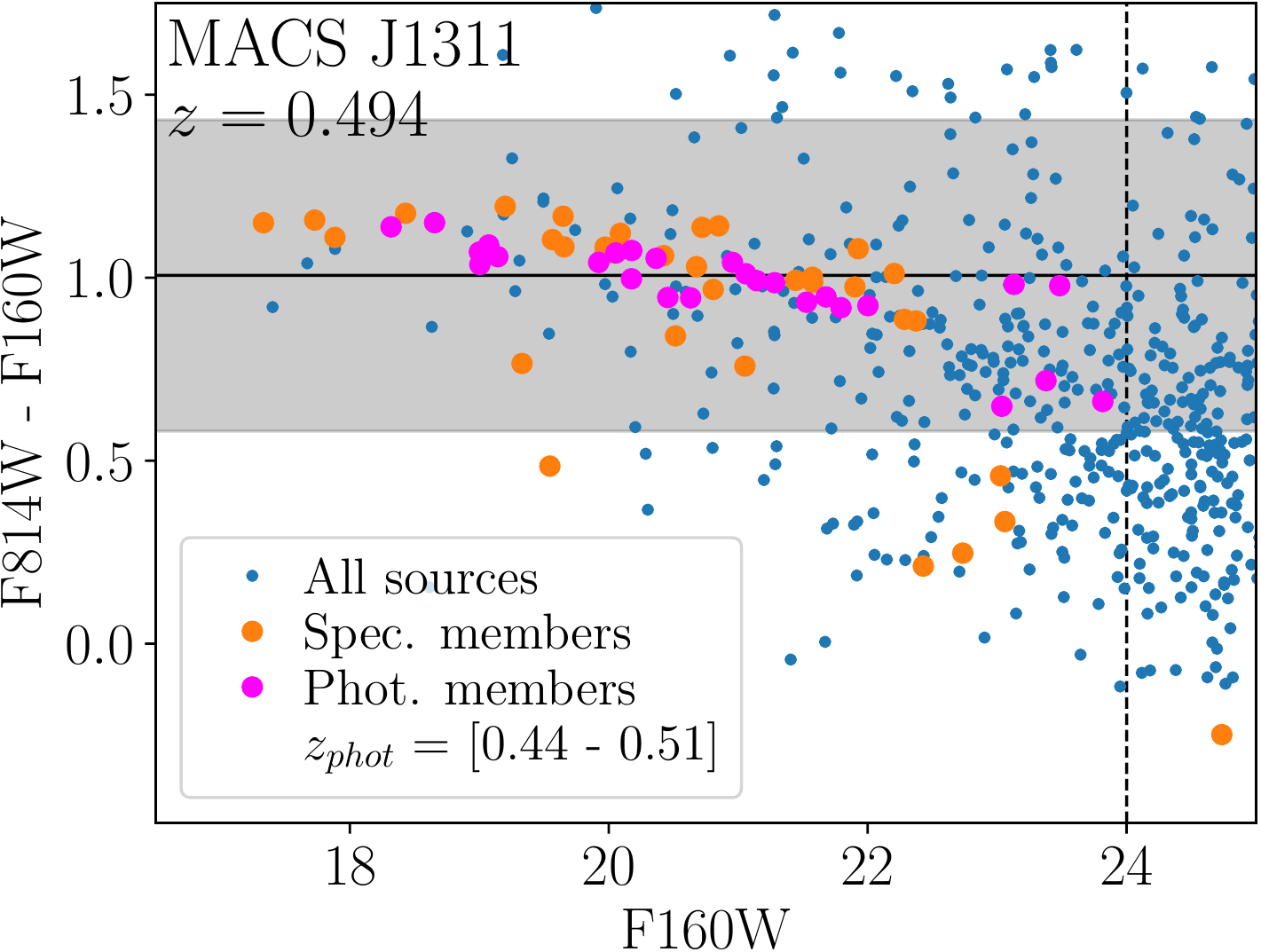}

  \caption{Colour-magnitude diagrams showing the red-sequence for all clusters analysed in this work. Only for RX~J2129 we use the filters F435W and F814W because of its low redshift, for all remaining clusters we use the F814W and F160W bands. Orange circles indicate the spectroscopically confirmed members. The grey region shows our colour cut and the photometric redshift range used to select photometric members is quoted in the legend (see Section \ref{sec:membership_selection} for more details). Both criteria are used to select the photometric members that are indicated by magenta circles. A summary of the membership selection is shown in Table \ref{tab:summary_members}.}
  \label{fig:member_selection}
\end{figure*}

For the total mass profiles of the galaxy members, we adopted the dual pseudoisothermal mass distribution \citep[dPIE,][]{2007arXiv0710.5636E,2010A&A...524A..94S} with zero ellipticity and core radius.
Moreover, the centres of the profiles were fixed to the centroids of the light distribution of the selected galaxy members (see Section \ref{sec:membership_selection}).
Therefore, each galaxy member has two free parameters, its central velocity dispersion ($\sigma^{\rm gals}_{{\rm v},i}$) and truncation radius ($r^{\rm gals}_{{\rm cut},i}$).
Since the total number of observables does not allow us to have two free parameters for each galaxy member, in our models we assumed a constant total mass-to-light ratio with the following scaling relations:
\begin{equation}
\sigma_{{\rm v},i}^{\rm gals} = \sigma_{\rm v}^{\rm gals}\left(\frac{L_i}{L_{\rm ref}} \right)^{0.25}, r_{{\rm cut},i}^{\rm gals} = r_{\rm cut}^{\rm gals}\left(\frac{L_i}{L_{\rm ref}} \right)^{0.5},
\label{eq:member_scale}
\end{equation}
where $L_{i}$ is the value of the luminosity of each member in the HST filter F160W and $L_{\rm ref}$ is the reference luminosity.
Hence, all galaxy members are described by the two parameters $\sigma_{\rm v}^{\rm gals}$ and $\rm r_{\rm cut}^{\rm gals}$.
We choose as reference luminosity the value corresponding to the rest-frame magnitude of $M_{\rm F160W} = -23$ in each cluster, which is close to $L^*$ for cluster galaxies at $z=0.4$ \citep[see e.g.][]{2017ApJ...848...37C, 2019arXiv190513236B}. 
This choice does not affect the lens modelling and the values of the normalization parameters (i.e. $\sigma_{\rm v}^{\rm gals}$ and $\rm r_{\rm cut}^{\rm gals}$) can be rescaled to any other reference luminosity by using Equation (\ref{eq:member_scale}).

By starting from the observed positions of the multiple images, we used the software {\tt lenstool} to obtain the best-fitting values of the free parameters of each model.
The software uses a $\chi^2$ function defined as
\begin{equation}
\chi^2\left(\vec{\Pi}\right) = \sum_{j=1}^{N_{\rm im}}{ \left(\frac{\left| \vec{x}_j^{\rm obs} -  \vec{x}_j^{\rm model}\left(\vec{\Pi}\right) \right|}{\sigma_j^{\rm obs}}\right)^{2} },
\label{eq:chi2}
\end{equation}
where $ \vec{x}_j^{\rm obs}$ and $\vec{x}_j^{\rm model}$ are the observed and model predicted positions of multiple images, respectively. 
The positional errors are given by $\sigma_{j}^{\rm obs}$ and $\vec{\Pi}$ is the vector containing the values of all model free parameters.
We used flat priors for all model free parameters ($\vec{\Pi}$) in every model.
The set of parameter values that minimises the $\chi^{2}$ function ($\chi^{2}_{\rm min}$) is called the best fit solution.
In addition to the $\chi^{2}_{\rm min}$ value, we also used the root-mean-square difference between the model predicted and observed positions ($\Delta_{\rm rms}$) of the multiple images to quantify the goodness of the fit of each model.

Given the different characteristics of the galaxy clusters in our sample, we explored a range of different models for each of them, from simple unimodal mass distributions to combinations of many different components in order to find the best fit solution.
First, we minimised the $\chi^{2}$ function for a single smooth mass component plus the galaxy members.
However, it is known that that in merging clusters, such as MACS~J0416 \citep{2017AA...600A..90C}, Abell~2744 \citep{2018MNRAS.473..663M} and Abell~370 \citep{2017MNRAS.469.3946L}, multiple smooth mass components are necessary to reproduce well the positions of all multiple images.
Therefore, we also optimised models with one and two extra smooth mass components, as well as models with an external shear component.
Since the BCG of a cluster undergoes formation and evolution processes that are different from those of the other galaxy members, we also tested models where the BCG parameters are free to vary and are not attached to those of the overall scaling relations described by Equations (\ref{eq:member_scale}).
We present the results of our lensing analyses in Section \ref{sec:results}.

\subsection{Membership selection}
\label{sec:membership_selection}

\begin{table*}[!]
\centering
\small
\caption{Summary of cluster member selection with F160W photometry.}
\begin{tabular}{l c c c c c c l} \hline \hline
Cluster & $z_{\rm cluster}$& $N_{\rm total}^a$ & $z_{\rm spec}^{\rm range}$ & $ N^{{\rm MUSE}\; b}_{\rm spec}$ & $N_{\rm cont.}^c$ & $N_{\rm miss.}^c$ \\
\hline
RX~J2129        & 0.234 & 70& $[0.217 - 0.250]$ & 32 & $1$ & $4$  \\
MACS~J1931      & 0.352 &120& $[0.334 - 0.370]$ & 56 & $0$ & $5$ \\
MACS~J0329      & 0.450 &106& $[0.431 - 0.470]$ & 50 & $4$ & $7$ \\
MACS~J2129      & 0.587 &138& $[0.566 - 0.608]$ & 90 & $0$ & $3$ \\
\hline
MACS~J1115$^{d}  $& 0.352 & 94& $[0.334 - 0.370]$ & 32 & $1$ & $22$ \\
MACS~J0429$^{d,e}$& 0.399 & 63& $[0.380 - 0.418]$ & 51 & $(e)$&$(e)$ \\
RX~J1347$^{d}  $  & 0.451 &114& $[0.432 - 0.470]$ & 33 & $3$ & $11$ \\
MACS~J1311$^{d}  $& 0.494 & 76& $[0.474 - 0.514]$ & 38 & $0$ & $12$ \\
\hline
\end{tabular}
\label{tab:summary_members}
\tablefoot{
\tablefoottext{a}{Total number of selected cluster members, i.e. spectroscopically confirmed and photometrically selected.}
\tablefoottext{b}{Number of members with spectroscopic confirmation from MUSE.}
\tablefoottext{c}{Number of interlopers and missing members are obtained from CLASH-VLT redshift measurements outside the MUSE field of view.
Spectroscopic interlopers (missing) members are removed (included) in the final version of the membership selection.}
\tablefoottext{d}{Clusters with shallow MUSE data to data ($<1$ hour) or with a small number of spectroscopically confirmed multiple images.}
\tablefoottext{e}{Not available since MACS~J0429 was not included in CLASH-VLT.}
}
\end{table*}

The cluster member selection is strongly based on our redshift measurements from MUSE in the very central regions and complemented with the CLASH-VLT spectroscopy and CLASH photometry in the outer regions.
For the photometric measurements, we used the CLASH public catalogues \citep{2012ApJS..199...25P, 2017MNRAS.470...95M}.
Firstly, we selected all sources from the MUSE spectroscopic catalogue located within $\pm 4000 \,{\rm km\,s^{-1}}$ from the cluster redshift (see Table \ref{tab:summary_members}, column $z_{\rm spec}^{\rm MUSE}$).
In Figure \ref{fig:member_selection}, we show the colour-magnitude diagrams for the HST filters F814W$-$F160W (except for RX~J2129 that has a low redshift value and for which we used F435W$-$F814W), where the cluster red sequences are well defined.
In order to select members with no spectroscopic confirmation (mainly outside the MUSE field of view), we used two criteria based on the empiric distribution of colours and photometric redshifts of confirmed members.
For the first criterion, we computed standard deviation of the colours in the aforementioned filters for each cluster, see the grey region in Figure \ref{fig:member_selection}.
The differences in the colour intervals observed in our cluster sample depend on the cluster redshift, the different regions covered by the MUSE pointings and cluster-to-cluster variance in the galaxy population in these small regions.
Moreover, the number of late-type galaxies is different in each cluster.
Specifically, RX~J2129, MACS~J1931, MACS~J2129, MACS~J0429 and MACS~J1311 have around 7\% - 10\% of late-type galaxies in the sample of spectroscopically confirmed members, whereas the other clusters have around 3\%.

For the second criterion, we used the photometric redshift distribution of spectroscopically confirmed members.
Since such distributions are fairly asymmetric and present some outliers, we used the 68\% confidence levels (after applying a sigma clipping) to define the lower and upper values of $z_{\rm phot}$ for members with no spectroscopic confirmation.
Therefore, we selected members with no MUSE measurements that are encompassed by both criteria (colour and $z_{phot}$ limits) and are brighter than $\rm mag_{F160W} = 24$.
We note that the distribution of photometrically selected galaxies in Figure \ref{fig:member_selection} closely follows the distribution of spectroscopic members, that is, the so called red-sequence.
A summary of the membership is presented in Table \ref{tab:summary_members}, where we quote the number of spectroscopically confirmed members by MUSE and the total number of members in our final selection.

To estimate the contamination and completeness of this selection, we used the available CLASH-VLT measurements (besides MACS~J0429 that was not included in this programme) in the regions outside the MUSE field of view.
Such comparison yields important information about missing members and provide lower limits for the number of interlopers.
In Table \ref{tab:summary_members}, we quote the number of photometric members wrongly identified and galaxies with CLASH-VLT redshifts that belong to the cluster (i.e. are within $\pm 4000 {\rm \,km\,s^{-1}}$ from the cluster redshift), but were not selected by our criteria.
The completeness of our sample is relatively high, around $\approx 80\% - 90\%$ thanks to the large number of spectroscopic confirmations of galaxy members.
We remark that members with CLASH-VLT confirmation, but not selected with our photometric criteria, are included in the strong lensing models.
Likewise, spectroscopic interlopers are removed from the final membership samples.
While VIMOS incomplete spectroscopic coverage always leads to some possible membership contamination, MUSE integral field spectroscopy ensures highly pure samples of cluster galaxies in its field of view.

In Figures \ref{fig:members} and \ref{fig:members_2}, we show the selected galaxy members used in our strong lens models.
We indicate the members with MUSE confirmation and galaxies selected using the colour - photo-z cuts (see Figure \ref{fig:member_selection}).
Members with spectroscopic confirmation from CLASH-VLT but not selected by our photometric criteria are also marked.
We note that within the MUSE field of view only a very small fraction of members does not have spectroscopic confirmation.
These galaxies are usually very faint or the contamination or confusion with nearby bright sources (caused by the atmospheric seeing) makes a secure spectroscopic confirmation difficult.

\subsection{Multiple-image identification}
\label{sec:multiple_image_identification}

In order to build reliable lensing models, we based our identification of multiple image families on spectroscopic confirmations. 
To identify multiple images of the same source, we selected entries in our MUSE and VIMOS (Rosati et al. in prep.) catalogues with similar redshifts (considering only measurements with QF greater than one) and inspected the HST images to verify if their positions, colours and parities are in agreement with what we expect from strong lensing theory.
We note that some Lyman-$\alpha$ emitters have no clear detection in the CLASH photometry, but their line profile, position and, in some cases, their spatial morphologies ensure a secure identification of a multiple image family.
In these cases, we used the centroid positions of the MUSE detections in the lens modelling.
With this first set of spectroscopic constraints, we built a preliminary version of strong lensing models and compute the model predicted positions of all other sources with measured redshifts behind the clusters.

When a model predicts that a source is multiply imaged, we checked the HST images for possible photometric counterparts near the model predicted positions.
In this step, we added to our set of multiple images only those with unambiguous photometric identification and with correct parity and colours.
This allowed us to include multiple images that are outside the MUSE field of view or too faint (given the different lensing magnification of each multiple image) to be spectroscopically confirmed.
The images selected in this way always belong to a family with measured redshift and are considered secure identifications.
In Figure \ref{fig:multiple_images}, we show the positions of all multiple images used in our models and in table \ref{tab:multiple_images} we list their positions, MUSE redshifts and previous measurements from the literature, when present.
Moreover, in Figure \ref{fig:specs}, we show the spectral features used to measure the multiple image redshifts from the MUSE spectra.

\begin{figure}
  \includegraphics[width = 1.0\columnwidth]{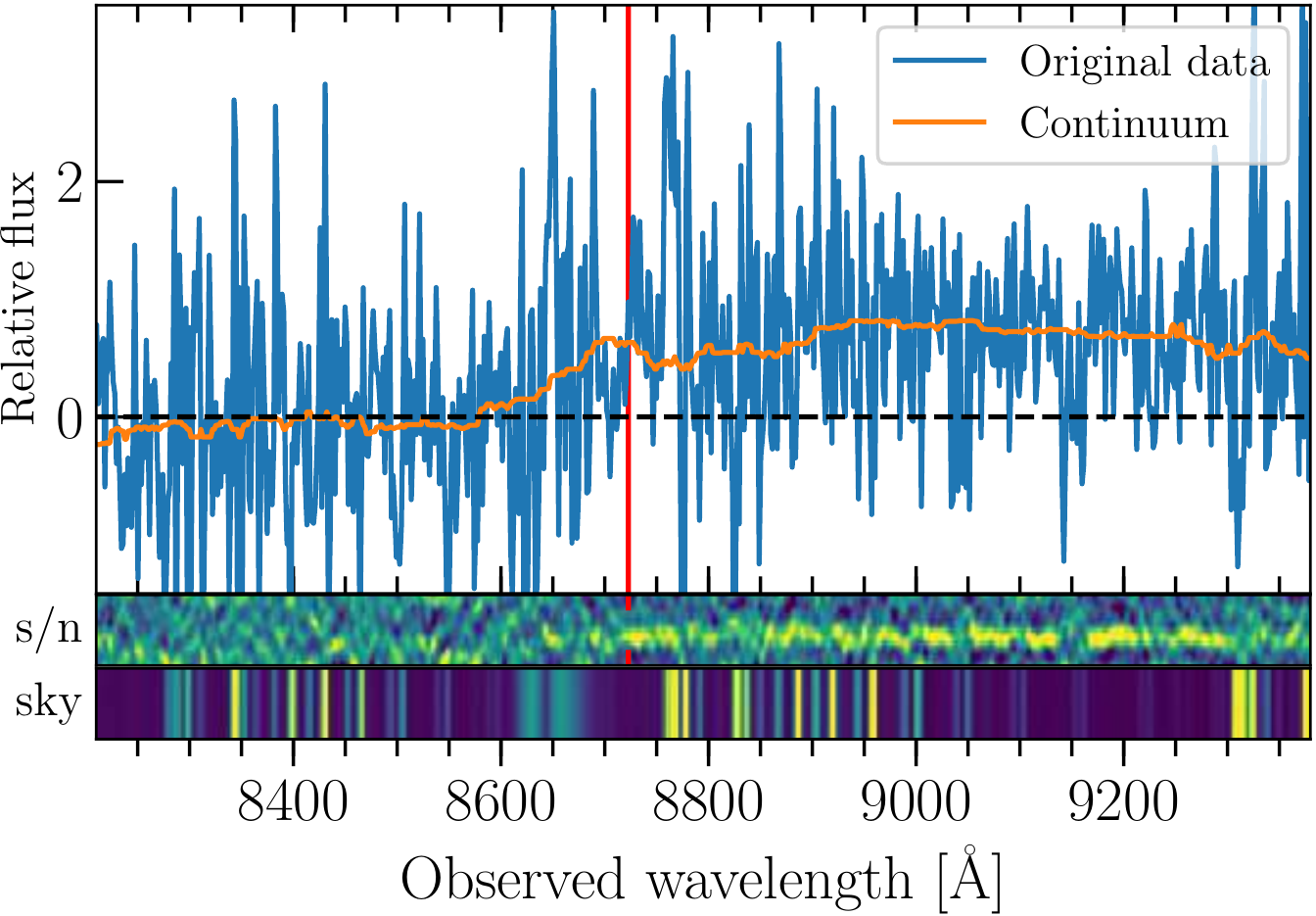}
  \caption{Top panel: ESO/FORS2 stacked spectrum of the multiple images belonging to family 9 in MACS~J0329 in arbitrary units, totaling 33 hours of exposure time on target. The original data and continuum are shown in blue and orange, respectively. The vertical red line indicates the Lyman-break position around $\approx 8720\AA$ and the dashed horizontal line shows the zero level. Middle and bottom panels: the two-dimensional signal-to-noise ratio of the data and the sky emission.}
  \label{fig:f9_m0329}
\end{figure}

We were able to confirm all previously published redshift measurements of the multiple images within the MUSE field of views, except for family 11 behind MACS~J2129.
This system was confirmed at $z=6.85$ by \citet{2016ApJ...823L..14H} using HST grism and Keck/DEIMOS data.
At this redshift, the Lyman-$\alpha$ line is outside the MUSE wavelength range, that extends out to $9350\AA$, corresponding to a maximum limit of $z\approx 6.69$ to detect this line.
Moreover, for family 9 behind MACS~J0329 \citep[see also][]{2012ApJ...747L...9Z}, we make use of archival ESO/FORS2 data to support our MUSE measurement of $z\approx6.1$, using the Lyman-break detection, and assign to this redshift a $\rm QF_{MUSE}=2$.
The ESO/FORS2 observations were obtained under the ESO programme ID 096.A-0650 (P.I. Vanzella) and we used the reduction methodology described in \citet{2011ApJ...730L..35V, 2014A&A...569A..78V}.
The multiple images 9b, 9c, and 9d were observed with an exposure time of 11 hours each, totalling 33 hours on the same source.
In Figure \ref{fig:f9_m0329}, we show the stacked one-dimensional spectrum and the two-dimensional signal to noise ratio and sky emission.
The spectrum shows a clear feature around $8720\AA$ that we associate to the break at $1215.7\;\AA$ rest-frame.
This break is in very good agreement with the MUSE data (see Figure \ref{fig:specs}) and supports our measurement.
In total, we present 27 new spectroscopic measurements of multiple image families and confirm other ten with previous redshift determinations.

We note that there are four clusters with a large number of spectroscopically confirmed families ($N_{\rm spec} > 5$, namely RX~J2129, MACS~J1931, MACS~J0329, and MACS~J2129), and four with a smaller number of strong lensing constraints (MACS~J1115, MACS~J0429, RX~J1347 and MACS~J1311).
For the second group, in order to have a larger number of observables to constrain the values of the free parameters of the models, we also included multiple image families without spectroscopic confirmation.
To do that, we inspected previous identifications available in the literature \citep[see e.g.][and references therein]{2015ApJ...801...44Z} and again select multiple images based on their positions, colours and parities.
For these families, we used the photometric redshift estimates from \citet{2017MNRAS.470...95M} to build priors for the strong lensing models.
We assumed flat priors for the redshift values, where the lower and upper limits were taken as the minimum and maximum of the 68\% confidence levels of all images of each source (excluding the cases where there is clear contamination from bright nearby objects).
In Table \ref{tab:multiple_images}, within brackets, we show the allowed redshift range of each family, that we used in the model optimization and sampling.

Because of this difference between the number of spectroscopic confirmations and the use of multiple image families selected from the photometric data only, we treated the two sample of clusters separately.
We refer to the sample of clusters with good constraints as 'gold' sample and explore a variety of parametrizations with increasing complexity.
Conversely, the small number of multiple images of the second sample, called 'silver',  did not allow us to test models with a large number of free parameters.
Therefore, we considered only a unimodal smooth mass component, except for RX~1347 that has two very bright central galaxies and is known to be a merging system \citep[see e.g.][]{1992A&A...256L..11S,2002MNRAS.335..256A,2018ApJ...866...48U}.

\begin{table*}[t!]
\centering
\small
\caption{Strong lensing model general information.}
\begin{tabular}{l c c c c c c r c} \hline \hline
Cluster & $z_{\rm cluster}$& $N_{\rm src}^{\rm spec}$ & $ N_{\rm src}^{\rm phot}$ & $N_{\rm img}$ & $z^{\rm spec}_{\rm range}$ & $ M^{{\rm SL},a}_{<200\,kpc} \left[ 10^{14} {\rm M_{\odot}}\right]$ & $M^{{\rm WL},c}_{\rm 200crit}\left[ 10^{14}\rm M_{\odot}\right]$ & $ R^{{\rm WL},c}_{\rm 200crit} \left[\rm Mpc\right]$ \\
\hline
RX~J2129        & 0.234 & 7 & 0 & 22 & $[0.68 - 3.43]$ & $\rm 1.19\pm0.01$ & $7.78 \pm 2.43$ & $1.76 \pm 0.18$\\
MACS~J1931      & 0.352 & 7 & 0 & 19 & $[1.18 - 5.34]$ & $\rm 1.70\pm0.02$ & $11.62 \pm 2.84$ & $1.92 \pm 0.16$ \\
MACS~J0329      & 0.450 & 9 & 0 & 23 & $[1.31 - 6.17]$ & $\rm 1.87\pm0.02$ & $12.70 \pm 2.19$ & $1.90 \pm 0.11$ \\
MACS~J2129      & 0.587 & 11& 0 & 38 & $[1.05 - 6.85]$ & $\rm 1.84\pm0.01$ & N/A & N/A \\
\hline
MACS~J1115$^{b}$& 0.352 & 2 & 1 & 9  & $[2.55 - 2.92]$ & $\rm 1.65\pm0.20$ & $17.91 \pm 3.81$ & $2.22 \pm 0.16$ \\
MACS~J0429$^{b}$& 0.399 & 2 & 1 & 11 & $[2.93 - 3.86]$ & $\rm 1.28\pm0.03$ & $8.88  \pm 1.70$ & $1.72 \pm 0.11$ \\
RX~J1347$^{b}$  & 0.451 & 4 & 4 & 20 & $[1.76 - 4.08]$ & $\rm 2.42\pm0.05$ & $35.40 \pm 5.05$ & $2.68 \pm 0.13$ \\
MACS~J1311$^{b}$& 0.494 & 1 & 2 & 8  & $[2.19]$        & $\rm 1.43\pm0.07$ & $4.6   \pm 0.3^{d}$  & $1.33 \pm 0.03^{d}$ \\

\hline
\end{tabular}
\label{tab:summary_bf}
\tablefoot{
\tablefoottext{a}{Total mass value projected within a circle with a radius of 200~kpc computed using our strong lensing reference models, as described in Table \ref{tab:summary_models_all}.}
\tablefoottext{b}{Clusters with shallow MUSE data ($<1$ hour) or with a small number of spectroscopically confirmed multiple images.}
\tablefoottext{c}{$M^{\rm WL}_{\rm 200crit}$ and $R^{\rm WL}_{\rm 200crit}$ are weak lensing measurements taken from \citet{2018ApJ...860..104U}.}
\tablefoottext{d}{Weak lensing values for MACS~J1311 are taken from \citet{2015ApJ...806....4M}, the most recent measurements for this cluster.}
}
\end{table*}

\section{Strong lensing model results}
\label{sec:results}

\begin{table*}[!htbp]
\centering
\small
\caption{Models summary.}
\begin{tabular}{l c c c c c c l}
\hline \hline
Model ID & N. par. & DOF & $\Delta_{\rm rms} ['']$ & $\rm \chi^2_{min}$  & BIC &AIC & Description \\
\hline
{\bf RXJ~2129-P1}         &\bf 8   &\bf 22  &\bf 0.20&\bf 3.39&\bf 34.7&\bf 27.6& One smooth elliptical PIEMD component\\
{\it RXJ~2129-gNFW1}      &\it 9   &\it 21  &\it 0.22&\it 4.40&\it 36.7&\it 30.1& One smooth elliptical gNFW component\\
RXJ~2129-P2               & 14  & 16  & 0.13& 1.56& 38.8& 38.7& Two smooth elliptical PIEMD components\\
RXJ~2129-P2-circular      & 12  & 18  & 0.14& 1.71& 37.0& 34.8& Second smooth component with circular symmetry\\
RXJ~2129-P1-shear         & 10  & 20  & 0.20& 3.37& 36.6& 31.6& Same as P1 plus an with external shear term\\
RXJ~2129-P1-BCG           & 10  & 20  & 0.19& 3.31& 36.6& 31.6& Same as P1 but with free circular BCG parameters\\
RXJ~2129-P1-BCG-shear     & 12  & 18  & 0.19& 3.30& 38.6& 35.6& Same as BCG but plus an external shear term\\
RXJ~2129-P1-BCGell        & 12  & 18  & 0.16& 2.37& 37.6& 35.1& Same as P1 but with free elliptical BCG parameters\\
\hline
MACS~J1931-P1             & 8   & 16  & 0.93& 41.0& 94.1& 57.5& One smooth elliptical PIEMD component\\
MACS~J1931-P2             & 14  & 10  & 0.34& 9.00& 43.3& 41.1& Two smooth elliptical PIEMD components\\
{\bf MACS~J1931-P2-circular}   &\bf 12  &\bf 12  &\bf 0.38&\bf 10.9&\bf 43.2&\bf 38.0& Second smooth component with circular symmetry\\
{\it MACS~J1931-gNFW2-curcular}&\it 13  &\it 11  &\it 0.39&\it 11.5&\it 44.7&\it 40.3& Main smooth component parametrised by gNFW\\
MACS~J1931-P1-shear       & 10  & 14  & 0.58& 25.4& 55.7& 41.3& Same as P1 plus an external shear term\\
MACS~J1931-P1-BCG         & 10  & 14  & 0.69& 36.2& 66.5& 46.7& Same as P1 but with free circular BCG parameters\\
MACS~J1931-P1-BCGell      & 12  & 12  & 0.55& 23.4& 55.7& 44.3& Same as P1 but with free elliptical BCG parameters\\
\hline
MACS~J0329-P1             & 8   & 20  & 1.74&  278&  310& 165 & One smooth elliptical PIEMD component\\
MACS~J0329-P2             & 14  & 14  & 0.32& 9.15& 47.3& 43.0& Two smooth elliptical PIEMD components\\
MACS~J0329-P2-circular    & 12  & 16  & 0.73& 49.2& 85.3& 59.0& Second smooth component with circular symmetry\\
MACS~J0329-P3-circular    & 18  & 10  & 0.26& 6.04& 48.2& 49.4& Same as P2 plus a 3rd circular component\\
MACS~J0329-P3             & 20  &  8  & 0.25& 5.63& 49.7& 53.2& Same as P2 plus a 3rd elliptical component\\
{\bf MACS~J0329-P2-shear}&\bf 16  &\bf 12  &\bf 0.24&\bf 5.18&\bf 45.3&\bf 45.0& Same as P2 plus an external shear term\\
{\it MACS~J0329-gNFW2-shear}&\it 17  &\it 11  &\it 0.22&\it 4.43&\it 45.5&\it 46.6& Main smooth component parametrised by gNFW\\
MACS~J0329-P2-BCG         & 16  & 12  & 0.31& 8.71& 48.8& 46.7& Same as P1 but with free circular BCG parameters\\
MACS~J0329-P2-BCGell      & 18  & 10  & 0.29& 9.00& 50.1& 50.4& Same as P1 but with free elliptical BCG parameters\\
\hline
MACS~J2129-P1             & 8   & 46  & 1.10& 180 & 226 & 123 & One smooth elliptical PIEMD component\\
MACS~J2129-P1-shear       & 10  & 44  & 1.05& 162 & 211 & 118 & Same as P1 plus an external shear term\\
MACS~J2129-P1-BCG         & 10  & 44  & 0.99& 149 & 197 & 111 & Same as P1 but with free circular BCG parameters\\
MACS~J2129-P1-BCGell      & 12  & 42  & 0.94& 134 & 184 & 108 & Same as P1 but with free elliptical BCG parameters\\
MACS~J2129-P2-circular    & 12  & 42  & 0.71& 77.4& 128 & 79.9& Second smooth component with circular symmetry\\
{\bf MACS~J2129-P2}       &\bf 14  &\bf 40  &\bf 0.56&\bf 47.1&\bf 99.5&\bf 68.7& Two smooth elliptical PIEMD components\\
{\it MACS~J2129-gNFW2}    &\it 15  &\it 39  &\it 0.83&\it 105&\it 158&\it 99.5& Main smooth component parametrised by gNFW\\
MACS~J2129-P2-circular-shear &14 & 40  & 0.65& 62.8& 115 & 76.5& Same as P2-circular plus an external shear term\\
MACS~J2129-P3-circular    & 16  & 38  & 0.64& 62.0& 116 & 80.1& Three smooth components with circular symmetry\\
\hline \hline
\multicolumn{8}{l}{Cluster sample with small number of constraints}  \\
\hline
MACS~J1115-P1             & 9   & 3   & 0.61& 13.2& 32.8& 28.7& One smooth elliptical PIEMD component\\
\hline
MACS~J0429-P1             & 9   & 7   & 0.32& 4.38& 26.1& 25.2& One smooth elliptical PIEMD component\\
\hline
RX~J1347-P2               & 18  & 6   & 1.05& 87.7& 127 & 88.9& Two smooth elliptical PIEMD\\
{\bf RX~J1347-P2-shear}   &\bf 20  &\bf 4   &\bf 0.36&\bf 10.1&\bf 51.1&\bf 54.1& Two smooth elliptical PIEMD plus an external shear term\\
\hline
MACS~J1311-P1             & 10  & 0   & 0.88& 24.9& 44.4& 36.1& One smooth elliptical PIEMD component\\
\hline \hline
\multicolumn{8}{l}{Previous models}  \\
\hline
Abell~1063                & 14  & 62  & 0.44& 44.5& 125 & 84.1& Updated model from \citet{2016AA...587A..80C}\\
MACS~J1206                & 22  & 88  & 0.44& 41.0& 225 & 143 & Model from \citet{2017AA...607A..93C}\\
MACS~J0416                & 26  & 104 & 0.59& 143 & 266 & 169 & Model from \citet{2017AA...600A..90C}\\
\hline
\end{tabular}
\label{tab:summary_models_all}
\tablefoot{For each model, we present the number of free parameters (N. par.), the degree of freedom (DOF), the root-mean-square difference between the model predicted and observed multiple image positions ($\Delta_{\rm rms}$), the minimum $\chi^2$ (see Equation \ref{eq:chi2}), the Bayesian and Akaike information criteria (BIC and AIC, respectively), and a short description of the model parametrization. Rows in boldface are our reference models, since they have the lowest BIC and AIC values. Model IDs in italics have the main smooth component parametrised by a gNFW mass profile.}
\end{table*}

As mentioned in Section \ref{sec:overall_description}, we used a bottom-to-top approach in order to find the mass model that better reproduces the positions of all multiple images.
We first considered simple models with one smooth PIEMD mass component plus galaxy members.
Since merging clusters or systems with high asymmetries in their mass distributions cannot reproduce the observed positions of all multiple images with only one smooth component \citep[see e.g.][]{2017MNRAS.469.3946L, 2017AA...607A..93C}, we gradually increased the complexity of our models by including additional smooth mass components and allowing values of the BCG parameters to vary independently from those of the other cluster members.
These different models are tested for our gold sample of clusters, whereas in the silver sample we do not have enough constraints.
Therefore, in these last four clusters we considered only simple models.

For each model, we computed the values of the best fitting $\Delta_{\rm rms}$ and $\chi^2_{\rm min}$, considering a positional error $\sigma_j^{\rm obs} = 0\arcsec.5$ for all multiple images.
We used this value of the positional errors to account for mass perturbers along the line of sight and limitations of parametric models.
This is in agreement with some theoretical predictions \citep{2012MNRAS.420L..18H} and estimates obtained from real data \citep{2018A&A...614A...8C}.
In Table \ref{tab:summary_models_all}, we summarise all models we optimised in our sample and give the numbers of free parameters ($N_{\rm par}$) and degrees of freedom ($\rm DOF\equiv number\,of\,constraints - N_{par}$).
Moreover, we also show the values of the Bayesian information criterion \citep[BIC,][]{schwarz1978} and of the Akaike information criterion \citep[AIC,][]{1974ITAC...19..716A}.
These two quantities are particularly important to select the best model, balancing between the goodness of the fit and the number of free parameters.
Therefore, the models with smaller BIC and AIC values were selected to be our reference models.
We note that both BIC and AIC values select the same reference model for each cluster.

For the reference models (indicated in bold in Table \ref{tab:summary_models_all}), we have run {\tt lenstool} also in the sampling mode to obtain the posterior distribution of each model parameter.
Although the values of the $\chi^2_{\rm min}$ are already close to those of the number of DOF, we rescaled the values of $\sigma^{\rm obs}_{j}$ in order to have $\chi^{2}_{\rm min}/{\rm DOF} = 1$ and obtain more realistic statistical uncertainties for the model parameters.
For each model, ten different sets of Markov Chain Monte Carlo (MCMC) are run until they reach a value of the so-called Gelman-Rubin test \citep{Brooks} lower than 1.2 for all model parameters, thus indicating the convergence of the chains.
We quote the median values and the the 68\%, 95\% and 99.7\% confidence level intervals of all model free parameters in Table \ref{tab:best_params}.

In addition to the best PIEMD models, we also sampled a model with a gNFW profile for the main smooth mass component, to quantify the dependence of the reconstructed total mass distribution on the specific choice of the smooth mass profile.
We remark that the pseudo-elliptical implementation of the gNFW profile in {\tt lenstool} might lead to unphysical projected mass density distributions (see Section \ref{sec:overall_description}).
We used these models only for the sake of comparison with our reference models.
In the following subsections, we briefly discuss the details of the strong lensing models of all eight clusters.

\begin{figure*}
  \centering
  \includegraphics[width = 0.504\columnwidth]{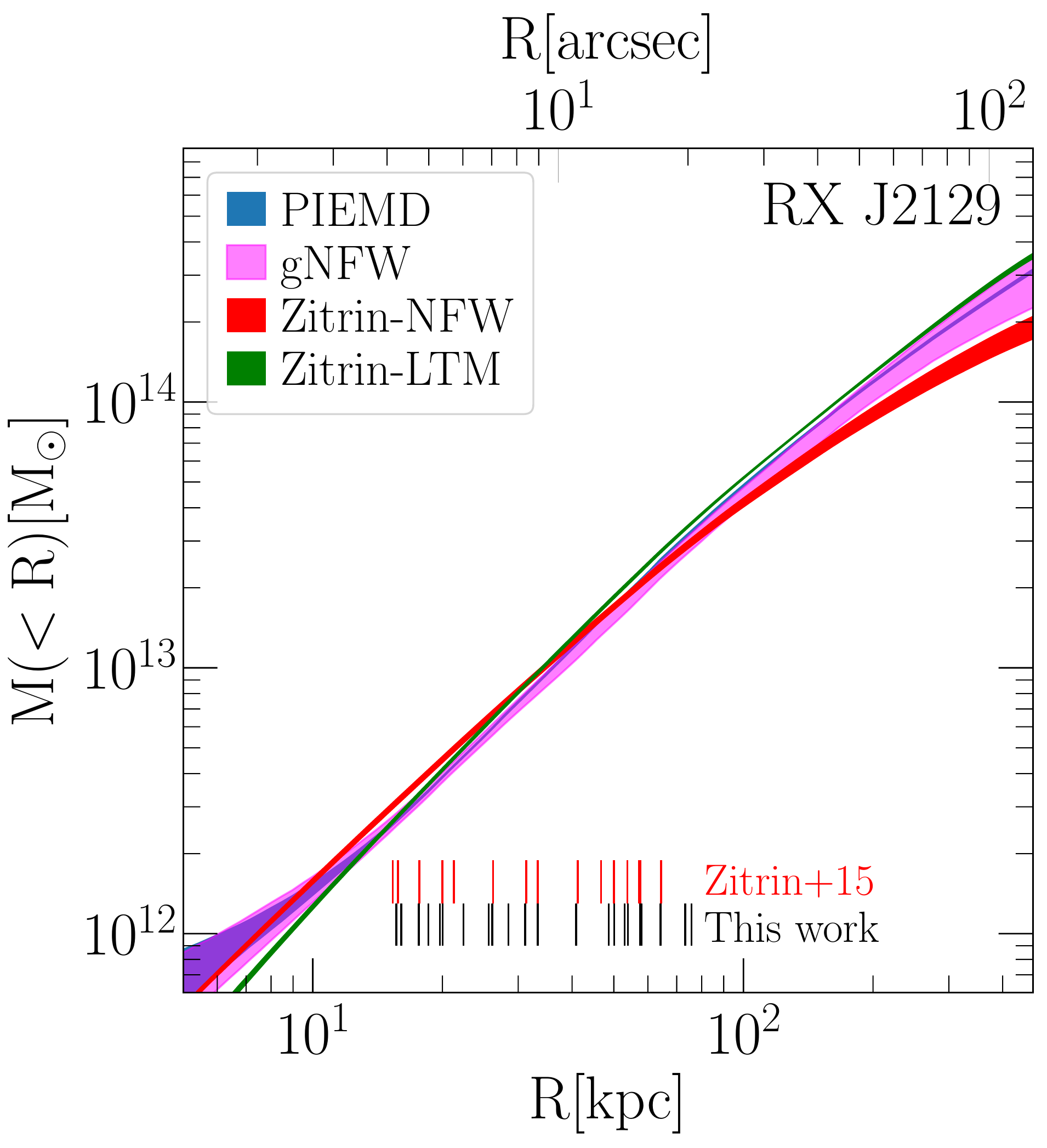}
  \includegraphics[width = 0.504\columnwidth]{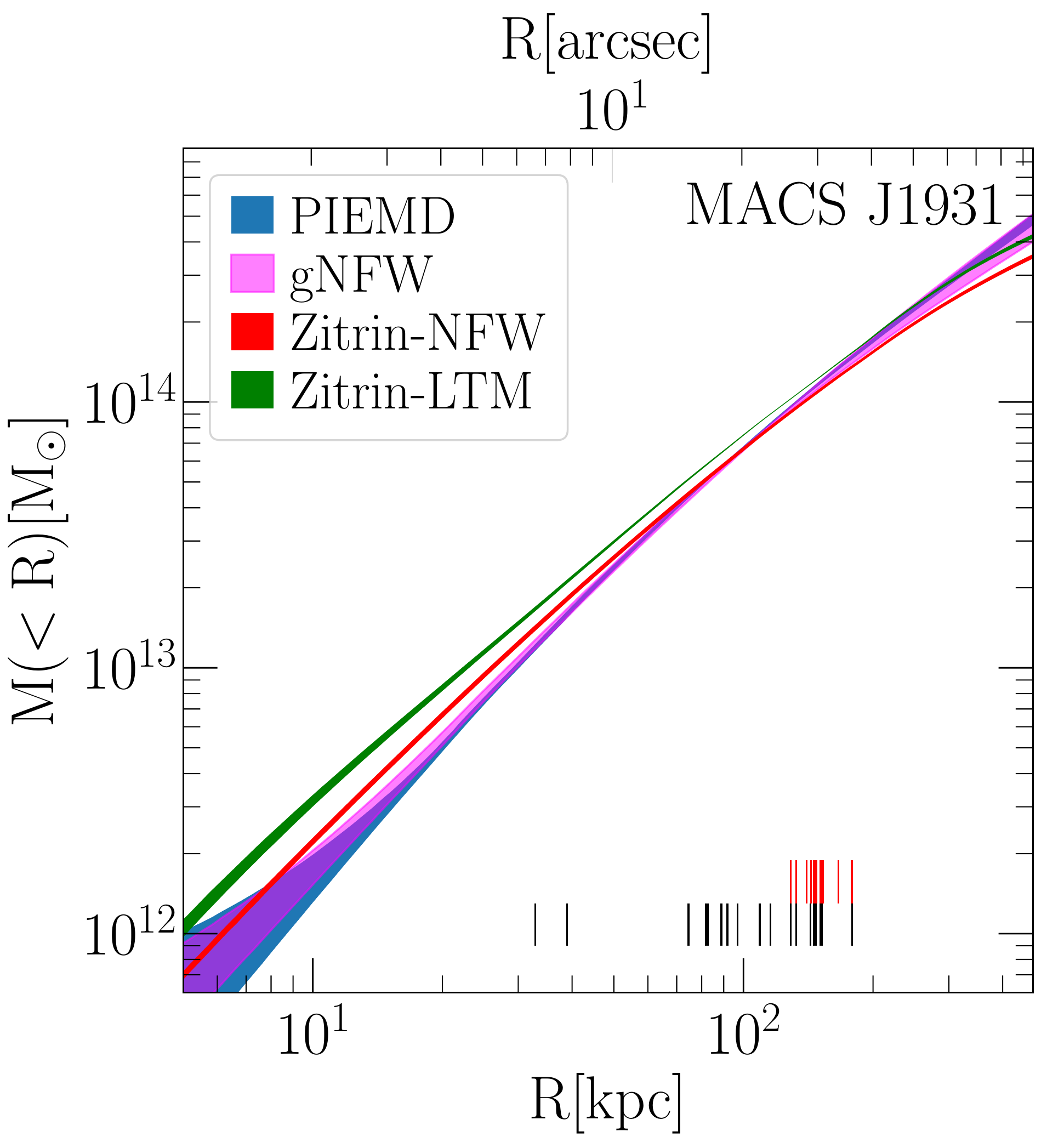}
  \includegraphics[width = 0.504\columnwidth]{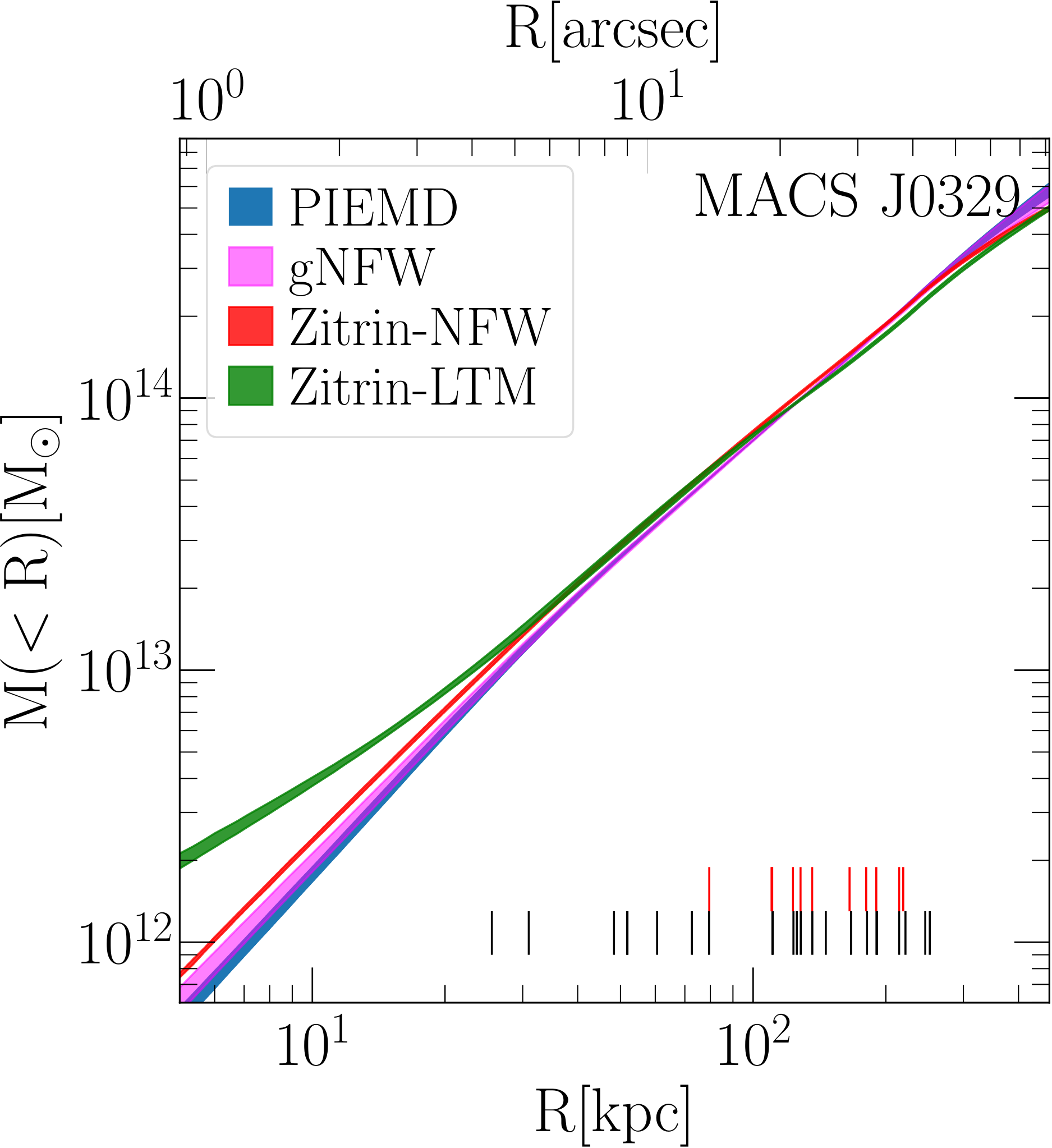}
  \includegraphics[width = 0.504\columnwidth]{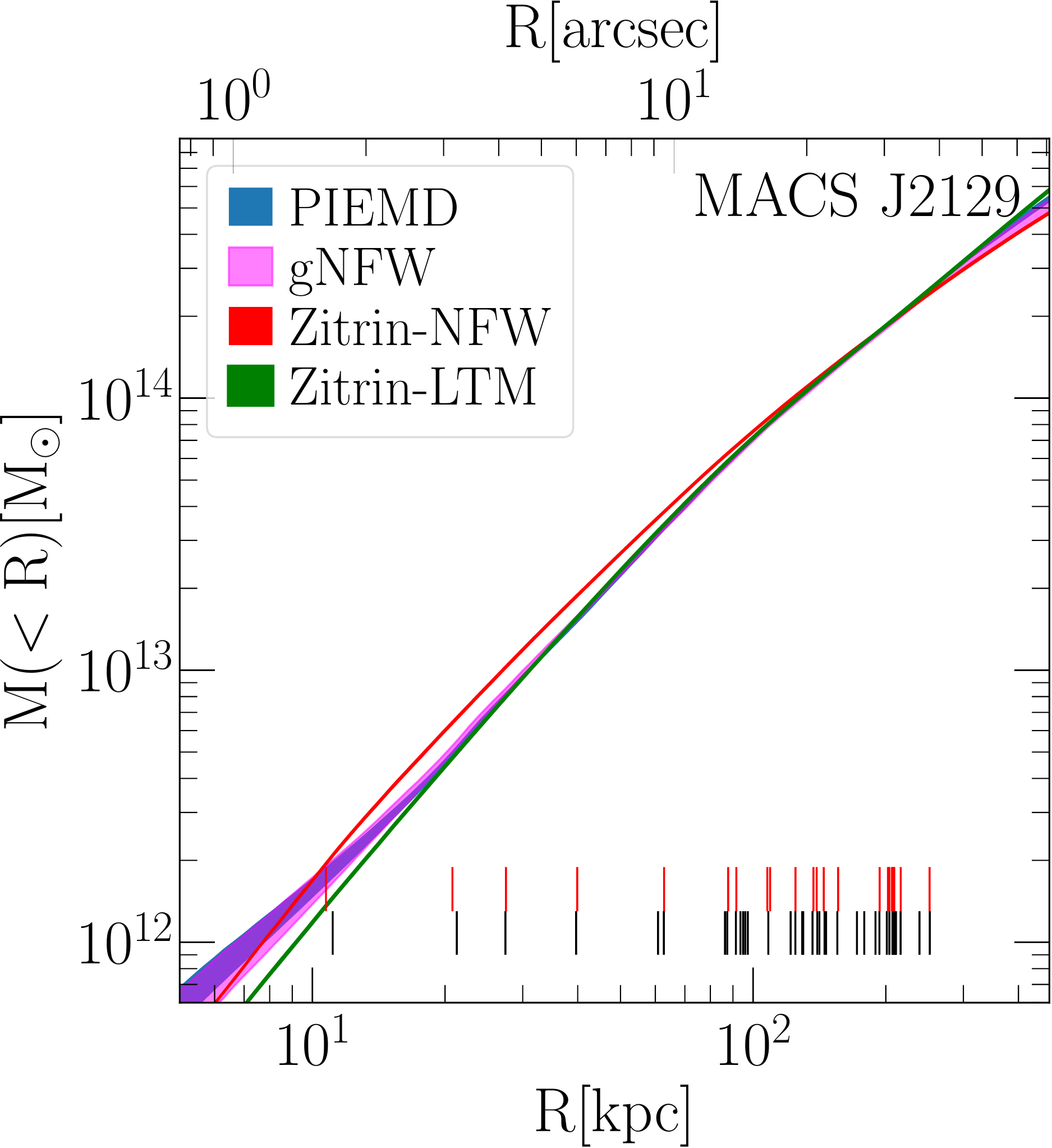}

  \vspace{0.5cm}

  \includegraphics[width = 0.504\columnwidth]{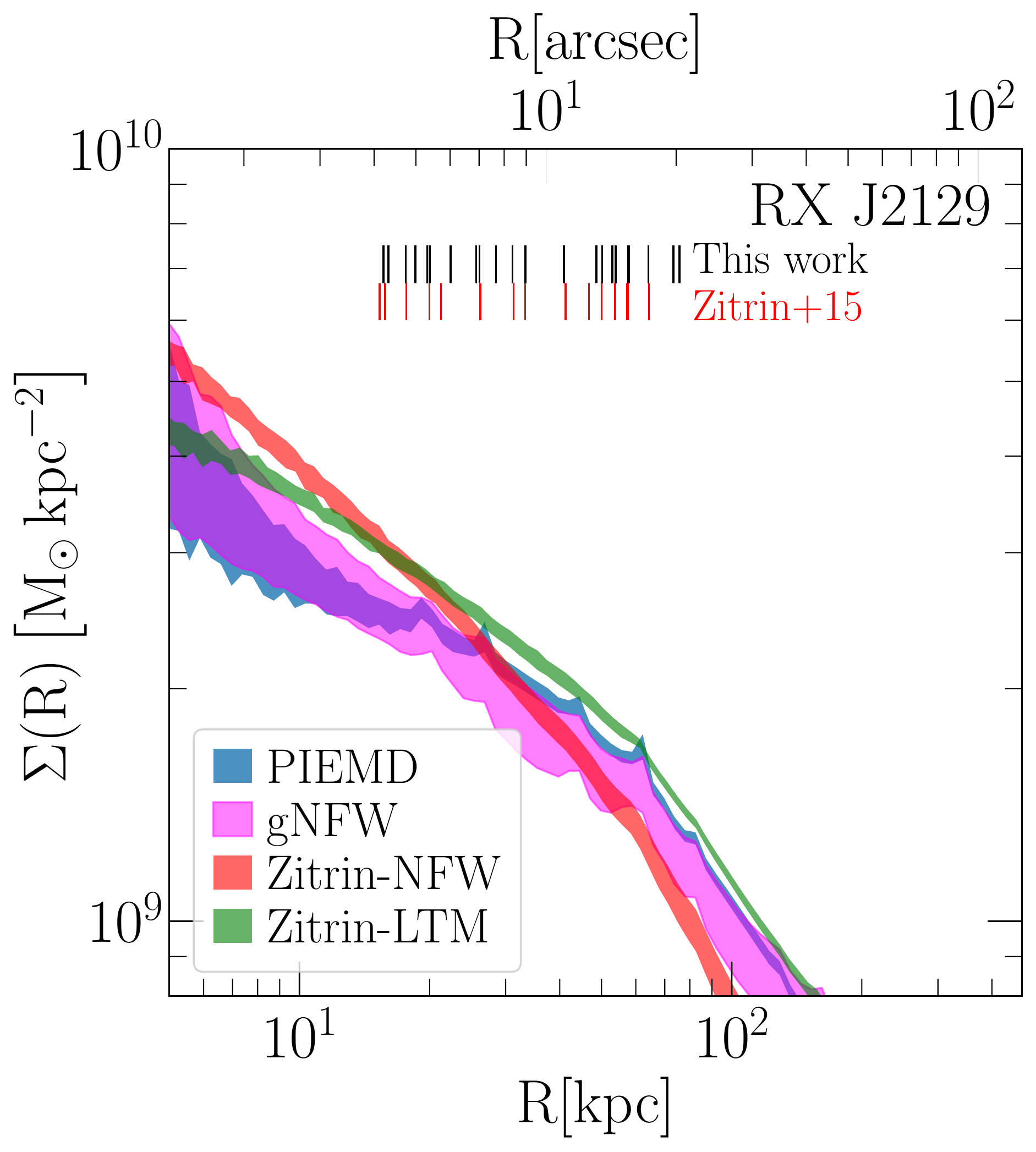}
  \includegraphics[width = 0.504\columnwidth]{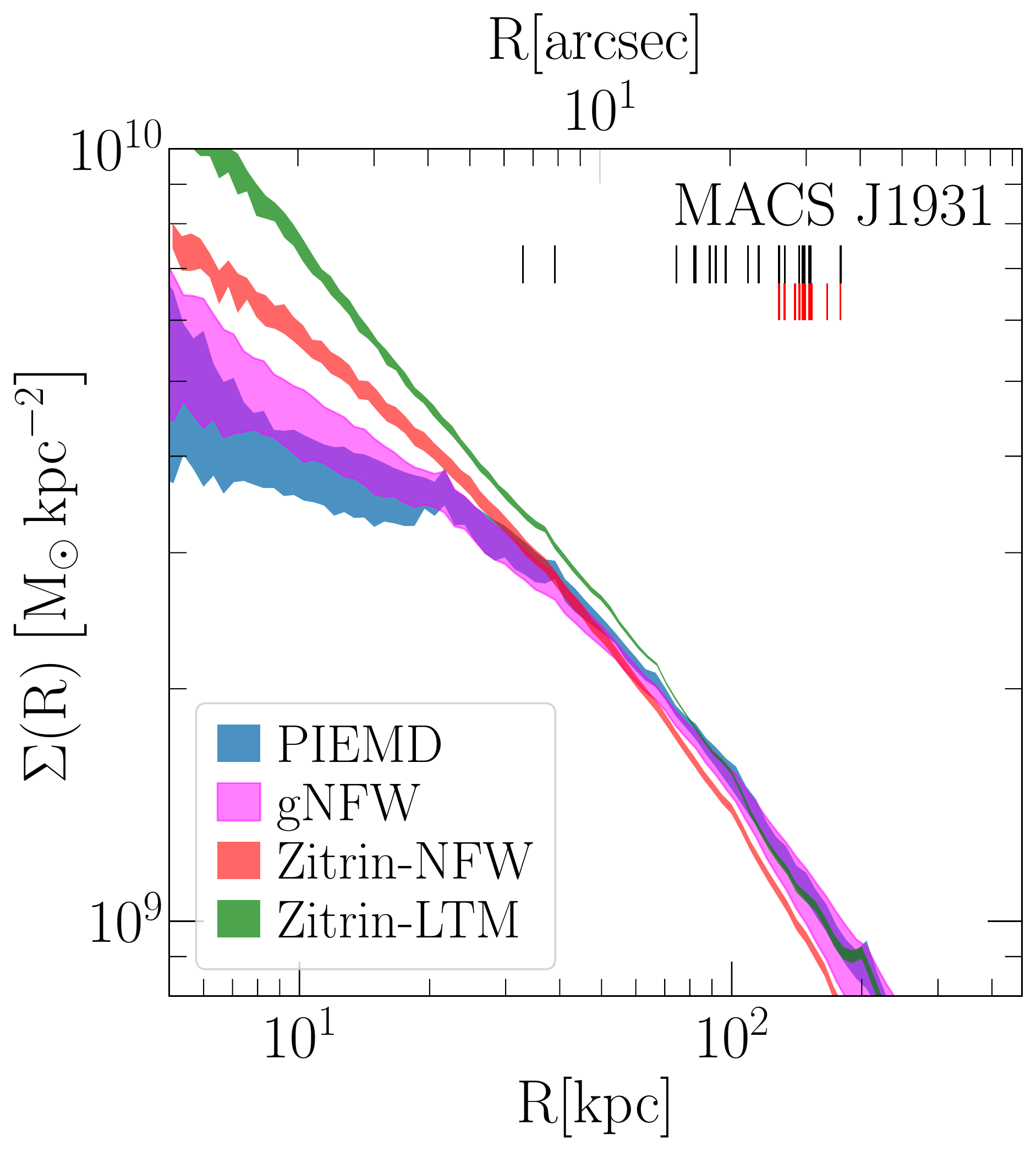}
  \includegraphics[width = 0.504\columnwidth]{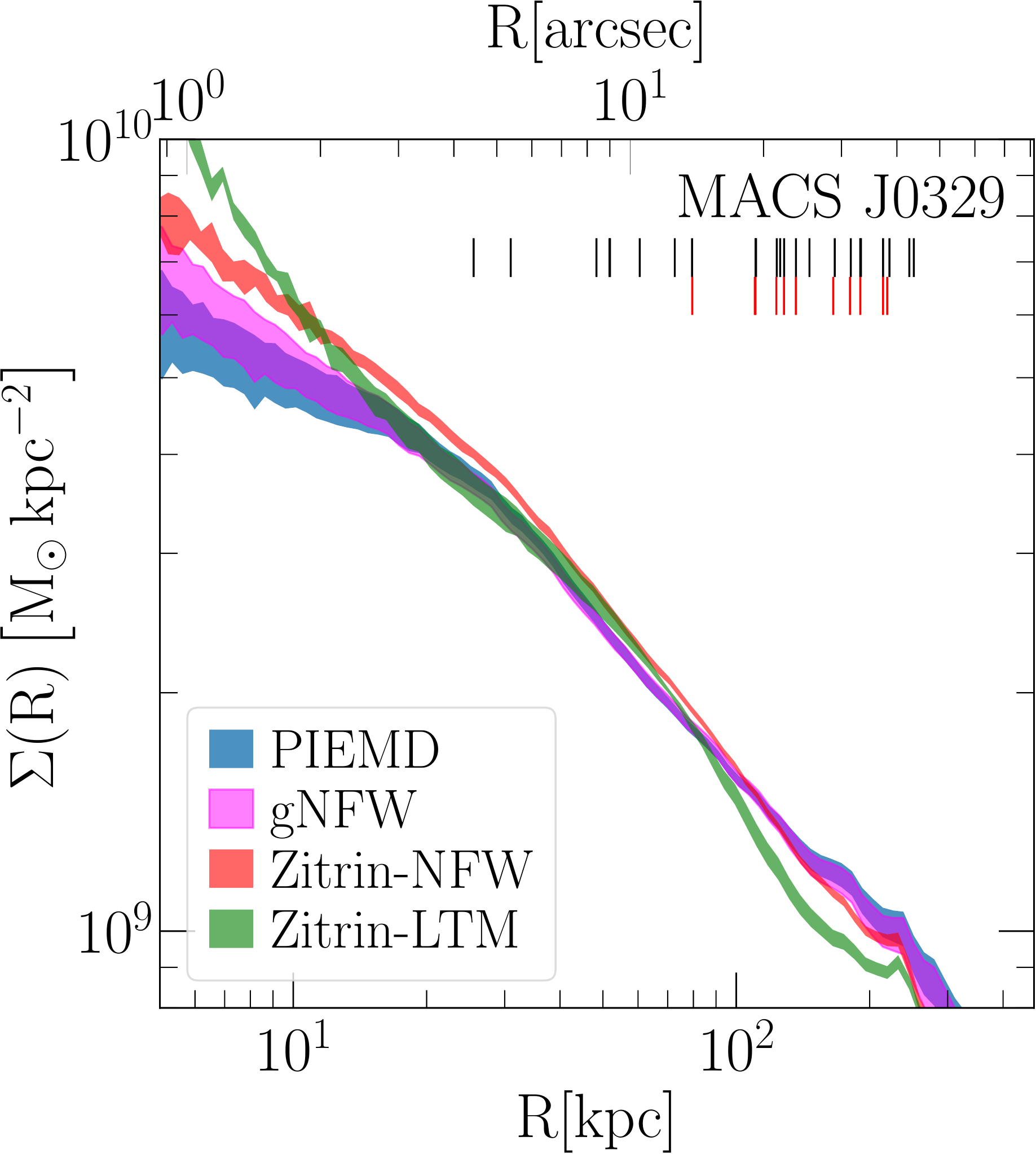}
  \includegraphics[width = 0.504\columnwidth]{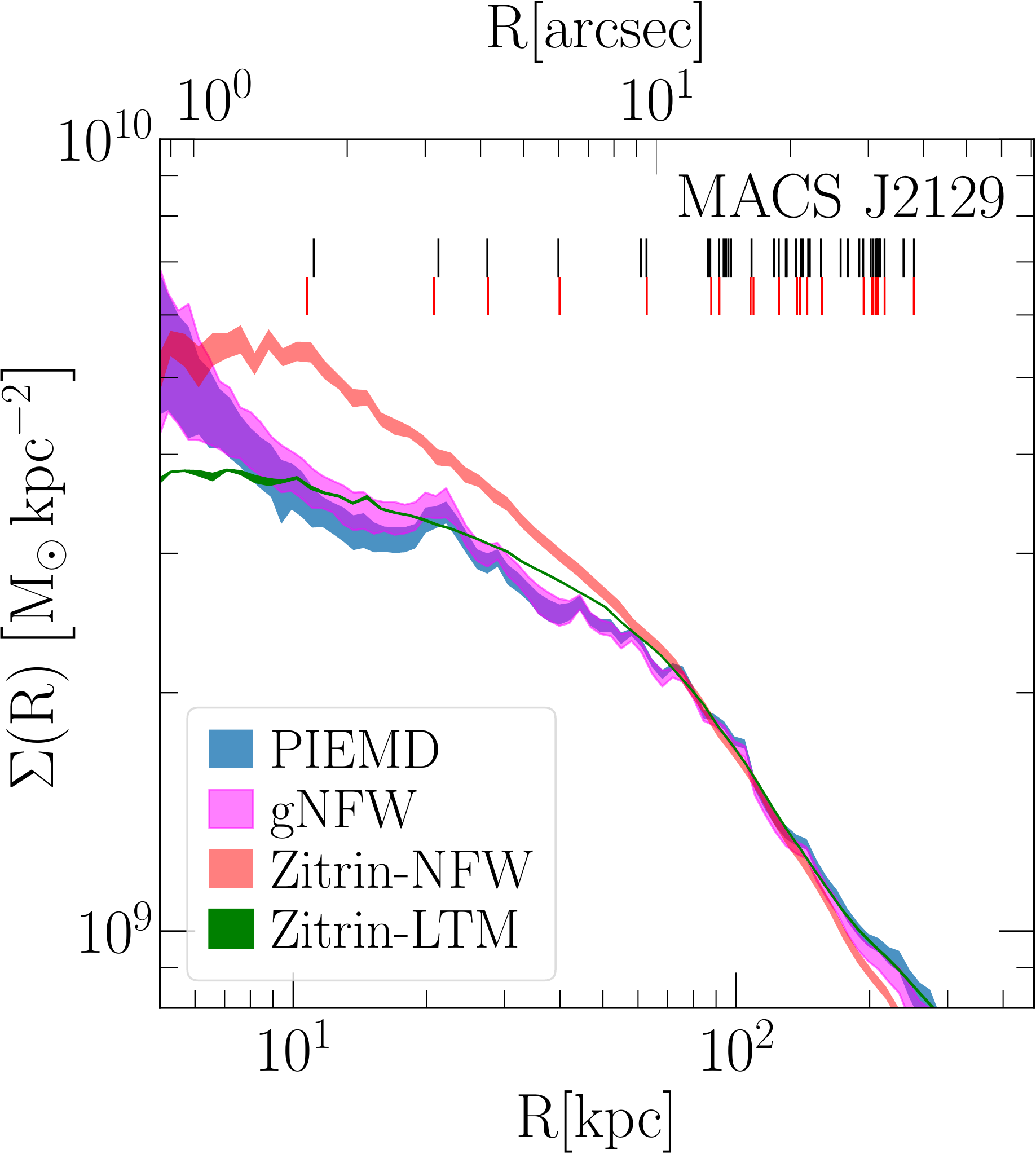}

  \caption{Top row: cumulative projected total mass profile out to $R=470$~kpc from our reference lensing models using the PIEMD (blue) and gNFW (magenta) profiles for the main smooth mass component. Red and green regions show the models from \citet{2015ApJ...801...44Z} with the two parametrizations NFW and LTM, respectively. Bottom row: Same for the total surface mass density profile. The areas correspond to the 95\% confidence level regions from 1000 random realizations of our models and 100 for the NFW and LTM models. Vertical lines indicate the distances from the BCG of the multiple images used to constraint the cluster total mass model in this work (black) and in \citet{2015ApJ...801...44Z} (red, mainly with no spectroscopic measurements). The position of the centre used to compute the profiles in each realization is given by the centre of mass estimated within a circle of 10$\arcsec$ radius from the BCG. We remark that the external smooth components of MACS~J1931 and MACS~J2129 do not affect the total mass distributions over the radial distances considered here.}
  \label{fig:mass_profile}
\end{figure*}

\begin{figure*}
  \centering

  \includegraphics[width = 0.504\columnwidth]{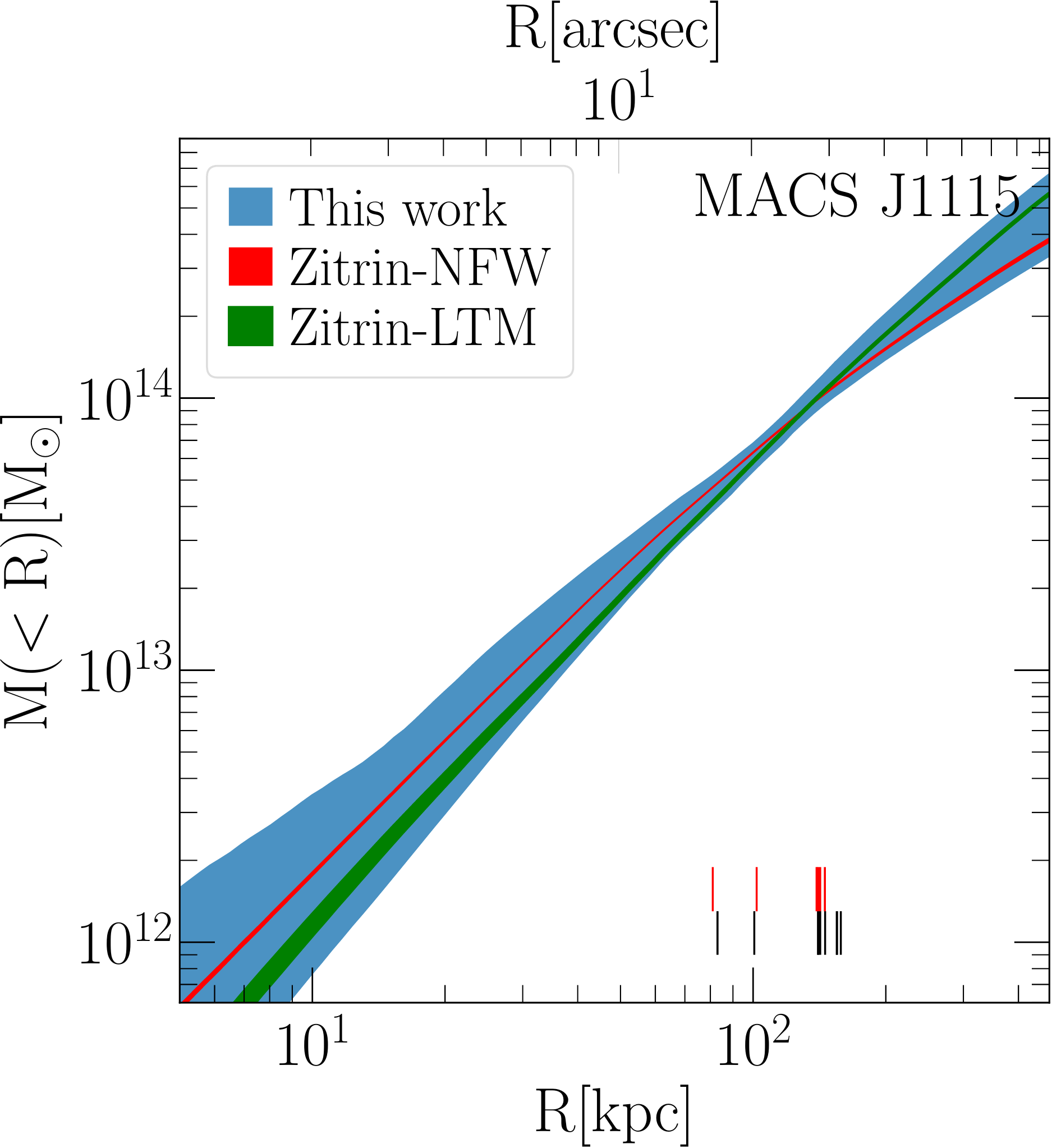}
  \includegraphics[width = 0.504\columnwidth]{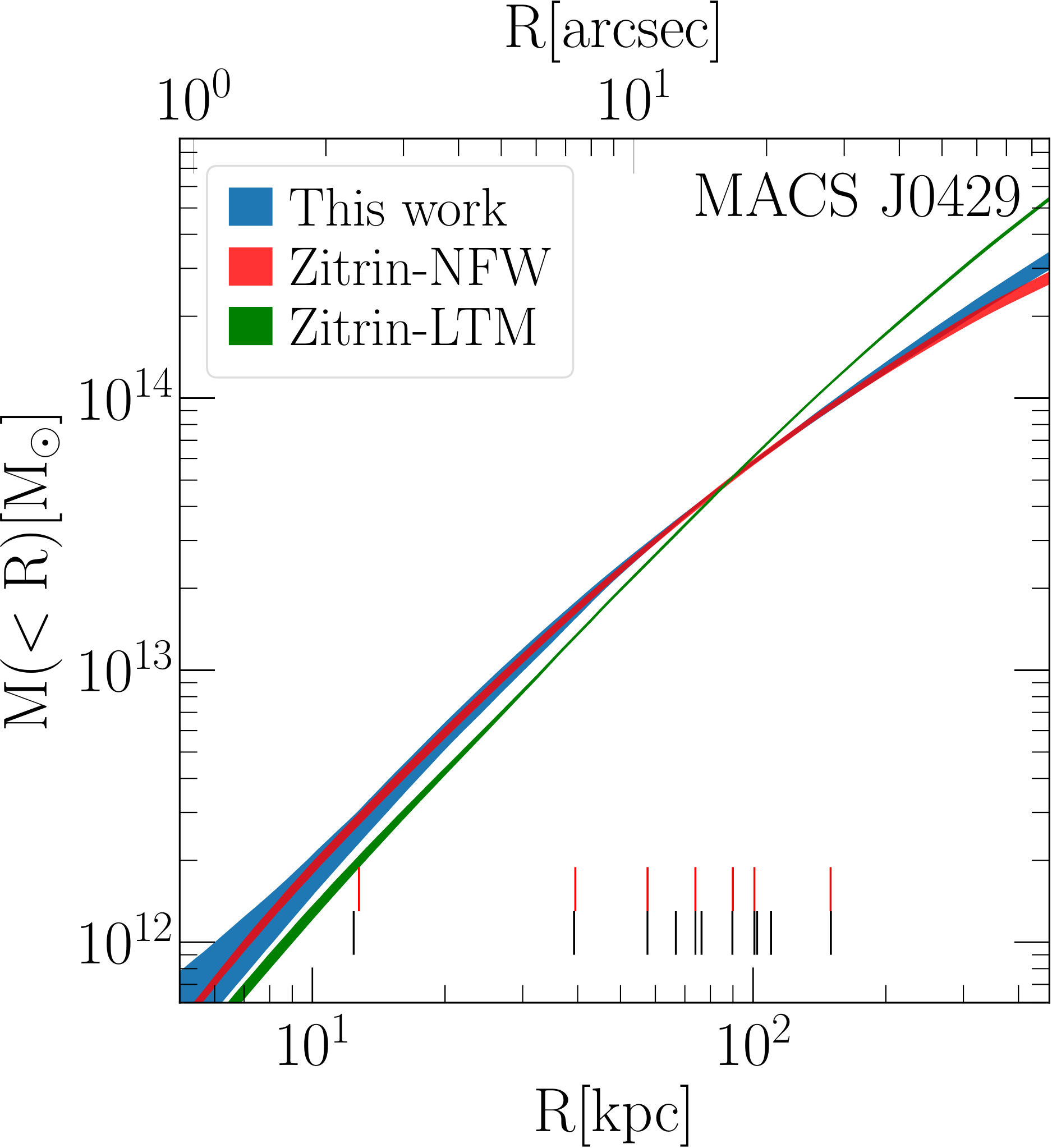}
  \includegraphics[width = 0.504\columnwidth]{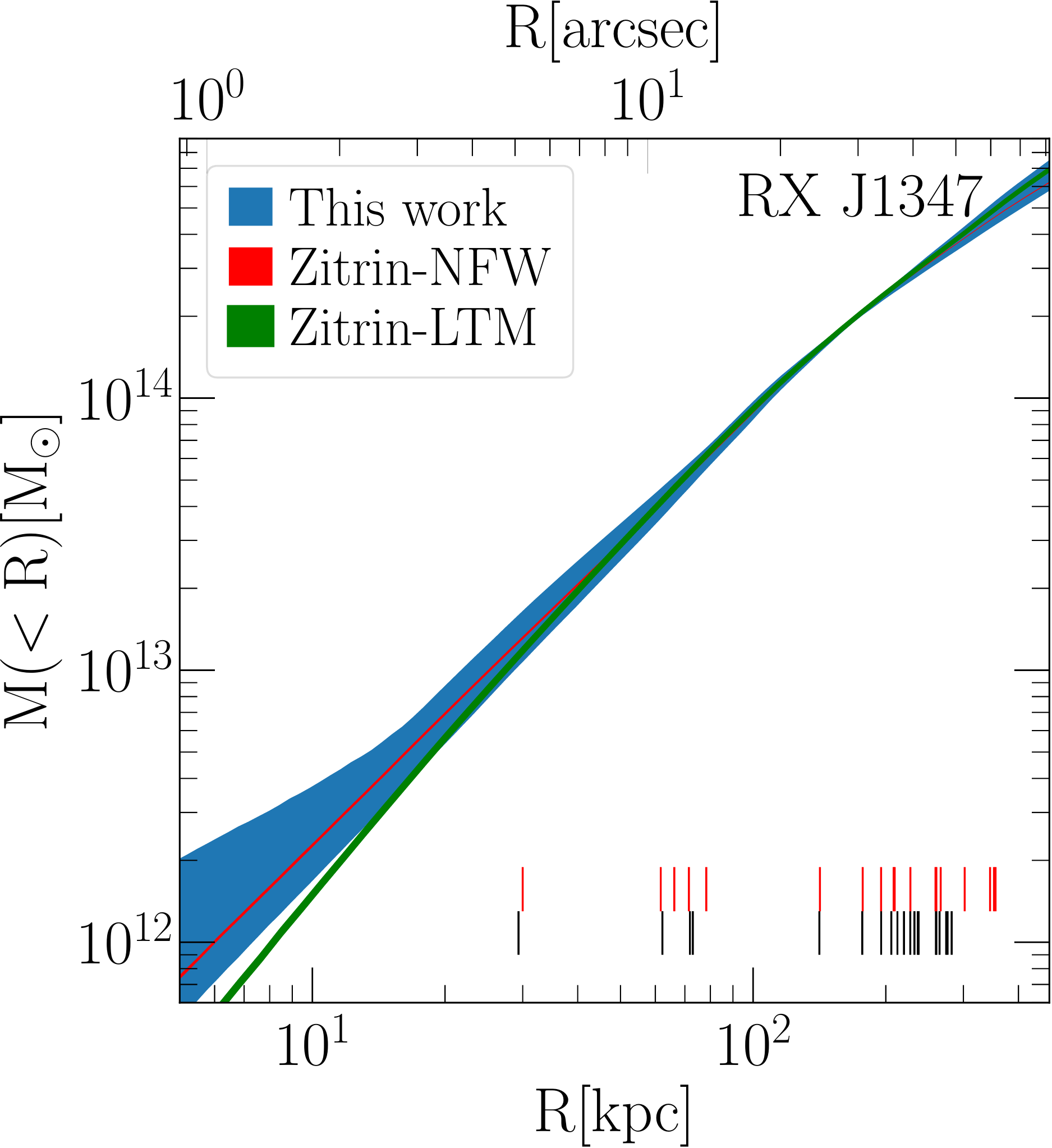}
  \includegraphics[width = 0.504\columnwidth]{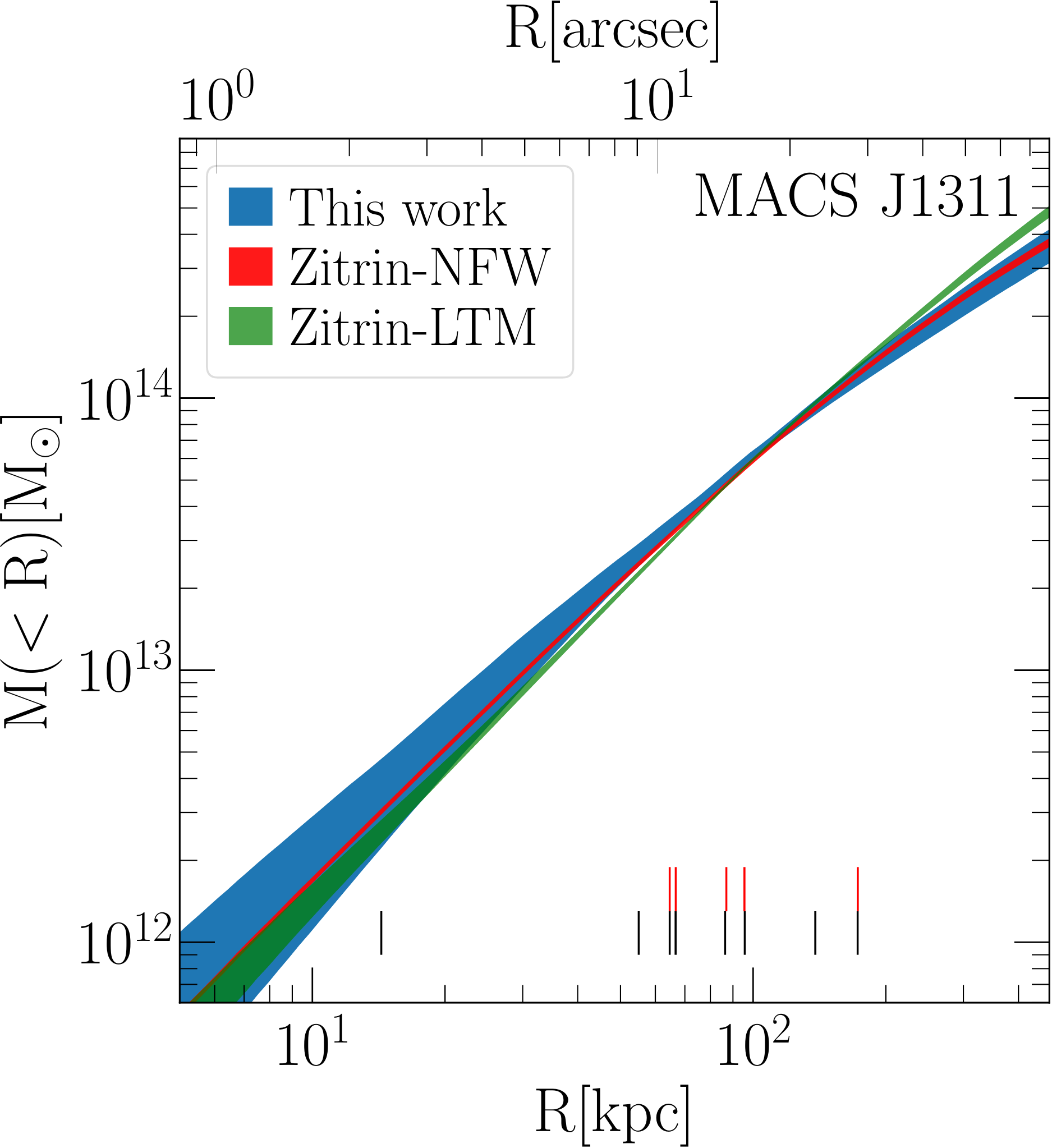}

  \vspace{0.25cm}

  \includegraphics[width = 0.504\columnwidth]{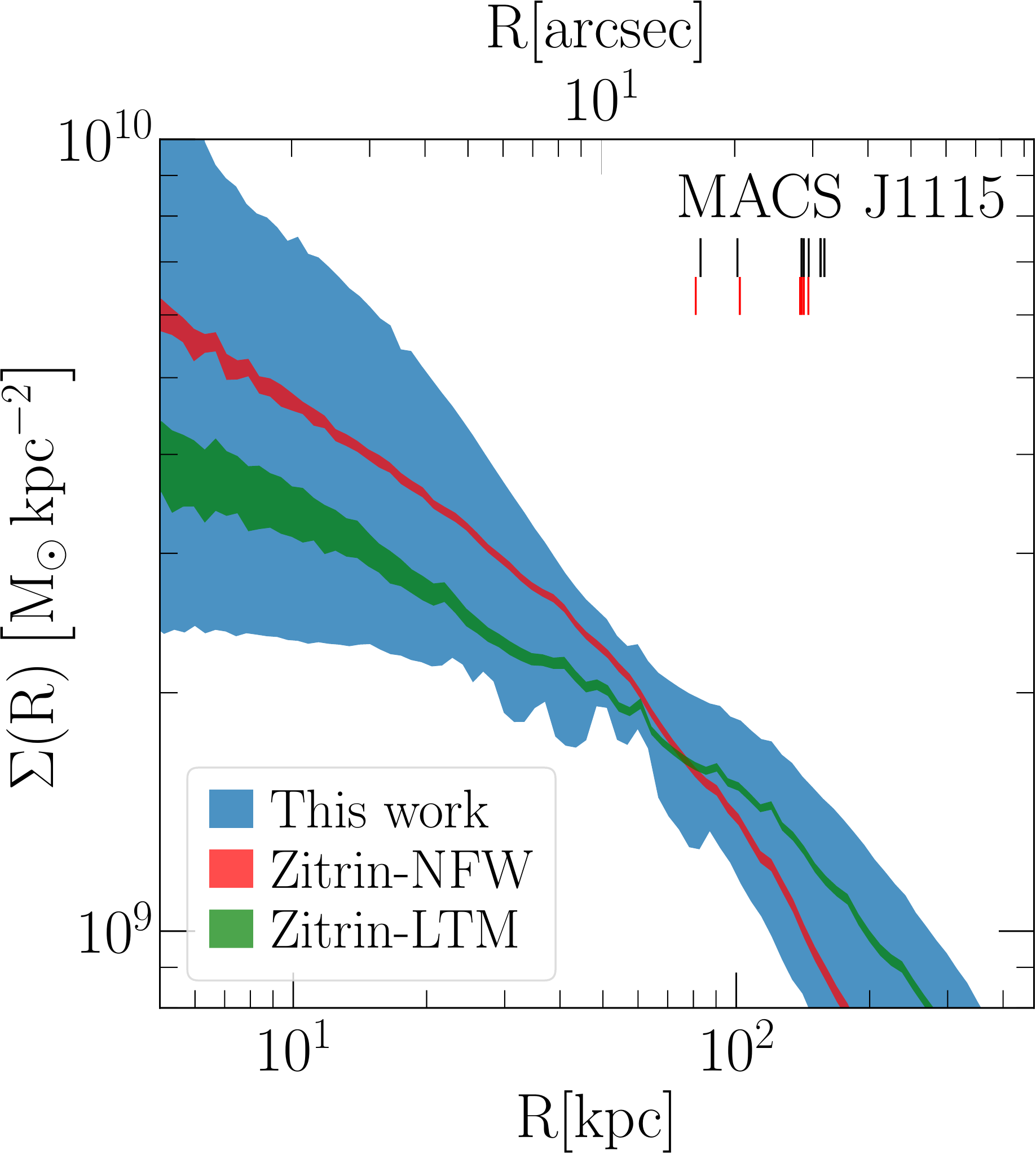}
  \includegraphics[width = 0.504\columnwidth]{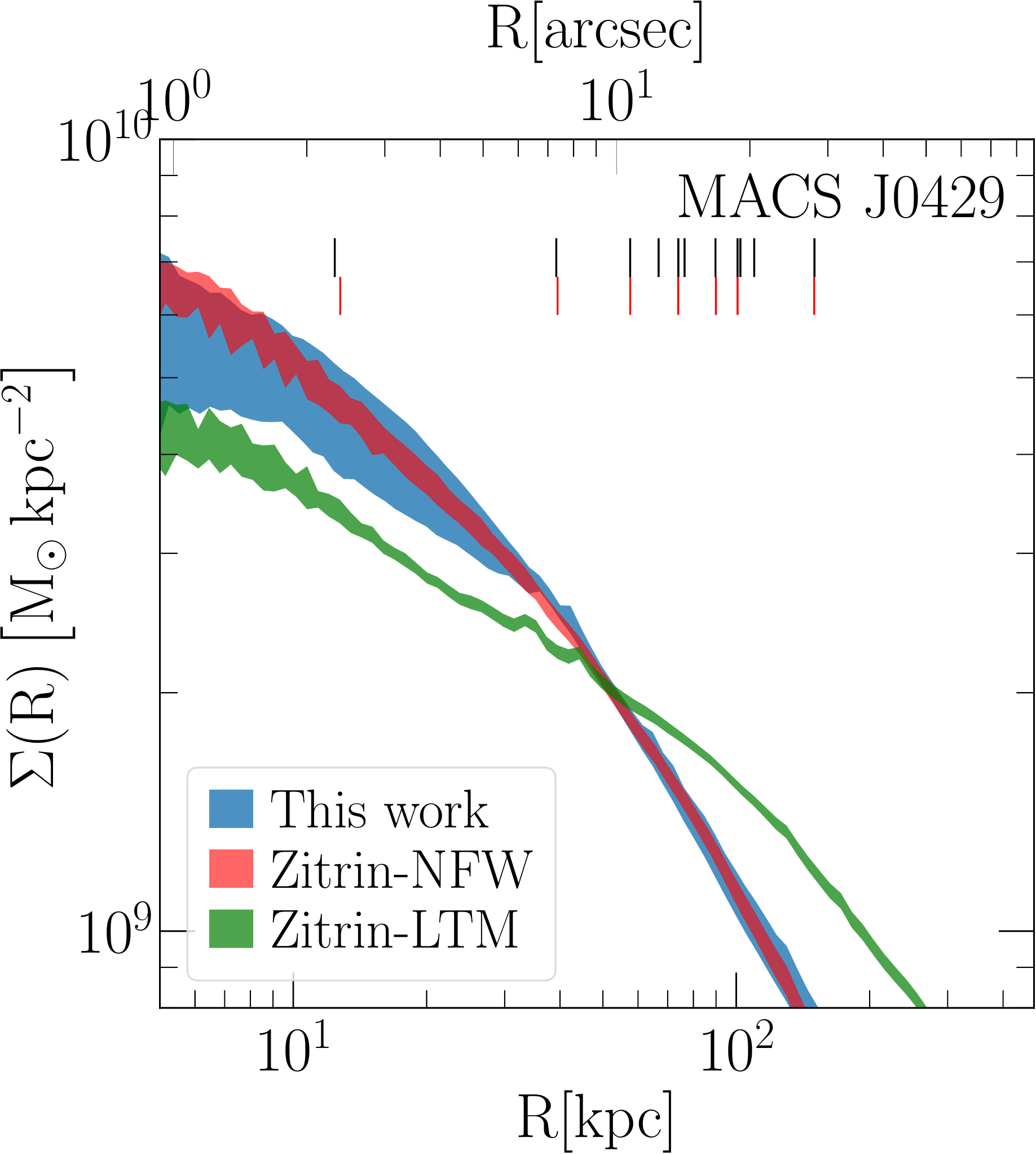}
  \includegraphics[width = 0.504\columnwidth]{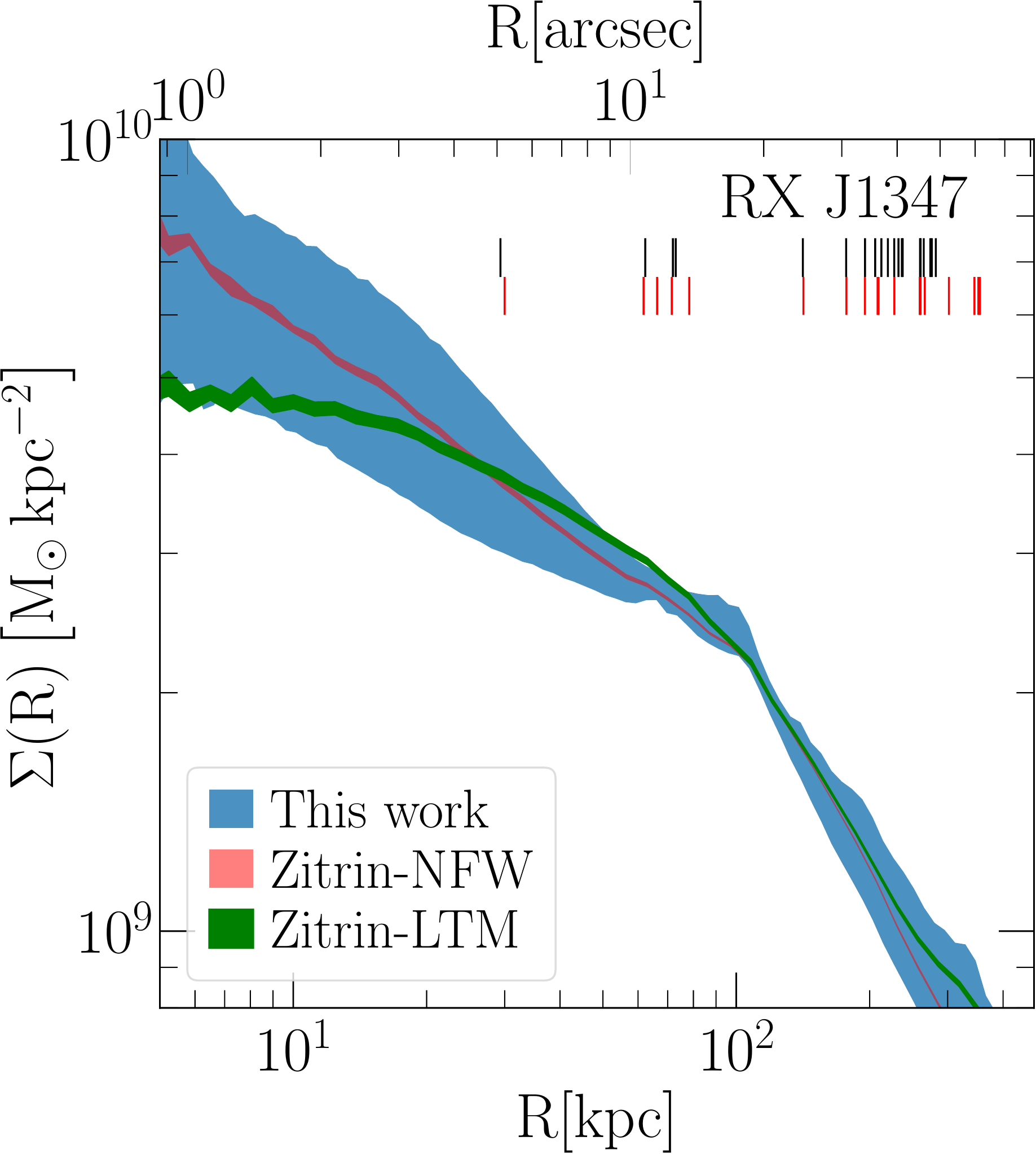}
  \includegraphics[width = 0.504\columnwidth]{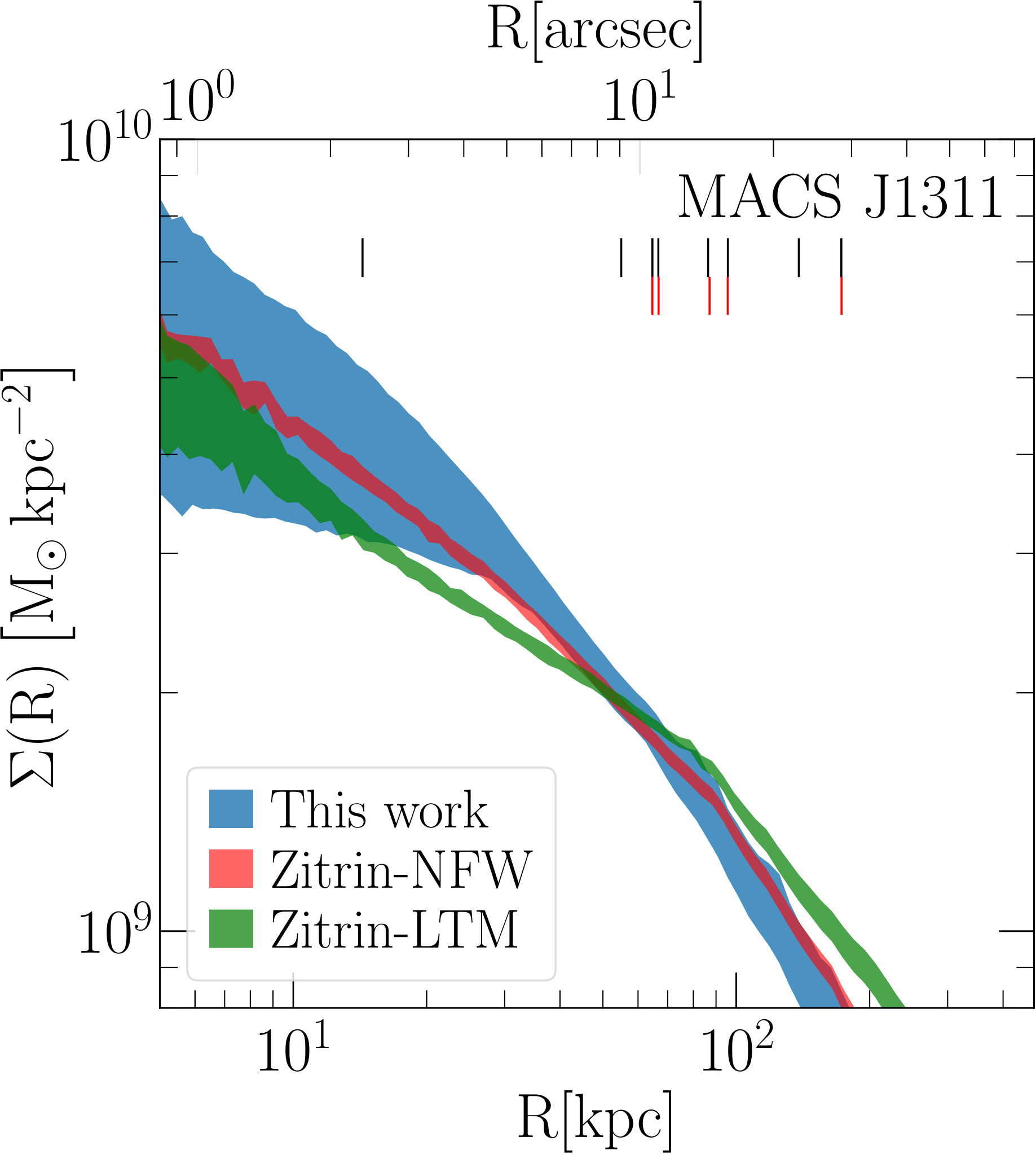}

  \caption{Same as in Figure \ref{fig:mass_profile}, but for the silver sample, i.e. the four clusters with lower number of strong lensing constraints.}
  \label{fig:mass_profile_second}
\end{figure*}

\subsection{RX~J2129}

The multiple images of RX~J2129 are located within $\approx 100$~kpc from the cluster centre and out to $z_{\rm src}=3.43$, the smallest redshift range for our gold lenses.
The positions of the galaxy members and of the multiple images (see Figures \ref{fig:members} and \ref{fig:multiple_images}) suggest a fairly regular total mass distribution, although some asymmetry in the intracluster light (ICL) towards the south-west region can be noted.

The observed positions of all 22 multiple images are well reproduced, with $\Delta_{\rm rms}=0\arcsec.2$, by a single smooth mass component plus the galaxy members.
The inclusion of extra smooth mass components reduces further the value of $\Delta_{\rm rms}$, but the increased values of the BIC and AIC do not justify these extra free parameters.
RX~J2129 is also the least massive cluster in our full sample, with $M^{\rm SL}(<200{\rm \,kpc}) \approx 1.2 \times 10^{14}{\rm M_{\odot}}$.
We note that in previous strong lensing analysis, presented in \citet{2015ApJ...801...44Z}, one multiple image belonging to family ID 2 was wrongly assigned. Multiple image 2a here (with $z_{\rm MUSE}=0.916$, see Table \ref{tab:multiple_images}; their image ID 5.3) was wrongly assigned to a nearby foreground object, located $\approx 4\arcsec$ from the correct one, and with $z_{\rm MUSE}=0.671$.

\subsection{MACS~J1931}
MACS~J1931 shows a strong BCG activity and an indication of a current infalling process of some galaxy members.
However, its multiple image positions and X-rays emission indicate a regular total mass distribution.
Our strong lens model is composed of two smooth mass components in addition to the galaxy members.
The first smooth component is located very close to the BCG centre and has a fairly low ellipticity value ($\varepsilon \approx 0.56$, see Table \ref{tab:best_params}).
A second smooth mass component with circular symmetry reduces significantly the best fitting $\Delta_{\rm rms}$ and is favoured by the information criteria.
This component is located relatively far ($\approx 300{\rm ~ kpc}$) from the cluster centre, however its position is not well determined by the available strong lensing constraints alone.

We remark that in previous works on HFF clusters the addition of extra smooth mass components within $\rm 500~kpc$ from the cluster centre \citep[][for Abell~2744]{2018MNRAS.473..663M} or an external shear component \citep[][for Abell~370]{2019MNRAS.485.3738L} was used in order to obtain better models and lower final multiple image $\Delta_{\rm rms}$.
In the case of Abell~2744, a detection of possible extra smooth components was discussed in \citet{2016MNRAS.463.3876J}, using weak lensing information.
We argue that the reduced values of the BIC and AIC parameters obtained with the inclusion of a second mass component in the model of MACS~J1931 might be an indication of the presence of an extra mass concentration.
However, a detailed weak lensing study in the corresponding region is not possible with the data currently available.

\subsection{MACS~J0329}

In this cluster, all multiple images are located around the south-east BCG (see Figure \ref{fig:multiple_images}), however the second bright galaxy located in the north-west indicates that the cluster is possibly undergoing a merging event.
In order to reproduce well the positions of all multiple images, a second smooth mass component must be included around this galaxy (indicated as NW in the top-right panel of Figure \ref{fig:multiple_images}).
This model improves significantly the final $\Delta_{\rm rms}$ and is accepted by the BIC and AIC criteria.
We note that no secure multiple image systems or photometrically selected candidates are present around the north-west BCG, making our strong lensing constraints weaker in that region.
Although an elongated gravitational arc is visible near the north-west BCG, this is not multiply lensed and does not add constraints to our models.

The BIC and AIC information criteria do not show clear preference between the parametrizations with and without an external shear term (model IDs MACSJ0329-P2 and  MACS J0329-P2-shear in Table \ref{tab:summary_models_all}).
We note that despite the differences in the two parametrizations, the final total mass and density profiles are nearly identical, therefore the specific choice between these two models will not affect our conclusions in this work.
Because of the lower values of the reduced $\chi^2$ and final $\Delta_{\rm rms}$, we select here as reference the model with the presence of an external shear term.

\subsection{MACS~J2129}
MACS~J2129 is the cluster with the largest set of multiple images (i.e. strong lensing constraints) in our sample.
Although the spectroscopic data have qualities similar to those of other clusters in this work, its higher redshift ($z\approx 0.6$) and total mass probably make it a more efficient gravitational lens.
The most recent strong lensing model, presented in \citet{2017MNRAS.466.4094M}, made use of eight multiple image families, for which six were spectroscopically confirmed by CLASH-VLT and GLASS.
The five new confirmations of multiple image families presented in this work highlights the high efficiency of MUSE over previous surveys in the cores of galaxy clusters.

The best-fit model is composed of two smooth mass components, one centred near the BCG and the other located $\approx 38 \arcsec$ ($\approx 250 {\rm ~ kpc}$) south from the cluster centre.
Interestingly, the position of this second smooth mass component lies close to a group of cluster members and an arclet (not multiply lensed) candidate.

In the gold sample, MACS~J2129 is the cluster with the highest value of $\Delta_{\rm rms}$.
Despite that, the reduced $\chi^2$ is 1.18, indicating a good fit.
The relatively high value of $\Delta_{\rm rms}$ ($=0\arcsec.56$) can be explained by the limitations of simple parametric models to reproduce an increasingly large number of multiple images.
For instance, the cluster MACS~J0416 is the cluster with the largest number of spectroscopically confirmed multiple images to date \citep{2017AA...600A..90C} and has the largest $\Delta_{\rm rms}$ in the full sample of this work (see Table \ref{tab:summary_models_all}).

\subsection{MACS~J1115}

As discussed in Section \ref{sec:multiple_image_identification}, four clusters have a small number ($<5$) of spectroscopic multiple image families.
These clusters form our silver sample and MACS~J1115 belongs to it.
MUSE observations were carried out off-centred with respect to the BCG and located $\approx 40 \arcsec$ towards the north-west direction, see Figure \ref{fig:members_2}.
In addition to the two spectroscopic families (one confirmed by MUSE and the other by CLASH-VLT), we included a family with three multiple images with secure identification from the HST data (see Figure \ref{fig:multiple_images}).
The positions of all multiple images is recovered by one single smooth mass component with $\Delta_{\rm rms}=0\arcsec.61$.
The low number of constraints does not allow us to test more complex mass model.

\subsection{MACS~J0429}

To model MACS~J0429, we used the constrains from three multiple image families, two of which have spectroscopic confirmation from MUSE.
The third family shows an Einstein Cross configuration with four clear detections in the HST images.
Three multiple images (IDs 3.b, 3.c and 3.d) have a relatively secure photometric redshift value in the range $[1.64 - 1.79]$, whereas the fourth (ID 3.a) is strongly contaminated by the BCG.
Our model with one smooth mass component can reproduce well the positions of all multiple images, with $\Delta_{\rm rms} = 0\arcsec.32$.

\subsection{RX~J1347}

RX~J1347 is a very massive cluster at $z_{\rm cluster}=0.451$ with $M_{\rm 200crit}=(3.54\pm0.51)\times 10^{15} {\rm M_{\odot}}$ from weak lensing only measurements (see Table \ref{tab:summary_bf}), and shows two bright central galaxies.
For the membership selection of this cluster we have also included five redshifts from \citet{2002ApJ...573..524C} and two from \citet{2012MNRAS.421.1949V}, in addition to the CLASH-VLT and GLASS measurements, in our spectroscopic sample.
By using X-ray and Sunyaev–Zel’dovich observations, \citet{2018ApJ...866...48U} argues that this cluster is likely to be undergoing the first passage of a major merging event and shows that the intra cluster medium has been perturbed by this event.
In \citet{2018ApJ...866...48U}, the most recent strong lensing model of this cluster is presented, using the software {\tt GLAFIC} \citep[][]{2010PASJ...62.1017O}.
The authors consider a set of strong lensing constraints very similar to that in \citet{2015ApJ...801...44Z}.
Moreover, they include a family containing six multiple images \citep[family ID 3 of][]{2018ApJ...866...48U}.
This family has no spectroscopic confirmation and for some images the lensing effect is dominated by the mass distribution on the scale of galaxy members.
Since the identification of some images cannot be considered secure due to significant differences between the model-predicted values of magnification and parity from the strong lensing model and those observable in the HST data, we decide not to include this family in our model.
In addition, we removed multiple images with dubious photometric identification.
As a result, we used 11 multiple images with photometric identification only (belonging to four families) in addition to the six multiple images with MUSE redshifts (three families) and one family (with three multiple images) with previous spectroscopy \citep{2008ApJ...681..187B, 2008A&A...481...65H}.
We note that further deep and wide MUSE observations will become available in the near future and will likely provide a larger number of strong lensing constraints, that will clarify any apparent inconsistency in the models.

As in the previous strong lensing models of this cluster, we have to consider two smooth mass components in order to reproduce the positions of all multiple images.
Moreover, we show that an external shear component improves significantly the best fitting model with no penalty to the BIC and AIC factors.
Although the final $\Delta_{\rm rms} = 0\arcsec.36$ is relatively small, the low number of DOF of this model suggests that the total mass distribution of this cluster might not be accurately described with the current data.

\subsection{MACS~J1311}

MACS~J1311 is the cluster with the smallest number of strong lensing constraints presented in this work, having spectroscopic confirmation for only one multiple image family.
The other two photometrically selected families show two and three multiple images.
Even in the simplest mass model composed of one smooth component plus galaxy members, the number of free parameters (eight for the mass distribution and two free redsfhits) is equal to the number of constraints.
This strong lensing model has $\Delta_{\rm rms}=0\arcsec.88$, the highest value in our sample.

\begin{figure*}
  \centering
  \includegraphics[width = 1.02\columnwidth]{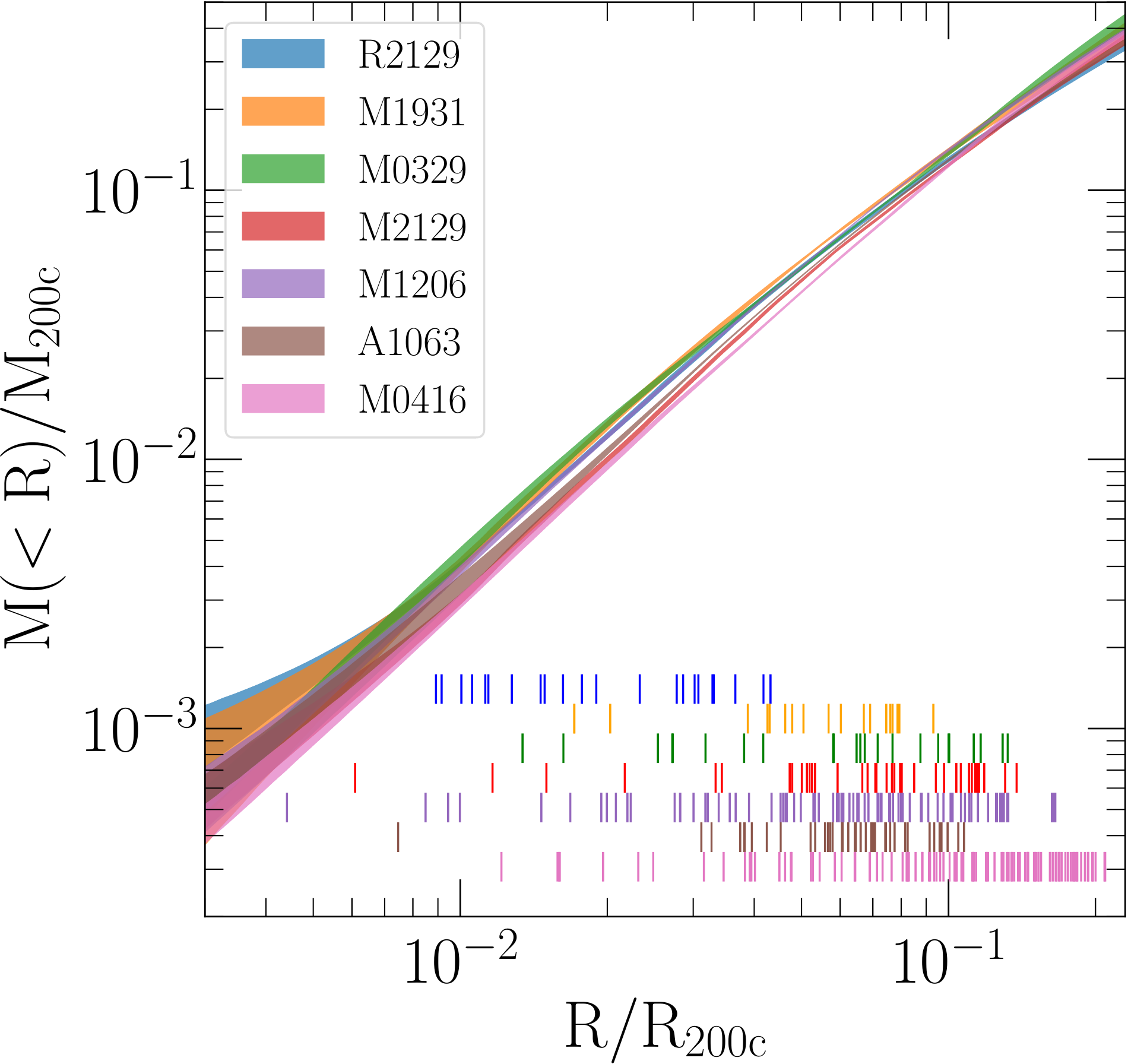}  
  \includegraphics[width = 1.0\columnwidth]{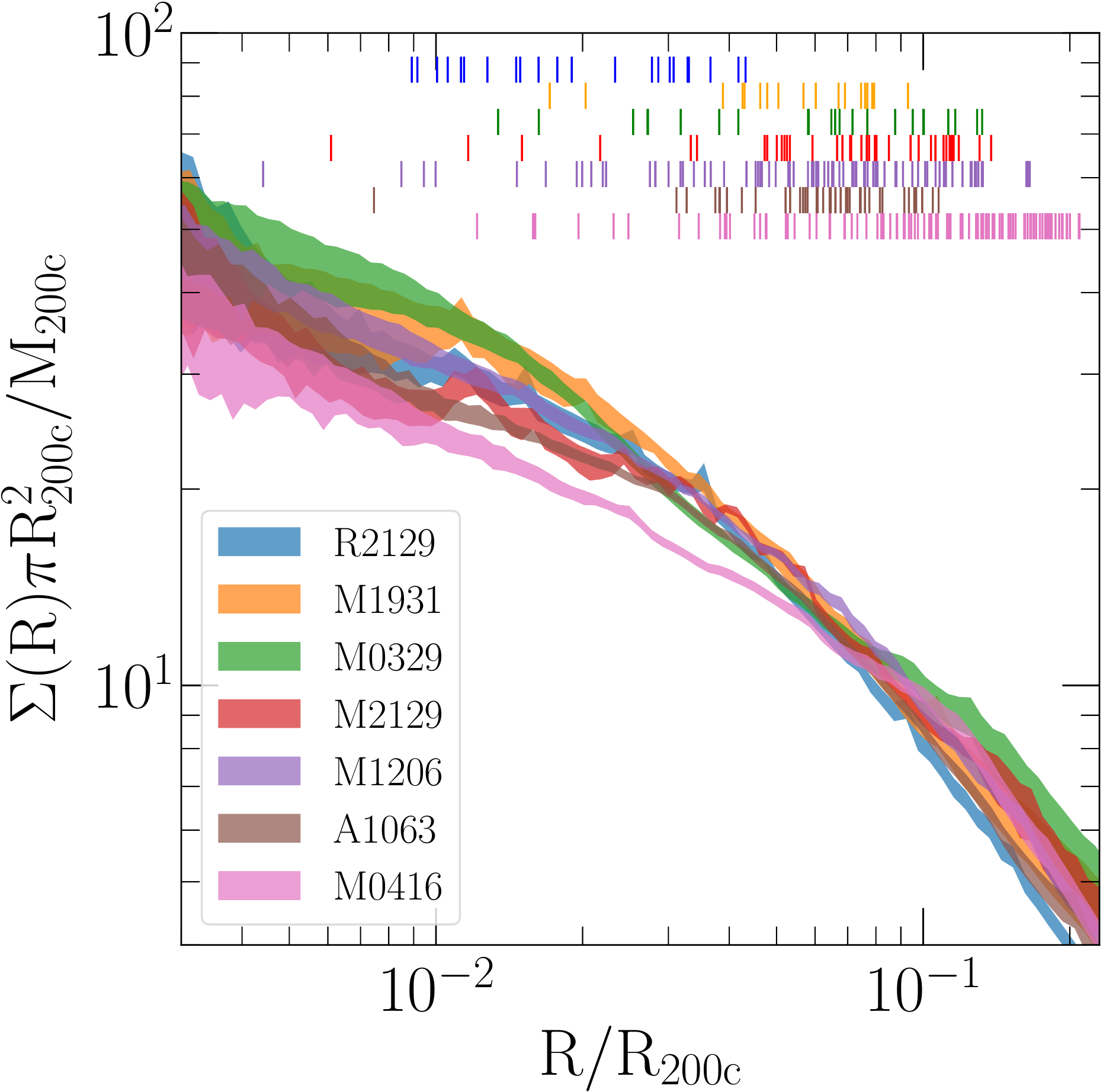}  

  \caption{Projected total mass (cumulative, left panel) and mass density (right panel) profiles rescaled by the quantities $M_{\rm 200c}$ and $R_{\rm 200c}$ from independent weak lensing measurements. In addition to the four galaxy clusters with the best strong lensing models presented in this work (i.e., the gold sample), we also show the profiles of MACS~J1206, Abell~1063 and MACS~J0416, the last one is known to be a prominent merging cluster. The vertical lines indicate the distances from the cluster centers of the multiple images belonging to the spectroscopically confirmed families.}
  \label{fig:projected_mass_density}
\end{figure*}

\subsection{Cluster sample mass distribution}

In Tab. \ref{tab:summary_models_all}, we see that three out of four of the gold clusters prefer a secondary smooth mass component and/or an external shear term.
Particularly two clusters, MACS~J1931 and MACS~J2129, have secondary smooth mass components relatively far from their luminosity centres and with no obvious association to bright galaxy members.
Although the inclusion of these extra components is supported by the information criteria, we cannot reconstruct their physical properties very accurately, given the strong lensing data presented here.
Unfortunately, both clusters lack detailed and deep weak lensing analyses.
In the case of MACS~J2129, this is not possible due the presence of a strong galactic cirrus in the field of view, and MACS~J1931 has the second smallest surface number density of back ground sources in the CLASH sample \citep[see Table 3 in ][]{2016ApJ...821..116U}.
Therefore, possible detections of substructures using weak lensing, similarly to the work by \citet{2016MNRAS.463.3876J, 2018MNRAS.481.2901J}, is very difficult in these two clusters.
We remark that in this work we focus on the inner total mass distribution of these clusters, that is, $R<200{\rm~kpc}$, where the contribution of these secondary mass components represents only a very small fraction of the total.
Moreover, we notice that the information criteria favour always cluster mass models where the BCG profiles are scaled according to the relations valid for all other cluster members.

In addition to the intrinsic limitations of parametric models to represent complex mass distributions in some irregular or merging clusters, mass concentrations along the line-of-sight might introduce extra deflections to the light of background sources and these are not taken into account in our work.
However, in \citet{2018A&A...614A...8C}, the detailed study of the effect of these extra perturbers in MACS~J0416 shows that, for a realistic case, the projected total mass density profile of the cluster is not significantly affected (see e.g. their Figure 18).
Another possible source of systematics in our modelling is some deviation of the adopted mass-to-light scaling relations of cluster members.
This effect has been also recently addressed in \citet{2019arXiv190513236B} by using robust stellar kinematics information of the cluster members.
When comparing our previous lens models of MACS~J1206, MACS~J0416 and Abell~1063 to the new models using prior information from the member kinematics, the authors show that the differences in the total mass distribution is of the order of a few percent and within the statistical errors, in the region where the multiple images are observed.
Therefore, although our modelling might be subject to some intrinsic systematics, their effect on the total mass profiles of the cluster sample should be very small.

In Figures \ref{fig:mass_profile} and \ref{fig:mass_profile_second}, we show the circularly averaged projected total mass (cumulative) and mass density profiles for each cluster for both the gold and silver samples.
For the gold sample, we also show the gNFW models, in addition to the reference PIEMD ones.
Remarkably, the different mass models provide very similar results in the regions where the multiple images are located.
This indicates that, given our good sample of strong lensing constraints, the recovered total mass distribution is mildly sensitive to the specific parametrization of the smooth mass component.
Moreover, we show for comparison the total mass distributions obtained by \citet{2015ApJ...801...44Z} with the PIEMDeNFW and light-traces-mass (LTM) methods.
The differences between the results obtained with these last two methods and those presented in this work might be ascribed primarily to the different sets of strong lensing constraints: \citet{2015ApJ...801...44Z} only had very small number of spectroscopically confirmed multiple images available.
The discrepancies between our and previous results are more evident in the very inner regions (i.e. at $R<30{\rm~kpc}$) and become less pronounced at large radii (i.e. at $R>250{\rm~kpc}$).
Finally, the smaller number of strong lensing constraints of the silver sample (Figure \ref{fig:mass_profile_second}) yields larger statistical errors on the recovered total mass distribution.
We were able to measure the total mass profiles of the clusters in this sample with a $\approx 2\% - 10\%$ statistical uncertainties within 200~kpc from the cluster centres (see Table~\ref{tab:summary_bf}).
However, these measurements of the silver sample are less precise and more likely to be biased than the ones of the gold clusters.

In order to compare the total mass profiles of the clusters in our gold sample, we used the values of $M_{\rm 200c}$ and $R_{\rm 200c}$ from independent weak lensing measurements \citep[][]{2018ApJ...860..104U} to rescale our total mass and mass density profiles.
We remark that weak lensing studies probe the outer regions of the clusters, with the innermost meaningful constraints typically located at $R \approx 300{\rm~kpc}$.
In Figure \ref{fig:projected_mass_density}, we show the projected total mass and density profiles of clusters in our sample, after rescaling the mass and physical scale of clusters by these two quantities.
We also include in this comparative analysis Abell~1063, MACS~J1206 and MACS~J0416 from \citet[][]{2016AA...587A..80C, 2017AA...600A..90C, 2017AA...607A..93C}.
Since the cluster MACS~J2129 does not have weak lensing measurements, we use an empirical scaling relation between $M^{\rm SL}\left(R<200{\rm~kpc}\right)$ and $M_{\rm 200crit}^{WL}$ obtained using the other six clusters to rescale its mass profiles.
All clusters have a relatively large number of strong lensing constraints in the region $10^{-2} \lesssim R/R_{\rm 200c} \lesssim 10^{-1}$ and we find that the shape of the one dimensional averaged projected total mass and mass density profiles are remarkably similar.
Even MACS~J0416, which is a highly asymmetric merging cluster, does not deviate significantly from the overall homologous profiles.
Within 10\% and 20\% of $R_{200c}$, we measure a mean projected total mass value for our seven clusters of $0.13$ and $0.32 \times M_{200c}$, respectively, finding a remarkably small scatter of 5-6~\%.
At these same radii, for the projected total mass density profiles, we find a mean value of $9.0$ and $4.7 \times M_{200c}~\pi^{-1}~R^{-2}_{200c}$, with a slightly larger scatter of 7\% and 9\%.
The observed trend is consistent with the predictions by \citet{2014ApJ...789....1D}, according to which dark-matter halos reveal a self-similar behaviour in their inner (or outer) structure when their mass profiles are expressed in units of spherical overdensity radii defined with respect to the critical (or mean) density of the Universe, especially $R_{\rm 200c}$ \citep[or $R_{\rm 200m}$, see also][]{2017ApJ...836..231U}.
Since the cluster total mass profiles reconstructed with a gNFW or a PIEMD main mass component are very similar (see Figure \ref{fig:mass_profile}), we conclude that the observed self-similarity does not depend on the specific modelling details.

\section{Conclusions}
\label{sec:conclusions}

In this work, we have performed a detailed strong lensing analysis of eight CLASH galaxy clusters, making use of extensive spectroscopic information from MUSE and complemented with CLASH-VLT.
Our cluster sample spans a range of masses of $M_{\rm 200c}^{\rm WL} = [5 - 35] \times 10^{14}{\rm~M_{\odot}}$ \citep[from weak lensing measurements,][]{2015ApJ...806....4M, 2018ApJ...860..104U} and redshifts $z_{\rm cluster} = 0.23 - 0.59$.
We used primarily MUSE spectroscopy to build a bona-fide set of strong lensing constraints (i.e. multiple image families), and to have a clean selection of galaxy cluster members.
Four lens clusters in our sample have spectroscopic confirmation of more than five multiple image families, defining our gold sample, whereas the other four clusters have a limited number of constraints (see Table \ref{tab:summary_bf}).
For the gold sample, we investigated different mass models with increased complexity and use the BIC and AIK information criteria in order to select our reference models.
Given the small number of strong lensing constraints for the silver sample, we used there only simple models to describe the cluster total mass distributions.
Although we can constrain the total mass profiles of these clusters with $\approx 2\% - 10\%$ statistical errors, these last lens models are less precise and more likely to be biased.
Therefore, our conclusions on the total mass distributions of the clusters studied in this work are focused only on the gold sample.
The main results can be summarised as follows:

\begin{itemize}

\item We built strong lensing models with a bottom-to-top approach, where we have first considered simple unimodal parametrizations and have gradually increased their complexity.
In order to choose the reference models with PIEMD profiles for the the smooth components, we used the BIC and AIK criteria to balance the goodness of a fit and the number of free parameters of a model.
In our gold sample, the models can reproduce well the positions of all multiple images, with values of $\Delta_{\rm rms}$ in the range of $0\arcsec.2 - 0\arcsec.6$.

\item When testing a pseudo-elliptical gNFW profile for the main smooth mass components, we found that the overall shapes of the reconstructed total mass and mass density profiles are similar to those obtained with a PIEMD profile.
This indicates that, with a sufficiently large number of strong lensing constraints, modelling details do not affect the general conclusions on the cluster total mass distributions, supporting the robustness of our results.

\item When comparing our cluster total mass profiles with those obtained in previous strong lensing analyses, we found some differences in the very inner regions ($R<30$~kpc).
This might be explained by the smaller sets of strong lensing constraints used in the past, based on a very limited number of spectroscopically confirmed multiple images.

\item Three out of four clusters in our gold sample require a secondary smooth mass component and/or an external shear term. 
Although the strong lensing constraints cannot provide detailed information on these components, because they are located in the cluster external regions ($R \gtrsim 200{\rm~kpc}$) where only a few multiple images are present, their inclusion in the models is favoured by the BIC and AIK criteria.
Given their projected distances from the cluster centres, these components do not affect significantly the inner total mass distributions of the clusters in our sample.

\item In order to compare the total mass profiles of the different clusters in our gold sample and in some of our previous analyses, we rescaled them using independent weak lensing measurements.
Remarkably, we found that all clusters have very similar one-dimensional projected total mass profiles with a small scatter of 5\% at $R=0.1R_{200c}$, including MACS~J0416 that is a clear merging cluster.
This is a noteworthy observational confirmation of the self-similarity of cluster-size halos predicted by cosmological simulations.

\end{itemize}

The high-quality strong lensing modelling presented in this work adds up to the sample of clusters with accurate total mass measurements in their inner regions (i.e. $R<200{\rm ~kpc}$) and constitutes an optimal sample to be compared to N-body and hydrodynamical simulations of clusters.
Investigations of the separate dark matter and baryonic mass components \citep[hot-gas and stars, see eg.,][]{2017ApJ...851...81A,2017ApJ...842..132B, 2018ApJ...864...98B} in a statistically significant sample of clusters might lead to interesting constraints on the physical nature of dark matter.
Moreover, the MUSE data presented in this work and the magnification maps produced by our analyses increase the number of gravitational telescopes with accurate lens models that can be used to study the very faint galaxy population at high redshift.

Finally, we make publicly available all MUSE spectroscopic redshift measurements in the electronic version of this paper.
The convergence, shear and magnification maps, as well as the {\tt lenstool} configuration files of the strong lensing models, are also available online\footnote{\url{https://www.astro.rug.nl/~caminha/SL_models}}.

\begin{acknowledgements}
  We thank the anonymous referee for the useful comments and suggestions that improved the manuscript.
  G.B.C, G.R. and K.I.C. acknowledge funding from the European Research Council through the award of the Consolidator Grant ID 681627-BUILDUP.
  C.G. acknowledges support by VILLUM FONDEN Young Investigator programme grant 10123. 
  A.M. acknowledges funding from the INAF PRIN-SKA 2017 programme 1.05.01.88.04.
  K.U. acknowledges support from the Ministry of Science and Technology of Taiwan (grant MOST 106-2628-M-001-003-MY3) and from Academia Sinica (grant AS-IA-107-M01).
  R.D. gratefully acknowledges support from the Chilean Centro de Excelencia en Astrof\'isica y Tecnolog\'ias Afines (CATA) BASAL grant AFB-170002
  This research made use of Astropy, \footnote{\url{http://www.astropy.org}} a community-developed core Python package for Astronomy \citep{2013A&A...558A..33A, 2018AJ....156..123A}.
  This work made use of the CHE cluster, managed and funded by ICRA/CBPF/MCTI, with financial support from FINEP (grant 01.07.0515.00 from CT-INFRA - 01/2006) and FAPERJ (grants E-26/171.206/2006 and E-26/110.516/2012).
  This work made use of data taken under the ESO programme IDs 186.A-0798, 095.A-0525, 096.A-0105, 096.A-0650, 097.A-0909 and 098.A-0590.

\end{acknowledgements}

\bibliographystyle{aa}
\bibliography{references}

\noindent {\bf \large Appendix A: Multiple images and reconstructed model parameters}
\label{ap:multiple_images}

\renewcommand{\thetable}{A.\arabic{table}}
\renewcommand{\thefigure}{A.1}

\longtab{
\setcounter{table}{0}
\begin{longtable}{lccccrlcc}
\caption{\label{tab:best_params} Median values and confidence levels of the cluster total mass distribution parameters from the MCMC analyses of the reference strong lensing models (see Table \ref{tab:summary_models_all}). }\\
\hline\hline
  ~ & Median & $68\%$ CL & $95\%$ CL & $99.7\%$ CL \\
\hline
\endfirsthead
\caption{continued.}\\
\hline\hline
  ~ & Median & $68\%$ CL & $95\%$ CL & $99.7\%$ CL \\
\hline
\endhead
\hline
\endfoot

\multicolumn{5}{c}{RX~J2129}  \\
\hline
\multicolumn{5}{l}{Smooth component 1} \\
\hline
$x$ ($\arcsec$) & $2.4$ & $_{-0.2}^{+0.3}$ & $_{-0.5}^{+0.5}$ &$_{-0.7}^{+0.9}$ \\
$y$ ($\arcsec$) & $-1.6$ & $_{-0.1}^{+0.1}$ & $_{-0.3}^{+0.2}$ &$_{-0.4}^{+0.3}$ \\
$\varepsilon$ & $0.67$ & $_{-0.02}^{+0.02}$ & $_{-0.03}^{+0.04}$ &$_{-0.05}^{+0.06}$ \\
$\theta$ ($^{\circ}$) & $-23.7$ & $_{-0.2}^{+0.2}$ & $_{-0.4}^{+0.4}$ &$_{-0.6}^{+0.7}$ \\
$r_{\rm core}$ ($\arcsec$) & $14.0$ & $_{-0.5}^{+0.6}$ & $_{-0.9}^{+1.4}$ &$_{-1.3}^{+2.5}$ \\
$\sigma_{\rm v}$ ($\rm km\;s^{-1}$) & $1079$ & $_{-8}^{+9}$ & $_{-16}^{+19}$ &$_{-23}^{+31}$ \\
\hline
\multicolumn{5}{l}{Galaxy members} \\
\hline
 $r_{\rm cut}^{\rm  M,\, gals}$ (kpc) & $  2.1$ & $ _{-1.2}^{+2.0}$ & $  _{-1.5}^{+4.4}$ &$  _{-1.6}^{+7.2}$ \\
$\sigma_{\rm v}^{\rm  M,\, gals}$ ($\rm km\;s^{-1}$) & $ 392$ & $ _{-85}^{+170}$ & $ _{-123}^{+307}$ &$ _{-144}^{+352}$ \\

\hline \hline
\multicolumn{5}{c}{MACS~J1931}  \\
\hline
\multicolumn{5}{l}{Smooth component 1} \\
\hline
$x$ ($\arcsec$) & $0.09$ & $_{-0.13}^{+0.13}$ & $_{-0.25}^{+0.25}$ &$_{-0.39}^{+0.39}$ \\
$y$ ($\arcsec$) & $0.22$ & $_{-0.31}^{+0.36}$ & $_{-0.60}^{+0.71}$ &$_{-0.87}^{+1.02}$ \\
$\varepsilon$ & $0.56$ & $_{-0.02}^{+0.02}$ & $_{-0.05}^{+0.04}$ &$_{-0.07}^{+0.05}$ \\
$\theta$ ($^{\circ}$) & $83.1$ & $_{-0.7}^{+0.7}$ & $_{-1.4}^{+1.3}$ &$_{-2.2}^{+2.0}$ \\
$r_{\rm core}$ ($\arcsec$) & $9.6$ & $_{-1.1}^{+0.8}$ & $_{-2.0}^{+1.6}$ &$_{-2.4}^{+2.5}$ \\
$\sigma_{\rm v}$ ($\rm km\;s^{-1}$) & $1199$ & $_{-26}^{+20}$ & $_{-54}^{+39}$ &$_{-74}^{+56}$ \\
\hline
\multicolumn{5}{l}{Smooth component 2} \\
\hline
$x$ ($\arcsec$) & $42$ & $_{-15}^{+21}$ & $_{-28}^{+46}$ &$_{-36}^{+57}$ \\
$y$ ($\arcsec$) & $-131$ & $_{-45}^{+45}$ & $_{-65}^{+70}$ &$_{-69}^{+82}$ \\
$r_{\rm core}$ ($\arcsec$) & $12$ & $_{-8}^{+9}$ & $_{-11}^{+12}$ &$_{-12}^{+13}$ \\
$\sigma_{\rm v}$ ($\rm km\;s^{-1}$) & $756$ & $_{-151}^{+181}$ & $_{-269}^{+379}$ &$_{-358}^{+529}$ \\
\hline
\multicolumn{5}{l}{Galaxy members} \\
\hline

 $r_{\rm cut}^{\rm  M,\, gals}$ (kpc) & $ 7$ & $ _{-6}^{+15}$ & $ _{-7}^{+32}$ &$ _{-7}^{+38}$ \\
$\sigma_{\rm v}^{\rm  M,\, gals}$ ($\rm km\;s^{-1}$) & $ 248$ & $ _{-101}^{+231}$ & $ _{-208}^{+492}$ &$ _{-237}^{+560}$ \\

\hline \hline
\multicolumn{5}{c}{MACS~J0329}  \\
\hline
\multicolumn{5}{l}{Smooth component 1} \\
\hline
$x$ ($\arcsec$) & $0.04$ & $_{-0.27}^{+0.26}$ & $_{-0.56}^{+0.52}$ &$_{-0.85}^{+0.76}$ \\
$y$ ($\arcsec$) & $0.02$ & $_{-0.24}^{+0.23}$ & $_{-0.49}^{+0.46}$ &$_{-0.74}^{+0.68}$ \\
$\varepsilon$ & $0.25$ & $_{-0.02}^{+0.02}$ & $_{-0.04}^{+0.05}$ &$_{-0.06}^{+0.07}$ \\
$\theta$ ($^{\circ}$) & $86$ & $_{-4}^{+4}$ & $_{-8}^{+7}$ &$_{-11}^{+11}$ \\
$r_{\rm core}$ ($\arcsec$) & $4.7$ & $_{-0.6}^{+0.6}$ & $_{-1.1}^{+1.2}$ &$_{-1.7}^{+1.9}$ \\
$\sigma_{\rm v}$ ($\rm km\;s^{-1}$) & $931$ & $_{-41}^{+42}$ & $_{-81}^{+84}$ &$_{-119}^{+123}$ \\
\hline
\multicolumn{5}{l}{Smooth component 2} \\
\hline
$x$ ($\arcsec$) & $40$ & $_{-4}^{+4}$ & $_{-7}^{+6}$ &$_{-10}^{+7}$ \\
$y$ ($\arcsec$) & $20$ & $_{-4}^{+4}$ & $_{-7}^{+6}$ &$_{-10}^{+7}$ \\
$\varepsilon$ & $0.52$ & $_{-0.11}^{+0.11}$ & $_{-0.23}^{+0.21}$ &$_{-0.32}^{+0.29}$ \\
$\theta$ ($^{\circ}$) & $71$ & $_{-7}^{+7}$ & $_{-14}^{+17}$ &$_{-20}^{+29}$ \\
$r_{\rm core}$ ($\arcsec$) & $29$ & $_{-5}^{+4}$ & $_{-10}^{+6}$ &$_{-14}^{+6}$ \\
$\sigma_{\rm v}$ ($\rm km\;s^{-1}$) & $1113$ & $_{-104}^{+84}$ & $_{-208}^{+147}$ &$_{-311}^{+197}$ \\
\hline
\multicolumn{5}{l}{External shear} \\
\hline
$\gamma_{\rm shear}$ & $0.07$ & $_{-0.02}^{+0.02}$ & $_{-0.05}^{+0.04}$ &$_{-0.07}^{+0.06}$ \\
$\theta$ ($^{\circ}$) & $-56$ & $_{-7}^{+7}$ & $_{-15}^{+17}$ &$_{-27}^{+135}$ \\
\hline
\multicolumn{5}{l}{Galaxy members} \\
\hline

 $r_{\rm cut}^{\rm  M,\, gals}$ (kpc) & $ 65$ & $ _{-23}^{+32}$ & $ _{-38}^{+54}$ &$ _{-46}^{+58}$ \\
$\sigma_{\rm v}^{\rm  M,\, gals}$ ($\rm km\;s^{-1}$) & $ 161$ & $ _{-8}^{+13}$ & $ _{-14}^{+31}$ &$ _{-18}^{+52}$ \\

\hline \hline
\multicolumn{5}{c}{MACS~J2129}  \\
\hline
\hline
\multicolumn{5}{l}{Smooth component 1} \\
\hline
$x$ ($\arcsec$) & $2.33$ & $_{-0.23}^{+0.22}$ & $_{-0.50}^{+0.46}$ &$_{-0.84}^{+0.73}$ \\
$y$ ($\arcsec$) & $1.62$ & $_{-0.14}^{+0.16}$ & $_{-0.25}^{+0.34}$ &$_{-0.35}^{+0.51}$ \\
$\varepsilon$ & $0.62$ & $_{-0.01}^{+0.01}$ & $_{-0.02}^{+0.03}$ &$_{-0.03}^{+0.05}$ \\
$\theta$ ($^{\circ}$) & $-5.5$ & $_{-0.3}^{+0.3}$ & $_{-0.7}^{+0.6}$ &$_{-1.1}^{+0.9}$ \\
$r_{\rm core}$ ($\arcsec$) & $11.1$ & $_{-0.4}^{+0.5}$ & $_{-0.8}^{+1.1}$ &$_{-1.2}^{+1.8}$ \\
$\sigma_{\rm v}$ ($\rm km\;s^{-1}$) & $1331$ & $_{-10}^{+9}$ & $_{-22}^{+17}$ &$_{-38}^{+26}$ \\
\hline
\multicolumn{5}{l}{Smooth component 2} \\
\hline
$x$ ($\arcsec$) & $6.0$ & $_{-1.2}^{+1.8}$ & $_{-2.8}^{+4.8}$ &$_{-5.3}^{+8.6}$ \\
$y$ ($\arcsec$) & $-39.1$ & $_{-3.5}^{+2.7}$ & $_{-8.4}^{+5.1}$ &$_{-16.0}^{+8.2}$ \\
$\varepsilon$ & $0.86$ & $_{-0.06}^{+0.03}$ & $_{-0.16}^{+0.04}$ &$_{-0.36}^{+0.04}$ \\
$\theta$ ($^{\circ}$) & $45$ & $_{-7}^{+6}$ & $_{-13}^{+12}$ &$_{-19}^{+19}$ \\
$r_{\rm core}$ ($\arcsec$) & $2.3$ & $_{-1.6}^{+3.1}$ & $_{-2.2}^{+8.0}$ &$_{-2.3}^{+15.8}$ \\
$\sigma_{\rm v}$ ($\rm km\;s^{-1}$) & $452$ & $_{-49}^{+70}$ & $_{-87}^{+166}$ &$_{-118}^{+323}$ \\
\hline
\multicolumn{5}{l}{Galaxy members} \\
\hline

 $r_{\rm cut}^{\rm  M,\, gals}$ (kpc) & $ 8$ & $ _{-2}^{+2}$ & $ _{-4}^{+5}$ &$ _{-5}^{+7}$ \\
$\sigma_{\rm v}^{\rm  M,\, gals}$ ($\rm km\;s^{-1}$) & $ 277$ & $ _{-17}^{+30}$ & $ _{-31}^{+85}$ &$ _{-42}^{+142}$ \\

\hline \hline
\multicolumn{5}{c}{MACS~J1115}  \\
\hline
\multicolumn{5}{l}{Smooth component 1} \\
\hline
$x$ ($\arcsec$) & $1.6$ & $_{-1.8}^{+0.8}$ & $_{-6.9}^{+1.4}$ &$_{-10.7}^{+2.0}$ \\
$y$ ($\arcsec$) & $2.1$ & $_{-2.4}^{+1.3}$ & $_{-8.3}^{+2.5}$ &$_{-11.7}^{+3.7}$ \\
$\varepsilon$ & $0.49$ & $_{-0.16}^{+0.17}$ & $_{-0.28}^{+0.34}$ &$_{-0.37}^{+0.40}$ \\
$\theta$ ($^{\circ}$) & $52$ & $_{-3}^{+2}$ & $_{-6}^{+3}$ &$_{-8}^{+4}$ \\
$r_{\rm core}$ ($\arcsec$) & $16$ & $_{-7}^{+8}$ & $_{-14}^{+17}$ &$_{-16}^{+29}$ \\
$\sigma_{\rm v}$ ($\rm km\;s^{-1}$) & $1300$ & $_{-144}^{+173}$ & $_{-242}^{+371}$ &$_{-302}^{+564}$ \\
\hline
\multicolumn{5}{l}{Redshifts} \\
\hline
$z_{3}$ & $2.8$ & $_{-0.6}^{+0.4}$ & $_{-1.0}^{+0.6}$ &$_{-1.4}^{+0.6}$ \\
\hline
\multicolumn{5}{l}{Galaxy members} \\
\hline

 $r_{\rm cut}^{\rm  M,\, gals}$ (kpc) & $ 28$ & $ _{-22}^{+66}$ & $ _{-27}^{+104}$ &$ _{-28}^{+111}$ \\
$\sigma_{\rm v}^{\rm  M,\, gals}$ ($\rm km\;s^{-1}$) & $ 152$ & $ _{-96}^{+254}$ & $ _{-136}^{+541}$ &$ _{-142}^{+623}$ \\

\hline \hline
\multicolumn{5}{c}{MACS~0429}  \\
\hline
\multicolumn{5}{l}{Smooth component 1} \\
\hline
$x$ ($\arcsec$) & $0.52$ & $_{-0.15}^{+0.16}$ & $_{-0.29}^{+0.33}$ &$_{-0.43}^{+0.52}$ \\
$y$ ($\arcsec$) & $0.00$ & $_{-0.28}^{+0.28}$ & $_{-0.61}^{+0.58}$ &$_{-1.10}^{+0.99}$ \\
$\varepsilon$ & $0.57$ & $_{-0.05}^{+0.04}$ & $_{-0.11}^{+0.08}$ &$_{-0.18}^{+0.12}$ \\
$\theta$ ($^{\circ}$) & $79.9$ & $_{-0.5}^{+0.5}$ & $_{-1.0}^{+1.0}$ &$_{-1.5}^{+1.5}$ \\
$r_{\rm core}$ ($\arcsec$) & $4.5$ & $_{-0.7}^{+1.1}$ & $_{-1.1}^{+3.0}$ &$_{-1.6}^{+5.1}$ \\
$\sigma_{\rm v}$ ($\rm km\;s^{-1}$) & $1023$ & $_{-17}^{+18}$ & $_{-39}^{+55}$ &$_{-63}^{+103}$ \\
\hline
\multicolumn{5}{l}{Redshifts} \\
\hline
$z_{3}$ & $1.72$ & $_{-0.05}^{+0.04}$ & $_{-0.08}^{+0.06}$ &$_{-0.09}^{+0.06}$ \\
\hline
\multicolumn{5}{l}{Galaxy members} \\
\hline

 $r_{\rm cut}^{\rm  M,\, gals}$ (kpc) & $ 29$ & $ _{-22}^{+32}$ & $ _{-28}^{+44}$ &$ _{-28}^{+46}$ \\
$\sigma_{\rm v}^{\rm  M,\, gals}$ ($\rm km\;s^{-1}$) & $ 128$ & $ _{-60}^{+87}$ & $ _{-90}^{+233}$ &$ _{-96}^{+401}$ \\

\hline \hline
\multicolumn{5}{c}{RX~J1347}  \\
\hline
\multicolumn{5}{l}{Smooth component 1} \\
\hline
$x$ ($\arcsec$) & $-0.9$ & $_{-1.6}^{+1.4}$ & $_{-2.6}^{+2.8}$ &$_{-3.1}^{+4.5}$ \\
$y$ ($\arcsec$) & $-1.9$ & $_{-1.2}^{+1.3}$ & $_{-1.8}^{+2.6}$ &$_{-2.3}^{+4.0}$ \\
$\varepsilon$ & $0.49$ & $_{-0.10}^{+0.11}$ & $_{-0.15}^{+0.19}$ &$_{-0.19}^{+0.28}$ \\
$\theta$ ($^{\circ}$) & $104$ & $_{-3}^{+4}$ & $_{-7}^{+7}$ &$_{-10}^{+10}$ \\
$r_{\rm core}$ ($\arcsec$) & $9$ & $_{-4}^{+6}$ & $_{-6}^{+8}$ &$_{-8}^{+10}$ \\
$\sigma_{\rm v}$ ($\rm km\;s^{-1}$) &$1244$ & $_{-123}^{+196}$ & $_{-219}^{+271}$ &$_{-307}^{+321}$ \\
\hline
\multicolumn{5}{l}{Smooth component 2} \\
\hline
$x$ ($\arcsec$) & $-17$ & $_{-2}^{+2}$ & $_{-3}^{+2}$ &$_{-4}^{+2}$ \\
$y$ ($\arcsec$) & $-2$ & $_{-2}^{+2}$ & $_{-3}^{+3}$ &$_{-3}^{+3}$ \\
$\varepsilon$ & $0.80$ & $_{-0.08}^{+0.07}$ & $_{-0.19}^{+0.10}$ &$_{-0.32}^{+0.10}$ \\
$\theta$ ($^{\circ}$) & $118$ & $_{-6}^{+7}$ & $_{-10}^{+13}$ &$_{-15}^{+18}$ \\
$r_{\rm core}$ ($\arcsec$) & $9$ & $_{-6}^{+22}$ & $_{-8}^{+37}$ &$_{-9}^{+41}$ \\
$\sigma_{\rm v}$ ($\rm km\;s^{-1}$) & $811$ & $_{-167}^{+131}$ & $_{-316}^{+286}$ &$_{-421}^{+481}$ \\
\hline
\multicolumn{5}{l}{External shear} \\
\hline
$\gamma_{\rm shear}$ & $0.10$ & $_{-0.02}^{+0.02}$ & $_{-0.03}^{+0.04}$ &$_{-0.05}^{+0.07}$ \\
$\theta$ ($^{\circ}$) & $81$ & $_{-11}^{+12}$ & $_{-19}^{+20}$ &$_{-26}^{+25}$ \\
\hline
\multicolumn{5}{l}{Redshifts} \\
\hline
$z_{5}$ & $2.13$ & $_{-0.10}^{+0.09}$ & $_{-0.16}^{+0.12}$ &$_{-0.17}^{+0.13}$ \\
$z_{6}$ & $3.08$ & $_{-0.95}^{+0.57}$ & $_{-2.26}^{+0.75}$ &$_{-2.61}^{+0.78}$ \\
$z_{7}$ & $2.48$ & $_{-0.11}^{+0.11}$ & $_{-0.15}^{+0.15}$ &$_{-0.16}^{+0.16}$ \\
$z_{8}$ & $4.37$ & $_{-0.94}^{+1.17}$ & $_{-1.70}^{+1.76}$ &$_{-2.21}^{+1.86}$ \\
\hline
\multicolumn{5}{l}{Galaxy members} \\
\hline

 $r_{\rm cut}^{\rm  M,\, gals}$ (kpc) & $ 30$ & $ _{-22}^{+54}$ & $ _{-27}^{+78}$ &$ _{-29}^{+83}$ \\
$\sigma_{\rm v}^{\rm  M,\, gals}$ ($\rm km\;s^{-1}$) & $ 213$ & $ _{-82}^{+140}$ & $ _{-167}^{+314}$ &$ _{-202}^{+427}$ \\

\hline \hline
\multicolumn{5}{c}{MACS~J1311}  \\
\hline
\multicolumn{5}{l}{Smooth component 1} \\
\hline
$x$ ($\arcsec$) & $-0.9$ & $_{-0.6}^{+0.6}$ & $_{-1.3}^{+1.1}$ &$_{-2.0}^{+1.6}$ \\
$y$ ($\arcsec$) & $-0.8$ & $_{-0.5}^{+0.8}$ & $_{-1.0}^{+1.9}$ &$_{-1.4}^{+3.6}$ \\
$\varepsilon$ & $0.35$ & $_{-0.07}^{+0.08}$ & $_{-0.14}^{+0.16}$ &$_{-0.19}^{+0.25}$ \\
$\theta$ ($^{\circ}$) &  $92$ & $_{-2}^{+3}$ & $_{-5}^{+6}$ &$_{-7}^{+10}$ \\
$r_{\rm core}$ ($\arcsec$) & $4.9$ & $_{-1.3}^{+1.4}$ & $_{-2.4}^{+2.9}$ &$_{-3.4}^{+4.5}$ \\
$\sigma_{\rm v}$ ($\rm km\;s^{-1}$) & $1015$ & $_{-74}^{+56}$ & $_{-171}^{+107}$ &$_{-276}^{+156}$ \\
\hline
\multicolumn{5}{l}{Redshifts} \\
\hline
$z_{2}$ & $6.12$ & $_{-0.26}^{+0.31}$ & $_{-0.35}^{+0.45}$ &$_{-0.37}^{+0.47}$ \\
$z_{3}$ & $2.27$ & $_{-0.26}^{+0.19}$ & $_{-0.49}^{+0.26}$ &$_{-0.67}^{+0.27}$ \\
\hline
\multicolumn{5}{l}{Galaxy members} \\
\hline

 $r_{\rm cut}^{\rm  M,\, gals}$ (kpc) & $ 48$ & $ _{-32}^{+49}$ & $ _{-46}^{+74}$ &$ _{-48}^{+78}$ \\
$\sigma_{\rm v}^{\rm  M,\, gals}$ ($\rm km\;s^{-1}$) & $ 147$ & $ _{-70}^{+76}$ & $ _{-102}^{+156}$ &$ _{-107}^{+362}$ \\

\end{longtable}
\tablefoot{The values of the velocity dispersion parameters ($\sigma_v$) are rescaled by the factor $\sqrt{2/3}$ as described in the {\tt lenstool} manual (see \href{http://projets.lam.fr/projects/lenstool/wiki/PIEMD}{http://projets.lam.fr/projects/lenstool/wiki/PIEMD}). The reference luminosities for the galaxy member parameters ($\sigma_{\rm v}^{\rm gals}$ and $\rm r_{\rm cut}^{\rm gals}$, see Equation \ref{eq:member_scale}) correspond to the the rest-frame magnitude of $M_{\rm F160W} = -23$ for each cluster.}
}

\longtab{
\sisetup{round-mode=places, output-decimal-marker={.}}
\setcounter{table}{1}
\begin{longtable}{l S[round-precision=6] S[round-precision=6] ccrlcc}
\caption{\label{tab:multiple_images} Information on the spectroscopically identified multiple images.}\\
\hline\hline
 ID & RA & {Dec} & $z_{\rm MUSE}$ & $z_{\rm previous}$ \\ 
\hline
\endfirsthead
\caption{continued.}\\
\hline\hline
 ID & RA & {Dec} & $z_{\rm MUSE}$ & $z_{\rm previous}$ \\ 
\hline
\endhead
\hline
\endfoot

\hline
\multicolumn{5}{c}{RX~J2129}  \\
\hline
R2129-1a & 322.4149133 & 0.09040428989 & 0.6786& ---           \\
R2129-1b & 322.4151812 & 0.08896061967 & 0.6786& ---           \\
R2129-1c & 322.4166369 & 0.08674799398 & 0.6786& ---           \\
\hline
R2129-2a & 322.4146277 & 0.09236277741 & 0.9160& ---           \\
R2129-2b & 322.4162941 & 0.08807537249 & 0.9160& ---           \\
R2129-2c & 322.4165779 & 0.08775984542 & 0.9160& ---           \\
\hline
R2129-3a & 322.4159496 & 0.09149525914 & 1.5194& ---           \\
R2129-3b & 322.4173403 & 0.09065614042 & ---   & ---           \\
R2129-3c & 322.4169443 & 0.09033158441 & ---   & ---           \\
R2129-3d & 322.4185603 & 0.08491578338 & 1.5194& ---           \\
\hline
R2129-4a & 322.4155603 & 0.09214539966 & 1.5202& ---           \\
R2129-4b & 322.4174785 & 0.09017938953 & ---   & ---           \\
R2129-4c & 322.4184168 & 0.08536534410 & 1.5202& ---           \\
\hline
R2129-5a & 322.4179655 & 0.09326575274 & ---   & ---           \\
R2129-5b & 322.4201775 & 0.08976076269 & 1.5210& 1.5222$^{a}$  \\
R2129-5c & 322.4203906 & 0.08830927034 & 1.5210& 1.5222$^{a}$  \\
\hline
R2129-6a & 322.4137584 & 0.09419516553 & 3.0815& ---           \\
R2129-6b & 322.4167336 & 0.08775787975 & 3.0815& ---           \\
R2129-6c & 322.4169857 & 0.08738699691 & 3.0815& ---           \\
\hline
R2129-7a & 322.4137271 & 0.09207931558 & 3.4270& ---           \\
R2129-7b & 322.4144306 & 0.08862720800 & 3.4270& ---           \\
R2129-7c & 322.4175389 & 0.08386292152 & 3.4270& ---           \\
\hline
\multicolumn{5}{c}{MACS~J1931}  \\
\hline
M1931-1a & 292.9556413 & -26.57417245 & 1.1784& ---           \\
M1931-1b & 292.9639604 & -26.57519341 & 1.1784& ---           \\
M1931-1c & 292.9521048 & -26.57618496 & 1.1784& ---           \\
M1931-1d & 292.9561196 & -26.57748170 & 1.1784& ---           \\
\hline
M1931-2a & 292.9578379 & -26.56857847 & 1.8347& 1.8437$^b$    \\
M1931-2b & 292.9607084 & -26.56919048 & 1.8347& 1.8437$^b$    \\
M1931-2c & 292.9496296 & -26.57059507 & 1.8347& ---           \\
\hline
M1931-3a & 292.9519796 & -26.58280409 & ---   & ---           \\
M1931-3b & 292.9554377 & -26.58366339 & 2.7069& ---           \\
\hline
M1931-4a & 292.9655056 & -26.58196223 & 4.0005& ---           \\
M1931-4b & 292.9529960 & -26.58357089 & 4.0005& 4.0000$^b$           \\
M1931-4c & 292.9539350 & -26.58380120 & 4.0005& 4.0000$^b$           \\
\hline
M1931-5a & 292.9531783 & -26.57133733 & 4.7448& ---           \\
M1931-5b & 292.9504751 & -26.57329848 & 4.7448& ---           \\
\hline
M1931-6a & 292.9583436 & -26.57407693 & 5.0785& ---           \\
M1931-6b & 292.9621377 & -26.57697810 & 5.0785& ---           \\
M1931-6c & 292.9607682 & -26.57863576 & 5.0785& ---           \\
\hline
M1931-7a & 292.9527961 & -26.57280898 & 5.3390& ---           \\
M1931-7b & 292.9511744 & -26.57444090 & 5.3390& ---           \\
\hline
\multicolumn{5}{c}{MACS~J0329}  \\
\hline
M0329-1a & 52.42812816 & -2.192448407 & ---   & ---           \\
M0329-1b & 52.42082751 & -2.196023678 & 1.3130& ---           \\
\hline
M0329-2a & 52.42570883 & -2.190441318 & 2.1446& ---           \\
M0329-2b & 52.42109085 & -2.191310894 & 2.1446& 2.1422$^b$    \\
M0329-2c & 52.42640130 & -2.198443743 & 2.1446& 2.1445$^b$    \\
M0329-2d & 52.41506800 & -2.200182875 & ---   & 2.1445$^b$    \\
\hline
M0329-3a & 52.42183629 & -2.187533753 & ---   & 2.7836$^b$    \\
M0329-3b & 52.41740283 & -2.190705376 & ---   & 2.7836$^b$    \\
M0329-3c & 52.41266865 & -2.197051094 & ---   & 2.7836$^b$    \\
\hline
M0329-4a & 52.42147510 & -2.194733492 & 2.9186& ---           \\
M0329-4b & 52.43038430 & -2.195987676 & 2.9186& ---           \\
\hline
M0329-5a & 52.42460593 & -2.196926235 & 3.8528& ---           \\
M0329-5b & 52.42546341 & -2.197447457 & 3.8528& ---           \\
\hline
M0329-6a & 52.42452717 & -2.196436725 & 4.5734& ---           \\
M0329-6b & 52.42614953 & -2.196909726 & 4.5734& ---           \\
\hline
M0329-7a & 52.42766923 & -2.207178871 & 5.6587& ---           \\
M0329-7b & 52.42720225 & -2.207653320 & 5.6587& ---           \\
\hline
M0329-8a & 52.43119727 & -2.191342771 & 6.0204& ---           \\
M0329-8b & 52.42021547 & -2.194539532 & 6.0204& ---           \\
\hline
M0329-9a & 52.42984385 & -2.188114905 & ---   & 6.17$^c$      \\
M0329-9b & 52.41737364 & -2.196010365 & ---   & 6.17$^c$      \\
M0329-9c & 52.41693078 & -2.197684600 & 6.1   & 6.17$^c$      \\
M0329-9d & 52.42182284 & -2.201280003 & ---   & 6.17$^c$      \\
\hline
\multicolumn{5}{c}{MACS~J2129}  \\
\hline
M2129-1a & 322.3548526 & -7.690679359 & 1.0480 & 1.047$^{b}$,1.040$^{h}$  \\
M2129-1b & 322.3547733 & -7.691599222 & 1.0480 & --- & \\
M2129-1c & 322.3554685 & -7.693276023 & 1.0480 & --- & \\
\hline
M2129-2a & 322.3567236 & -7.685551318 & 1.3568 & --- & \\
M2129-2b & 322.3562323 & -7.691723071 & 1.3568 & --- & \\
M2129-2c & 322.3572259 & -7.694321533 & 1.3568 & --- & \\
\hline
M2129-3a & 322.3539358 & -7.687586260 & ---    & --- & \\
M2129-3b & 322.3533247 & -7.691187481 & 1.3585 & --- & \\
M2129-3c & 322.3544685 & -7.694413516 & ---    & --- & \\
\hline
M2129-4a & 322.3579635 & -7.685881890 & 1.3634 & 1.365$^{b}$,1.372$^{h}$ & \\
M2129-4b & 322.3596767 & -7.690845569 & ---    & --- & \\
M2129-4c & 322.3592595 & -7.690947126 & 1.3634 & 1.372$^{h}$ & \\
M2129-4d & 322.3571199 & -7.691084524 & 1.3634 & 1.364$^{b}$,1.373$^{h}$ & \\
M2129-4e & 322.3576424 & -7.691140237 & 1.3634 & 1.370$^{h}$ & \\
M2129-4f & 322.3586238 & -7.694884030 & 1.3634 & 1.367$^{h}$ & \\
\hline
M2129-5a & 322.3613111 & -7.685896466 & ---    & --- & \\
M2129-5b & 322.3624812 & -7.691420301 & ---    & --- & \\
M2129-5c & 322.3625937 & -7.693602639 & 1.4519 & --- & \\
\hline
M2129-6a & 322.3502349 & -7.688817805 & 2.2427 & 2.236$^{b}$,2.240$^{h}$ & \\
M2129-6b & 322.3501313 & -7.689457921 & 2.2427 & 2.236$^{b}$,2.240$^{h}$ & \\
M2129-6c & 322.3509908 & -7.695782768 & ---    & 2.239$^{b}$ & \\
\hline
M2129-7a & 322.3616733 & -7.683622879 & 3.1059 & --- & \\
M2129-7b & 322.3648793 & -7.690103438 & 3.1059 & --- & \\
M2129-7c & 322.3633399 & -7.697055909 & 3.1059 & --- & \\
\hline
M2129-8a & 322.3665058 & -7.686902595 & 3.1100 & 2.237$^{b}$ & \\
M2129-8b & 322.3669527 & -7.688234891 & ---    & --- & \\
M2129-8c & 322.3666659 & -7.695243802 & ---    & --- & \\
\hline
M2129-9a & 322.3545492 & -7.685185222 & 3.9006 & --- & \\
M2129-9b & 322.3527854 & -7.688410699 & 3.9006 & --- & \\
M2129-9c & 322.3573628 & -7.699774072 & ---    & --- & \\
\hline
M2129-10a & 322.3586055 & -7.684904850 & 4.4086 & --- & \\
M2129-10b & 322.3616458 & -7.688096597 & 4.4086 & --- & \\
M2129-10c & 322.3541953 & -7.688761478 & 4.4086 & --- & \\
M2129-10d & 322.3569890 & -7.689230874 & 4.4086 & 4.411$^{b}$ & \\
M2129-10e & 322.3603558 & -7.700940300 & ---    & --- & \\
\hline
M2129-11a & 322.3539317 & -7.681649317 & ---    & 6.846$^{d}$ & \\
M2129-11b & 322.3509348 & -7.693330050 & ---    & 6.846$^{d}$ & \\
M2129-11c & 322.3532384 & -7.697444677 & ---    & 6.846$^{d}$ & \\

\hline
\multicolumn{5}{c}{MACS~J1115}  \\
\hline
M1115-1a  & 168.966823 & 1.506877 & ---   & ---           \\
M1115-1b  & 168.961854 & 1.505248 & ---   & ---           \\
M1115-1c  & 168.958464 & 1.500308 & ---   & $2.5520^{b}$  \\
\hline
M1115-2a  & 168.964480 & 1.507233 & 2.9175& ---           \\
M1115-2b  & 168.962530 & 1.506565 & 2.9175& ---           \\
M1115-2c  & 168.957635 & 1.500751 & 2.9175& ---           \\
\hline
M1115-3a  & 168.973224 & 1.502403 & ---   & $[0.367-3.411]^{e}$ \\
M1115-3b  & 168.966512 & 1.494089 & ---   & ''         \\
M1115-3c  & 168.963670 & 1.493729 & ---   & ''         \\
\hline
\multicolumn{5}{c}{MACS~J0429}  \\
\hline
M0429-1a & 67.398809 & -2.881399 & 2.9286& ---           \\
M0429-1b & 67.405683 & -2.883986 & 2.9286& ---           \\
M0429-1c & 67.394847 & -2.885733 & 2.9286& ---           \\
M0429-1d & 67.400178 & -2.888594 & 2.9286& ---           \\
\hline
M0429-2a & 67.400749 & -2.885197 & 3.8665& ---           \\
M0429-2b & 67.392409 & -2.886273 & 3.8665& ---           \\
M0429-2c & 67.402549 & -2.886853 & 3.8665& ---           \\
\hline
M0429-3a & 67.399814 & -2.883126 & --- & $[1.637-1.785]^{e}$           \\
M0429-3b & 67.404684 & -2.885985 & --- & ''           \\
M0429-3c & 67.395423 & -2.887432 & --- & ''           \\
M0429-3d & 67.401022 & -2.888857 & --- & ''           \\
\hline
\multicolumn{5}{c}{RX~J1347}  \\
\hline
R1347-1a & 206.871201 & -11.760319 & ---   & $1.7638^{b,f}$        \\
R1347-1b & 206.872629 & -11.761609 & ---   & $1.7638^{b,f}$        \\
R1347-1c & 206.882279 & -11.764403 & ---   & $1.7638^{b,f}$        \\
\hline
R1347-2a & 206.885022 & -11.744131 & 3.6804& ---           \\
R1347-2b & 206.891017 & -11.753639 & 3.6804& ---           \\
\hline
R1347-3a & 206.888217 & -11.744547 & 3.6814& ---           \\
R1347-3b & 206.889938 & -11.746559 & 3.6814& ---           \\
\hline
R1347-4a & 206.888299 & -11.752540 & 4.0790& $4.083^{g}$         \\
R1347-4b & 206.880519 & -11.754408 & 4.0790& ---           \\
\hline
R1347-5a & 206.8695013 & -11.74730166 & --- & $[1.965-2.265]^{e}$   \\
R1347-5b & 206.8850766 & -11.74837918 & --- & ''            \\
R1347-5c & 206.8786695 & -11.75333785 & --- & ''            \\
R1347-5d & 206.8874583 & -11.75747984 & --- & ''            \\
R1347-5e & 206.8775624 & -11.75934602 & --- & ''            \\
\hline
R1347-6a & 206.882883 & -11.741196 & ---   & $[0.151-3.860]^{e}$   \\
R1347-6b & 206.884048 & -11.741763 & ---   & ''           \\
\hline
R1347-7a & 206.878328 & -11.749177 & ---   & $[2.321-2.636]^{e}$           \\
R1347-7b & 206.878227 & -11.749630 & ---   & ''   \\
\hline
R1347-8a & 206.875988 & -11.741197 & ---   & $[0.669-6.228]^{e}$   \\
R1347-8b & 206.882313 & -11.742182 & ---   & ''           \\
\hline
\multicolumn{5}{c}{MACS~1311}  \\
\hline
M1311-1a & 197.755540 & -3.176000 & 2.1867& 2.1893$^{b}$  \\
M1311-1b & 197.756897 & -3.177289 & ---   & 2.1850$^{b}$  \\
M1311-1c & 197.763604 & -3.179186 & 2.1867& 2.1883$^{b}$  \\
\hline
M1311-2a & 197.765361 & -3.177361 & ---   & $[5.755-6.594]^{e}$          \\
M1311-2b & 197.755199 & -3.179719 & ---   & ''           \\
\hline
M1311-3a & 197.758730 & -3.174978 & ---   & $[1.219-2.542]^{e}$           \\
M1311-3b & 197.761644 & -3.178933 & ---   & ''           \\
M1311-3c & 197.759620 & -3.180957 & ---   & ''           \\

\hline
\end{longtable}
\tablefoot{
\tablefoottext{a}{Spectroscopic redshifts from \citet{2013ApJ...772..141B}.}
\tablefoottext{b}{Spectroscopic redshifts from CLASH-VLT (Rosati et al. in prep.).}
\tablefoottext{c}{Spectroscopic redshift from Vanzella et al. in prep.}
\tablefoottext{d}{Spectroscopic confirmation from \citet{2016ApJ...823L..14H}.}
\tablefoottext{e}{Photometric redshift limits from \citet{2017MNRAS.470...95M}.}
\tablefoottext{f}{Independent spectroscopic confirmations from \citet{2002ApJ...577..133R,2008ApJ...681..187B, 2008A&A...481...65H}.}
\tablefoottext{g}{Spectroscopic redshift from \citet{2002ApJ...573..524C}.}
\tablefoottext{h}{Spectroscopic redshift obtained from the public GLASS catalogues \citet{2015ApJ...812..114T}.}}
}

\begin{figure*}

   \includegraphics[width = 0.666\columnwidth]{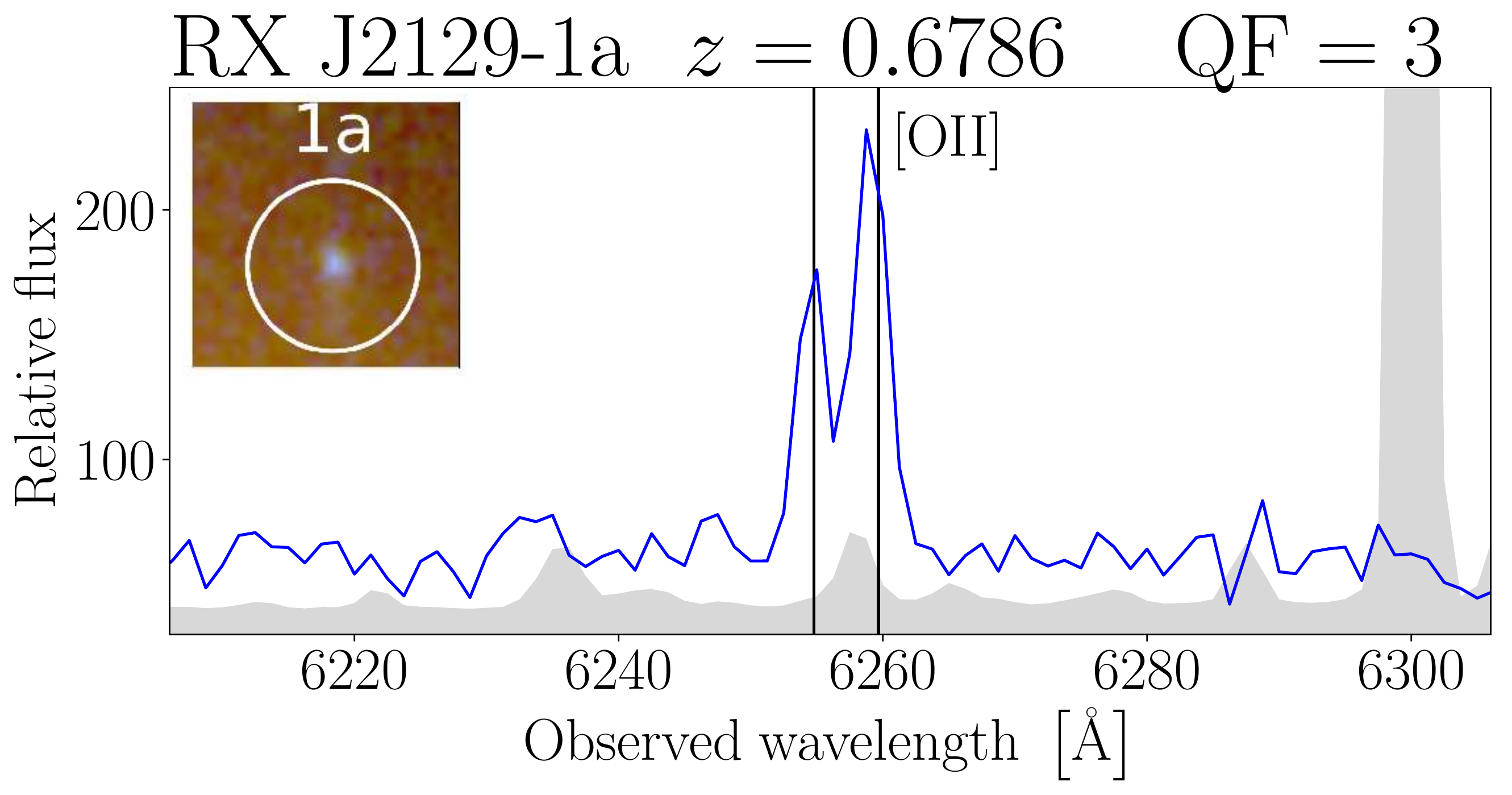}
   \includegraphics[width = 0.666\columnwidth]{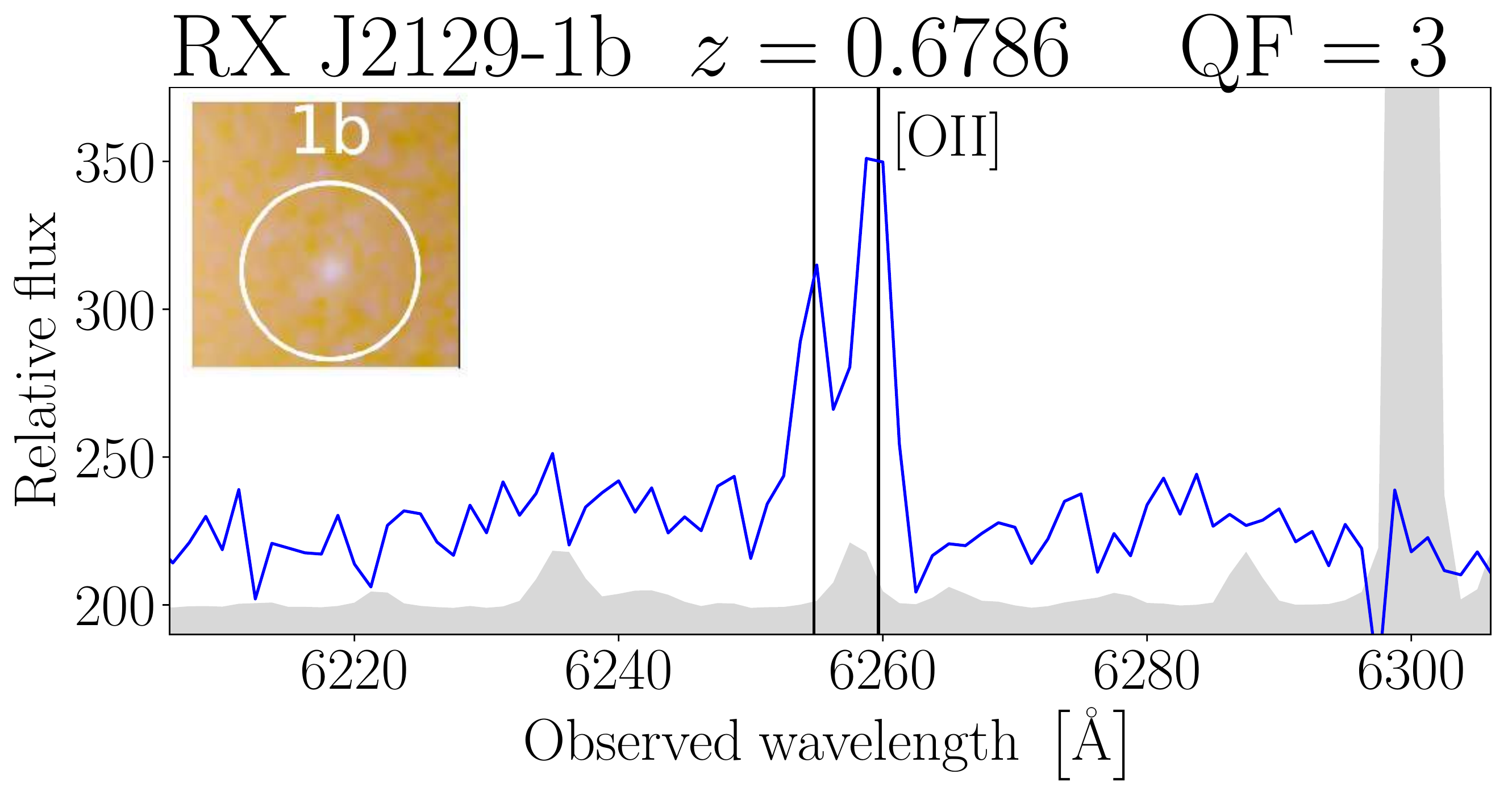}
   \includegraphics[width = 0.666\columnwidth]{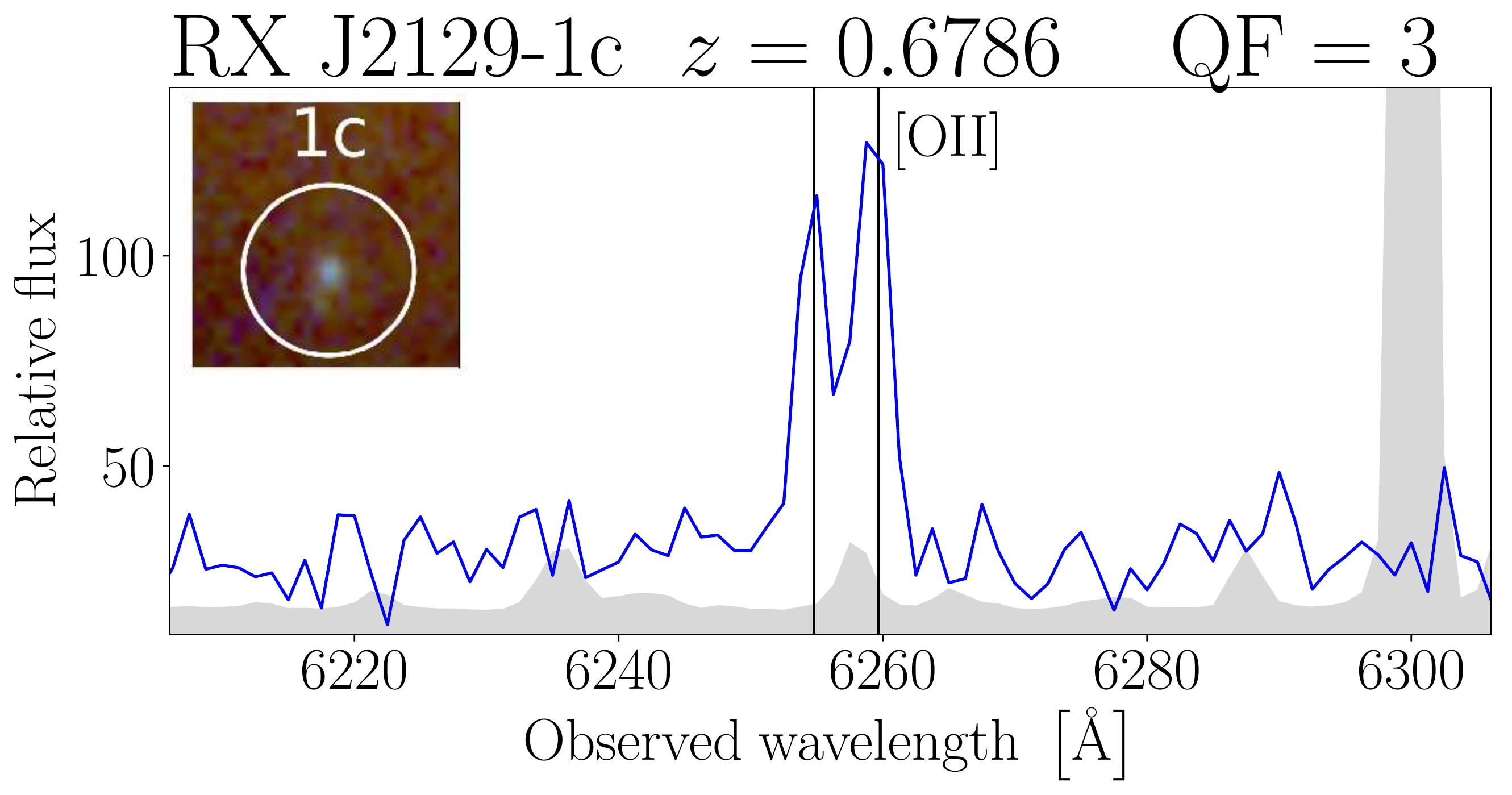}
   \includegraphics[width = 0.666\columnwidth]{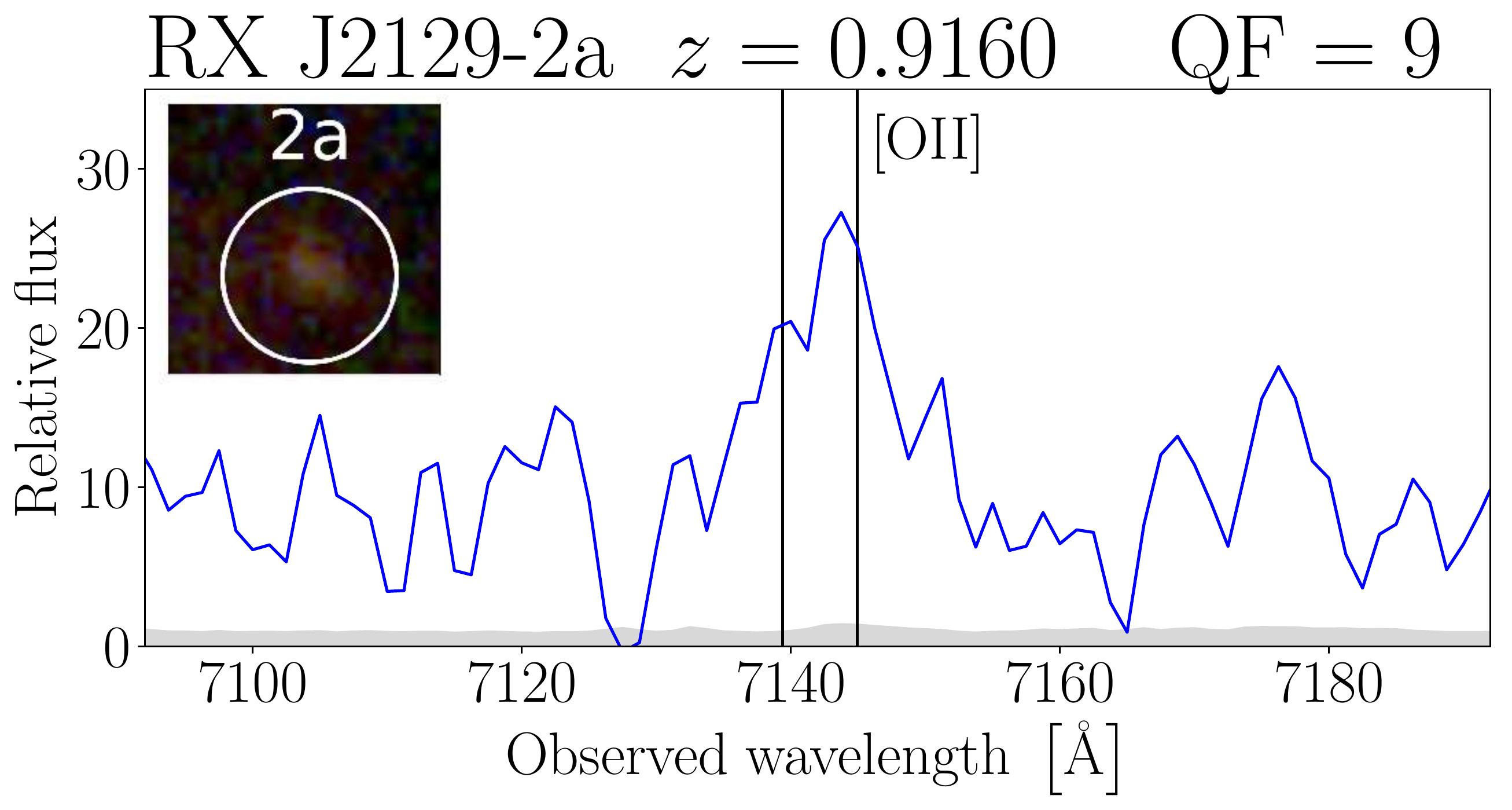}
   \includegraphics[width = 0.666\columnwidth]{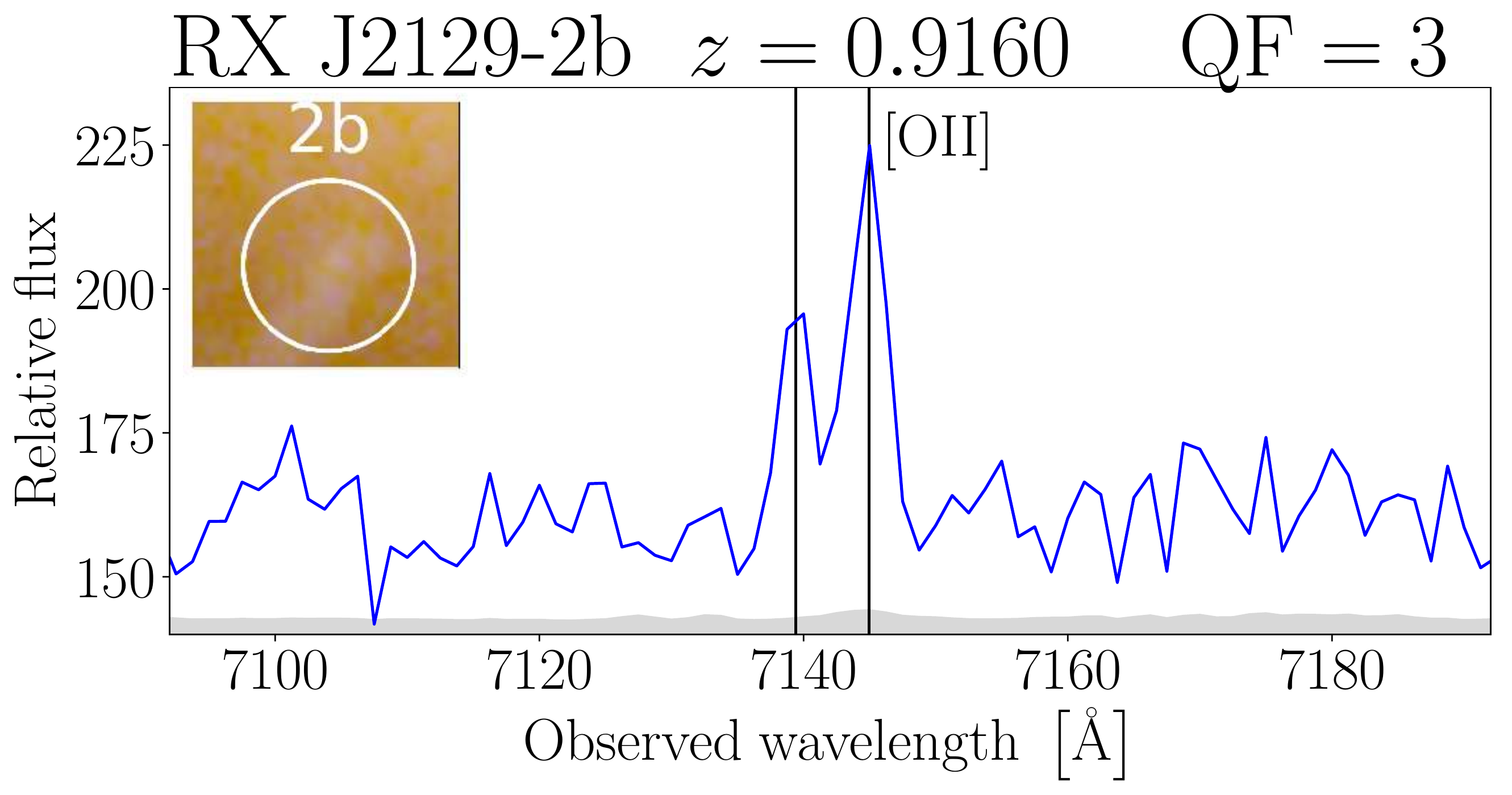}
   \includegraphics[width = 0.666\columnwidth]{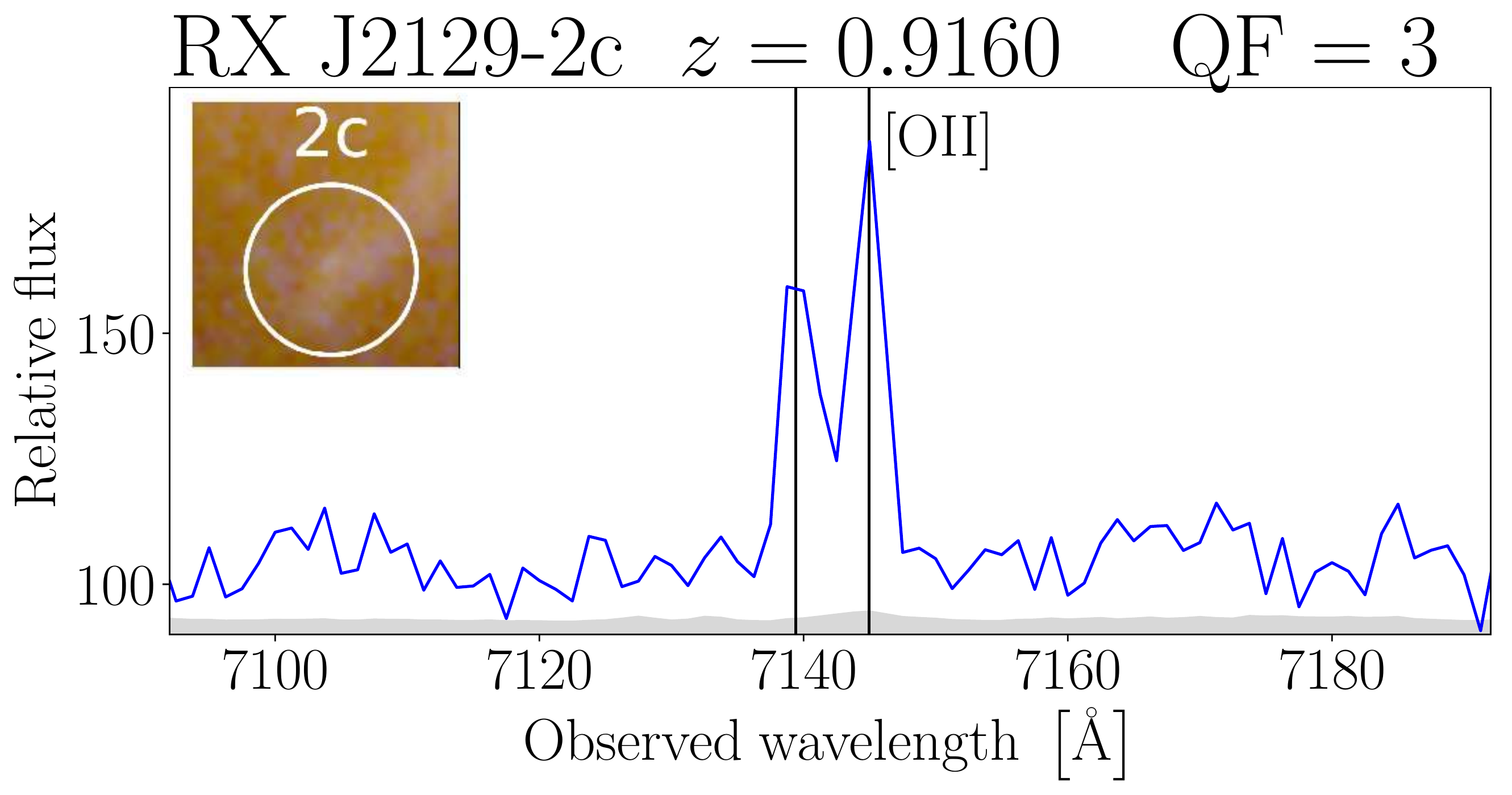}
   \includegraphics[width = 0.666\columnwidth]{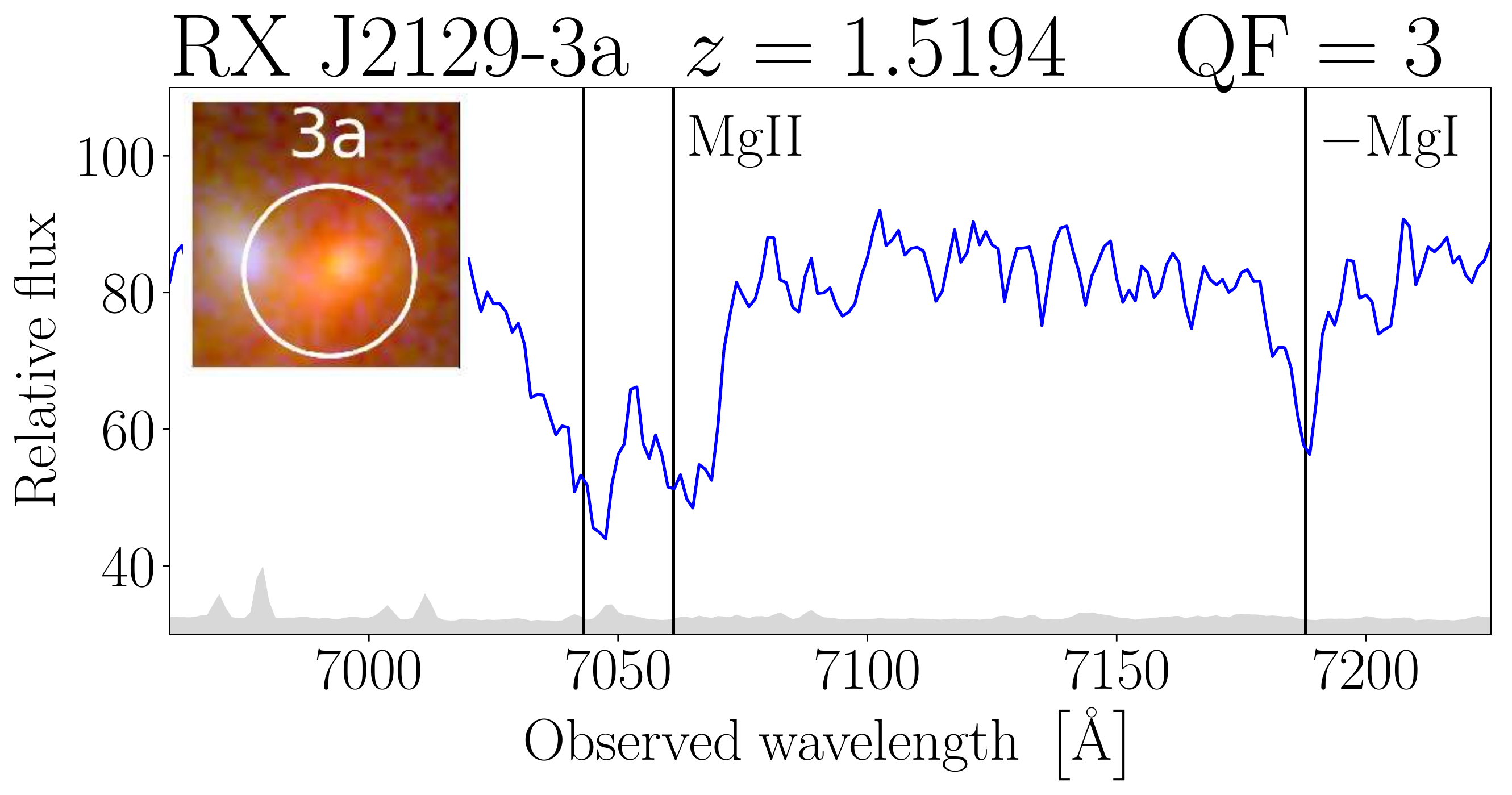}
   \includegraphics[width = 0.666\columnwidth]{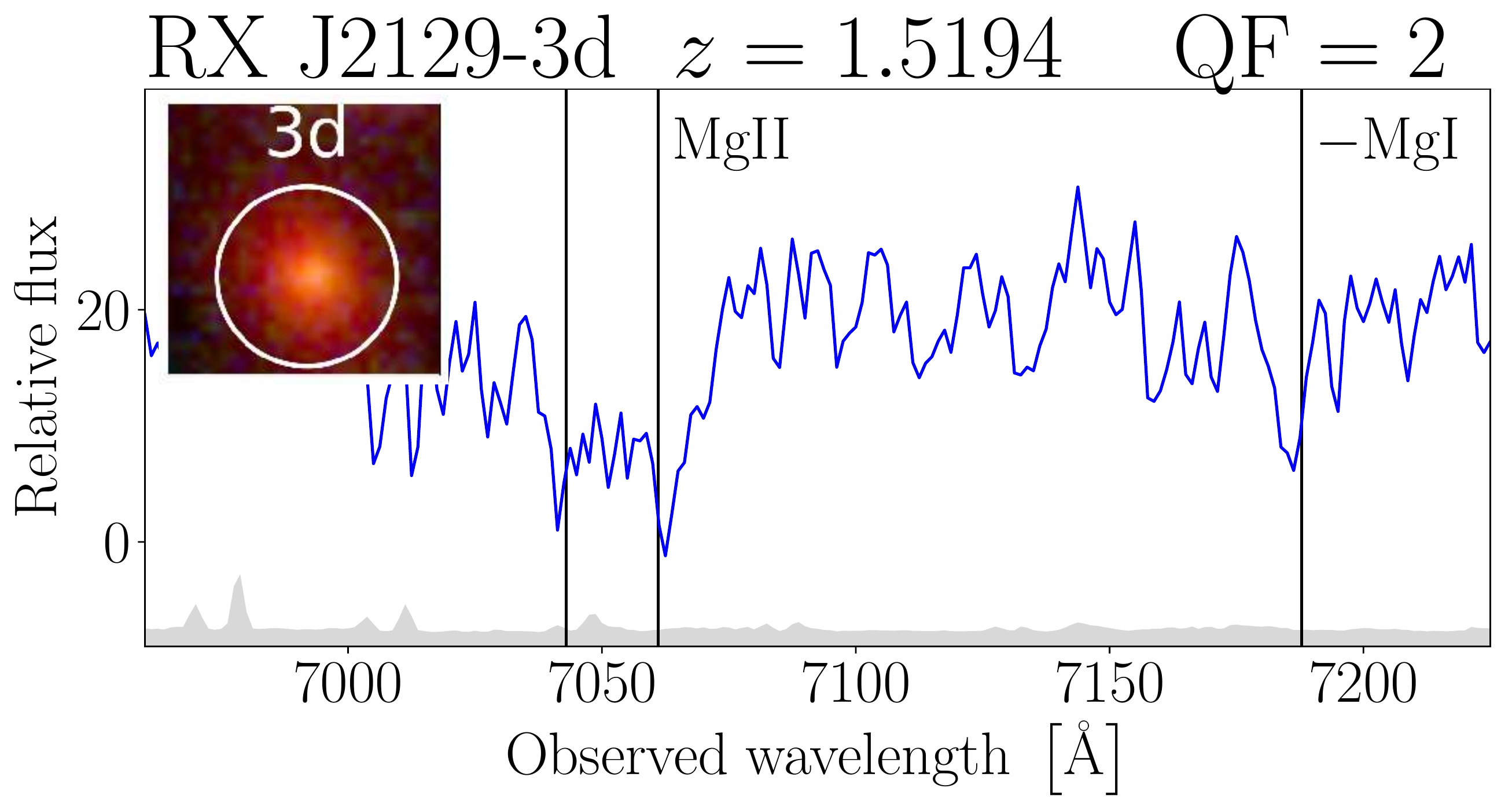}
   \includegraphics[width = 0.666\columnwidth]{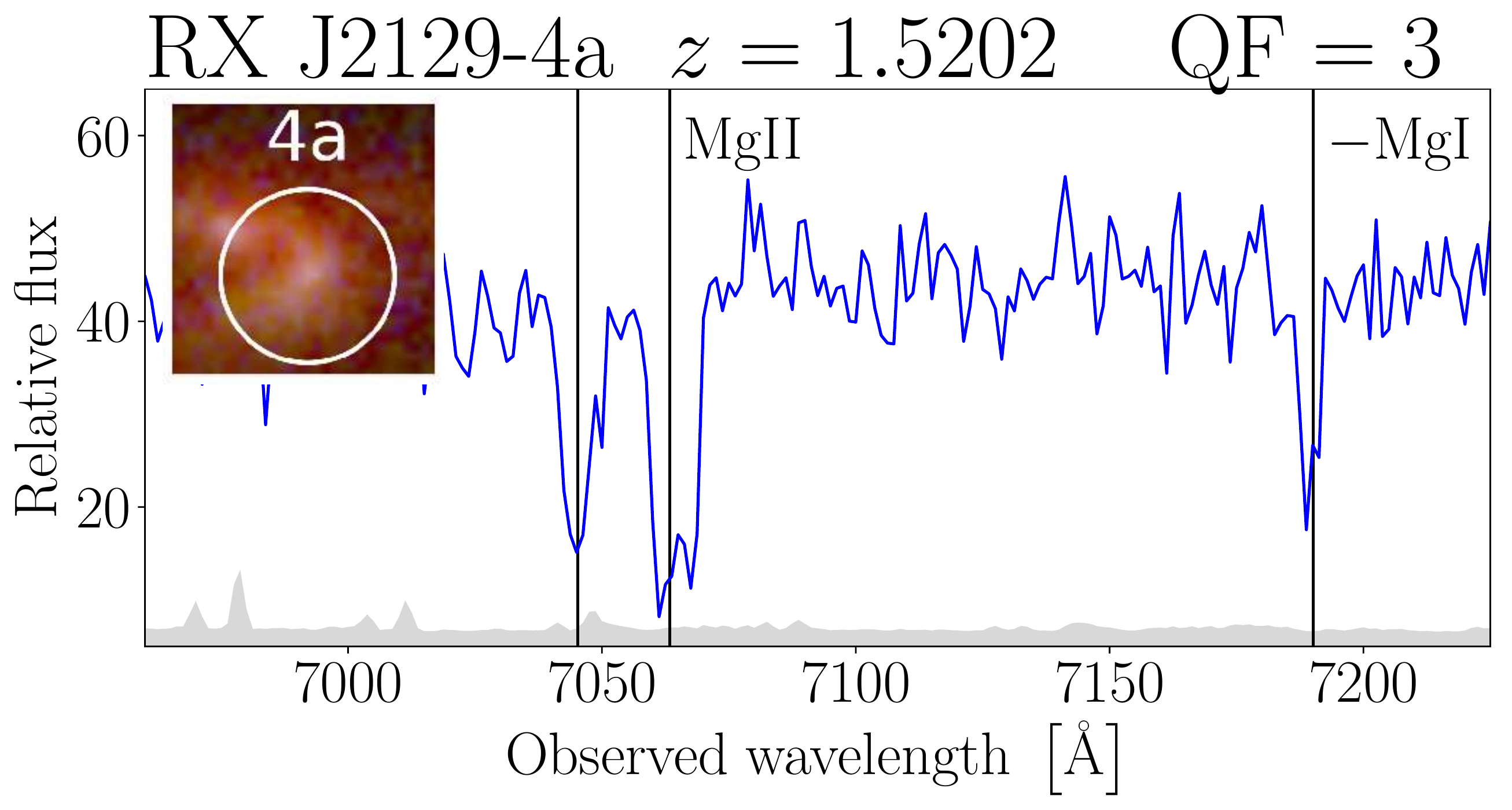}
   \includegraphics[width = 0.666\columnwidth]{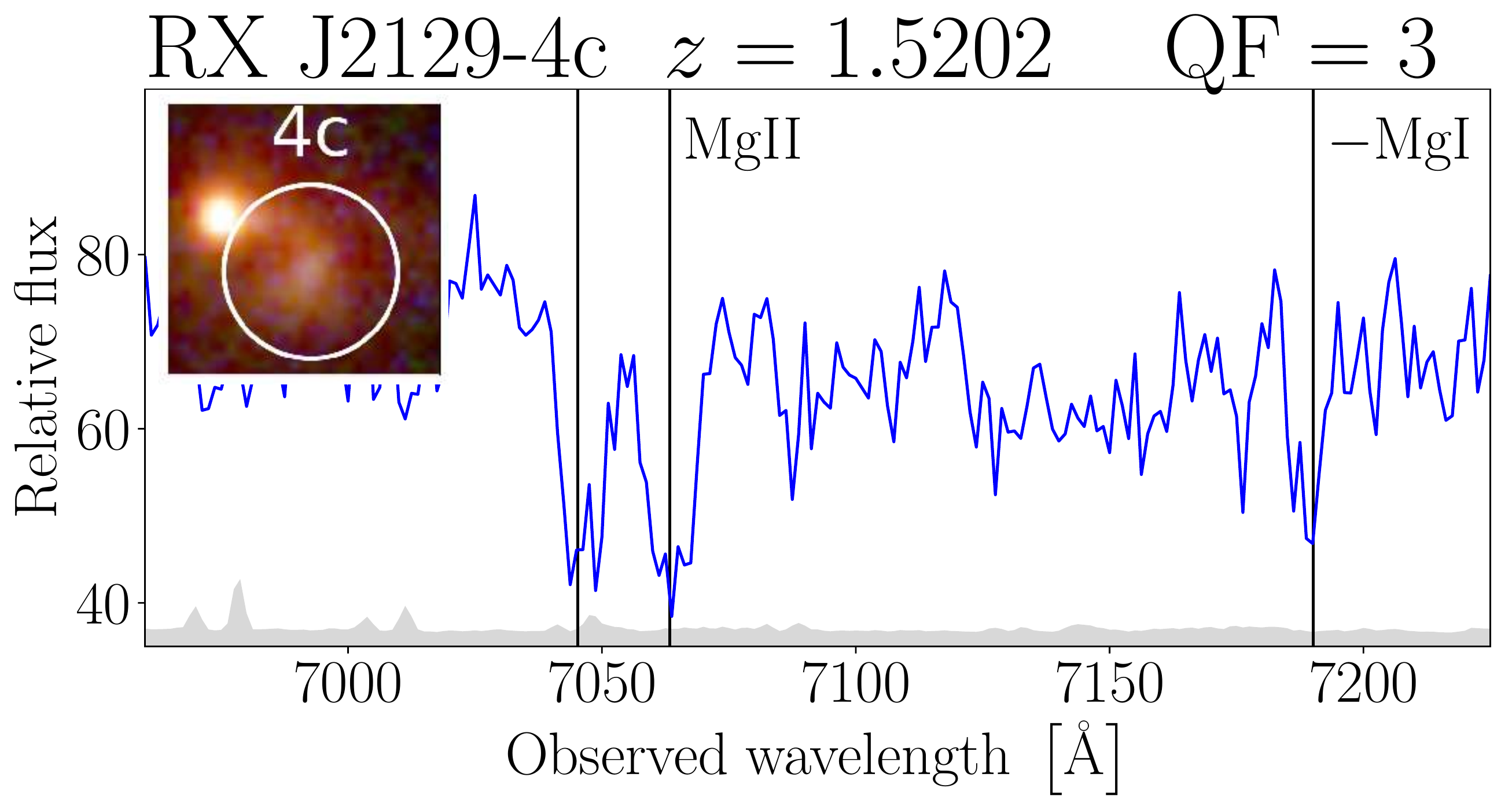}
   \includegraphics[width = 0.666\columnwidth]{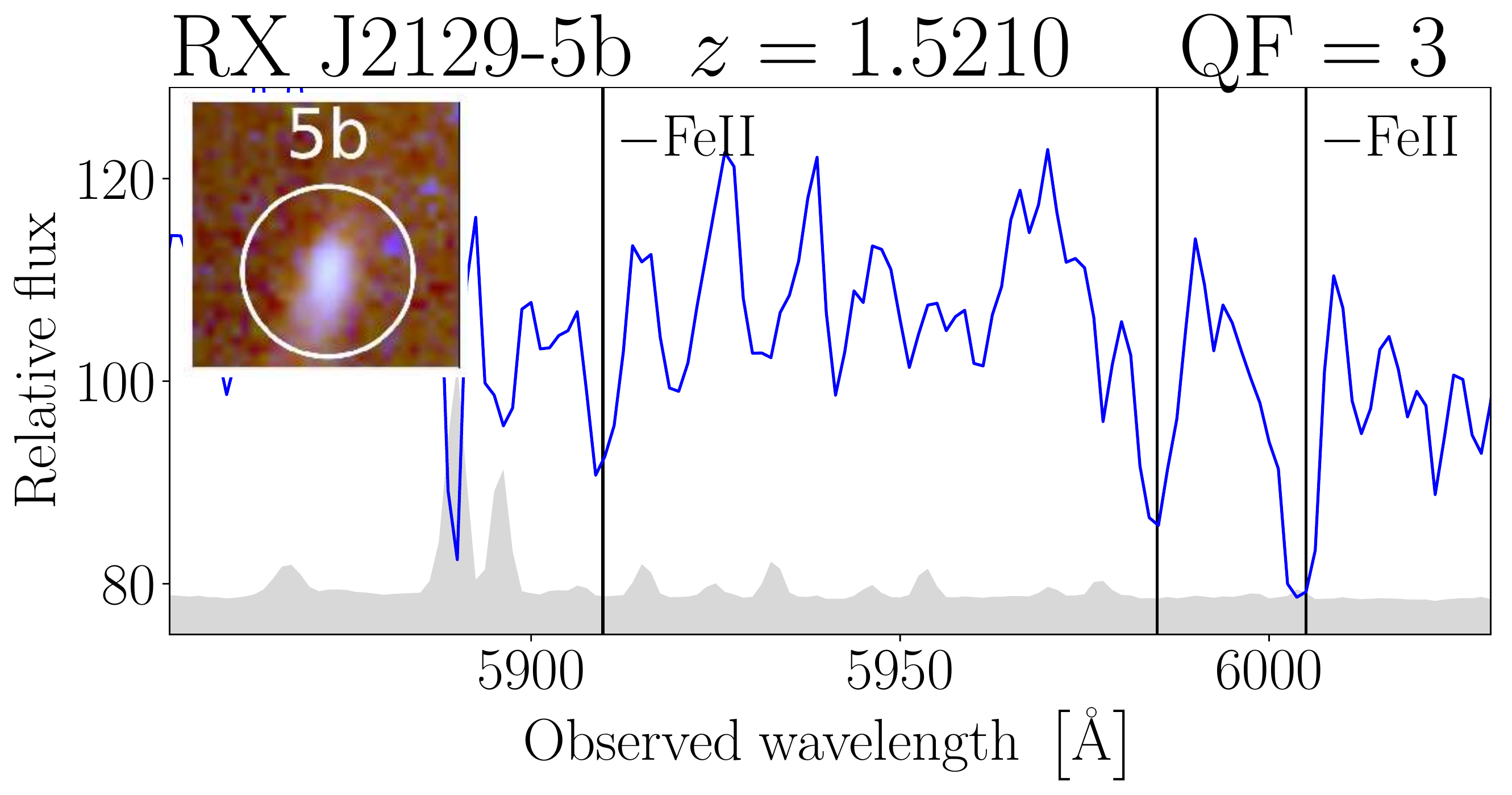}
   \includegraphics[width = 0.666\columnwidth]{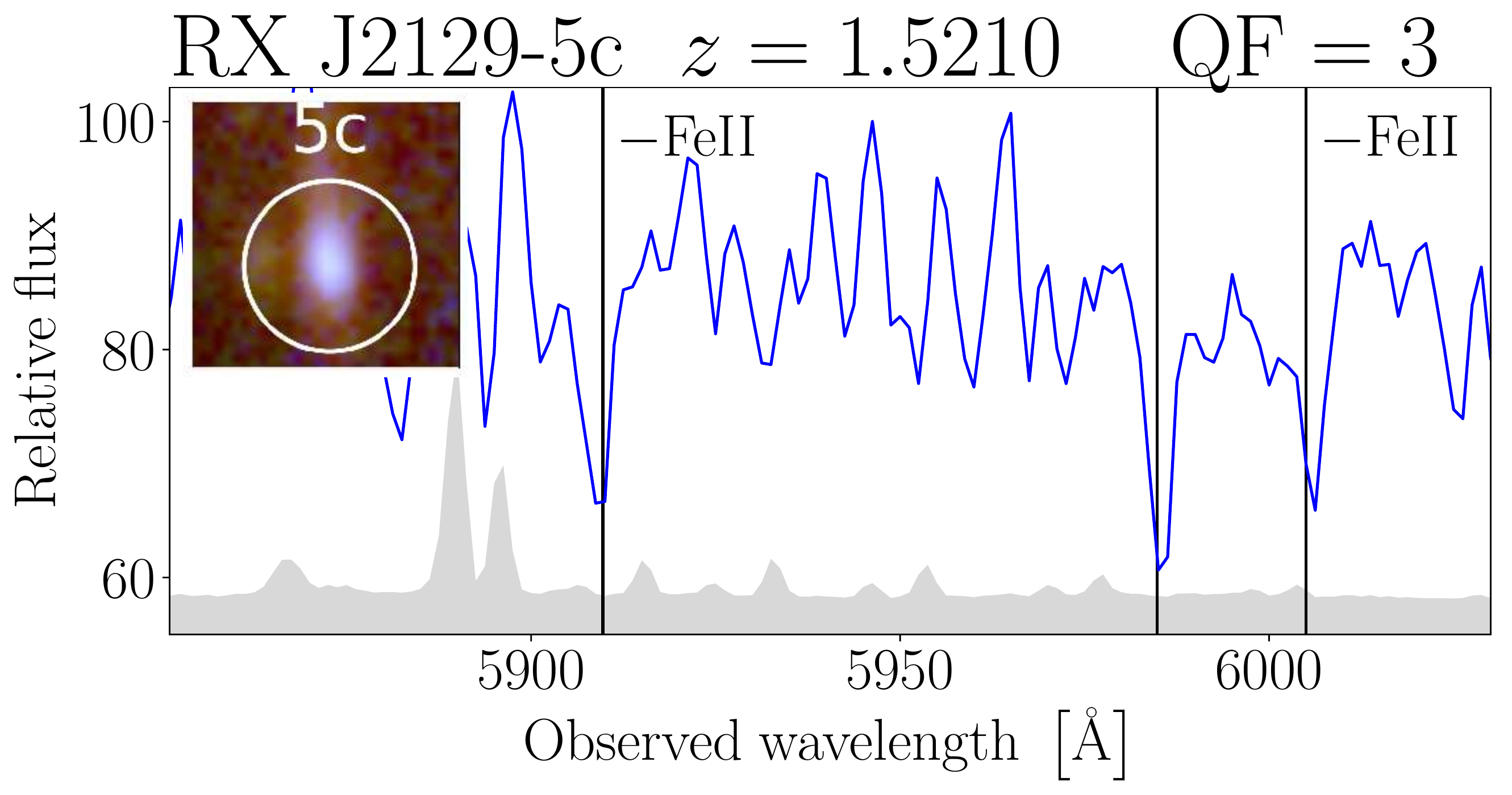}
   \includegraphics[width = 0.666\columnwidth]{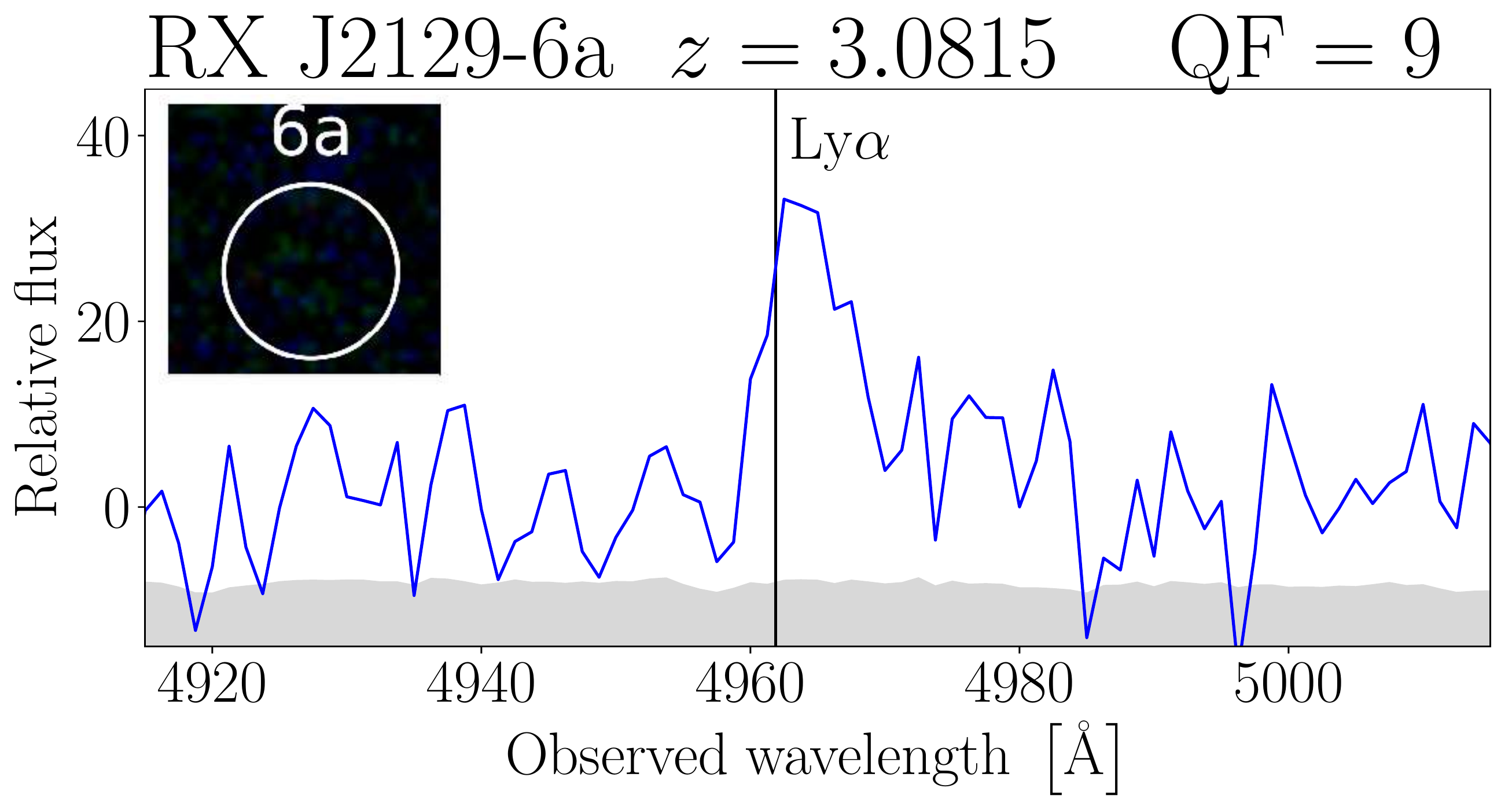}
   \includegraphics[width = 0.666\columnwidth]{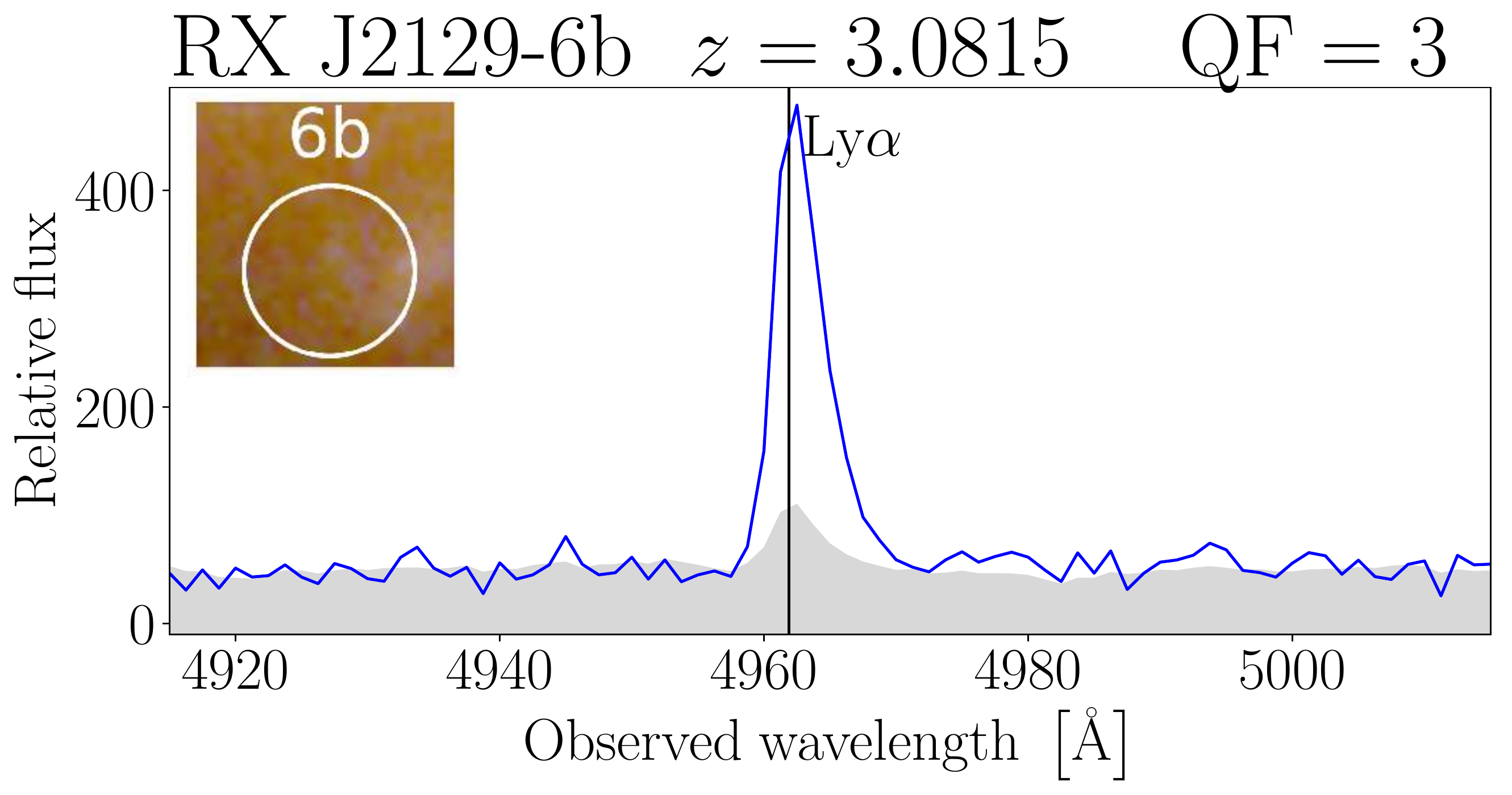}
   \includegraphics[width = 0.666\columnwidth]{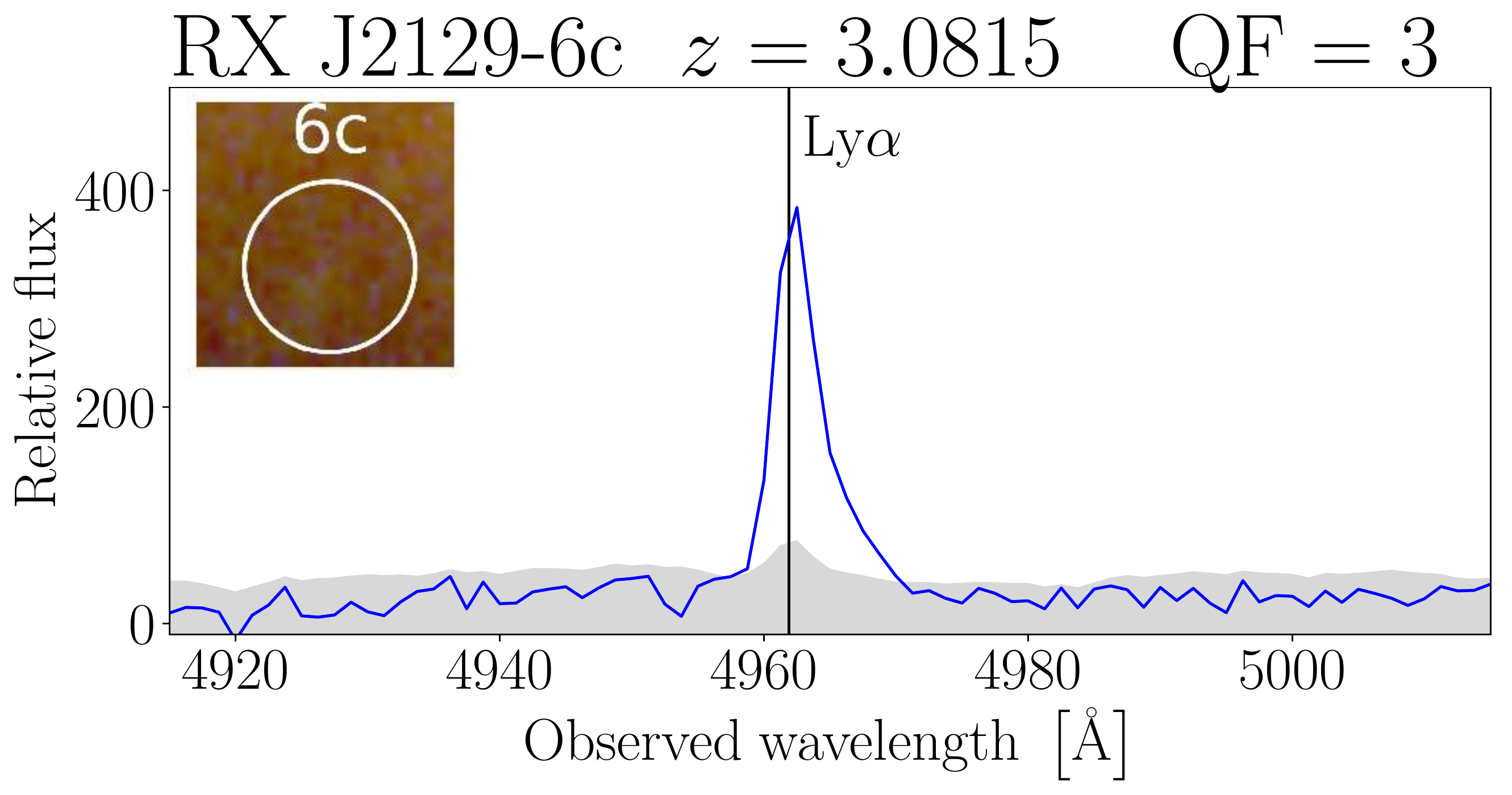}
   \includegraphics[width = 0.666\columnwidth]{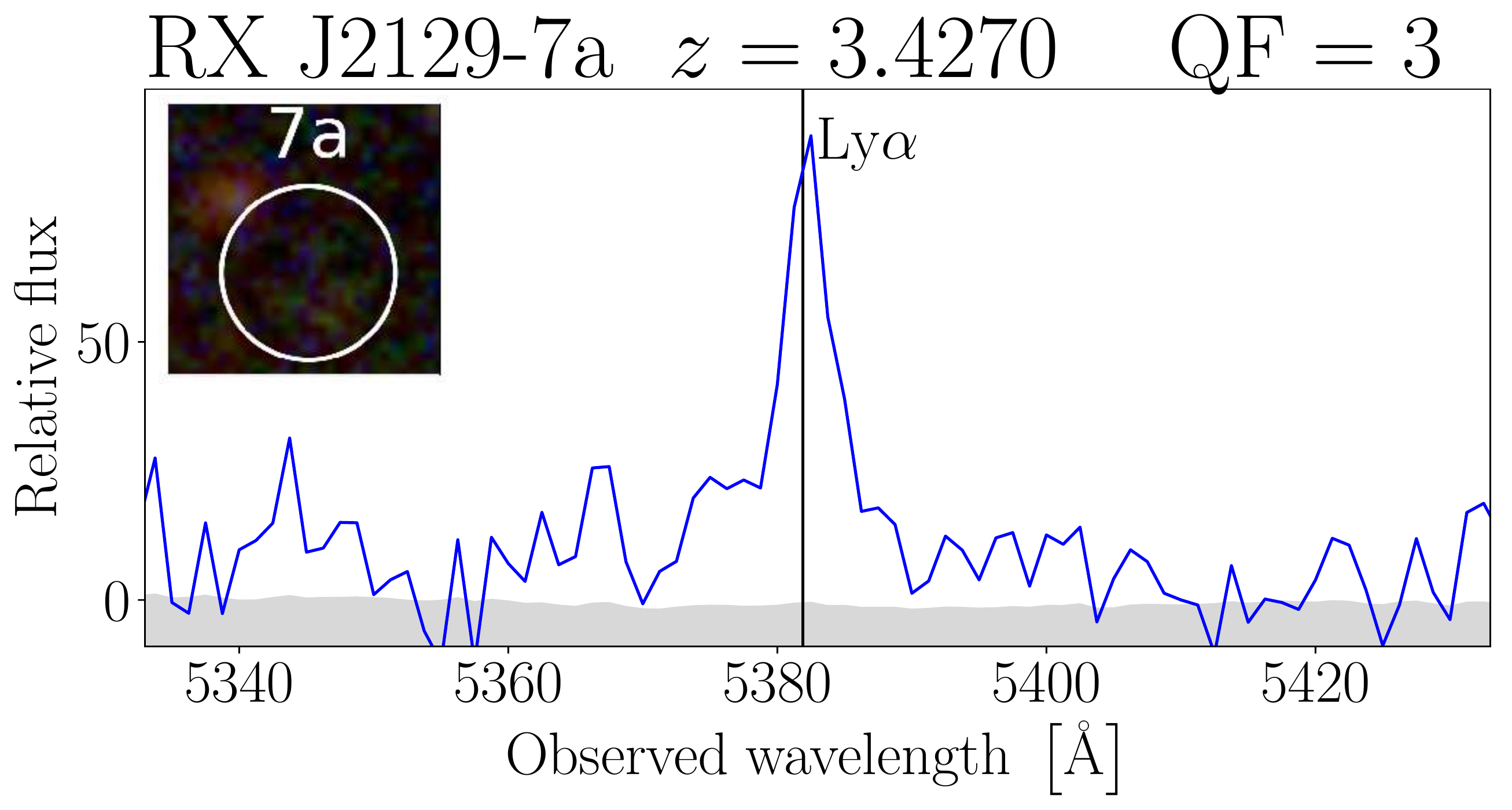}
   \includegraphics[width = 0.666\columnwidth]{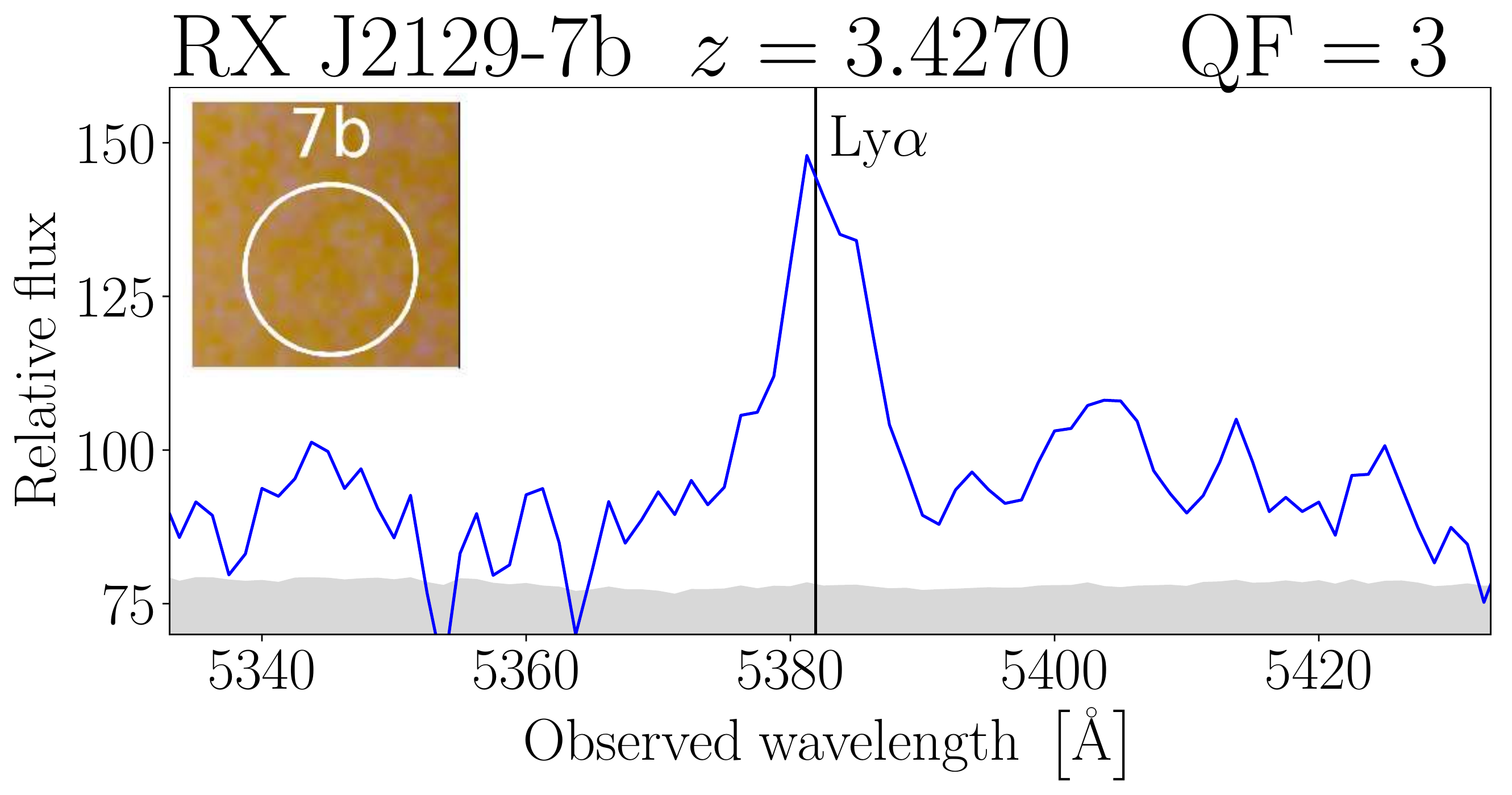}
   \includegraphics[width = 0.666\columnwidth]{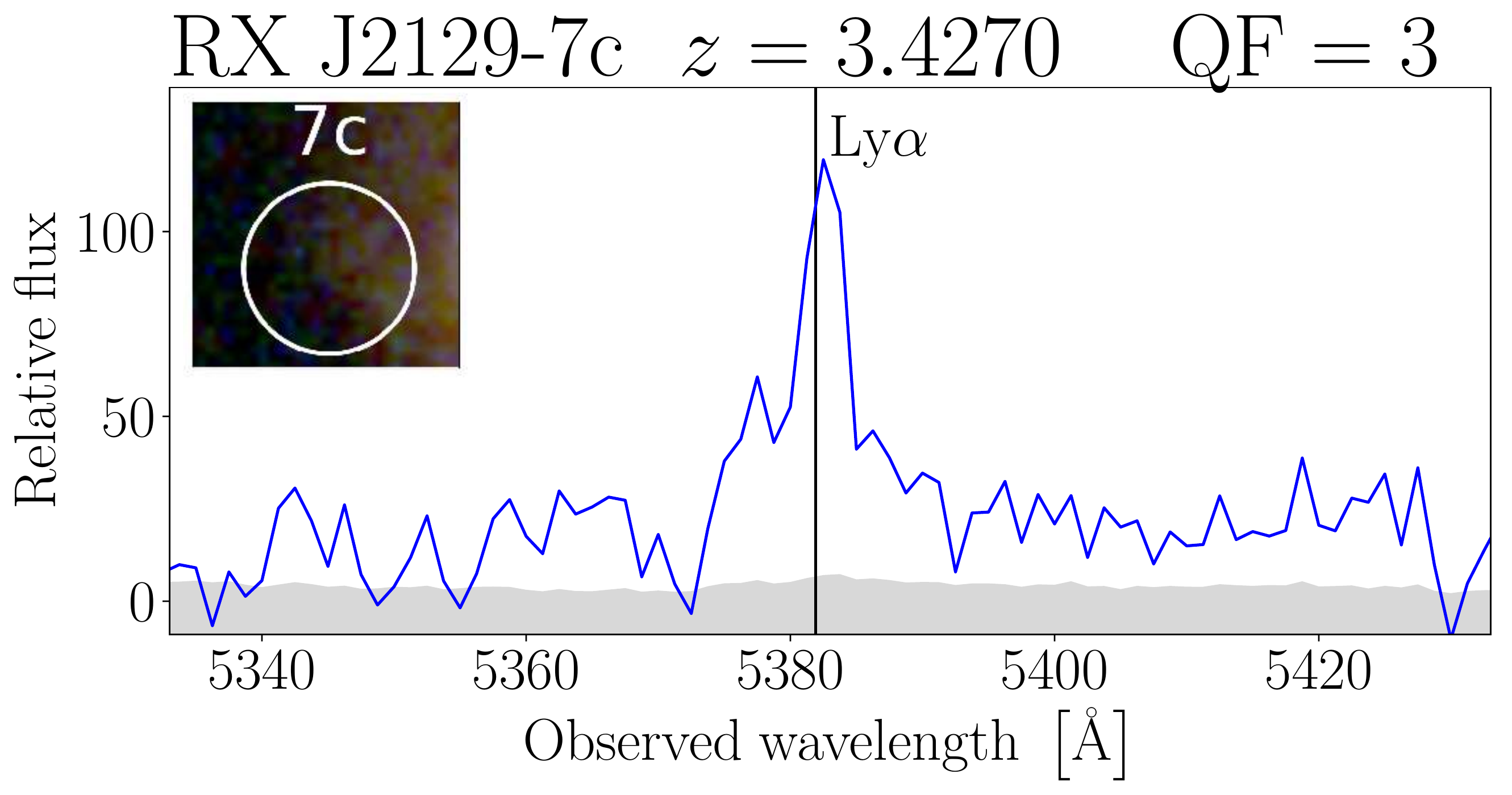}
   \includegraphics[width = 0.666\columnwidth]{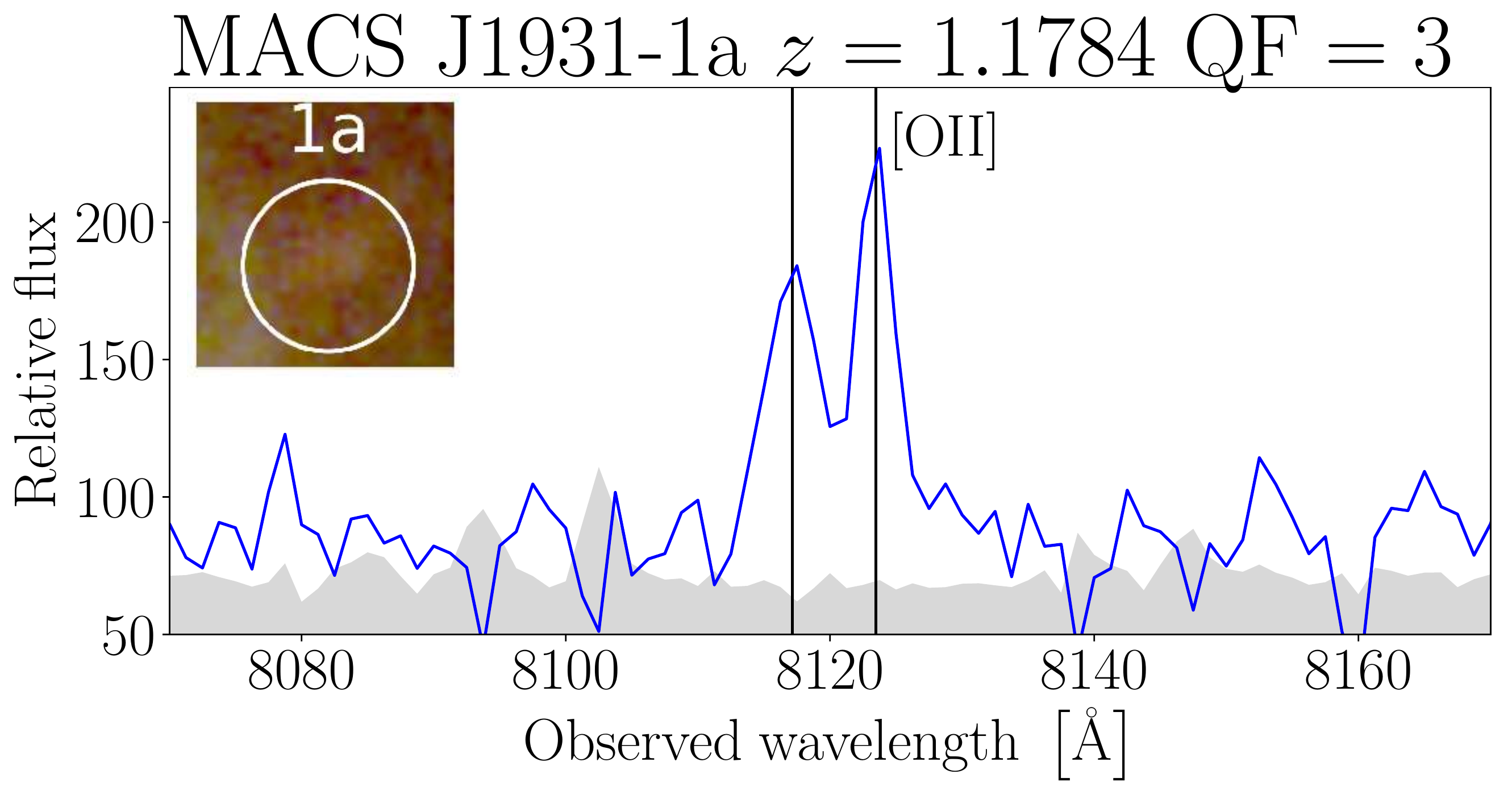}
   \includegraphics[width = 0.666\columnwidth]{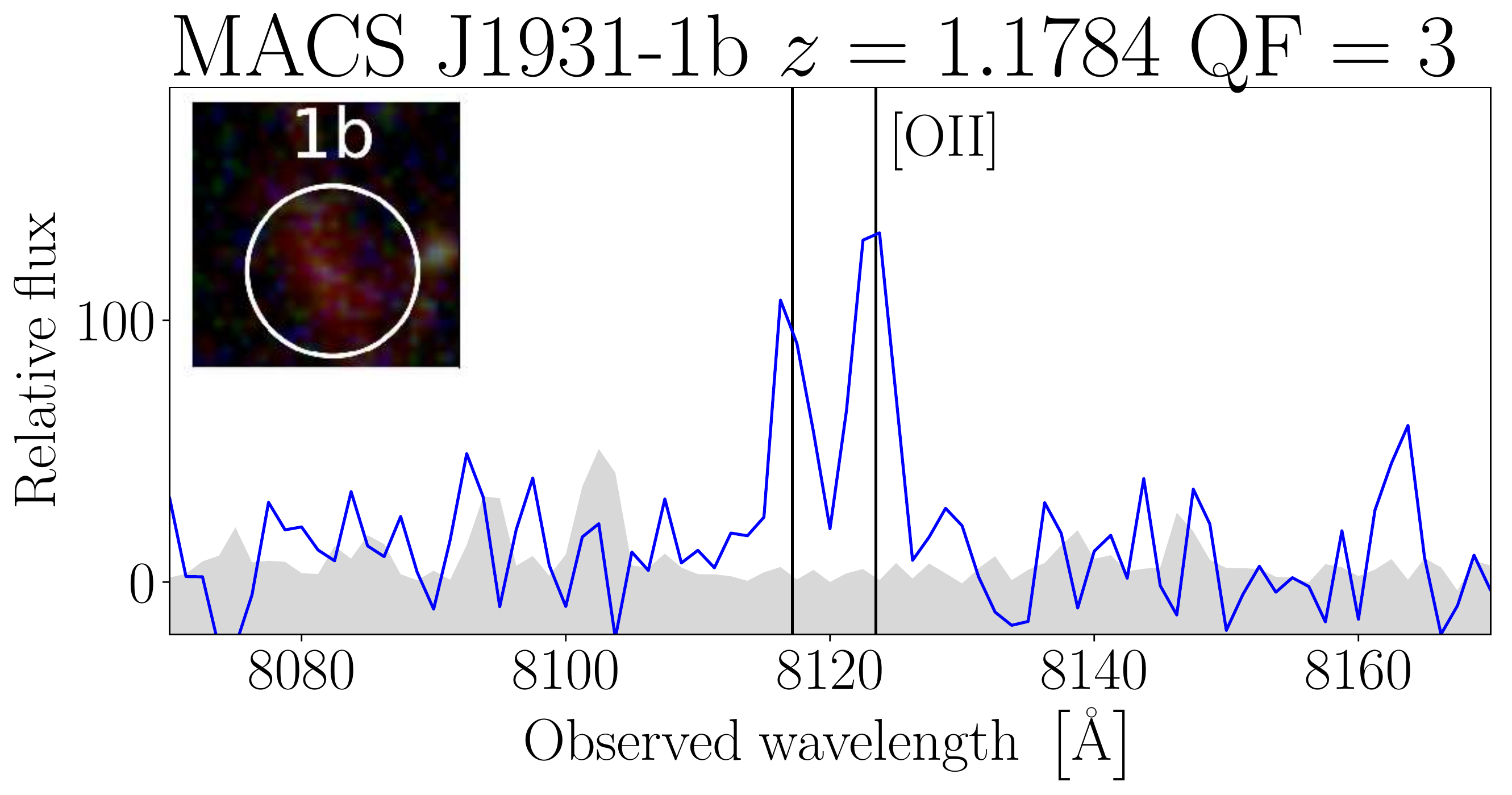}
   \includegraphics[width = 0.666\columnwidth]{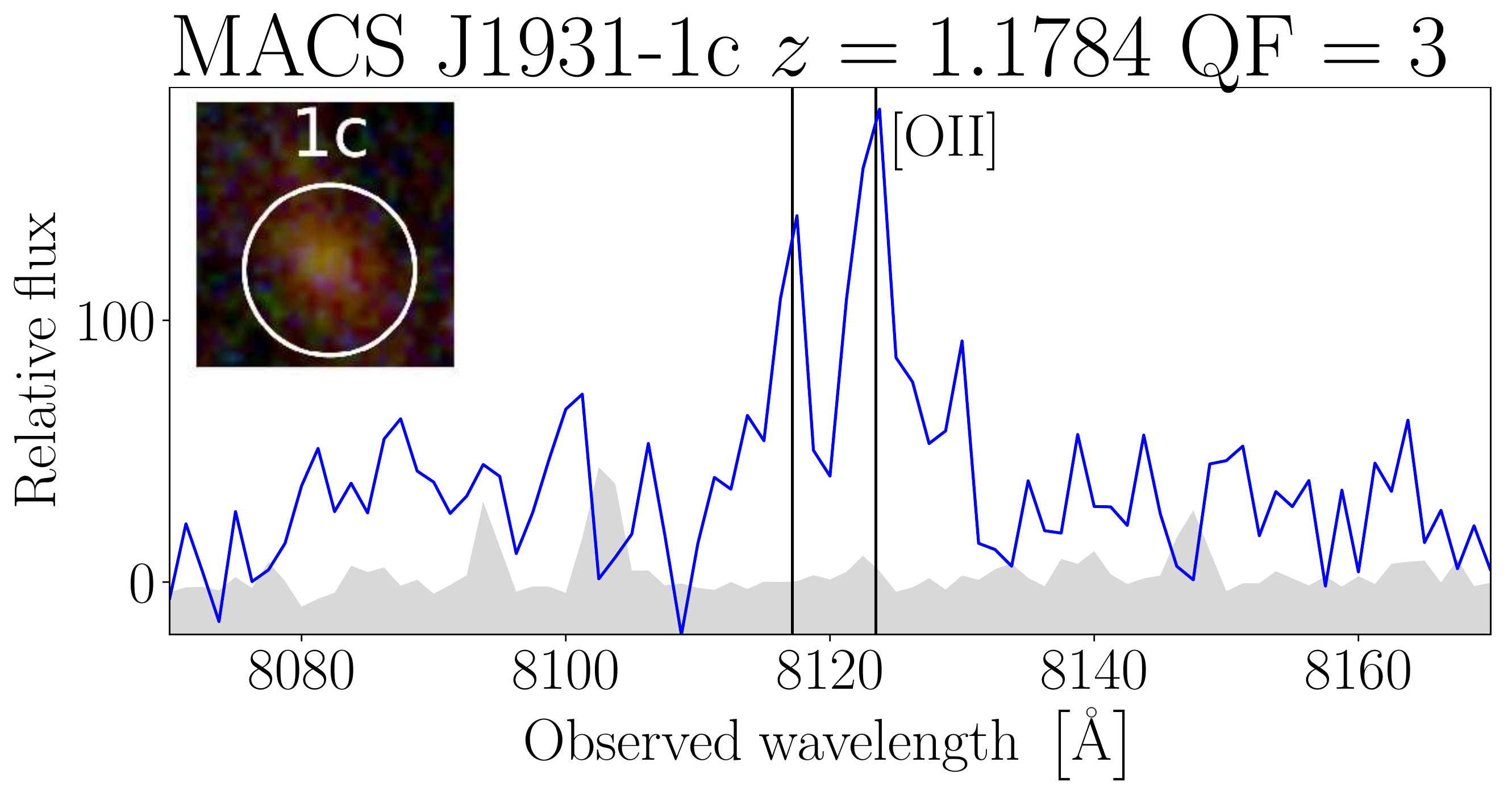}

   \caption{MUSE spectra of multiply lensed background sources. Vertical black lines indicate the positions of spectral features used to estimate the redshifts. The grey area shows the rescaled variance obtained from the data reduction pipeline; the flux is given in units of $\rm 10^{-20}\, erg\, s^{-1}\, cm^{-2}\, \AA^{-1}$. The image boxes are extracted from the CLASH colour images and have $2\arcsec$ side. The white circles show the HST counterparts and, for the cases with no apparent photometric counterparts, they are centred at the position of the MUSE peak emission.}
  \label{fig:specs}
\end{figure*}

\begin{figure*}
\setcounter{figure}{\value{figure}-1}
   \includegraphics[width = 0.666\columnwidth]{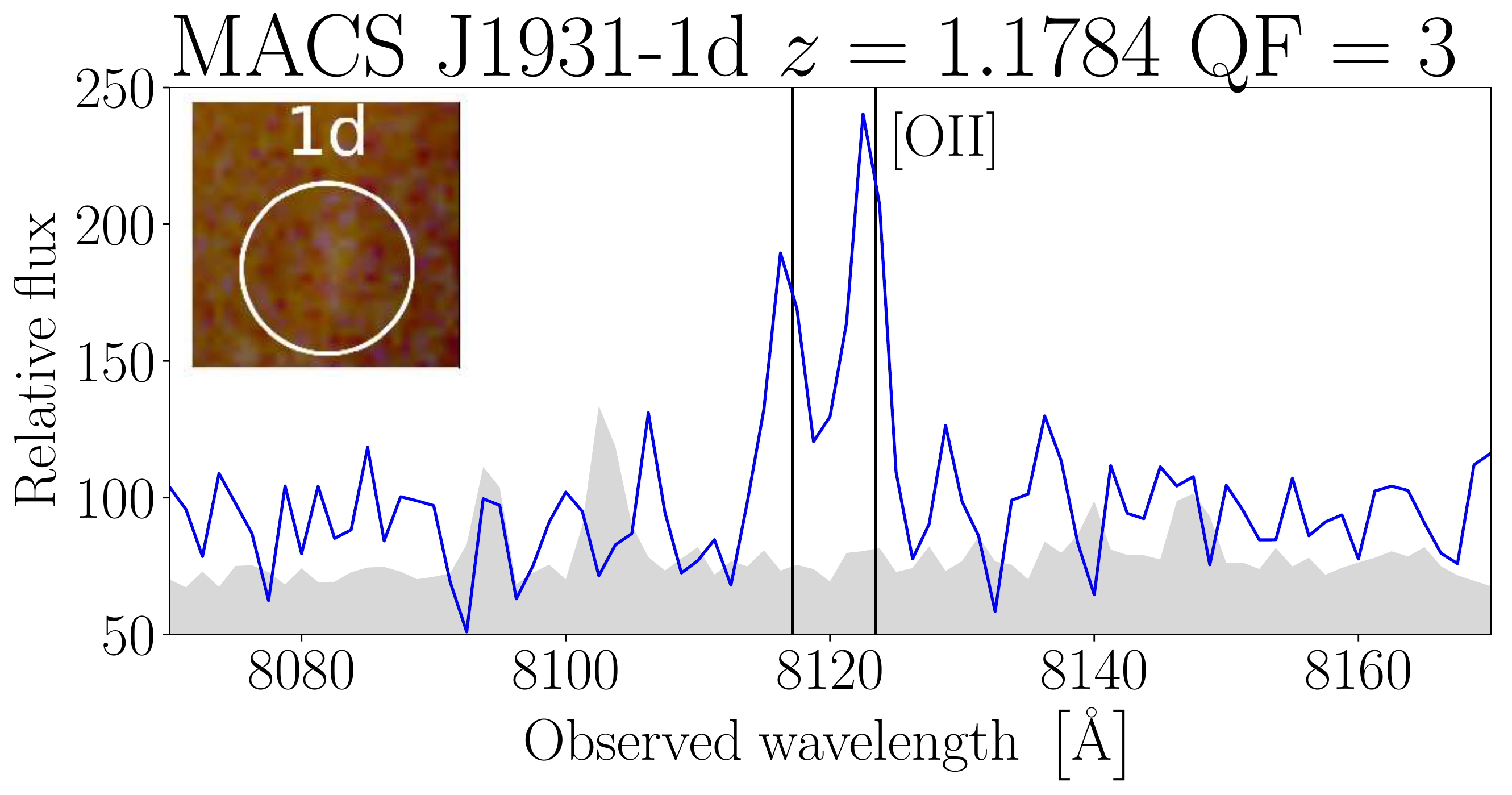}
   \includegraphics[width = 0.666\columnwidth]{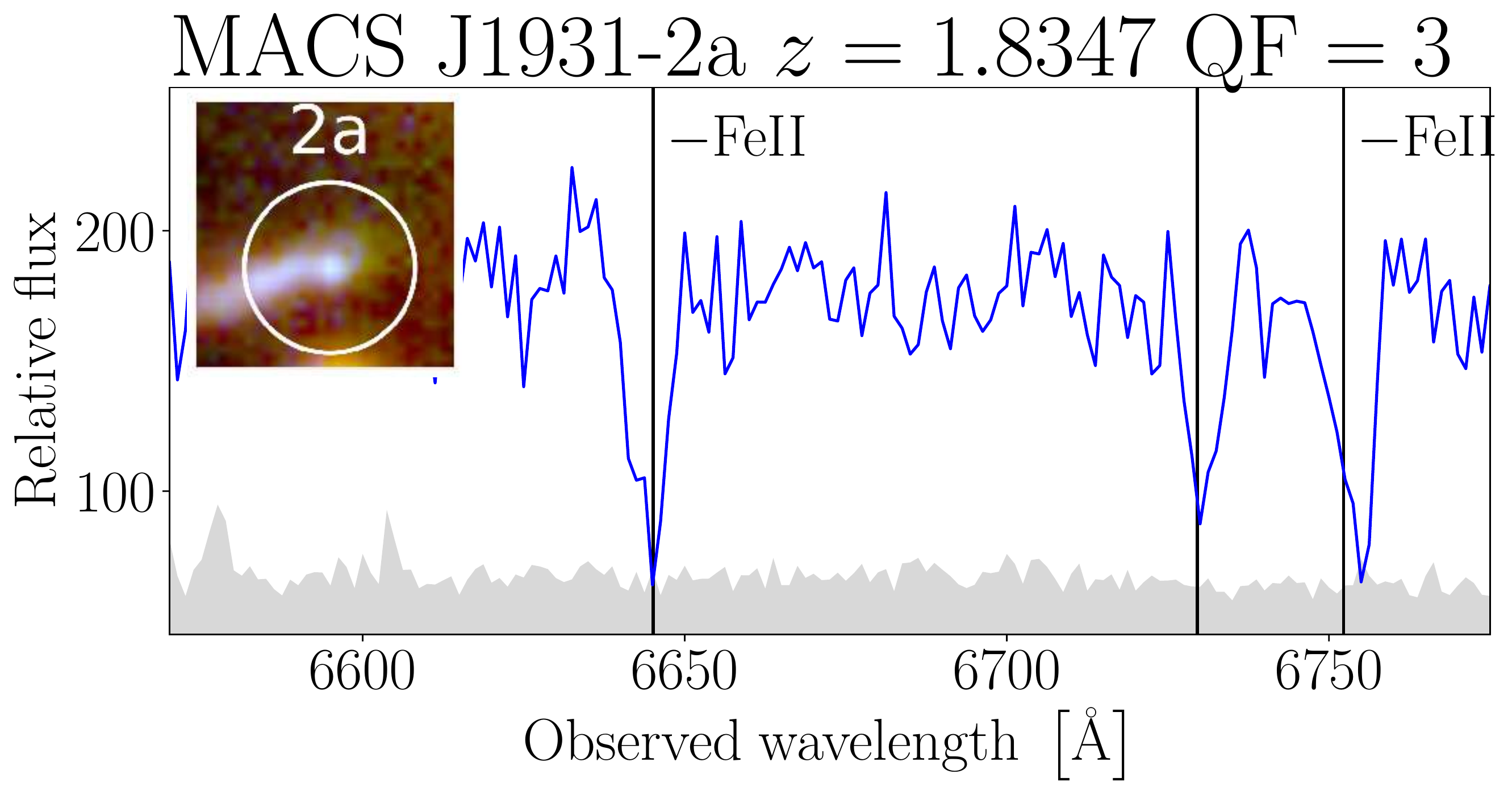}
   \includegraphics[width = 0.666\columnwidth]{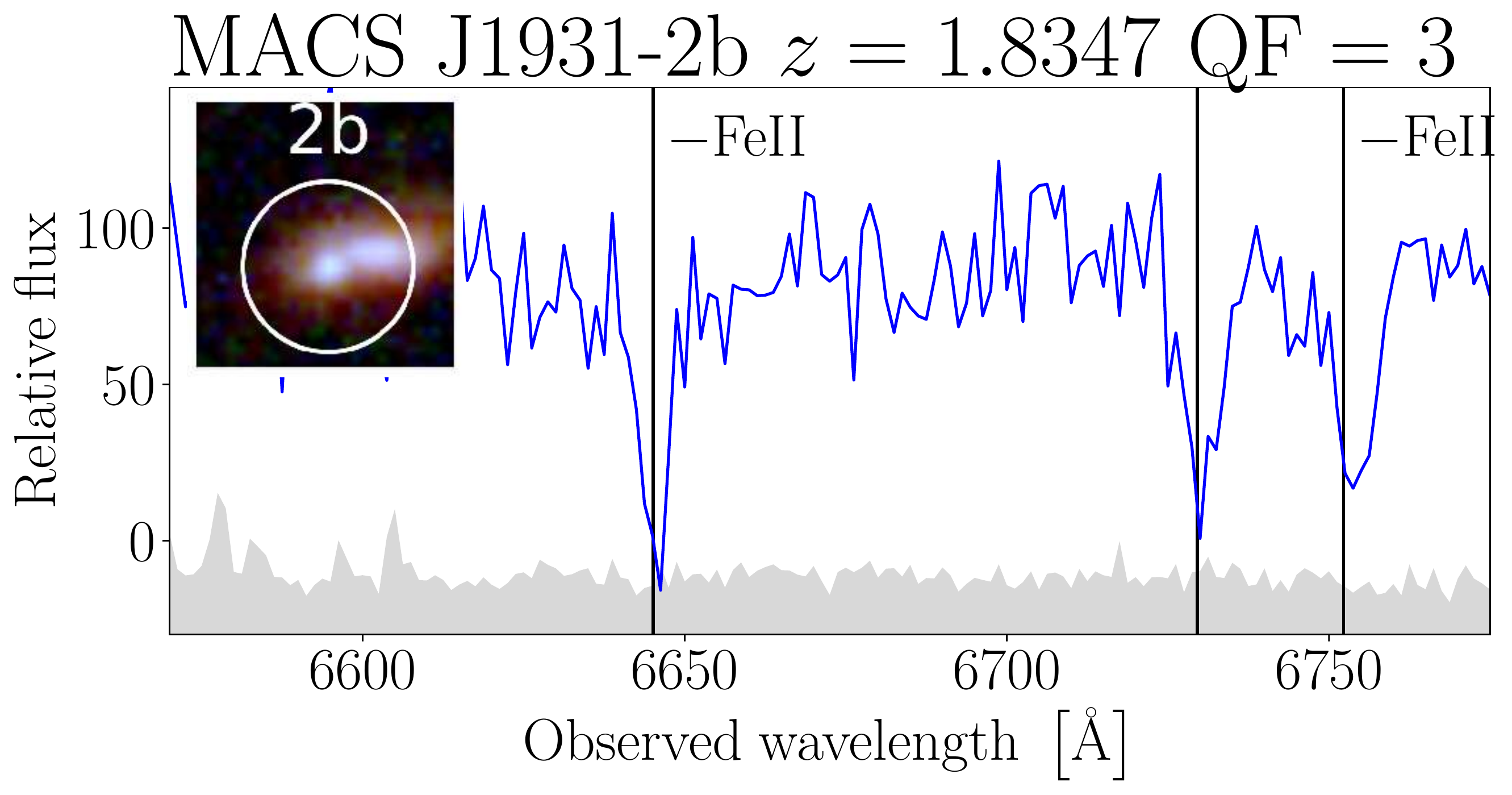}
   \includegraphics[width = 0.666\columnwidth]{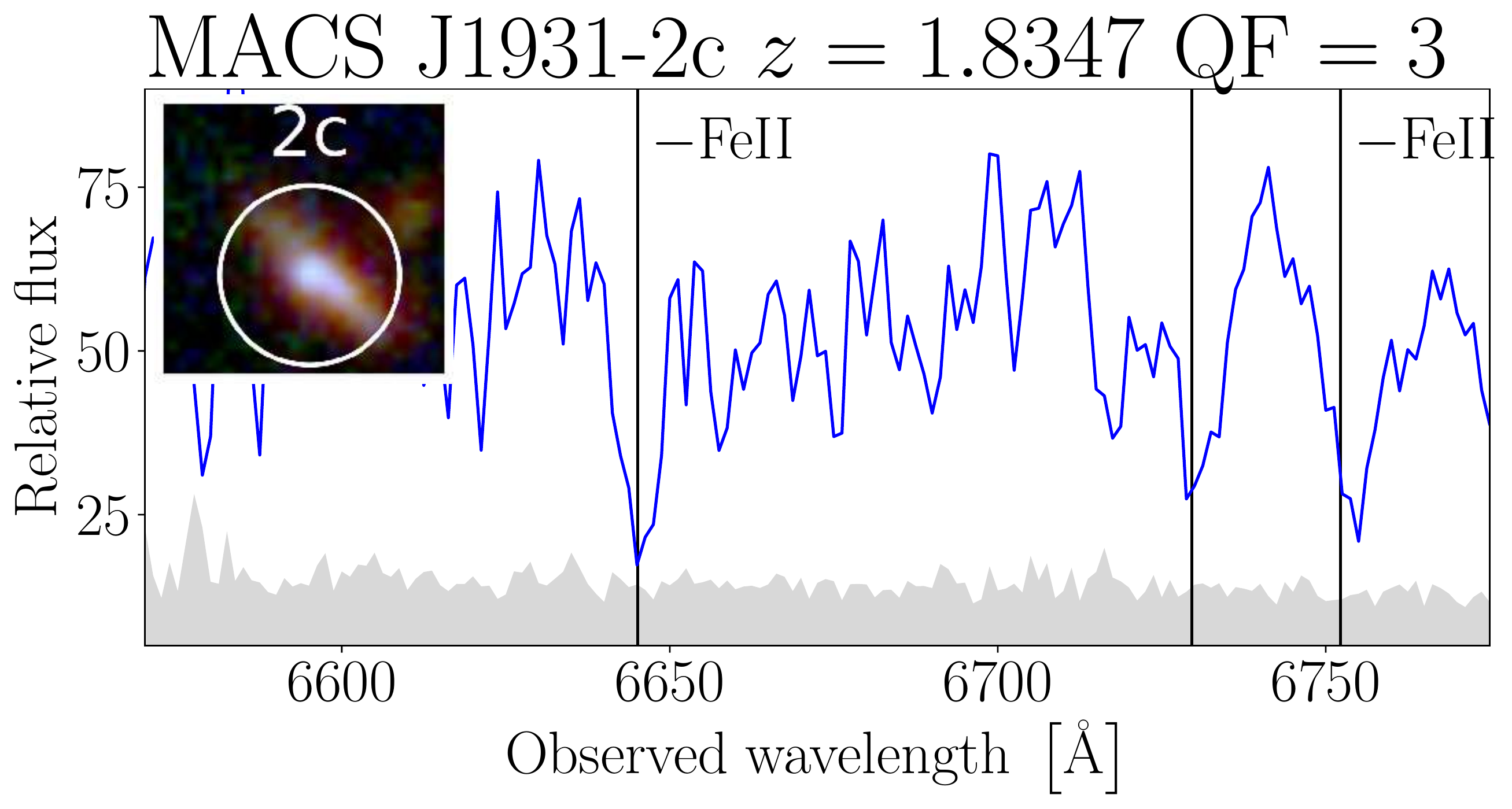}
   \includegraphics[width = 0.666\columnwidth]{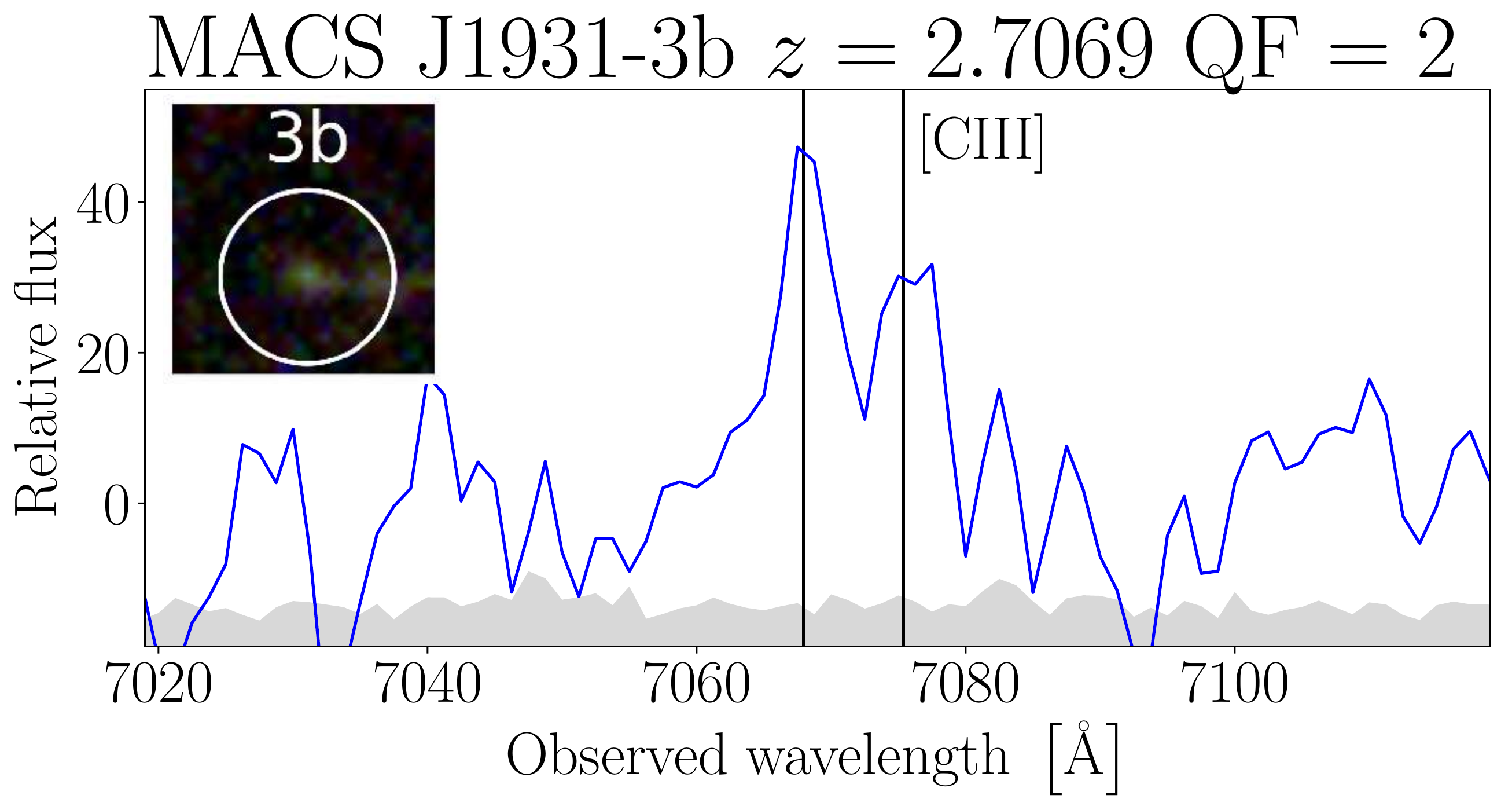}
   \includegraphics[width = 0.666\columnwidth]{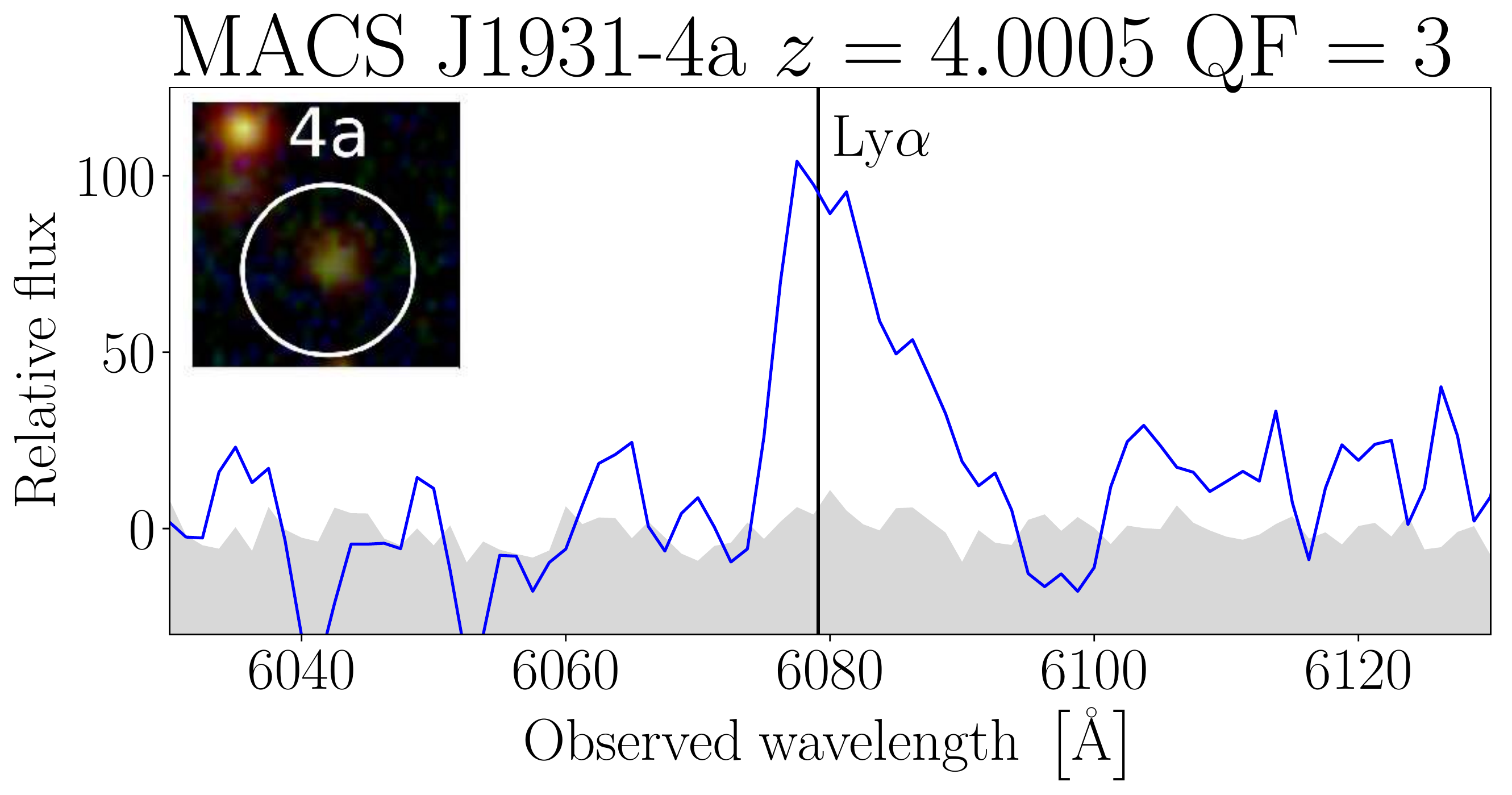}
   \includegraphics[width = 0.666\columnwidth]{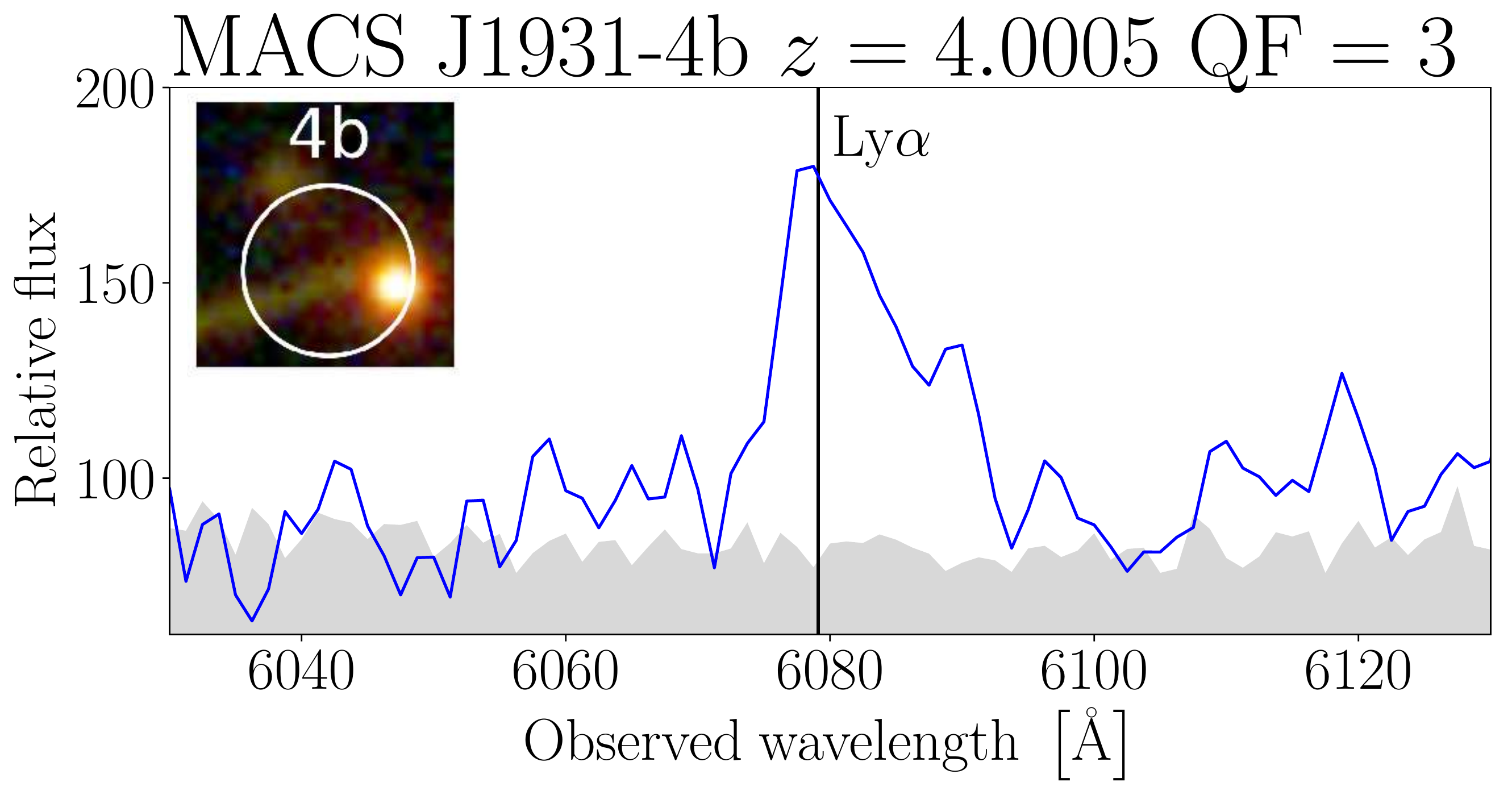}
   \includegraphics[width = 0.666\columnwidth]{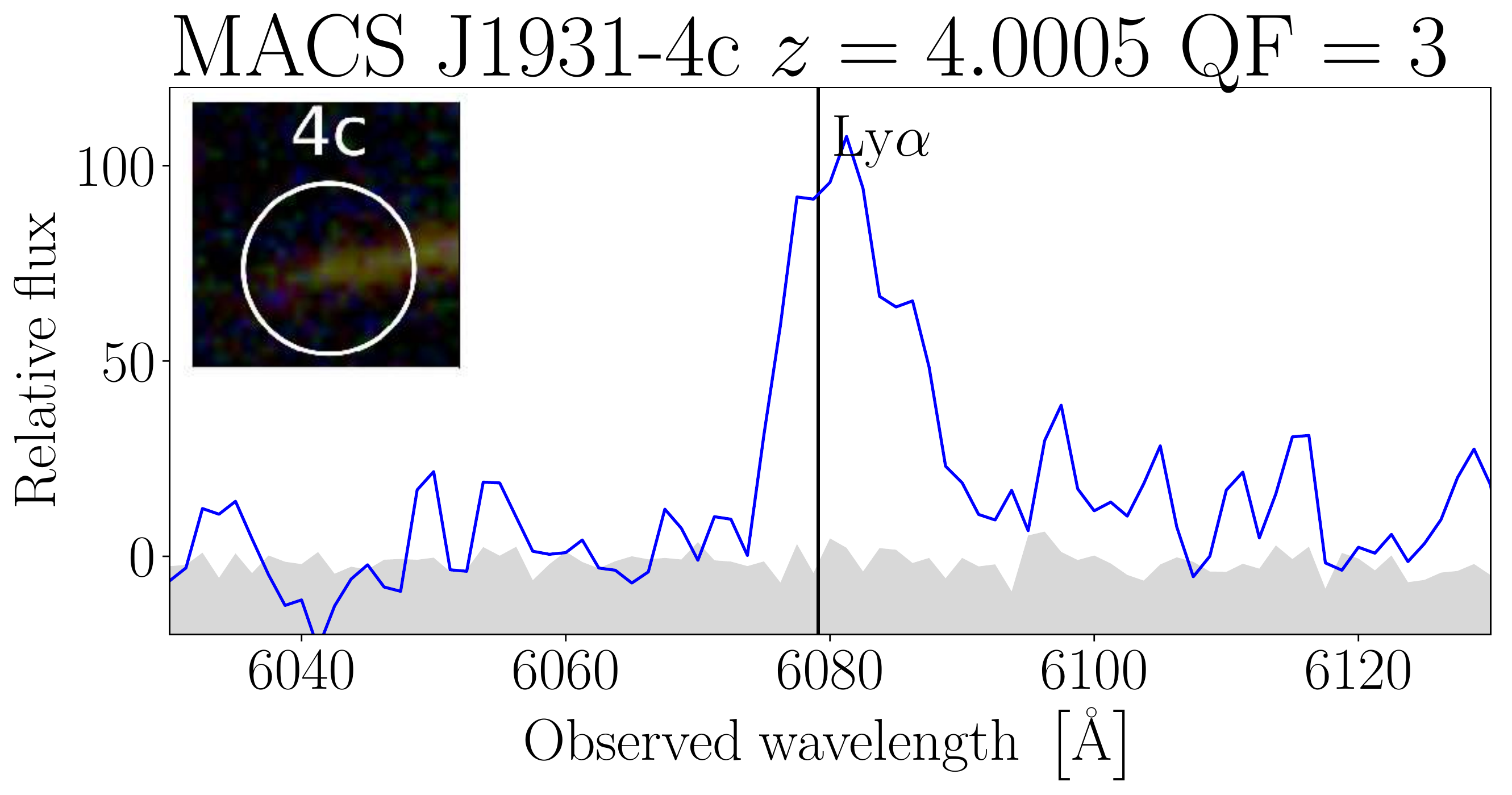}
   \includegraphics[width = 0.666\columnwidth]{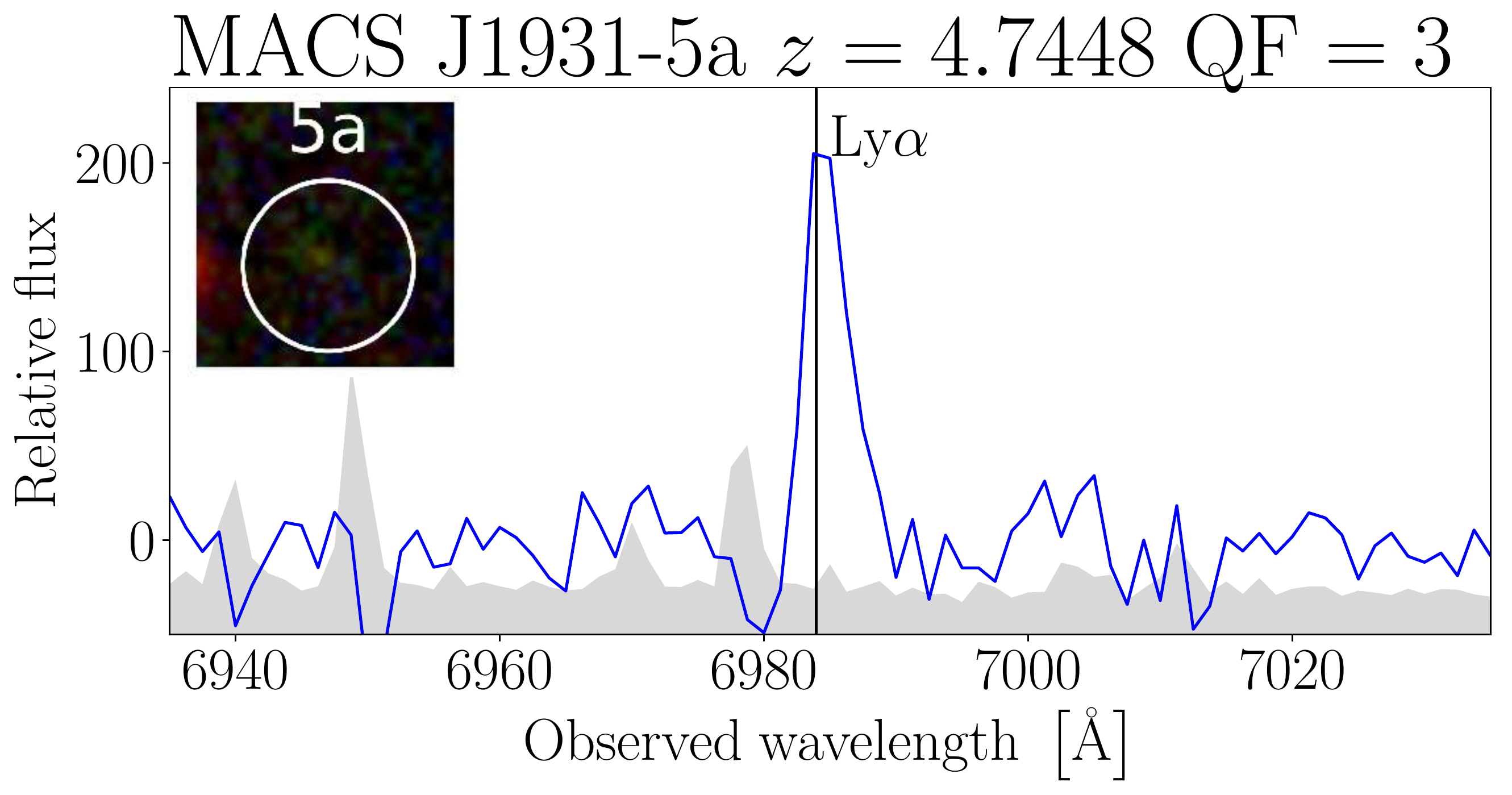}
   \includegraphics[width = 0.666\columnwidth]{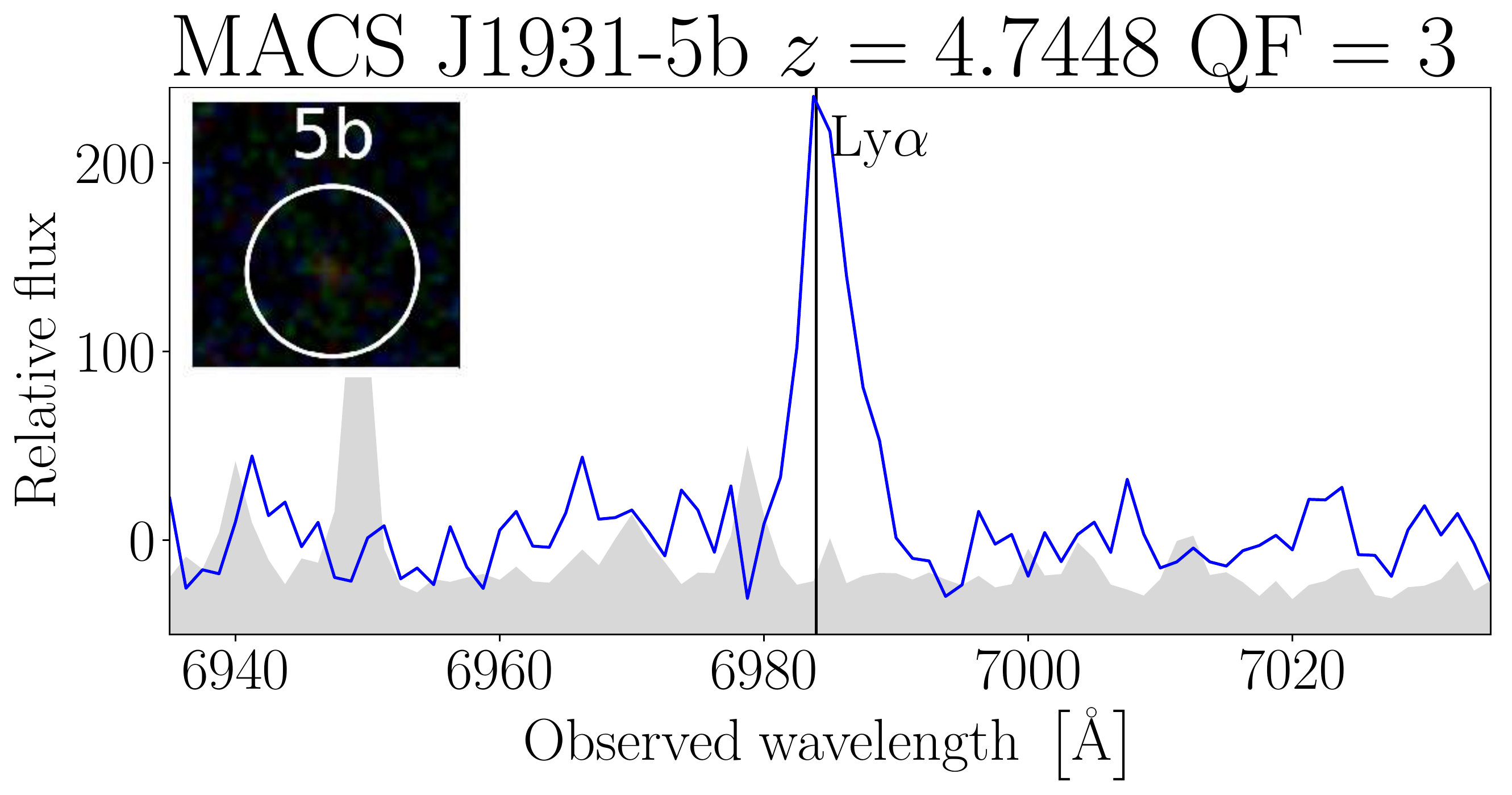}
   \includegraphics[width = 0.666\columnwidth]{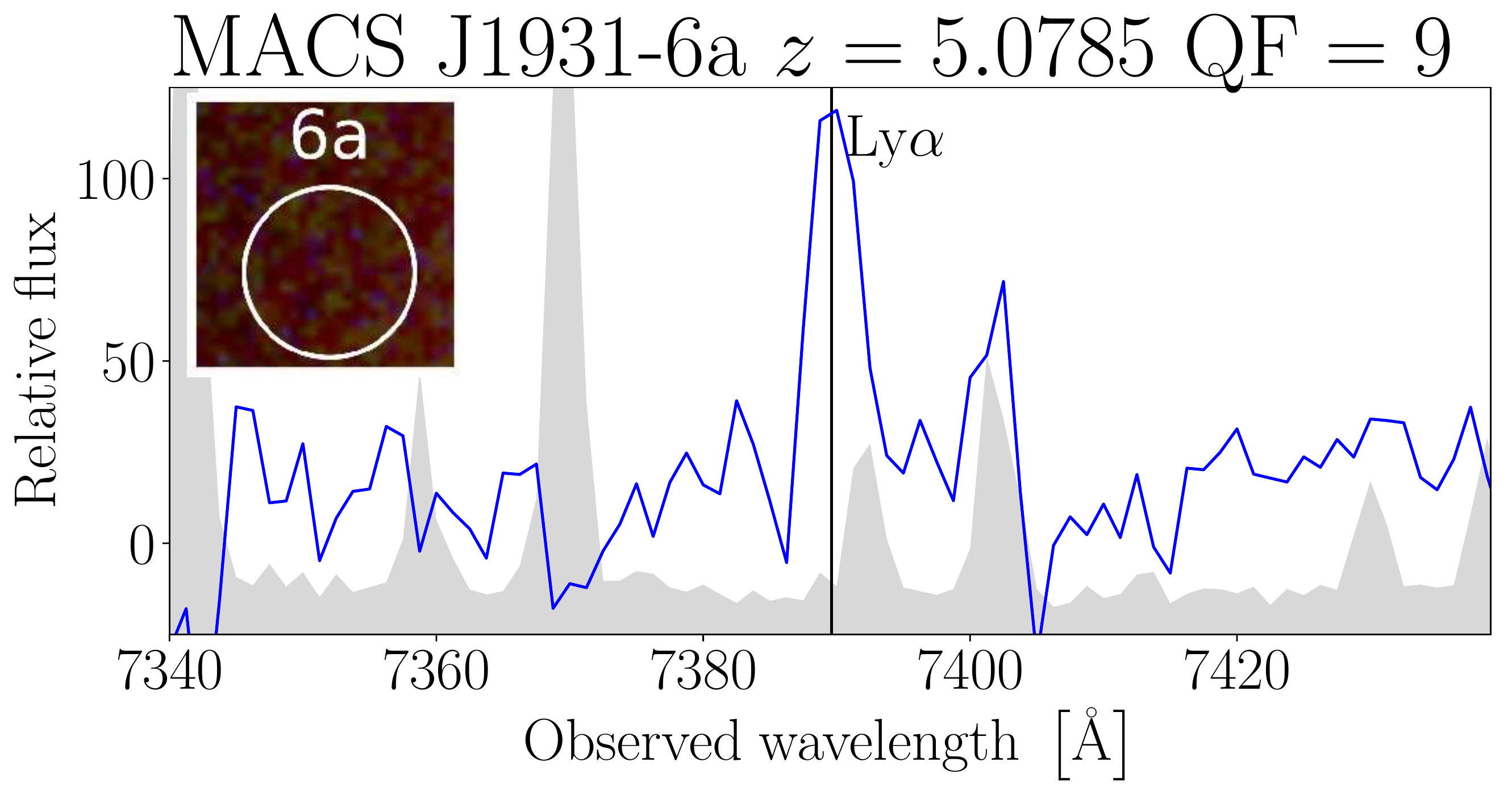}
   \includegraphics[width = 0.666\columnwidth]{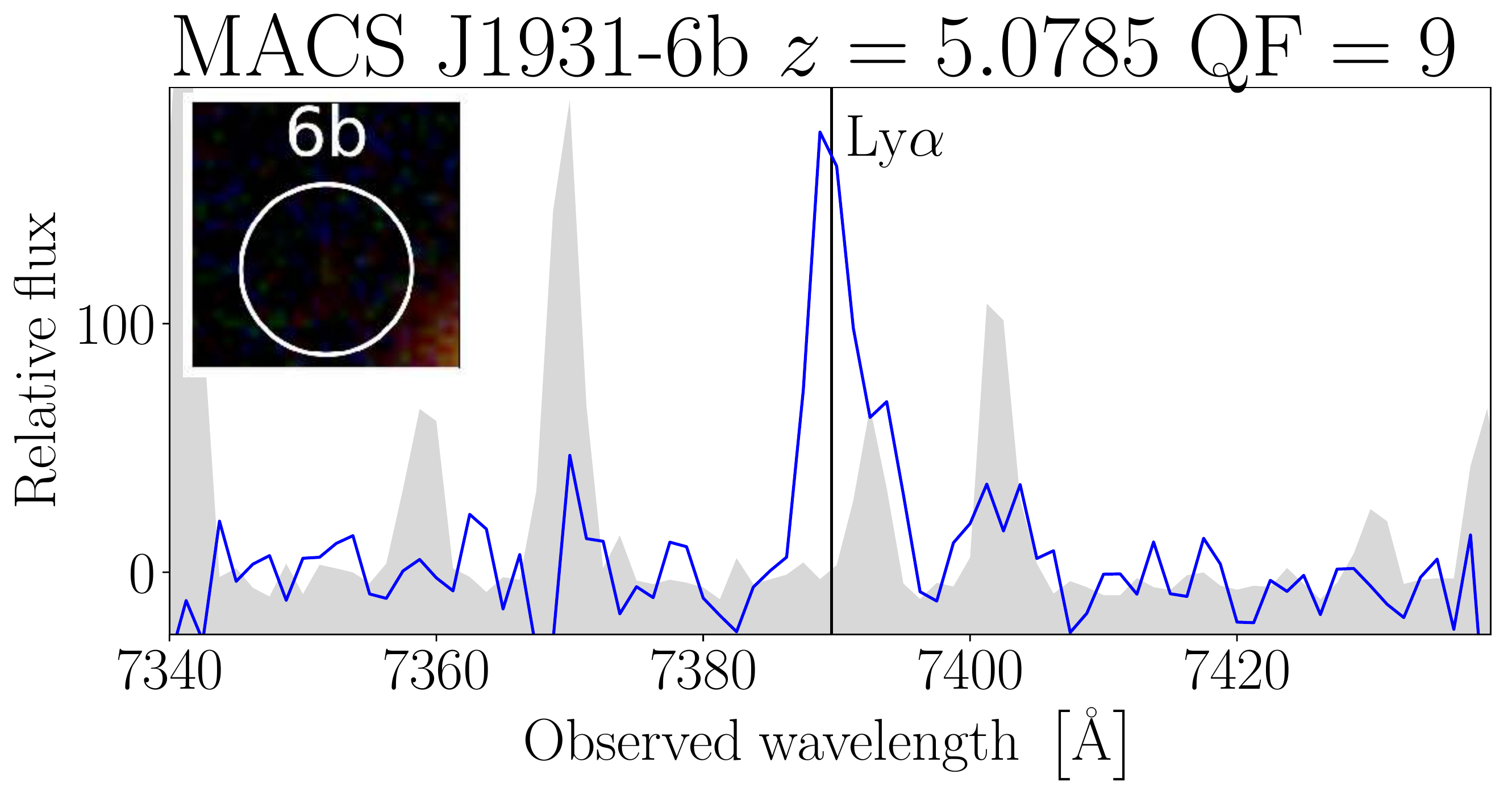}
   \includegraphics[width = 0.666\columnwidth]{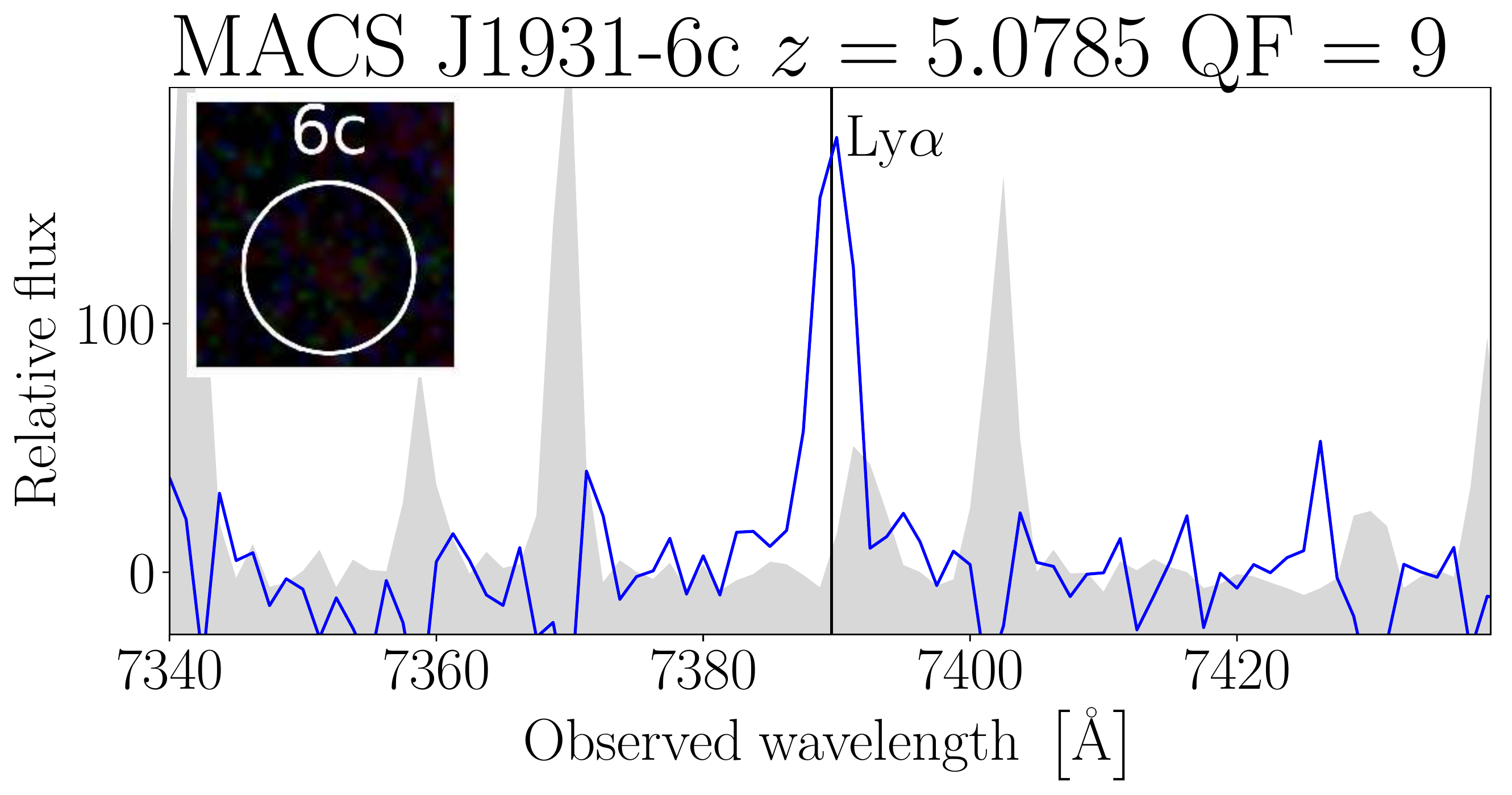}
   \includegraphics[width = 0.666\columnwidth]{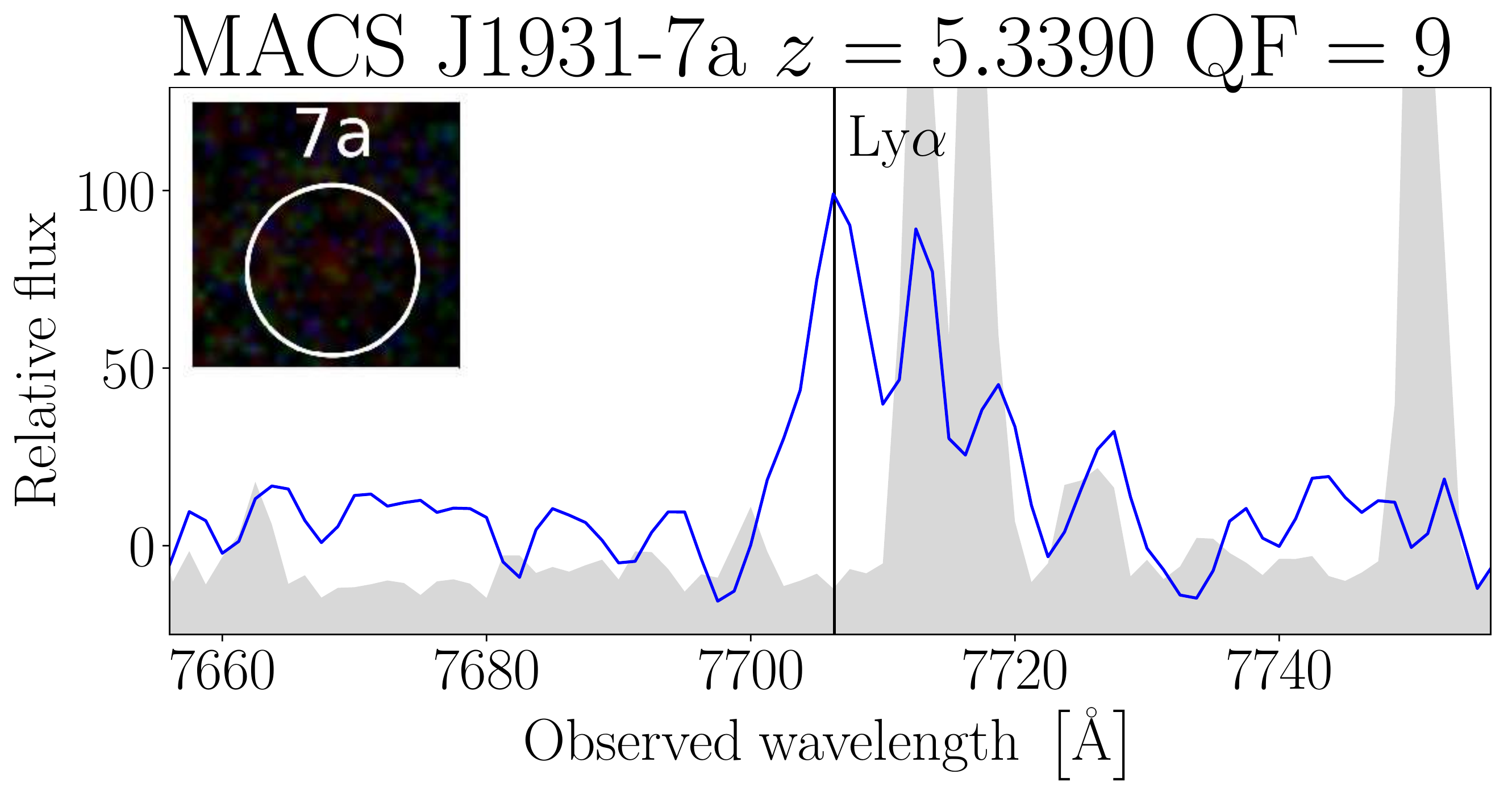}
   \includegraphics[width = 0.666\columnwidth]{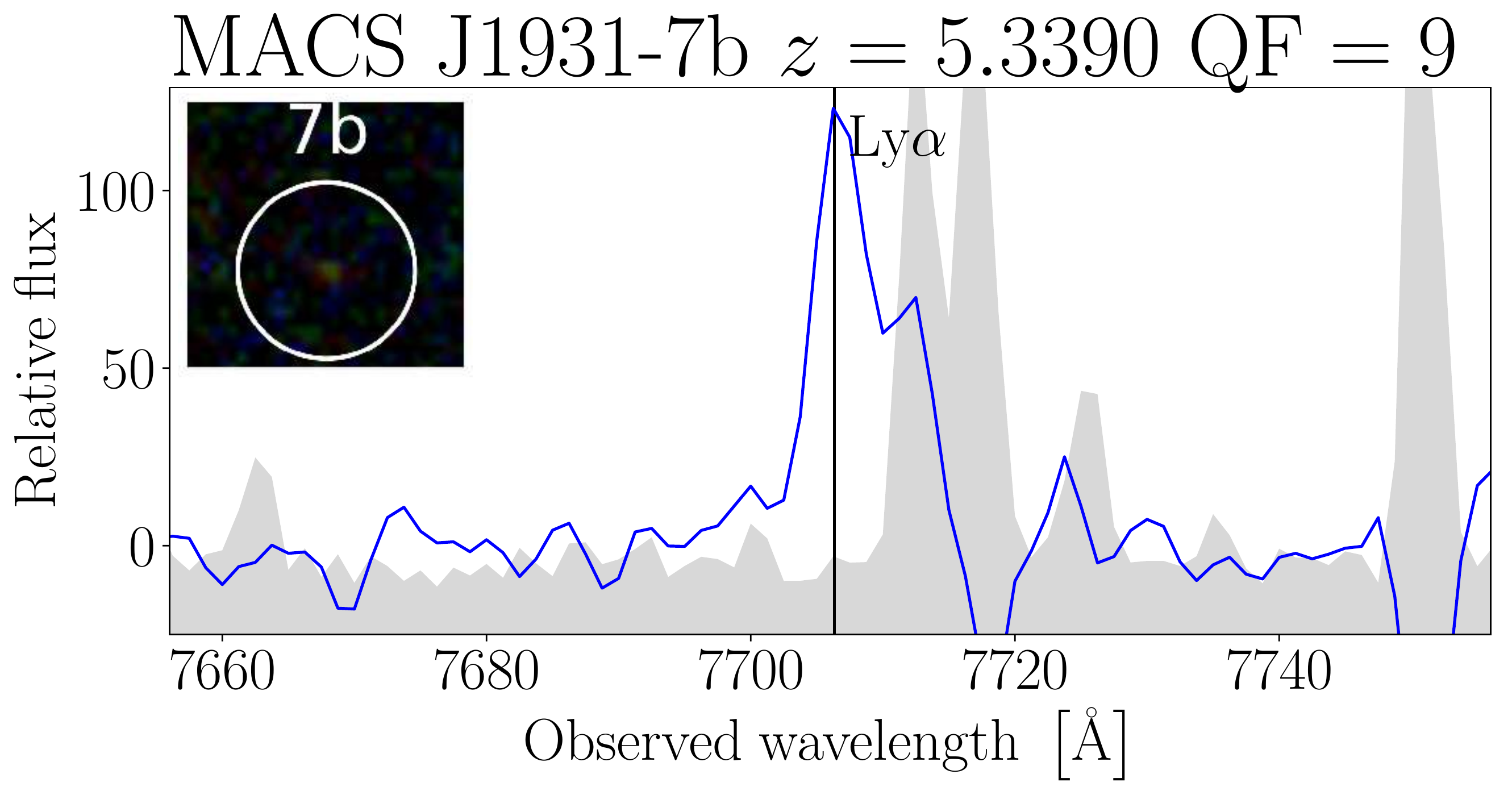}
   \includegraphics[width = 0.666\columnwidth]{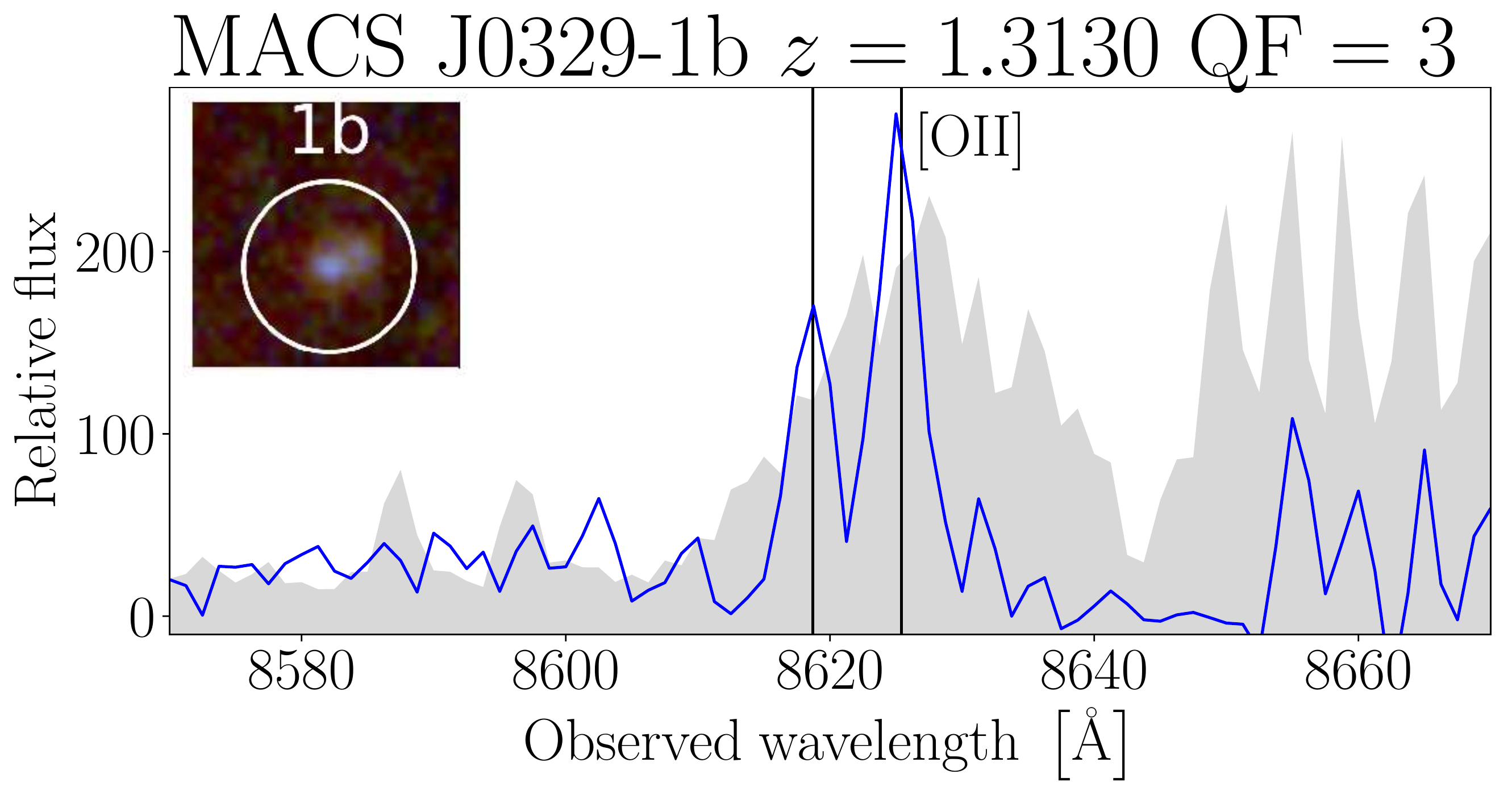}
   \includegraphics[width = 0.666\columnwidth]{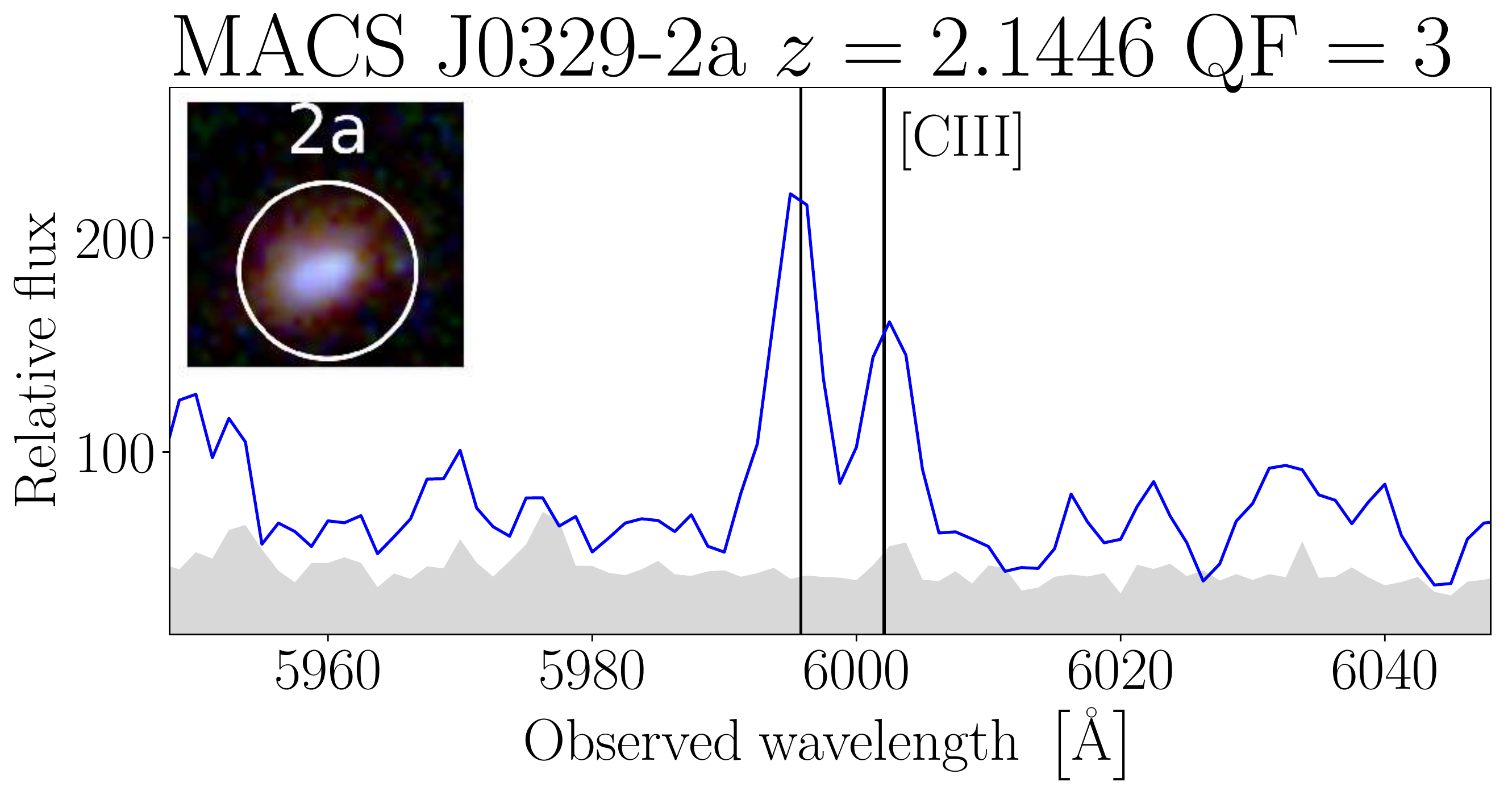}
   \includegraphics[width = 0.666\columnwidth]{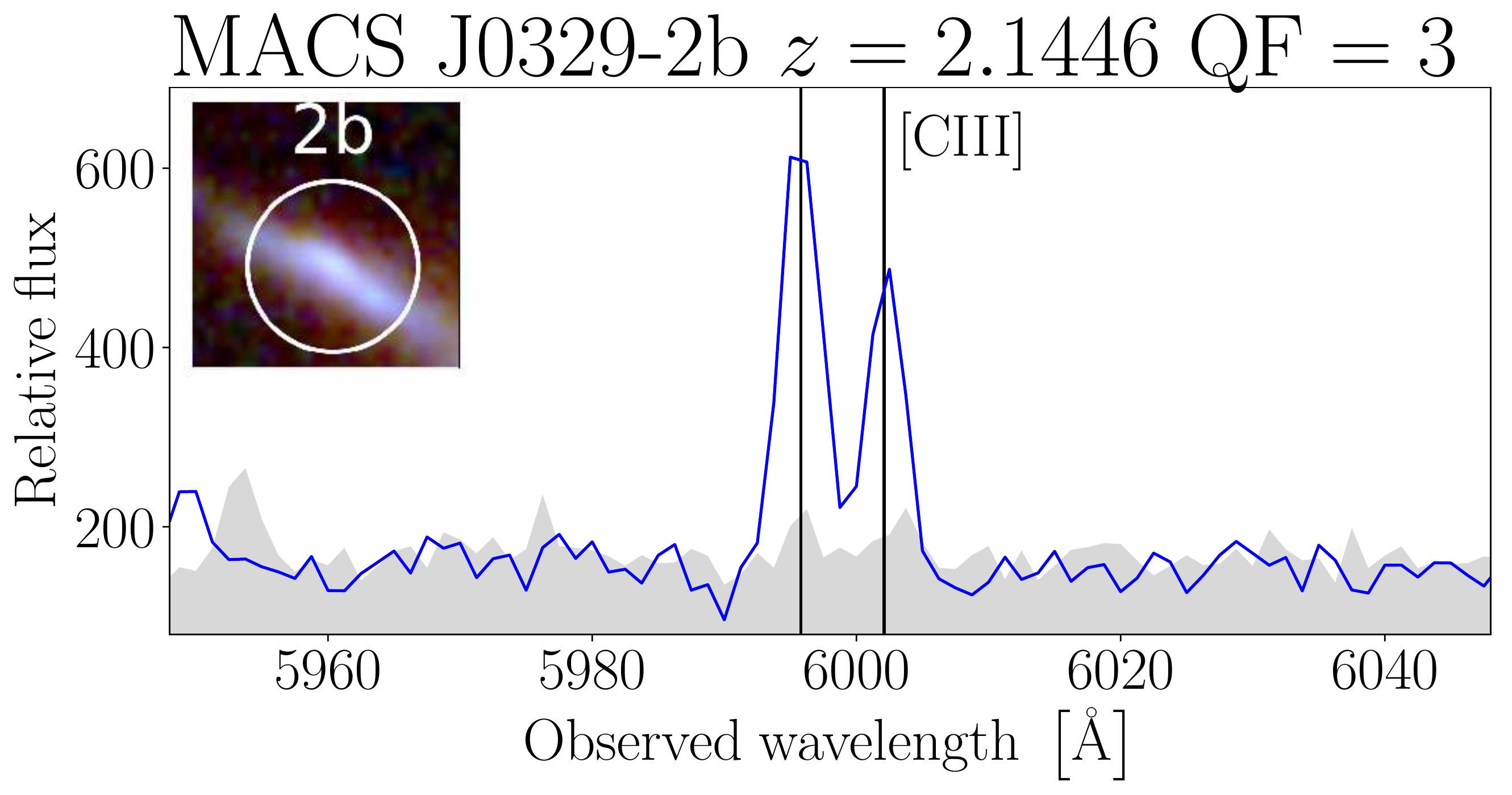}
   \includegraphics[width = 0.666\columnwidth]{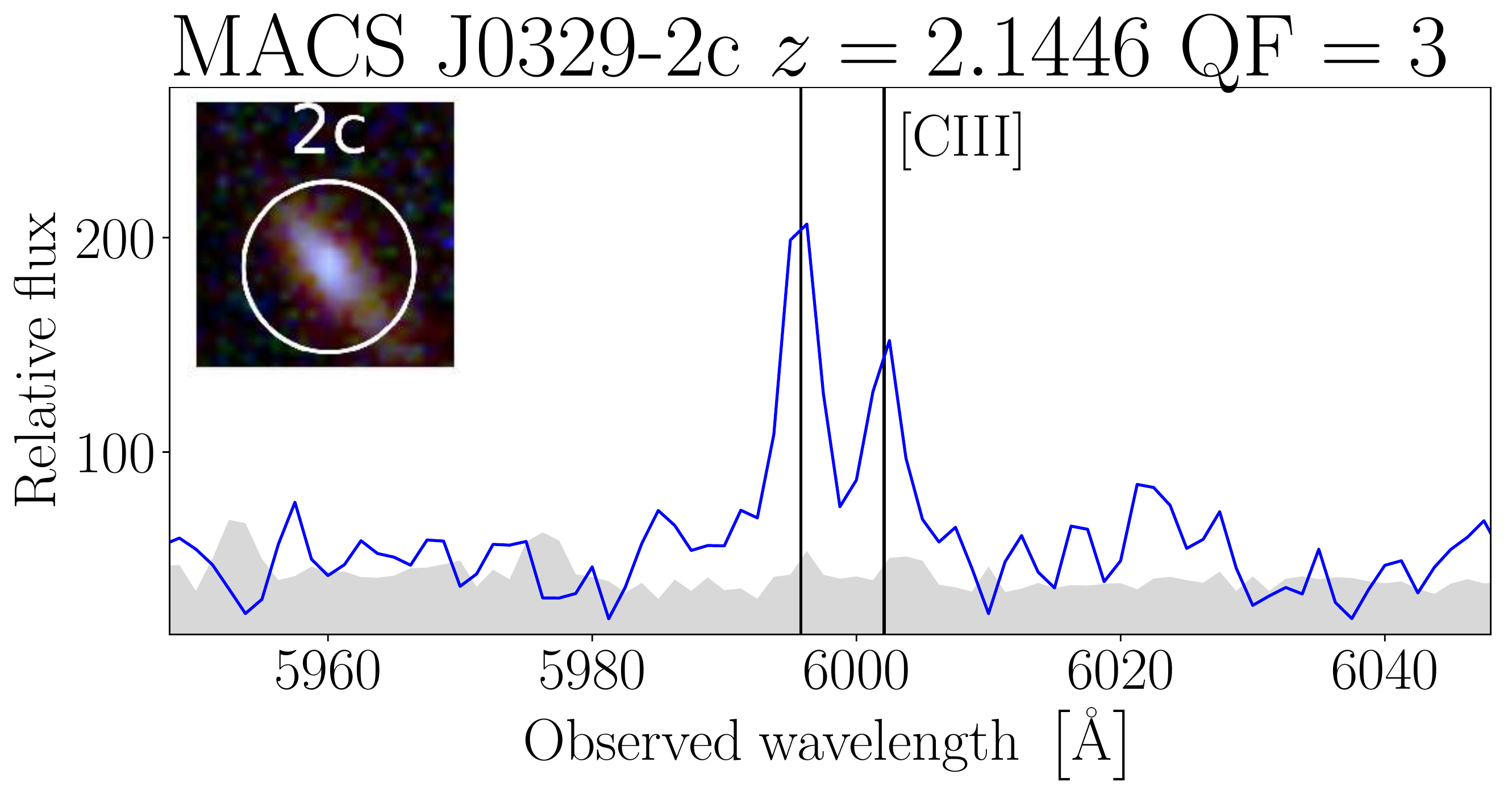}
   \includegraphics[width = 0.666\columnwidth]{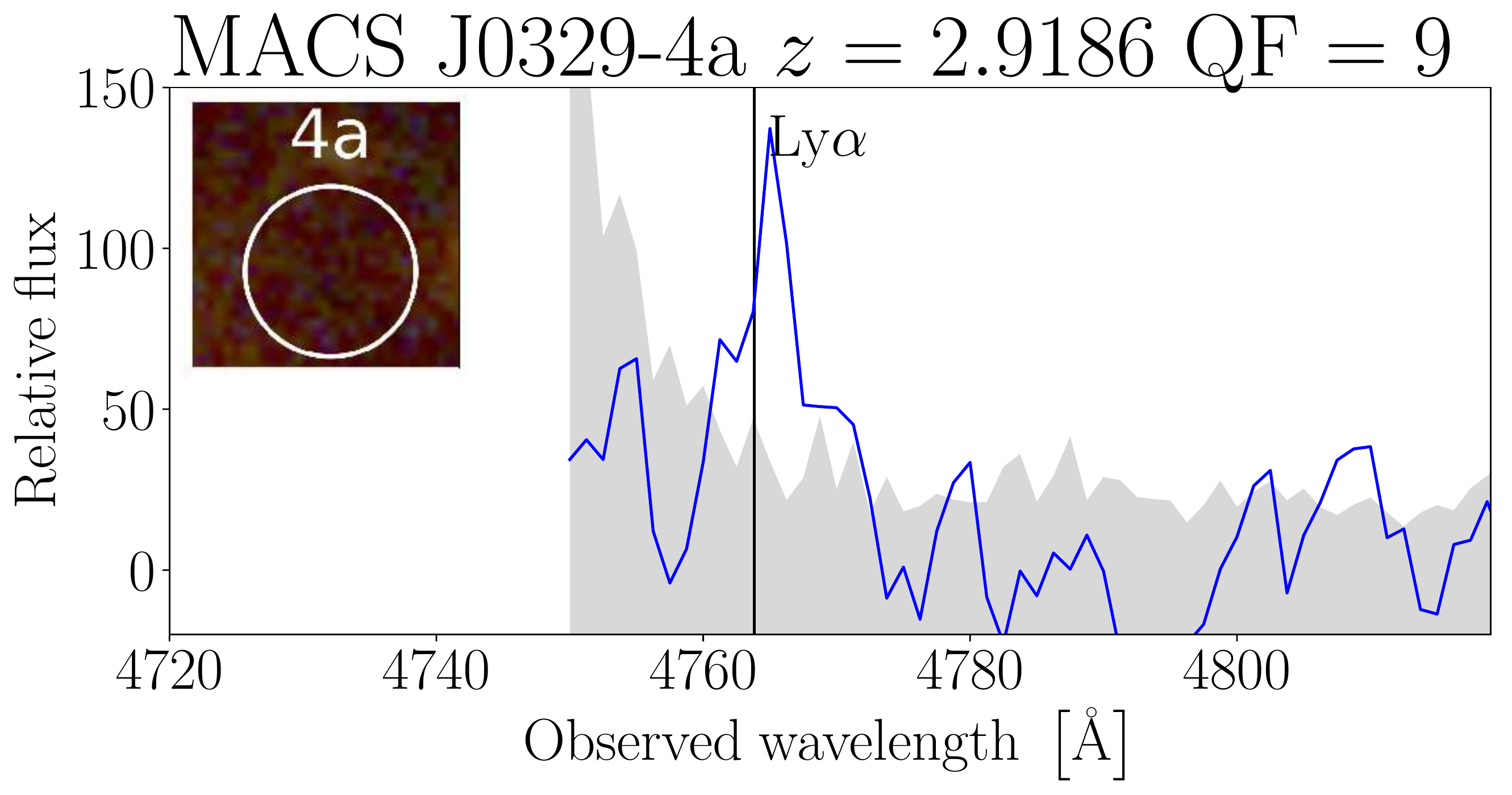}
   \includegraphics[width = 0.666\columnwidth]{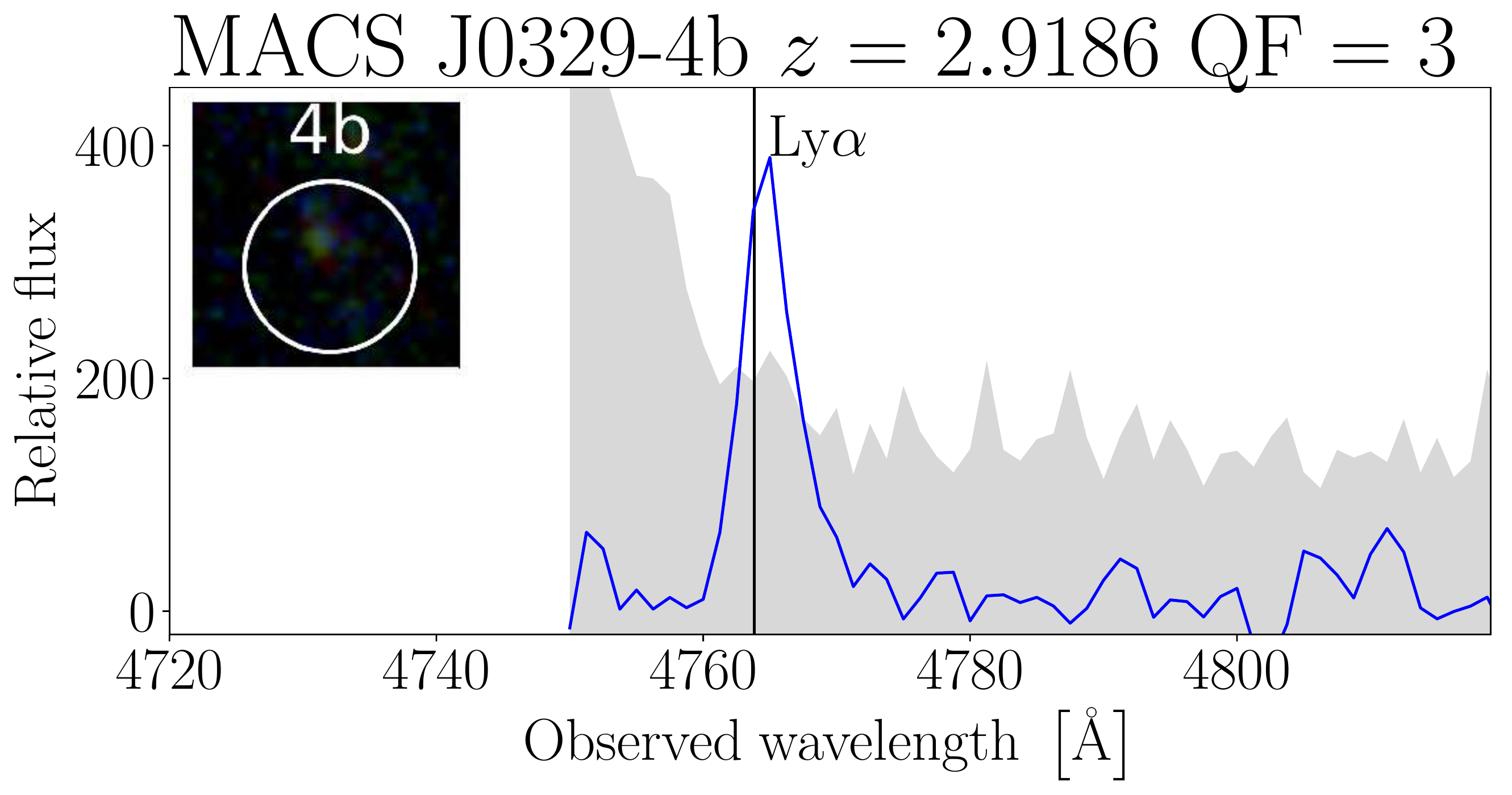}
   
  \caption{(Continued)}
  \label{fig:specs}
\end{figure*}

\begin{figure*}
\setcounter{figure}{\value{figure}-1}
   \includegraphics[width = 0.666\columnwidth]{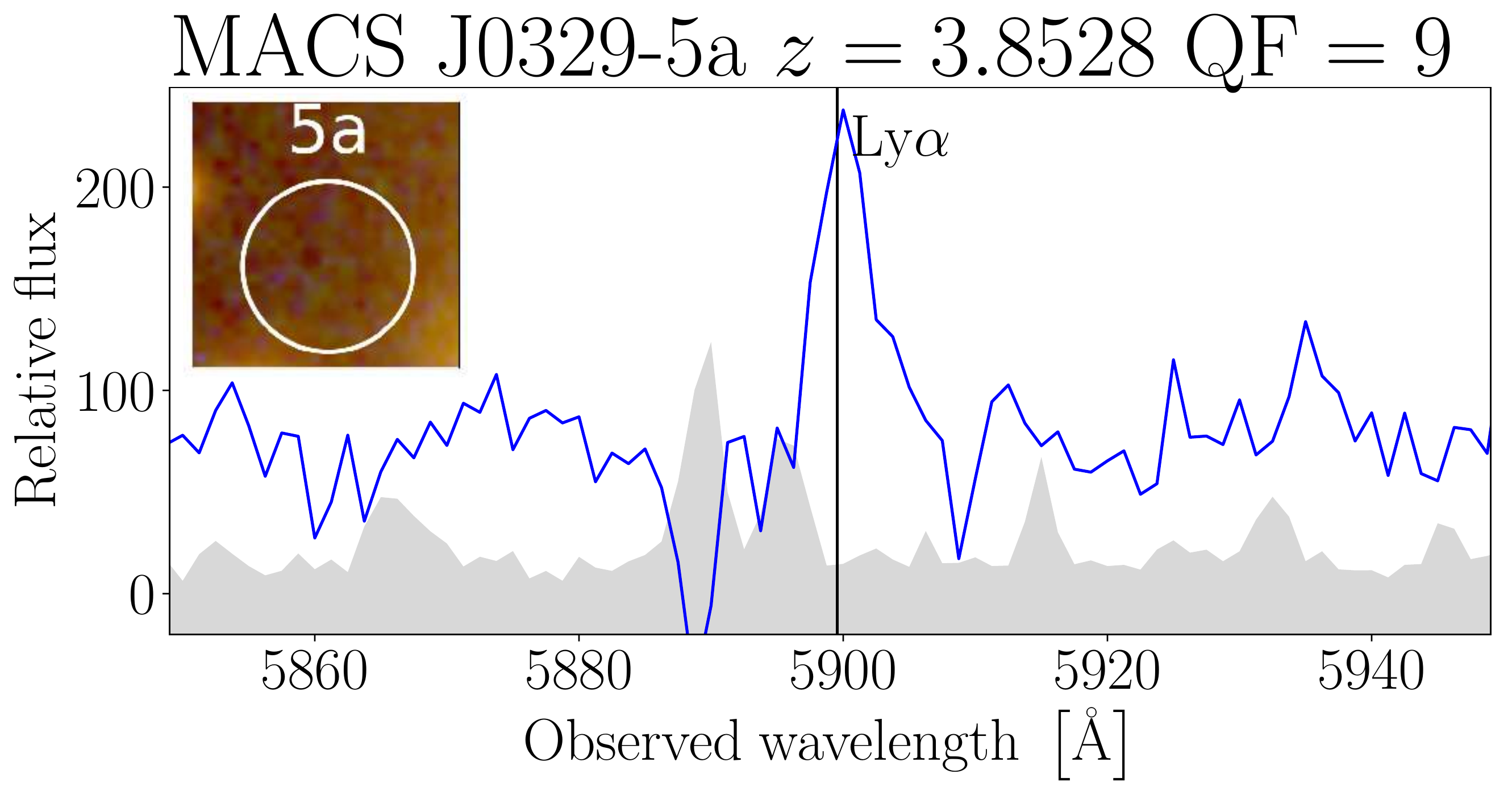}
   \includegraphics[width = 0.666\columnwidth]{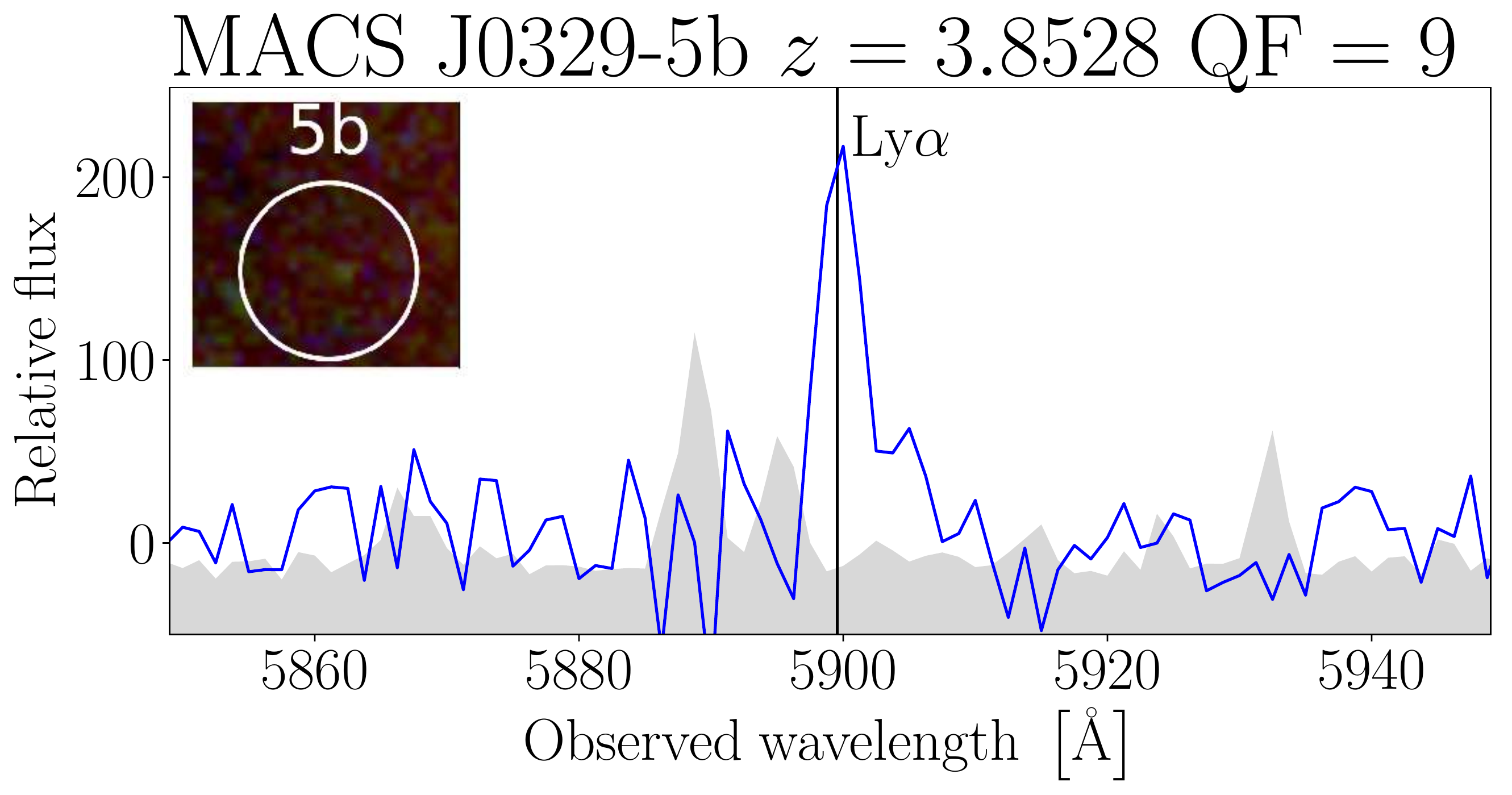}
   \includegraphics[width = 0.666\columnwidth]{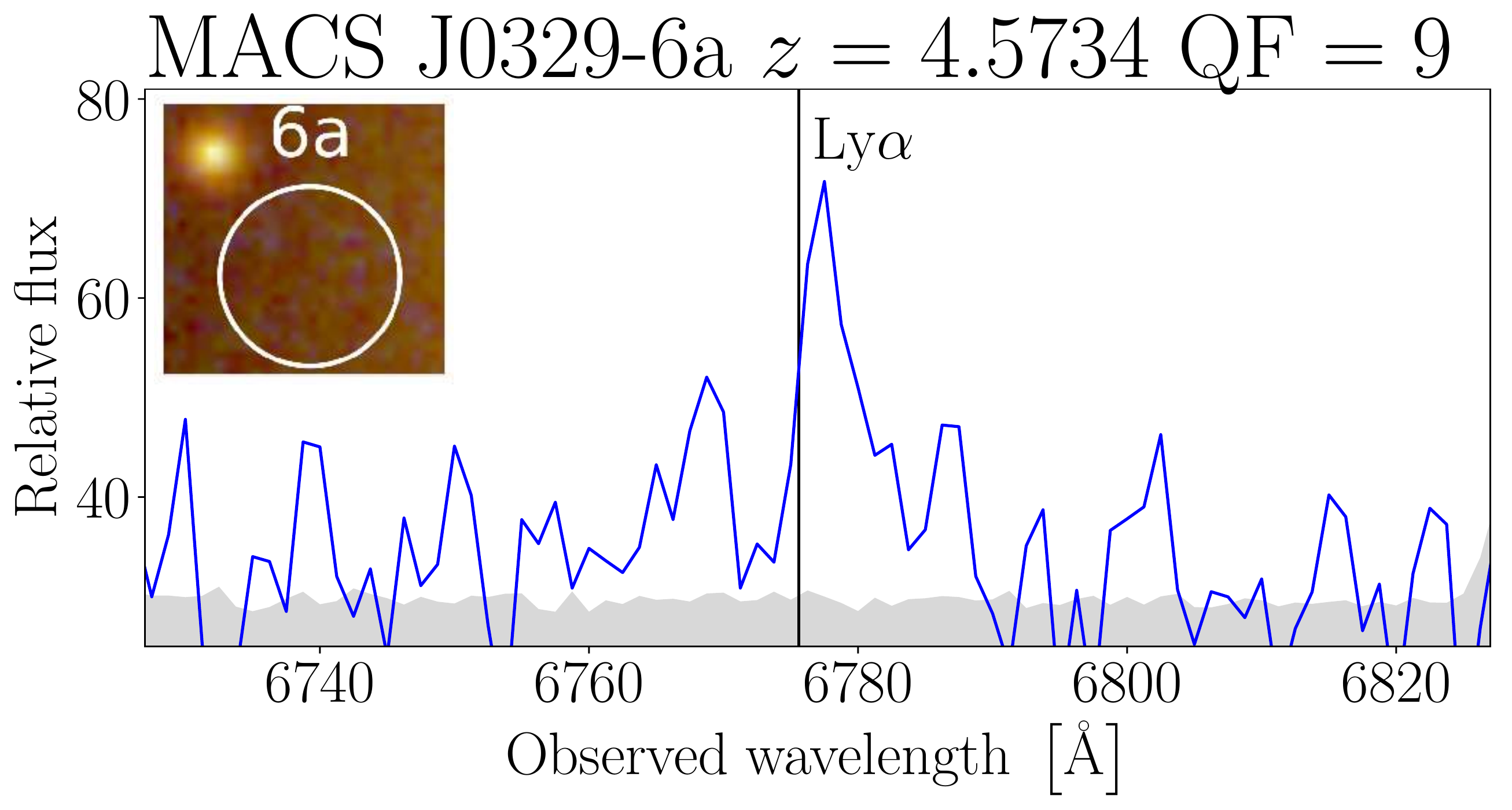}
   \includegraphics[width = 0.666\columnwidth]{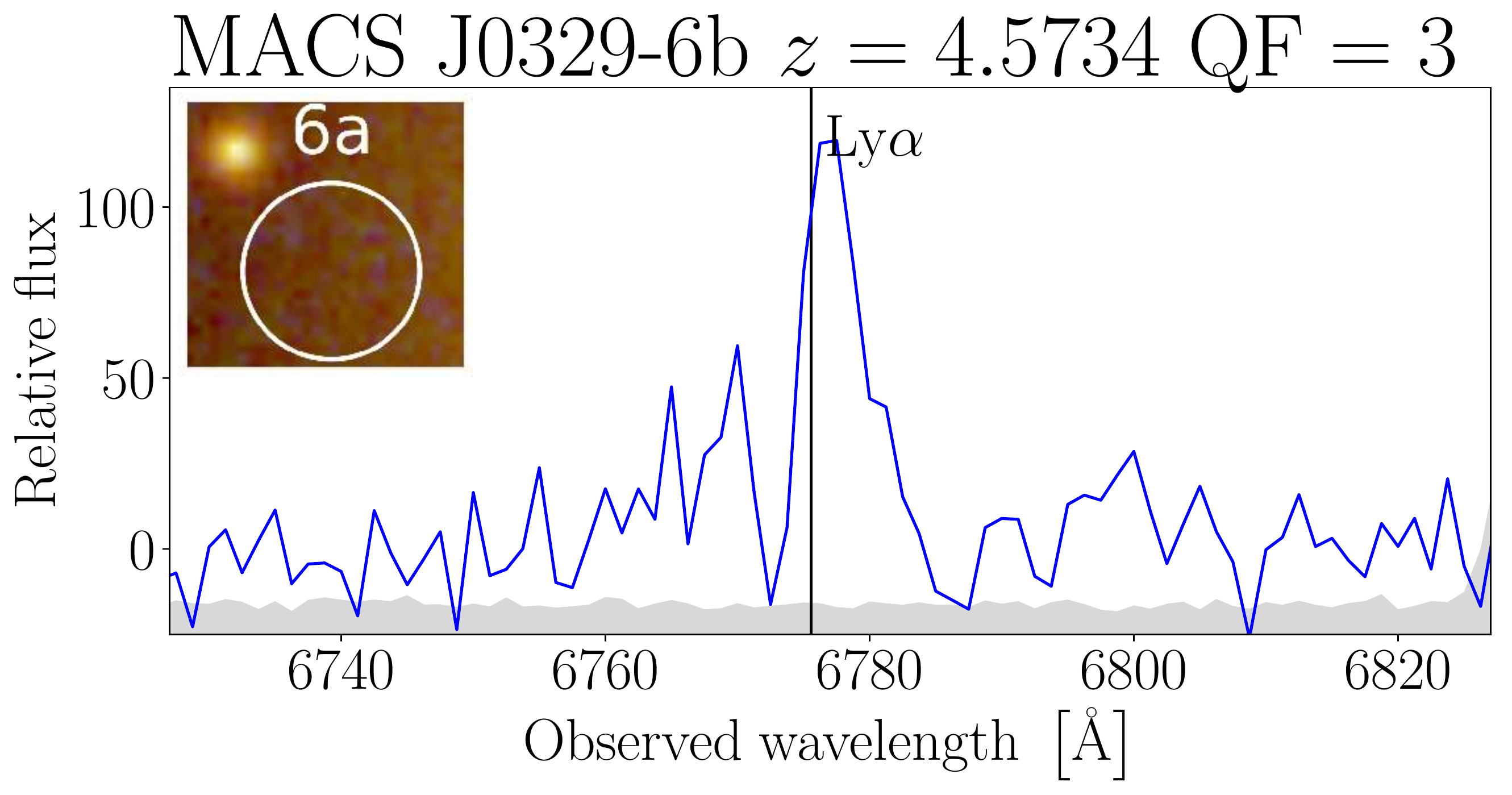}
   \includegraphics[width = 0.666\columnwidth]{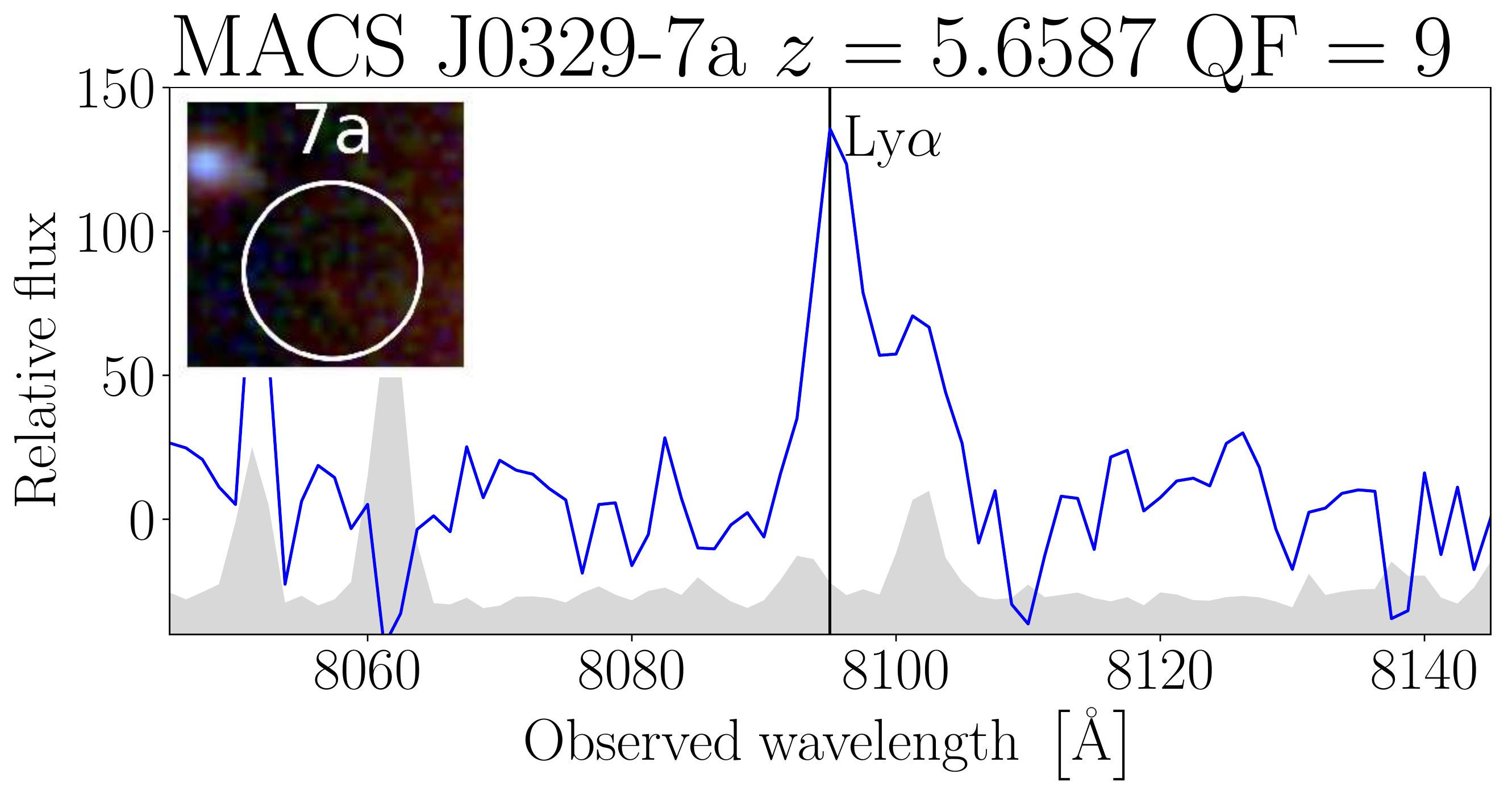}
   \includegraphics[width = 0.666\columnwidth]{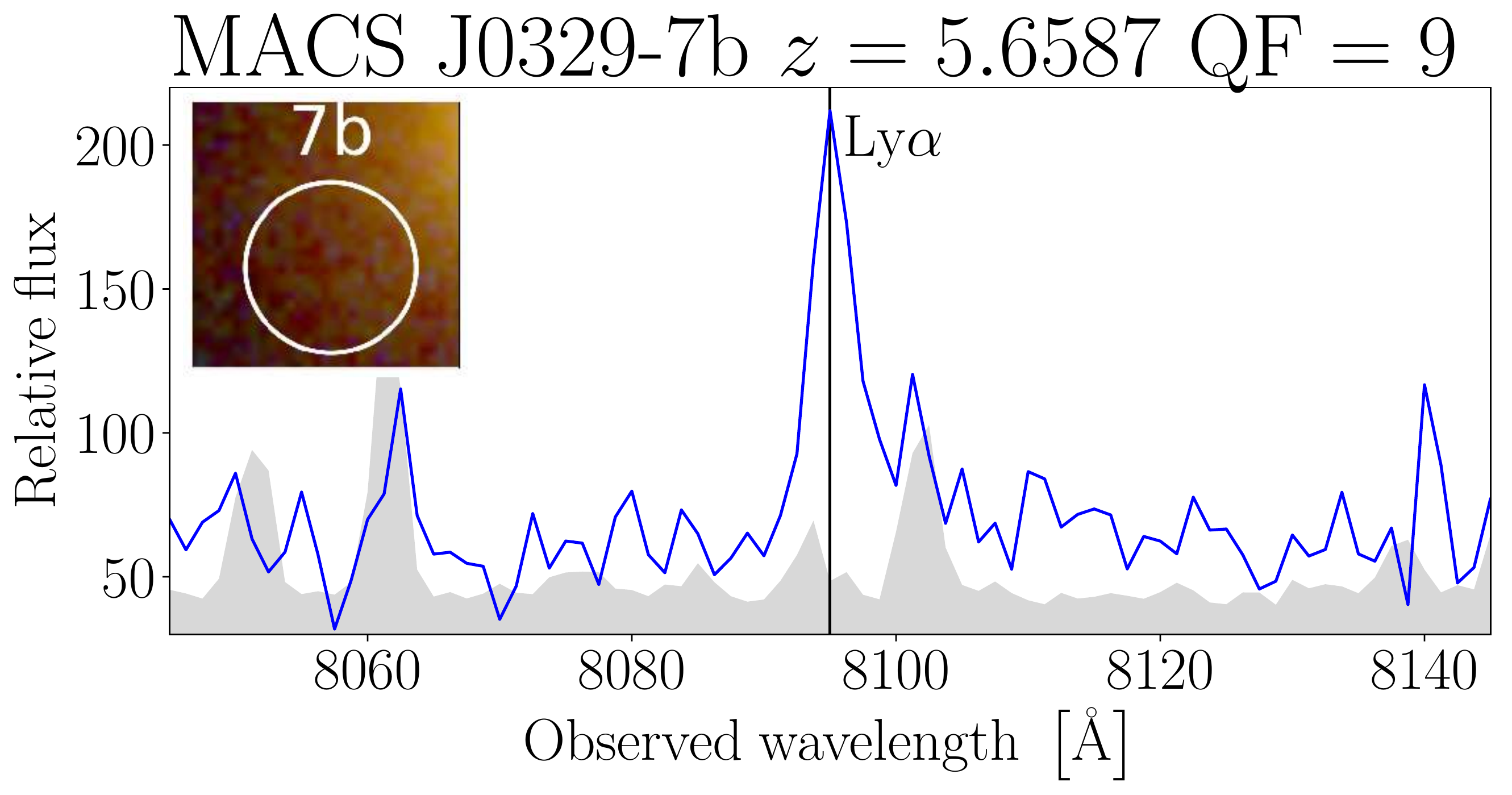}
   \includegraphics[width = 0.666\columnwidth]{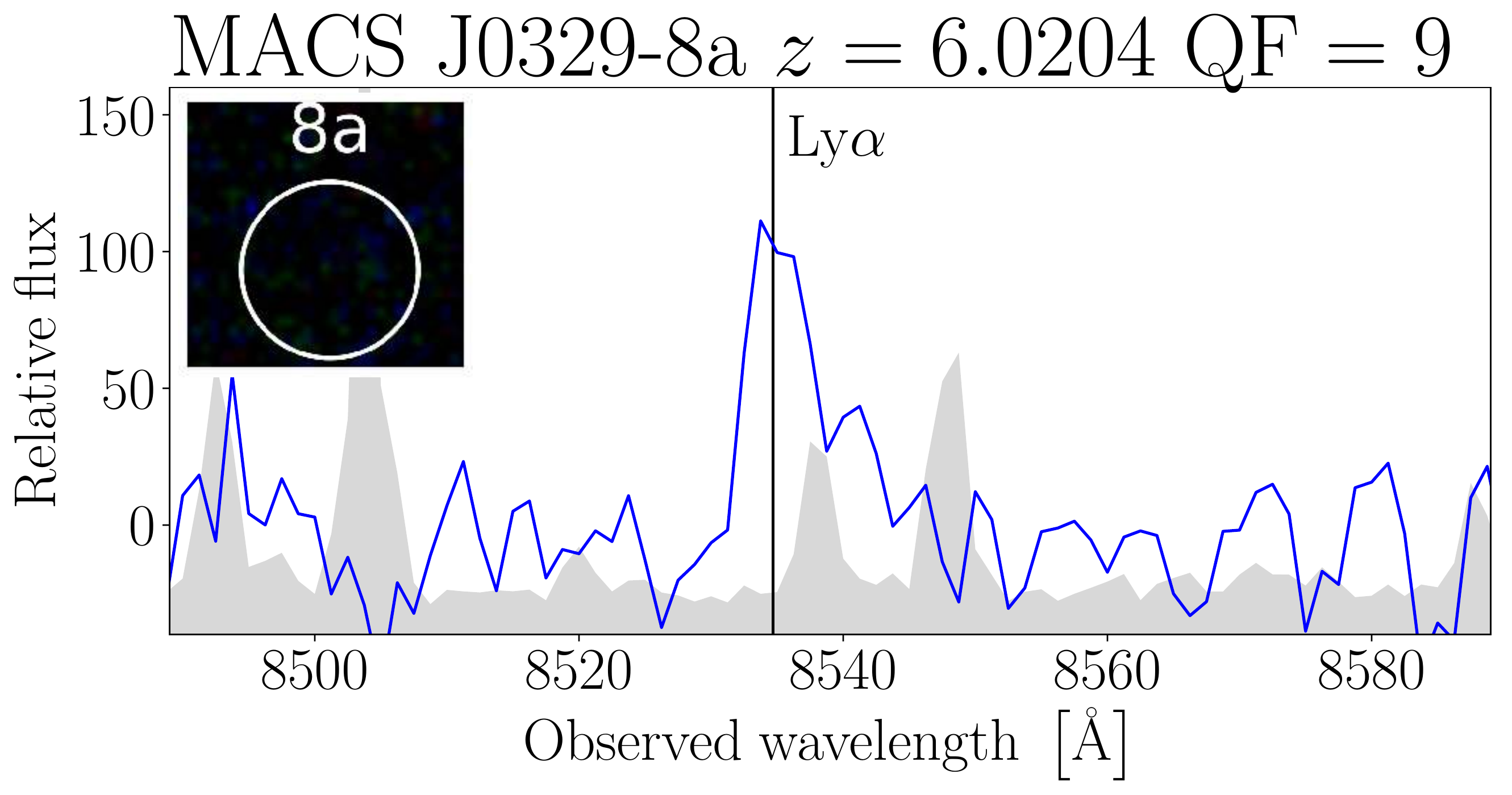}
   \includegraphics[width = 0.666\columnwidth]{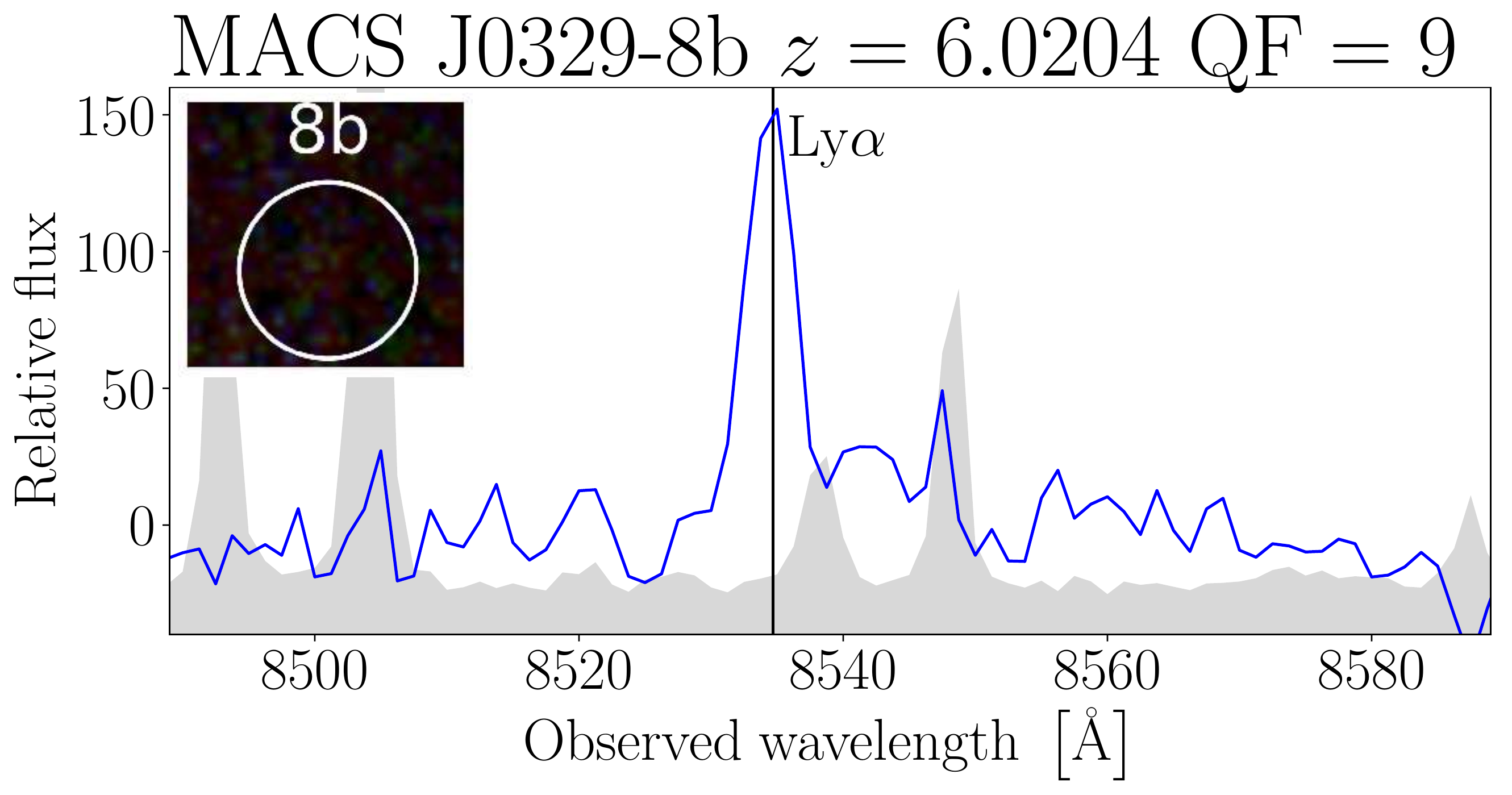}
   \includegraphics[width = 0.666\columnwidth]{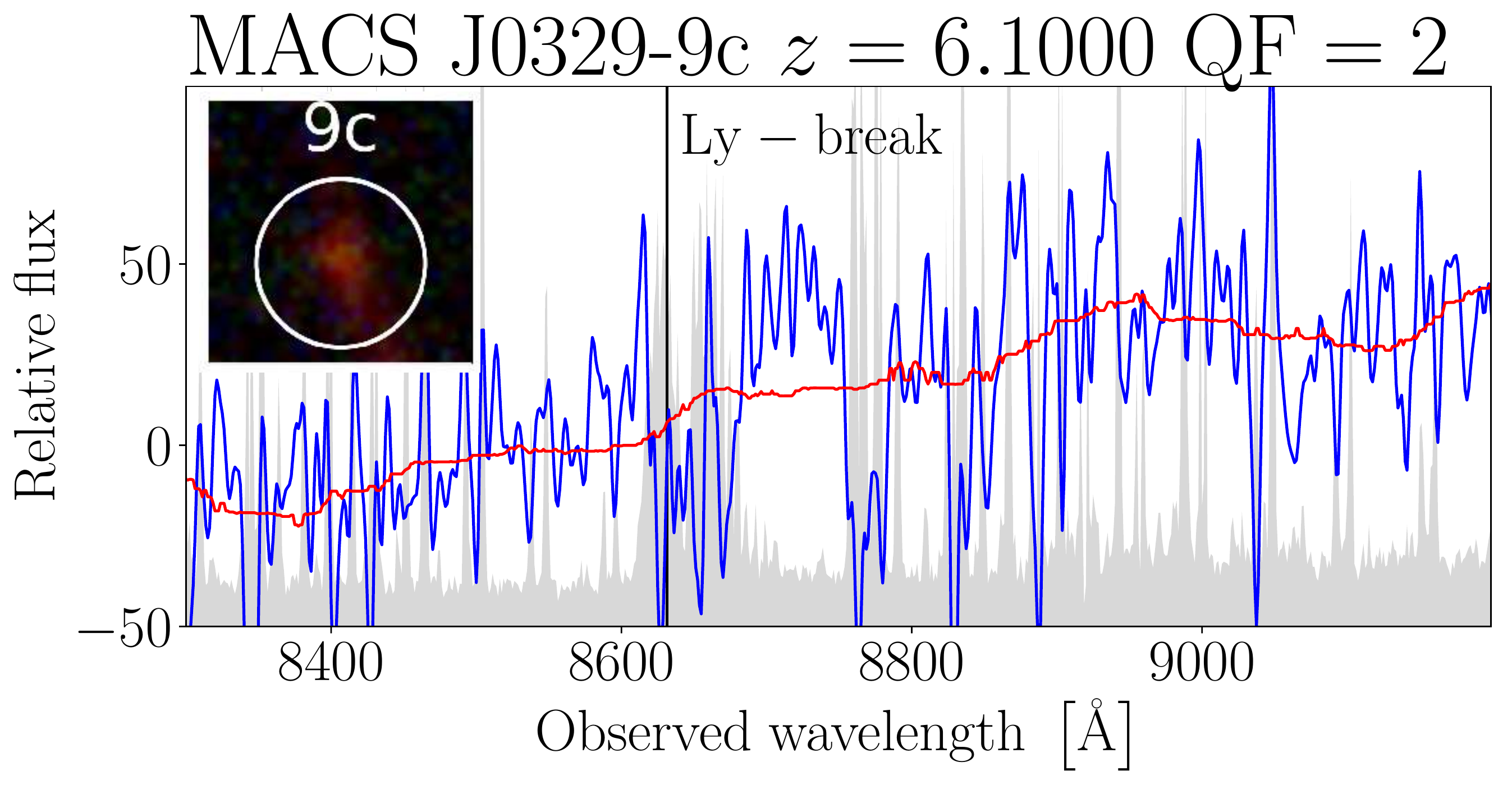}
   \includegraphics[width = 0.666\columnwidth]{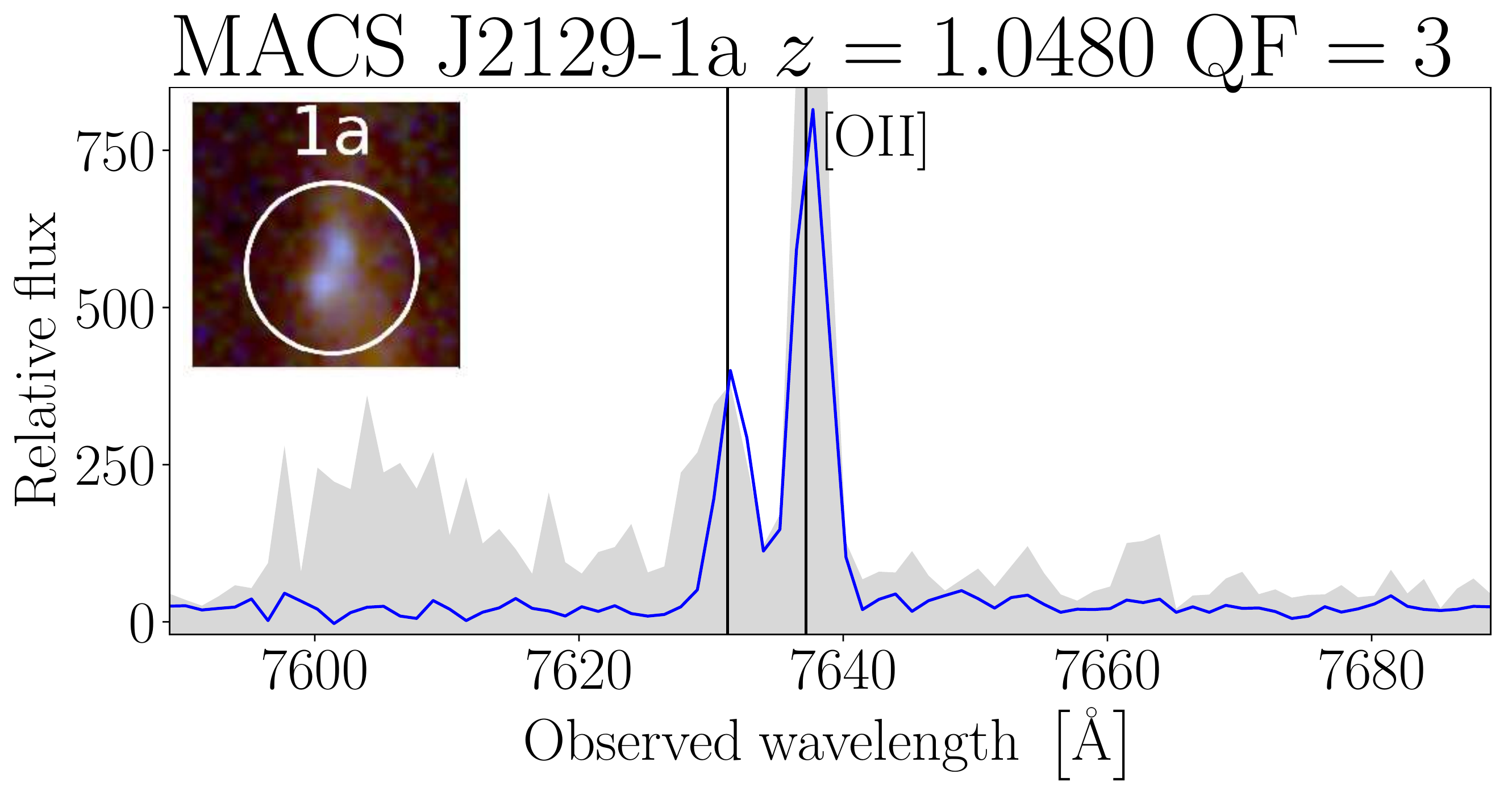}
   \includegraphics[width = 0.666\columnwidth]{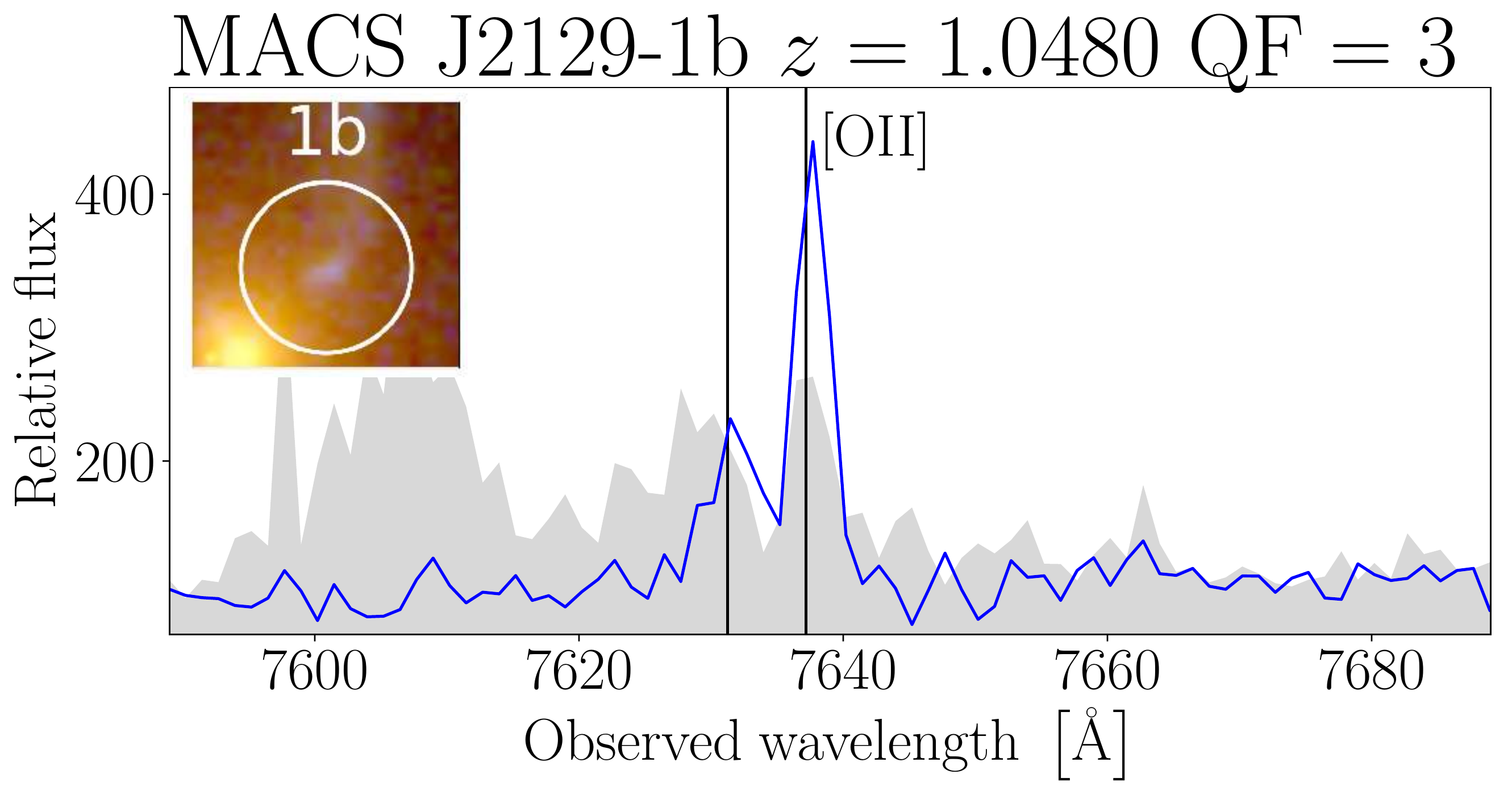}
   \includegraphics[width = 0.666\columnwidth]{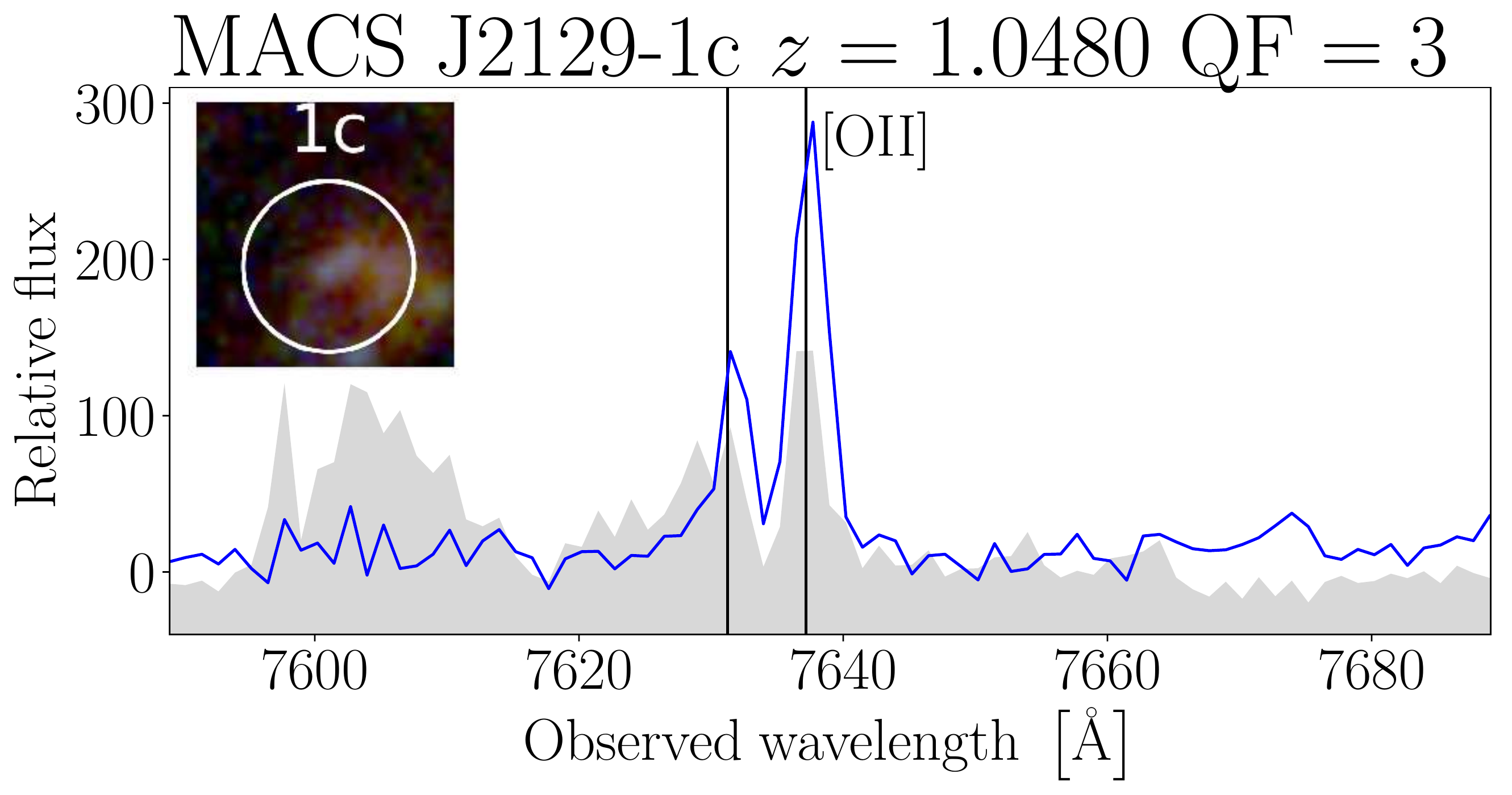}
   \includegraphics[width = 0.666\columnwidth]{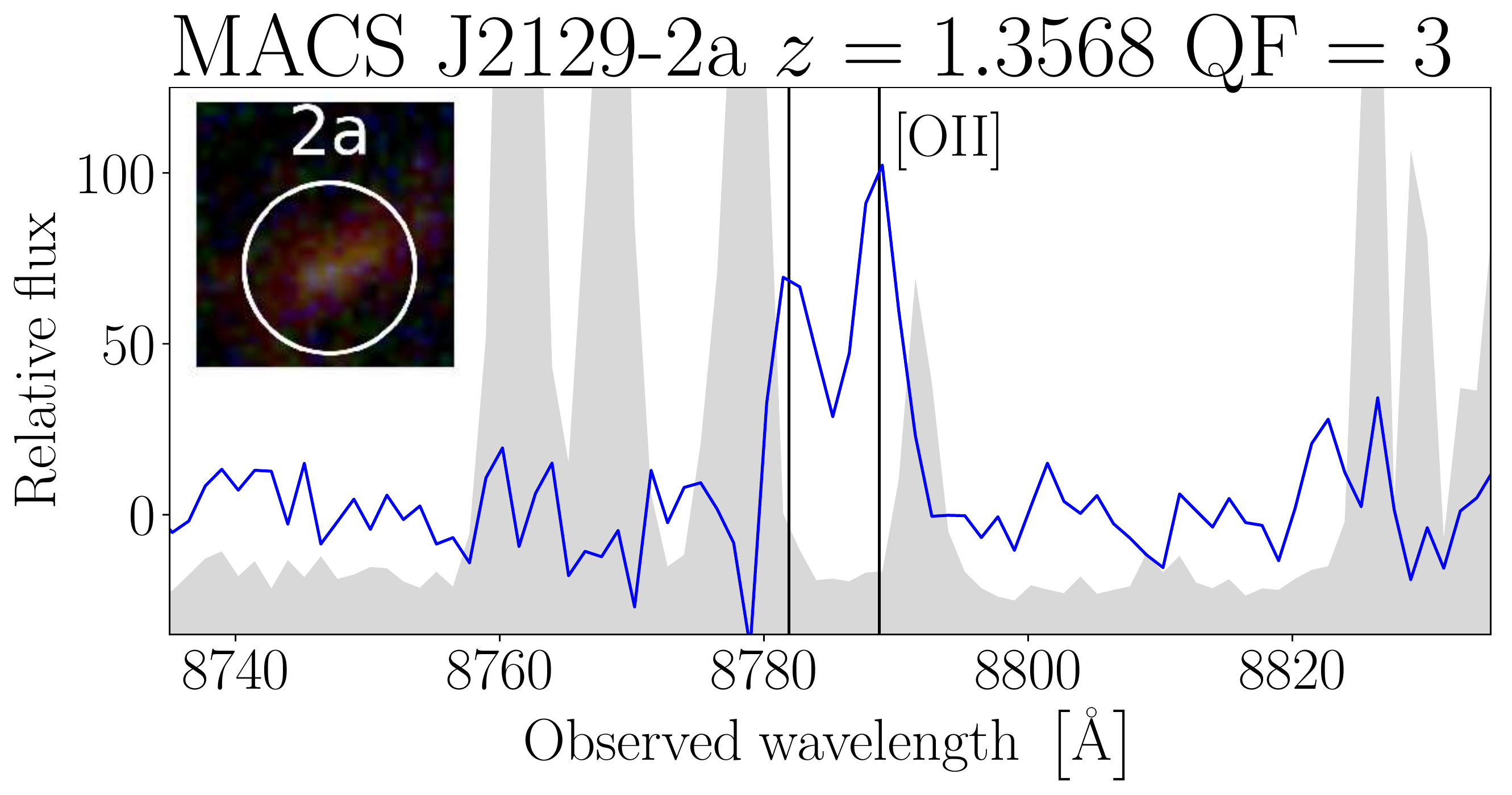}
   \includegraphics[width = 0.666\columnwidth]{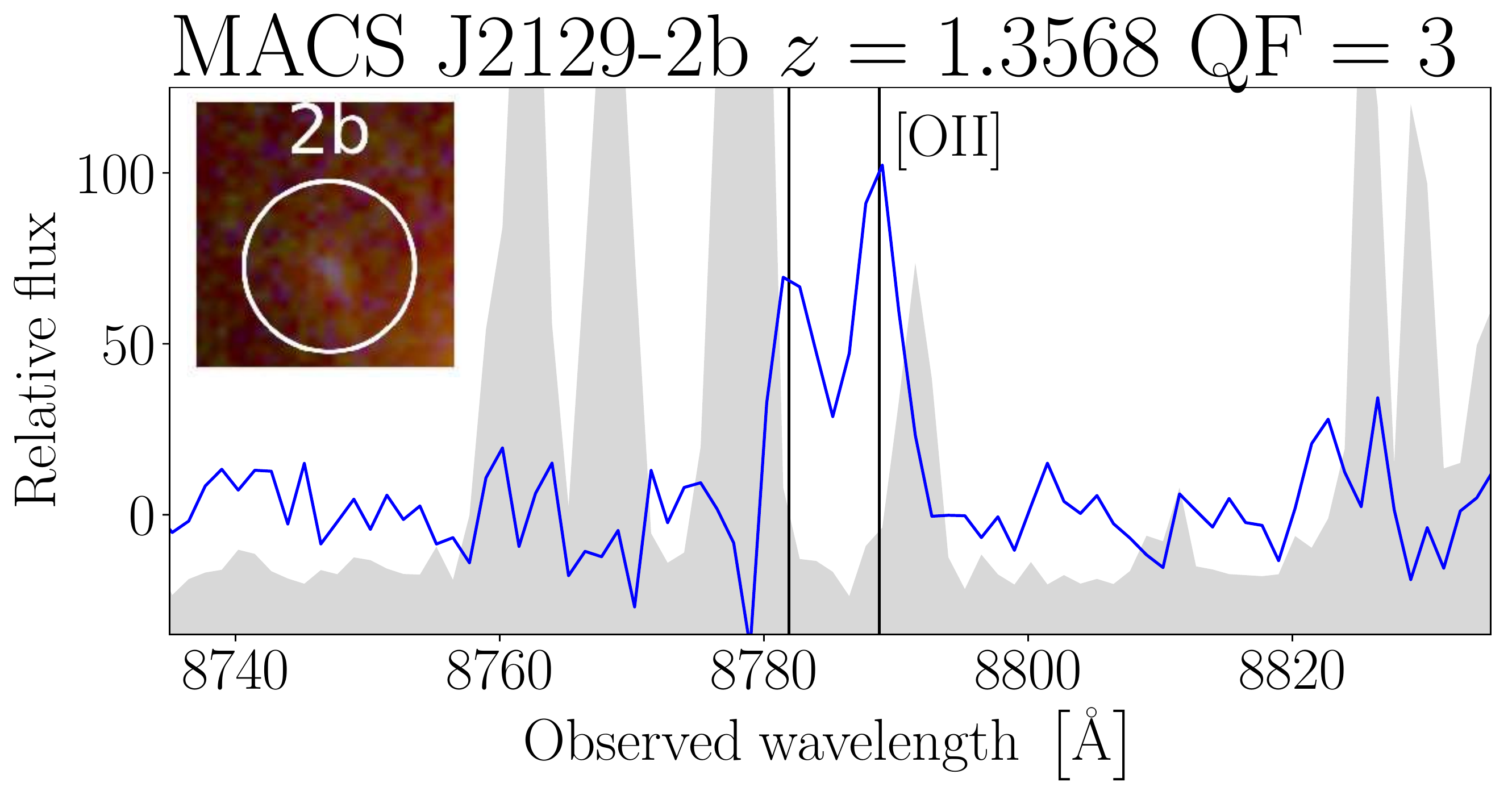}
   \includegraphics[width = 0.666\columnwidth]{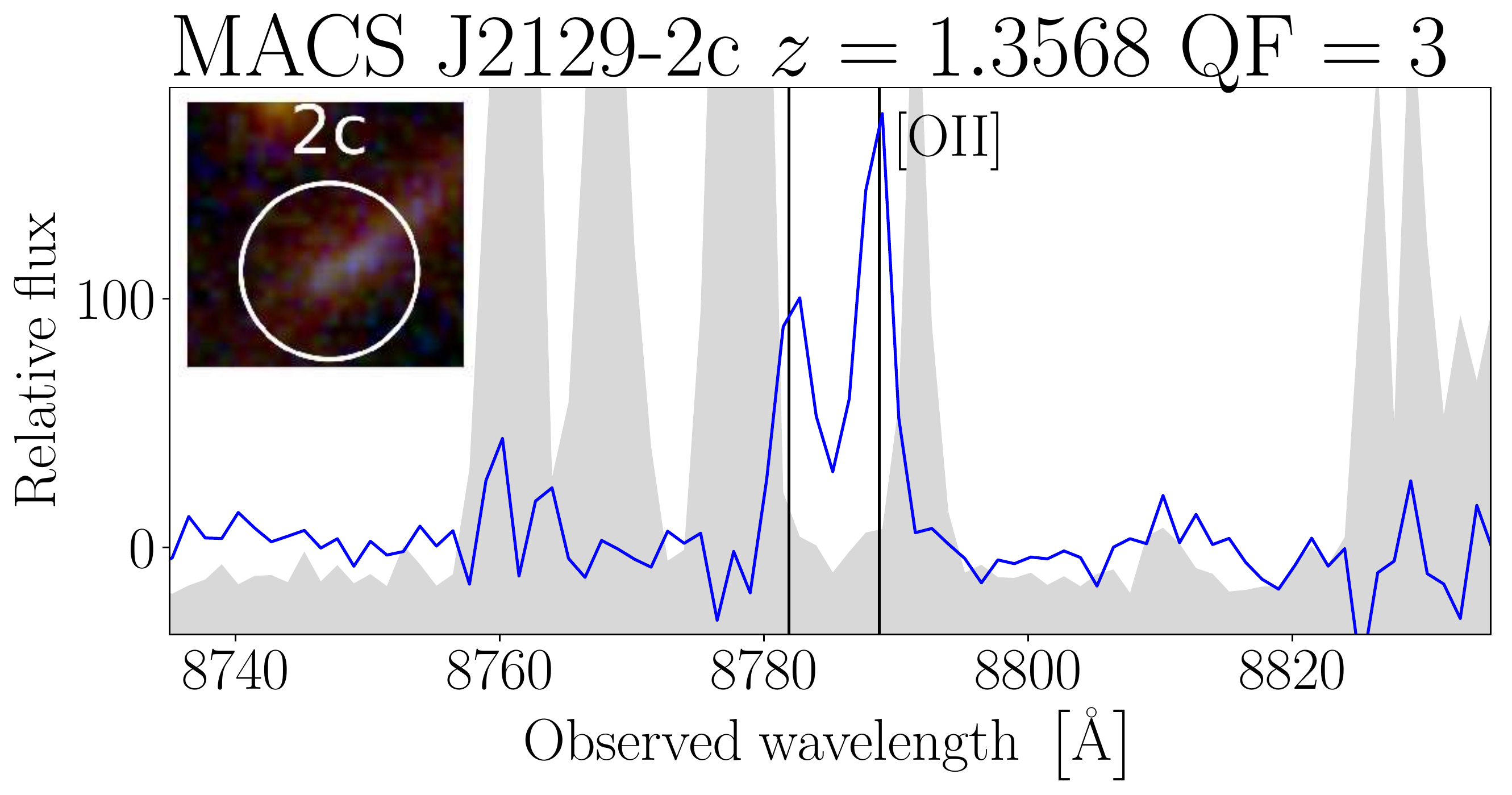}
   \includegraphics[width = 0.666\columnwidth]{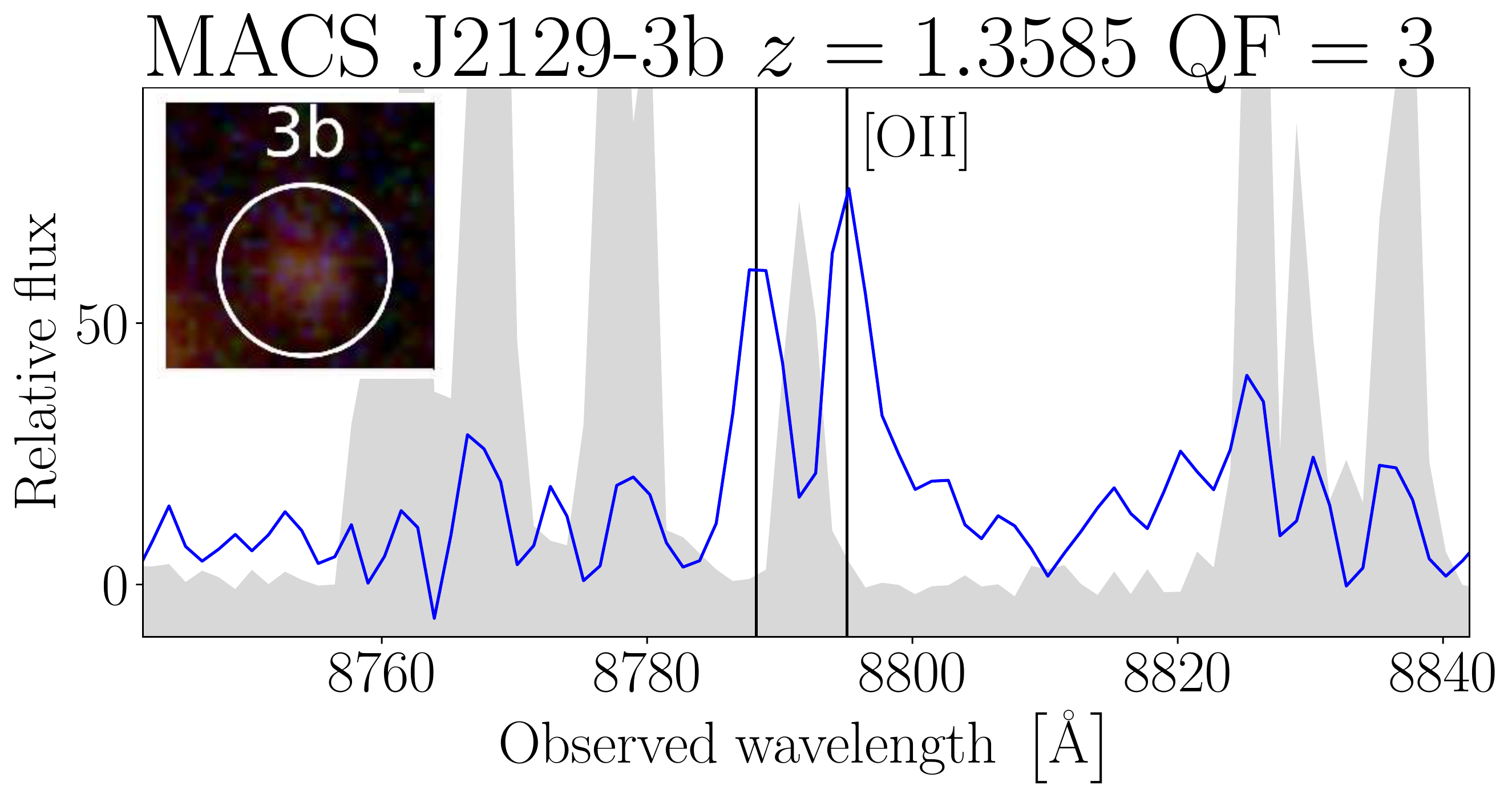}
   \includegraphics[width = 0.666\columnwidth]{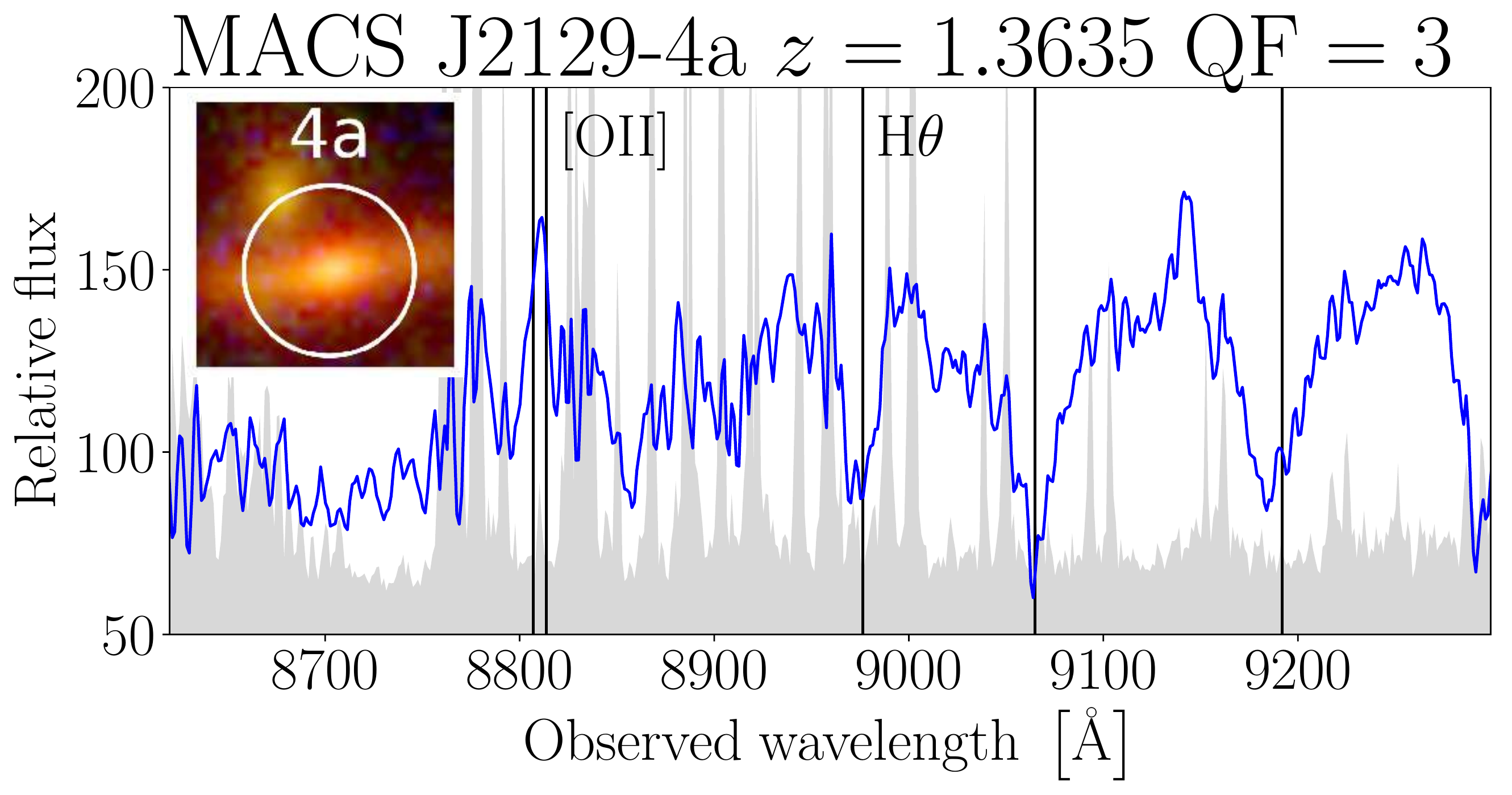}
   \includegraphics[width = 0.666\columnwidth]{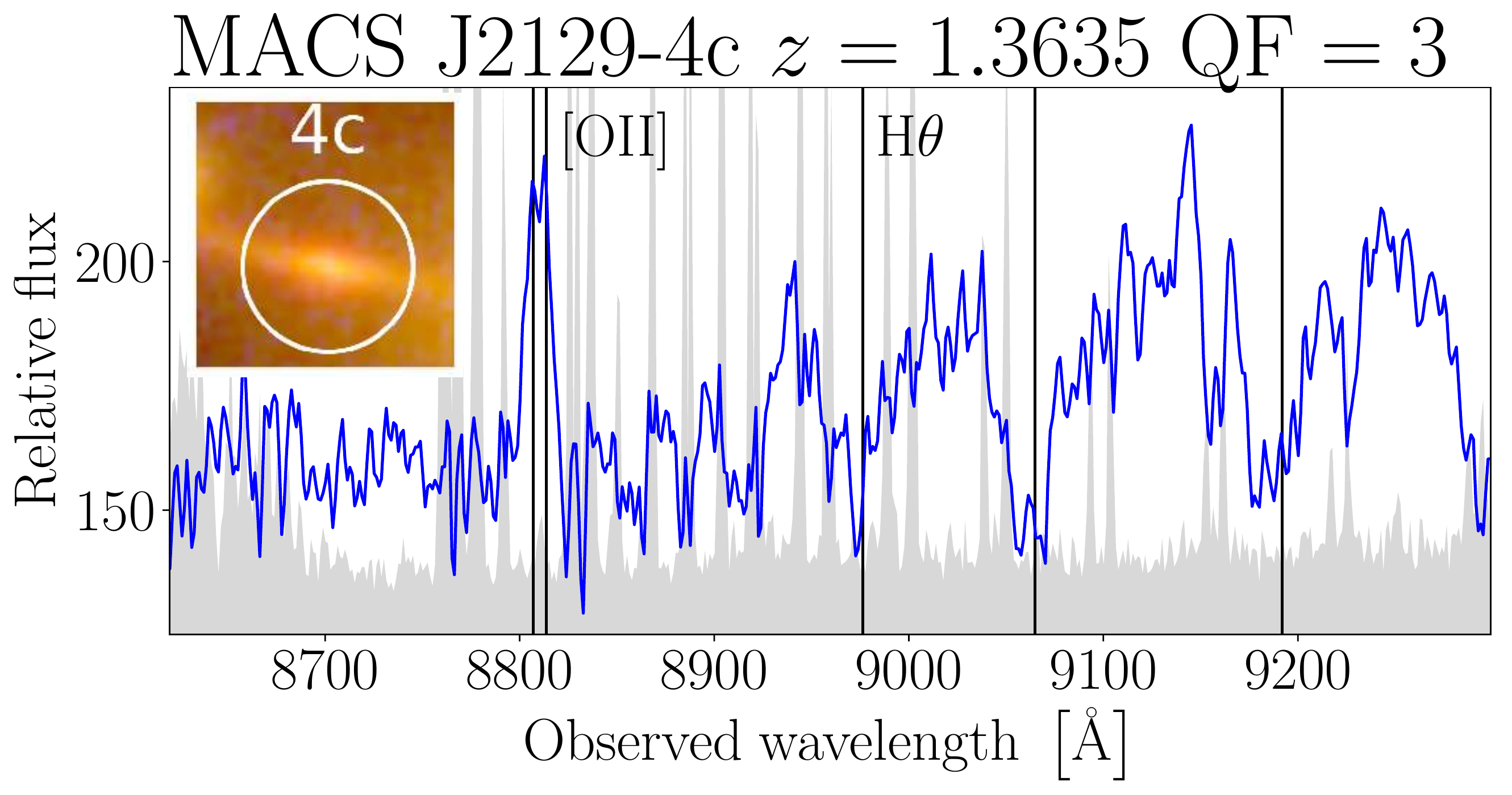}
   \includegraphics[width = 0.666\columnwidth]{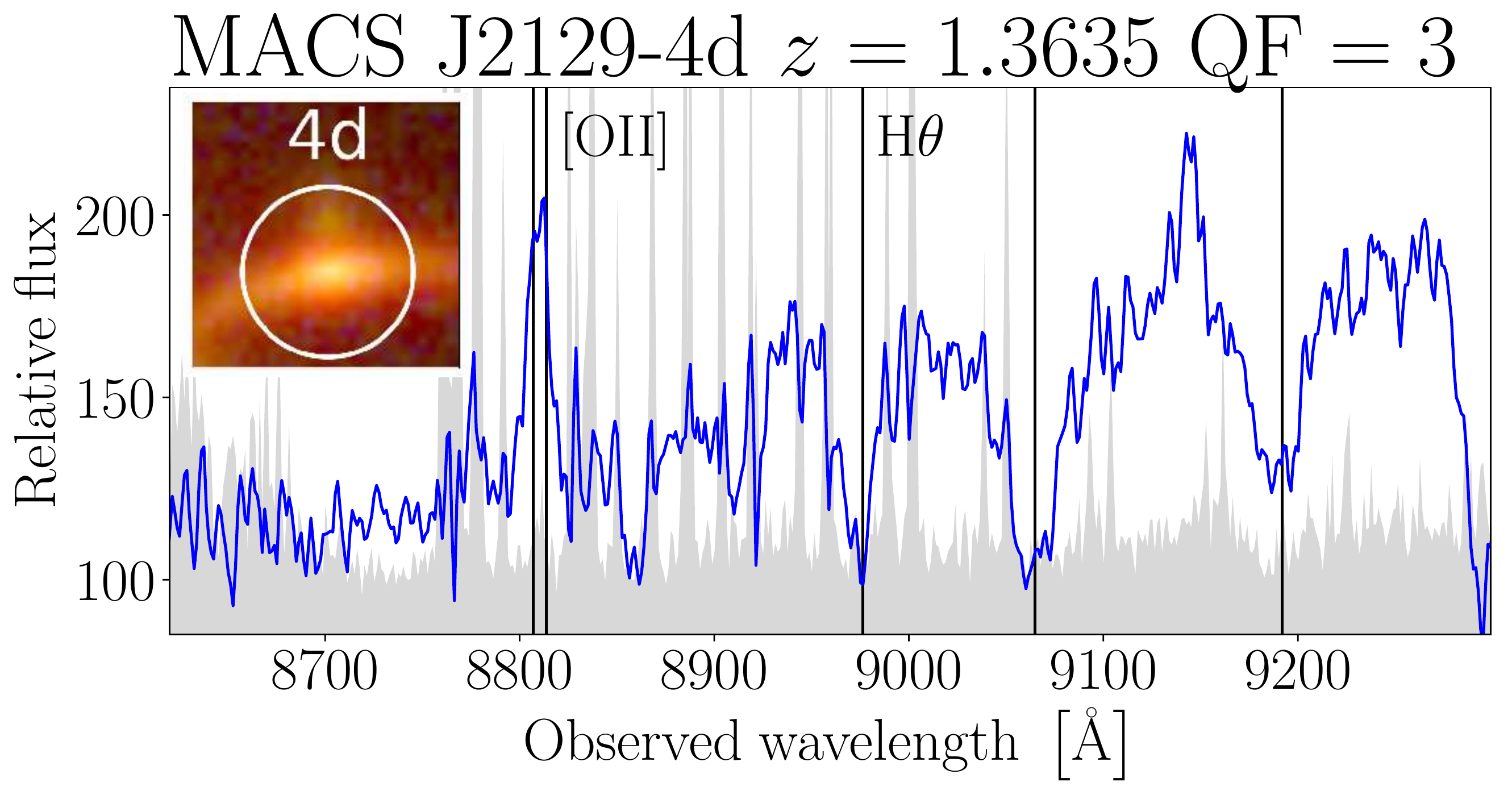}
   \includegraphics[width = 0.666\columnwidth]{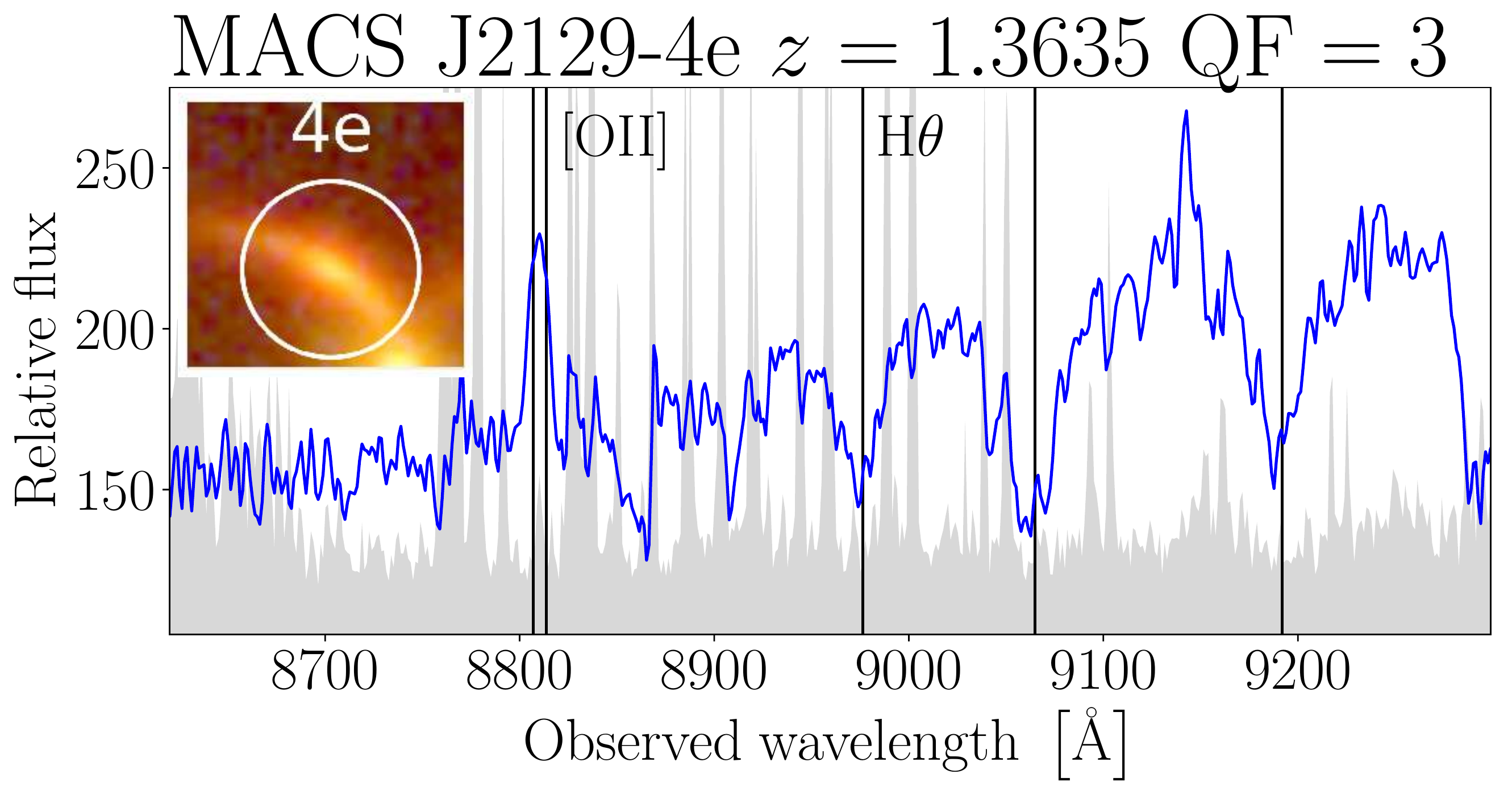}
   \includegraphics[width = 0.666\columnwidth]{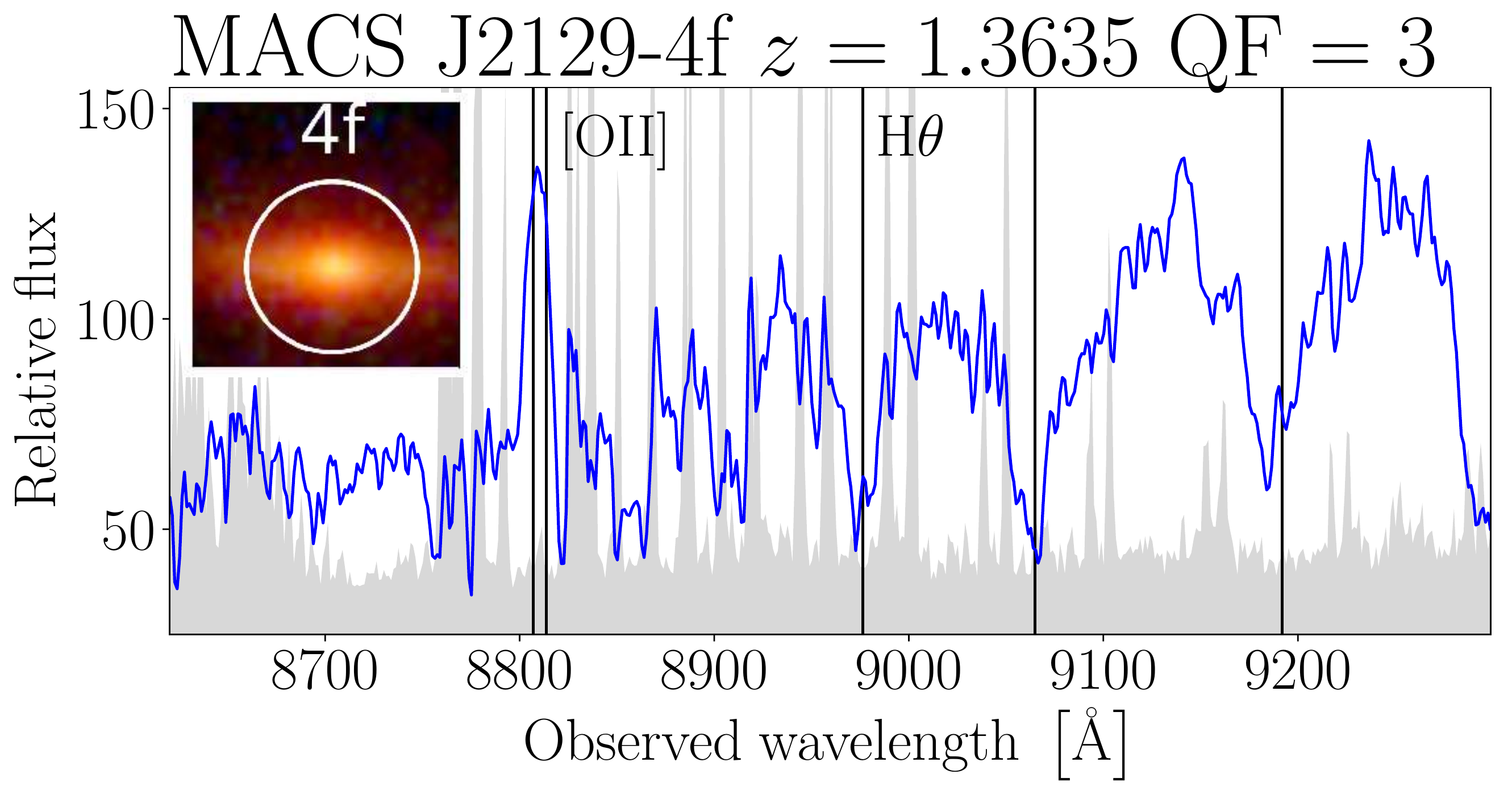}
   
  \caption{(Continued) We remark that multiple image MACS J0329-9c has an independent redshift measurement from Vanzella et al. (in prep.).}
  \label{fig:specs}
\end{figure*}

\begin{figure*}
\setcounter{figure}{\value{figure}-1}

   \includegraphics[width = 0.666\columnwidth]{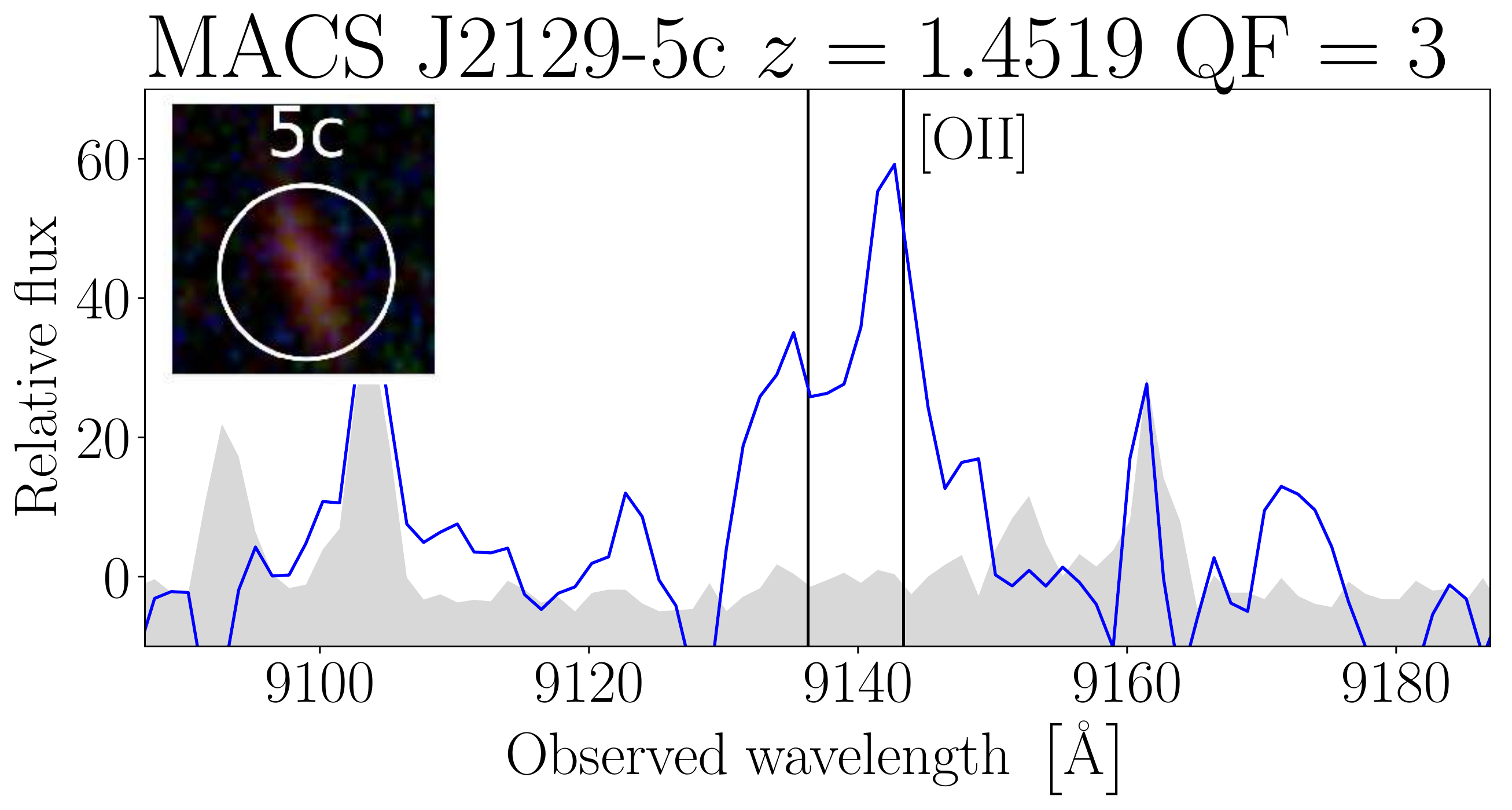}
   \includegraphics[width = 0.666\columnwidth]{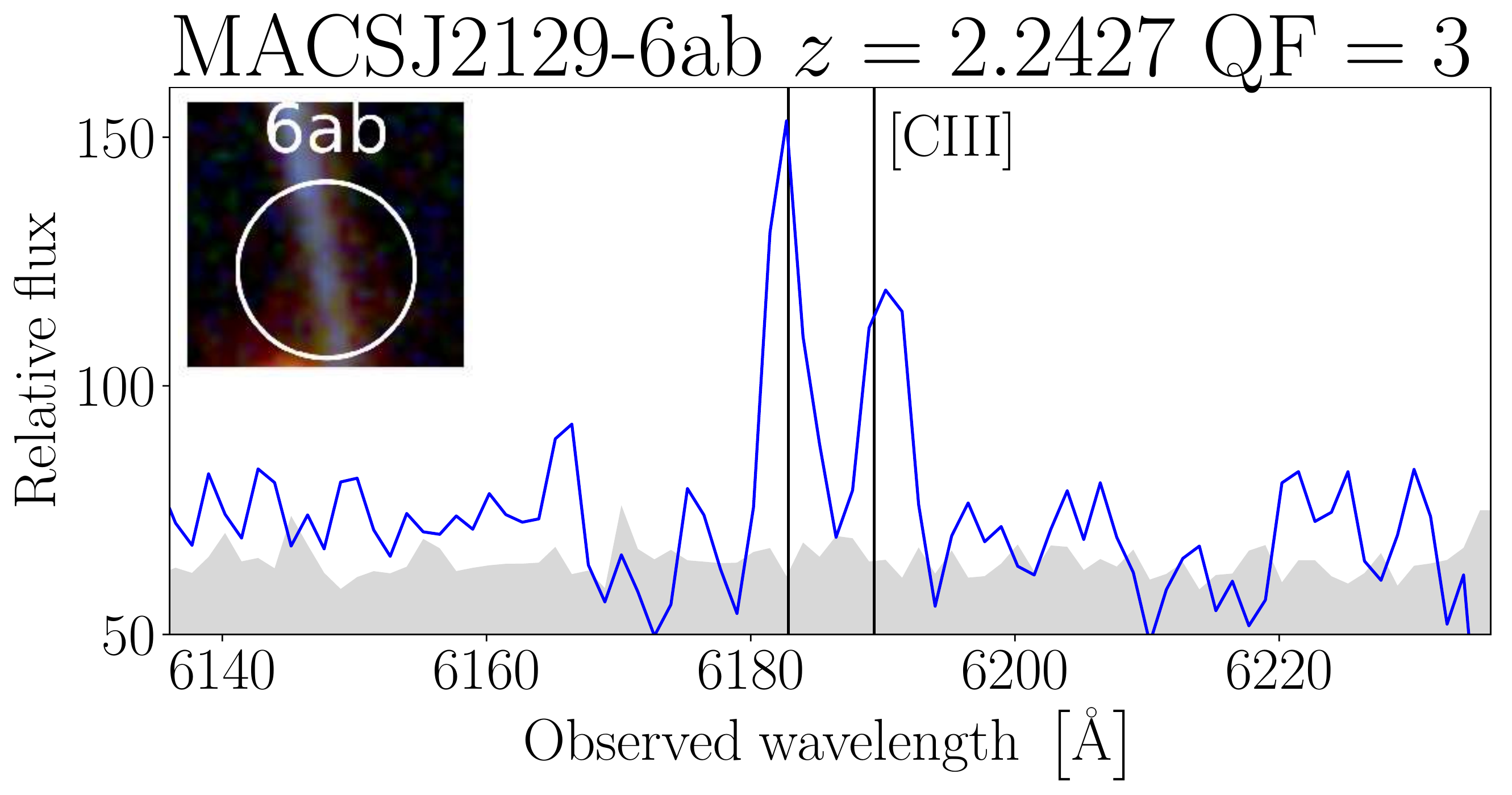}
   \includegraphics[width = 0.666\columnwidth]{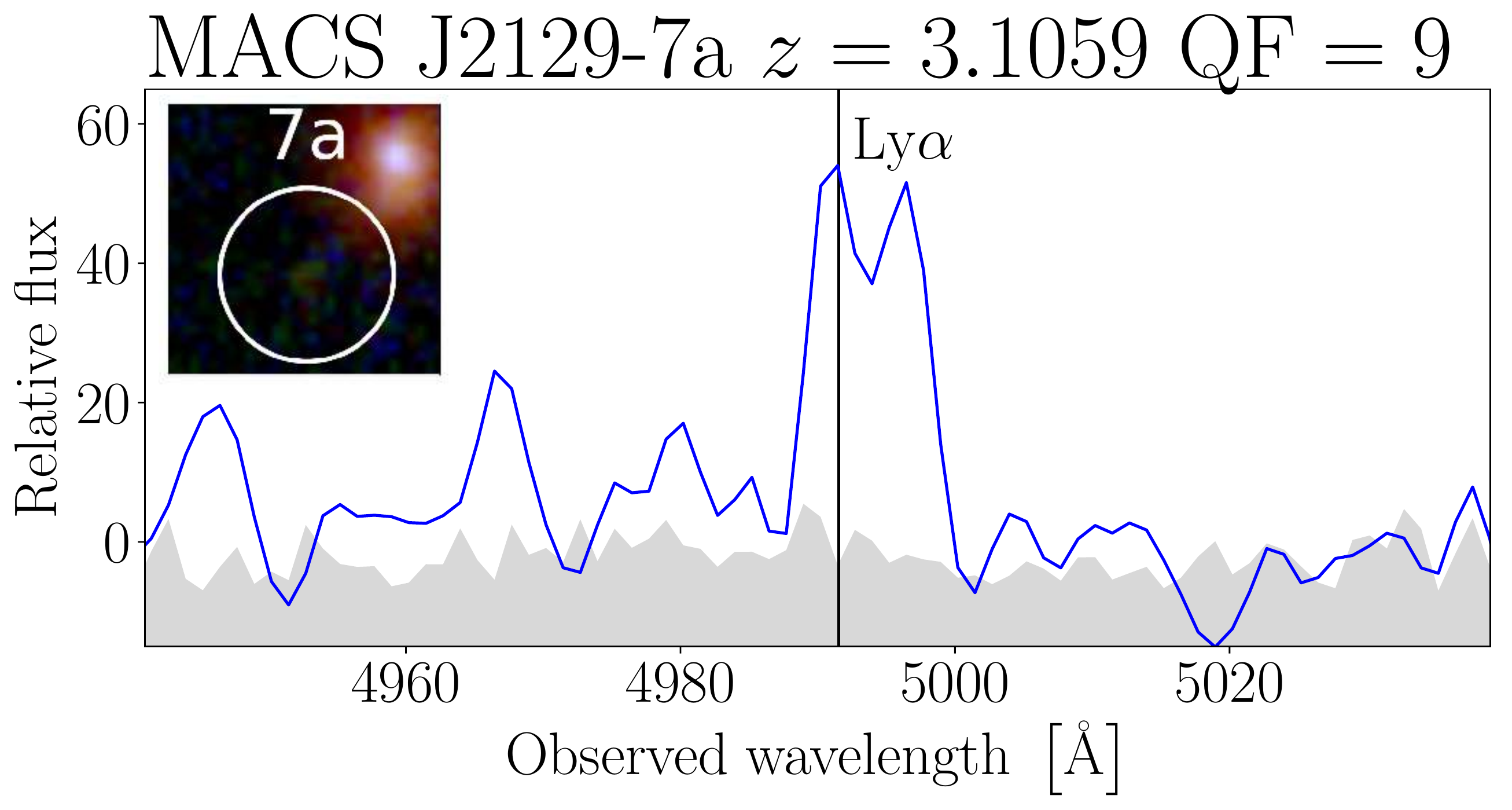}
   \includegraphics[width = 0.666\columnwidth]{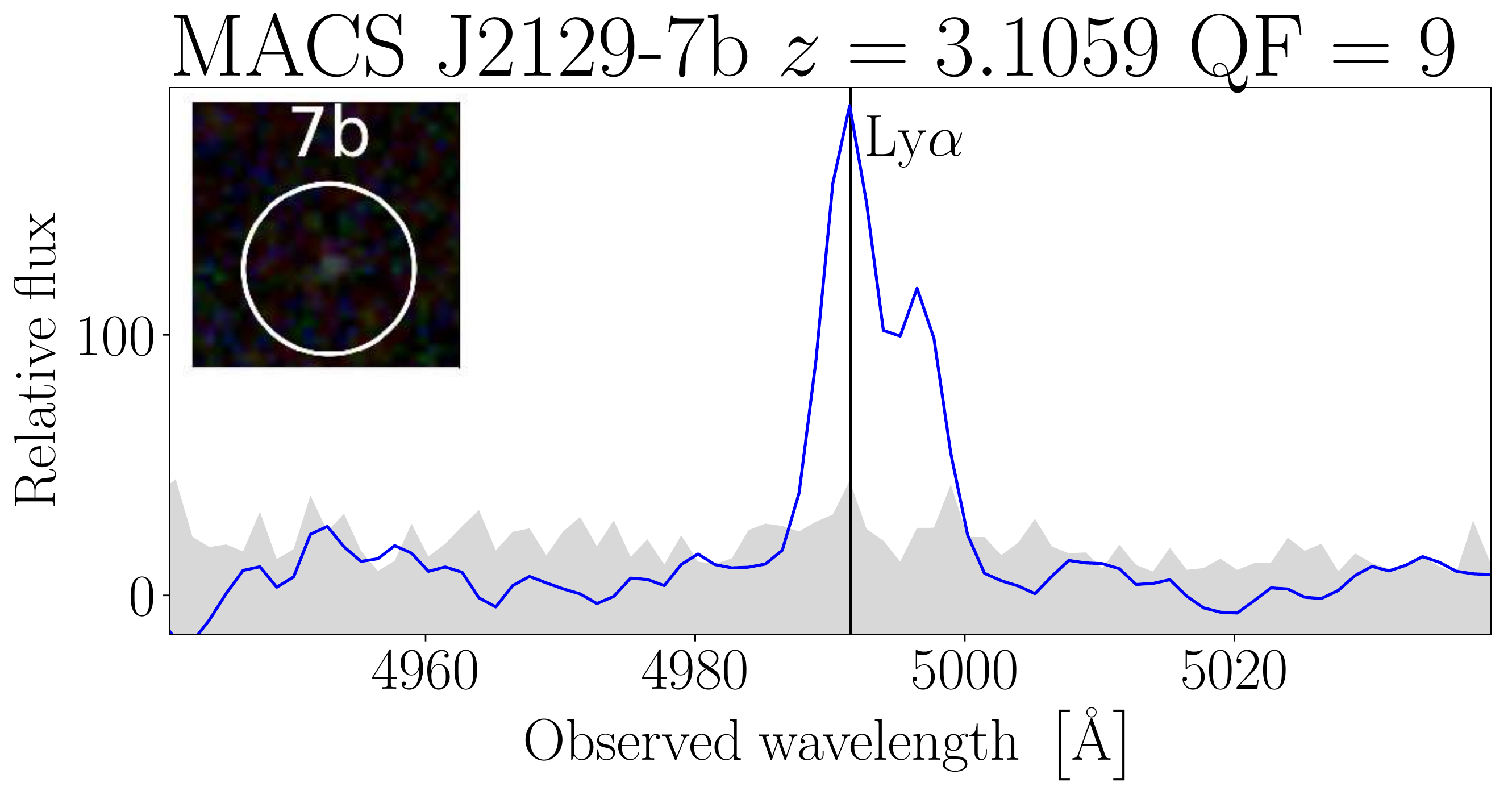}
   \includegraphics[width = 0.666\columnwidth]{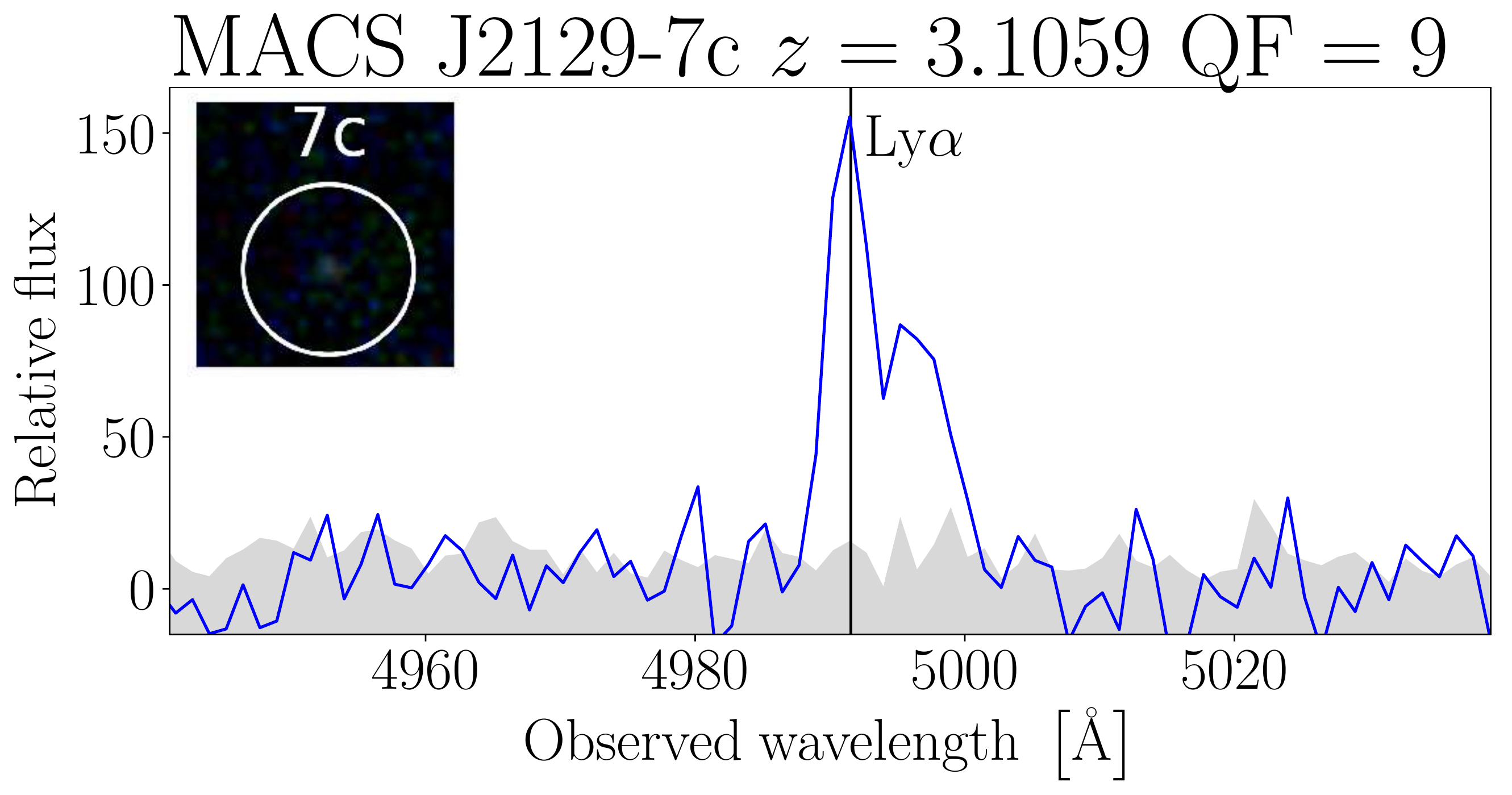}
   \includegraphics[width = 0.666\columnwidth]{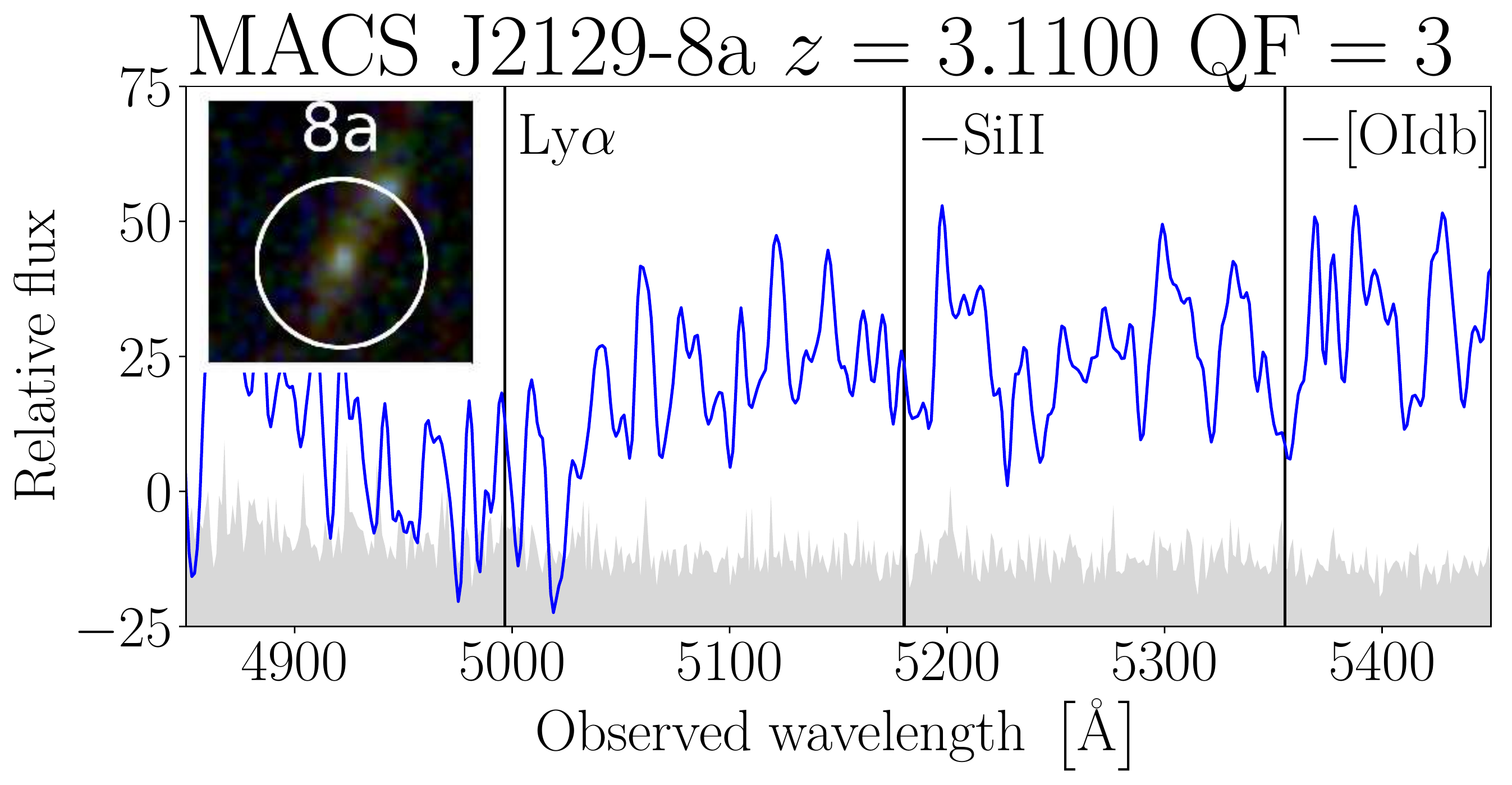}
   \includegraphics[width = 0.666\columnwidth]{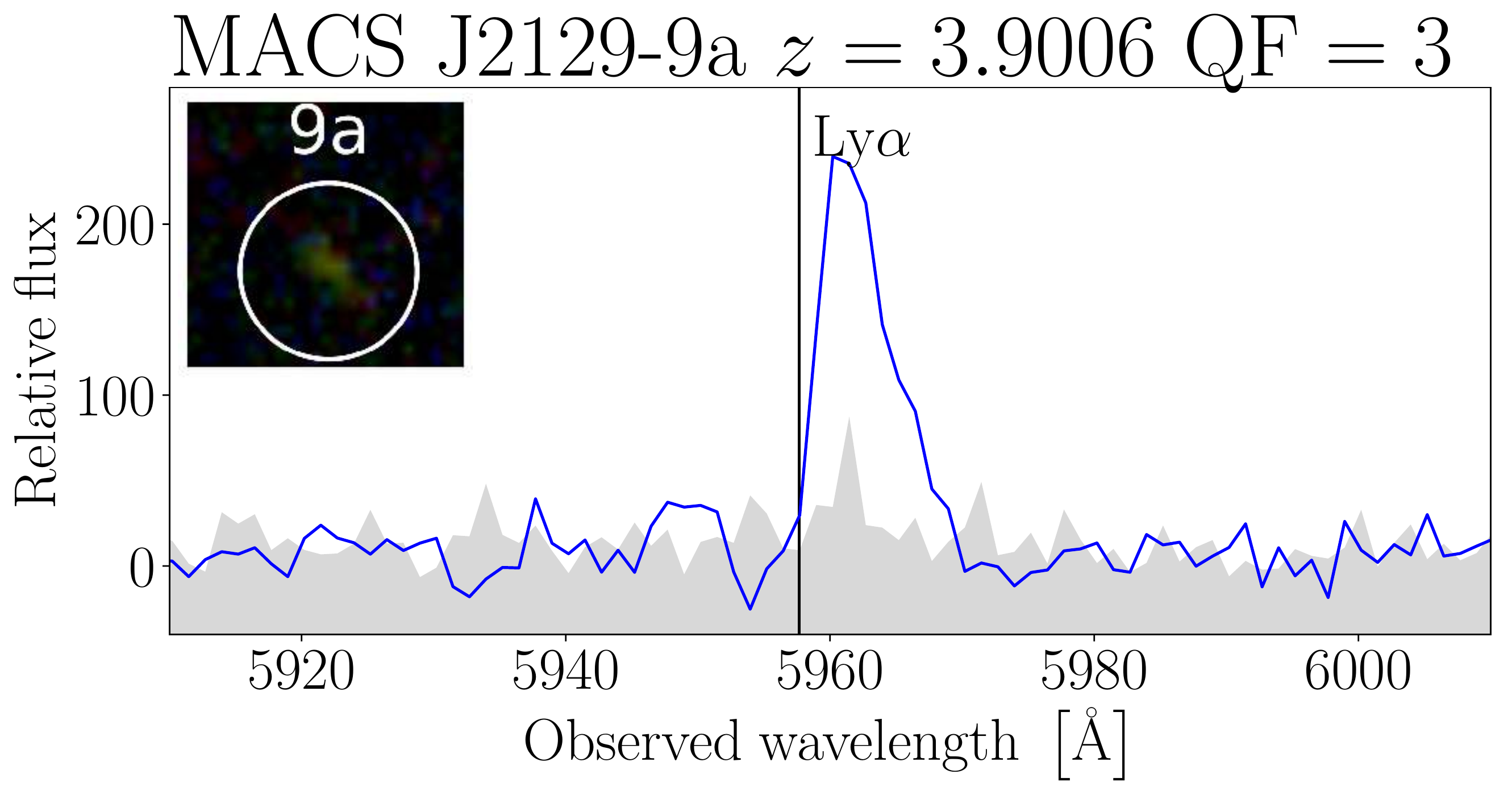}
   \includegraphics[width = 0.666\columnwidth]{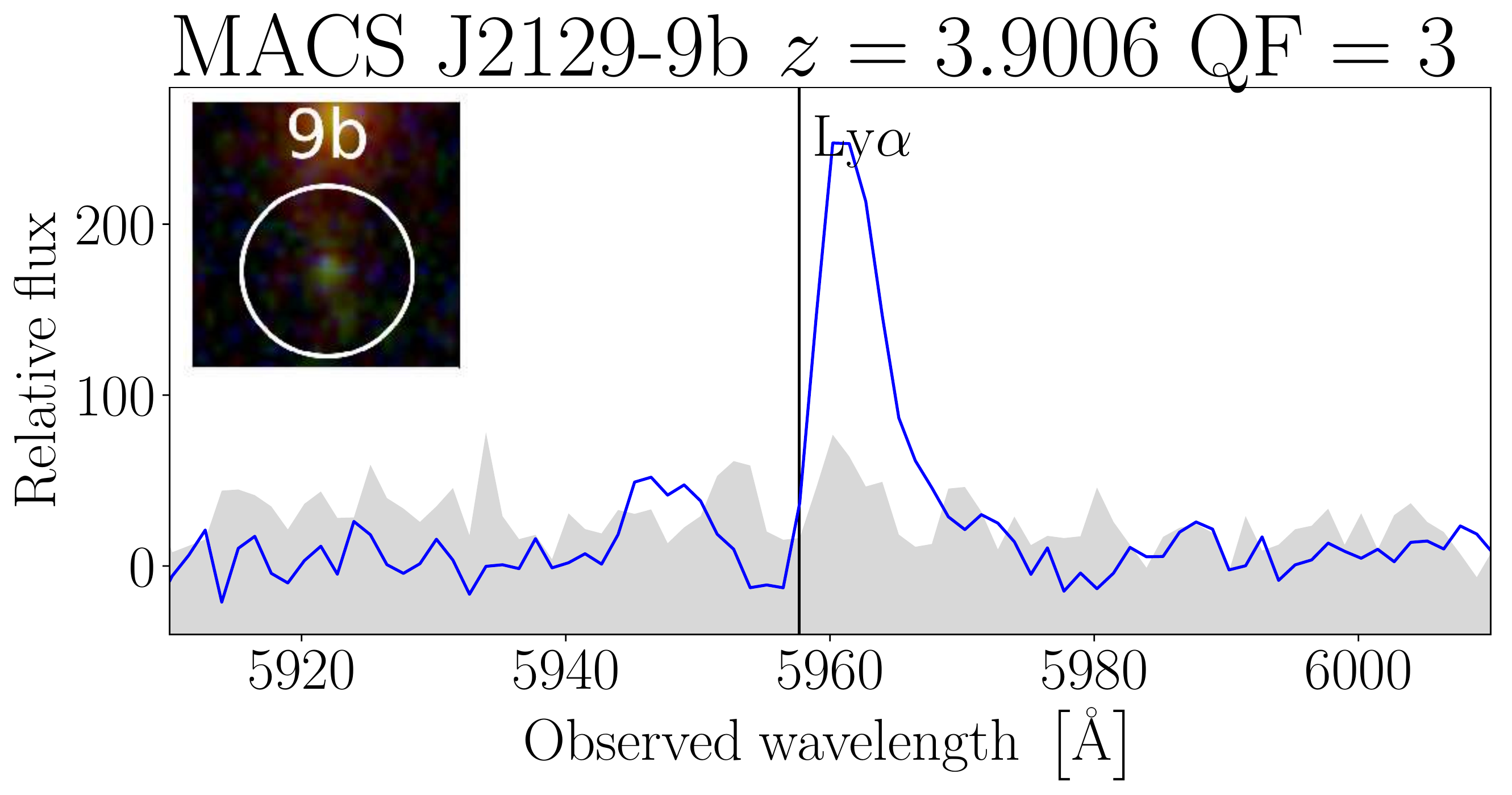}
   \includegraphics[width = 0.666\columnwidth]{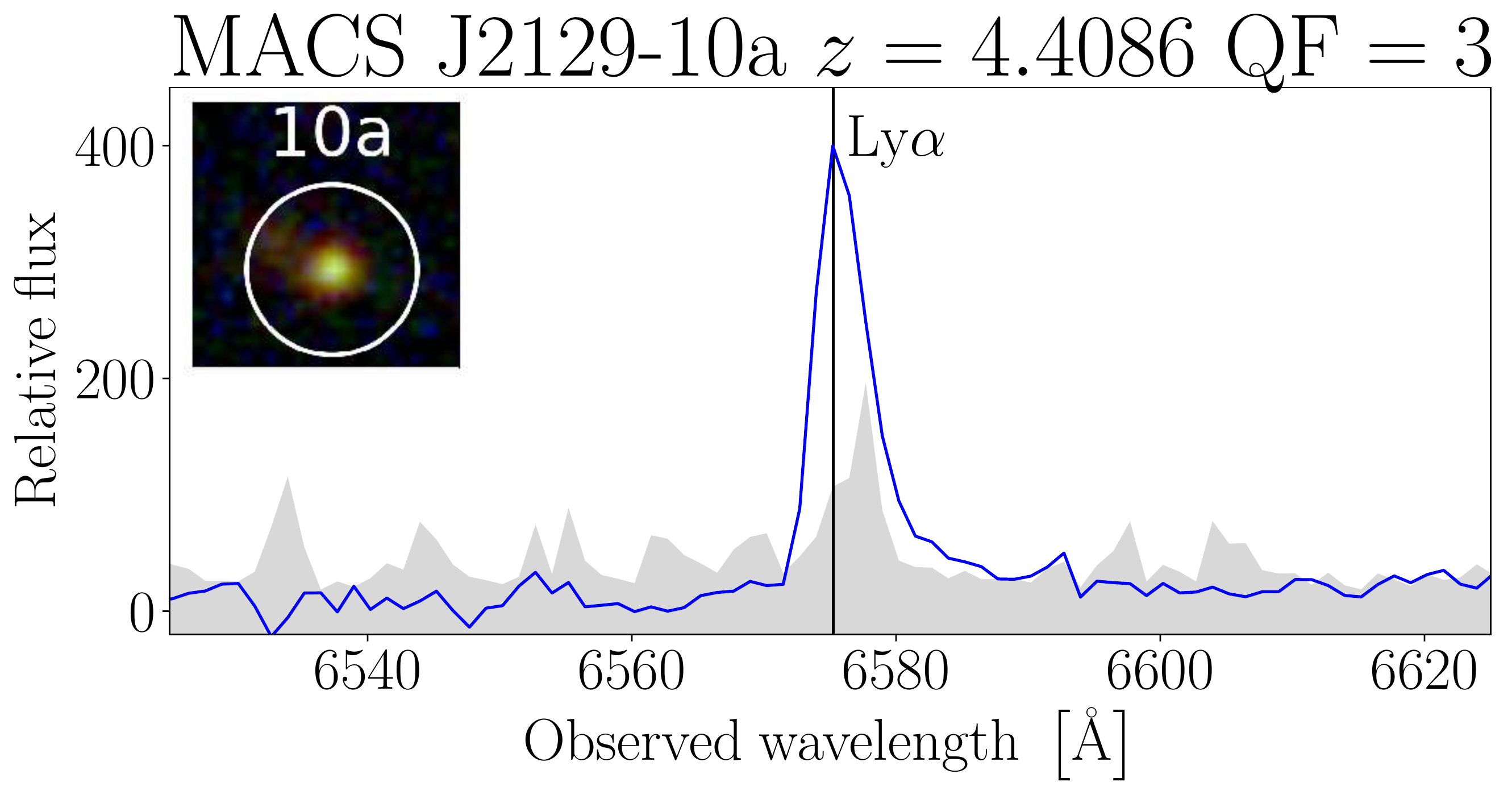}
   \includegraphics[width = 0.666\columnwidth]{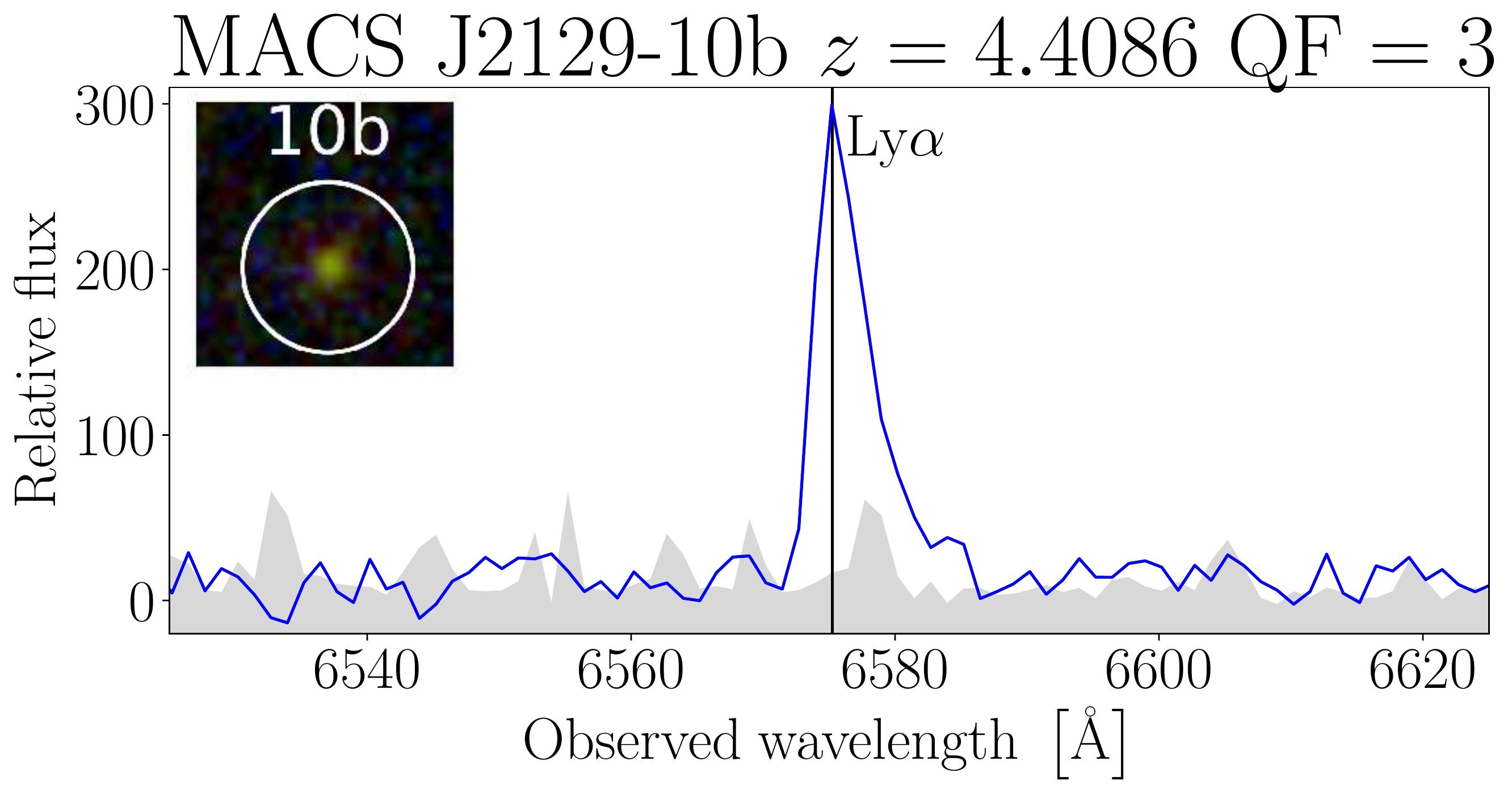}
   \includegraphics[width = 0.666\columnwidth]{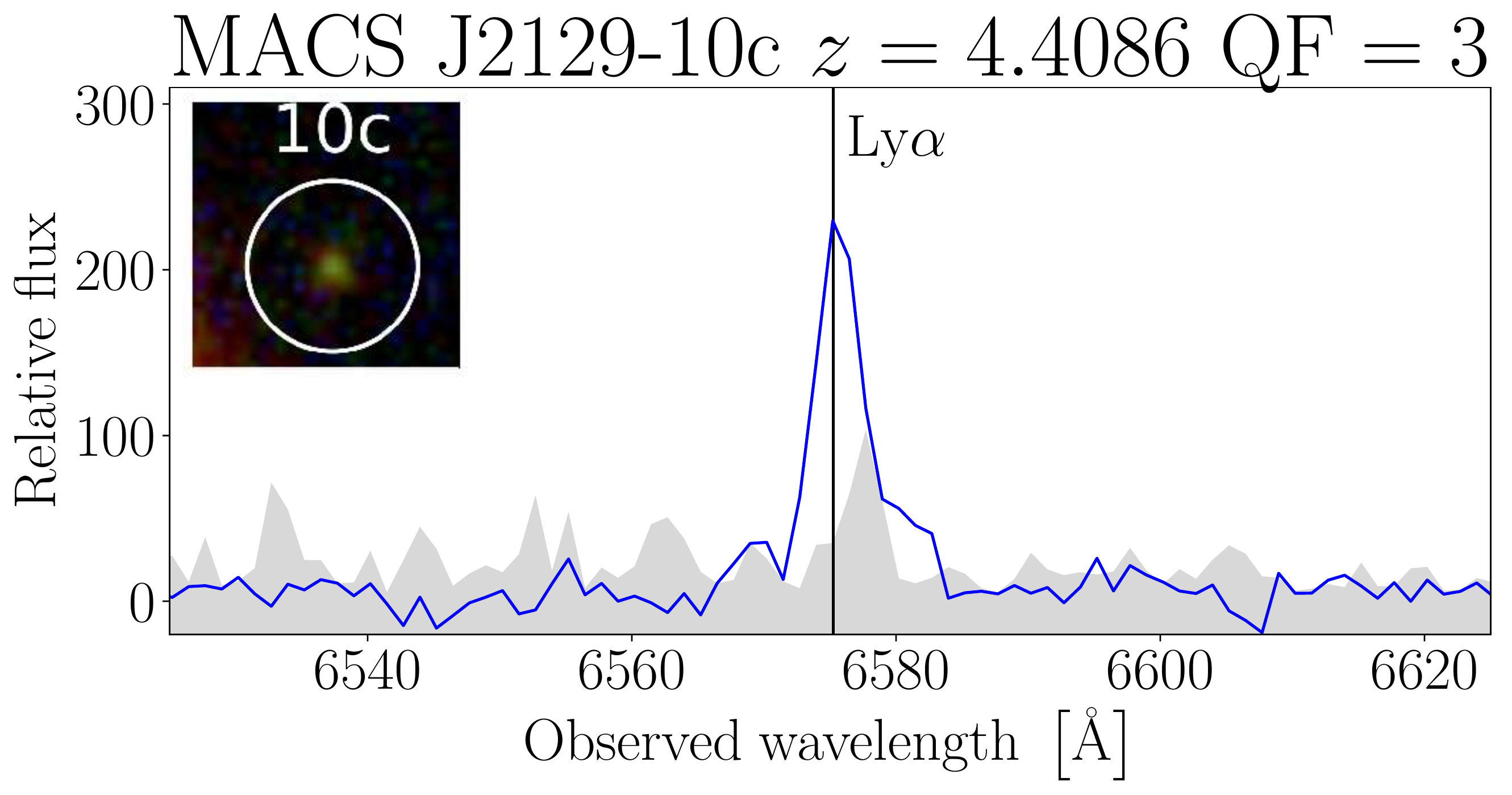}
   \includegraphics[width = 0.666\columnwidth]{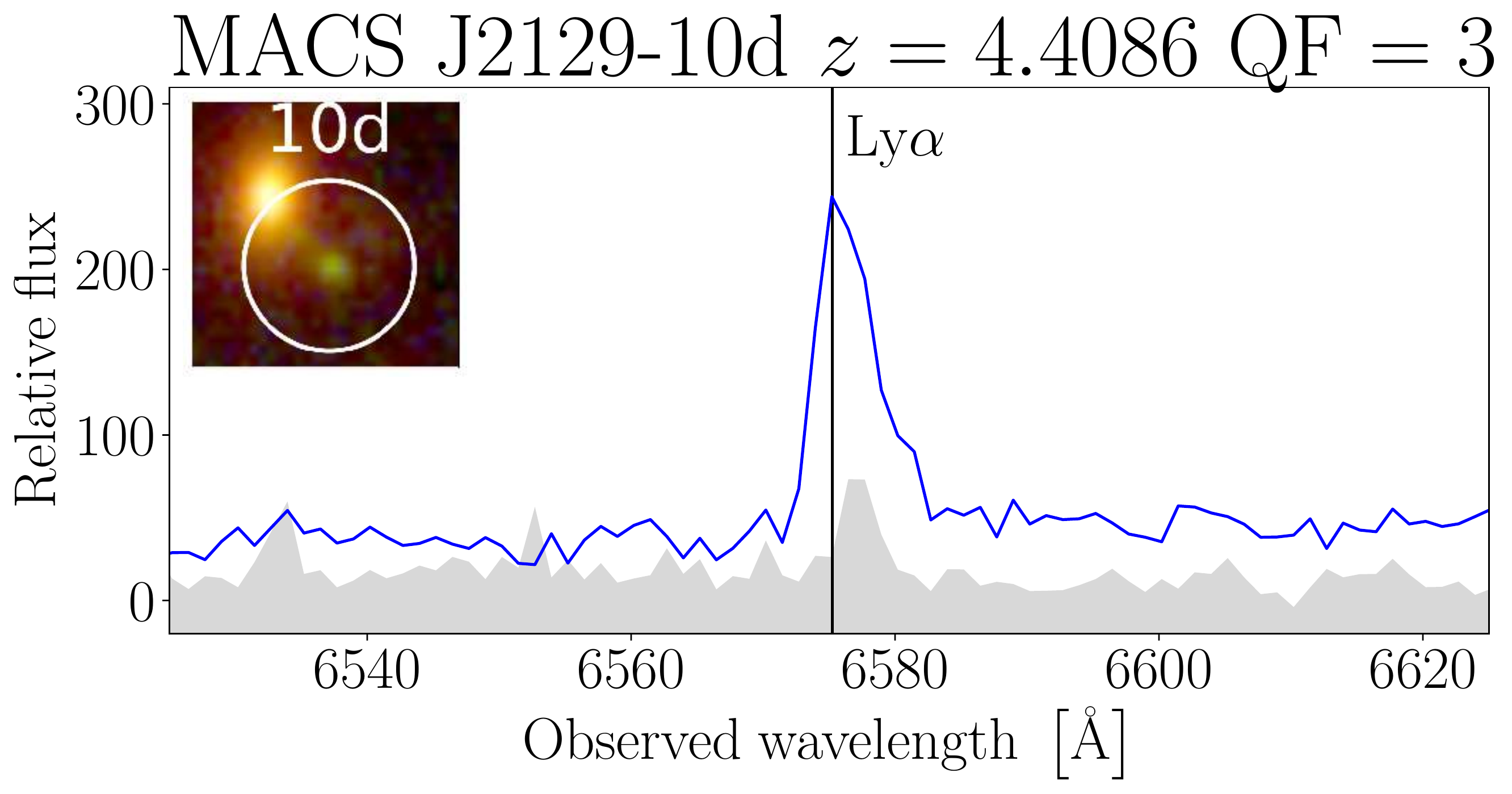}
   \includegraphics[width = 0.666\columnwidth]{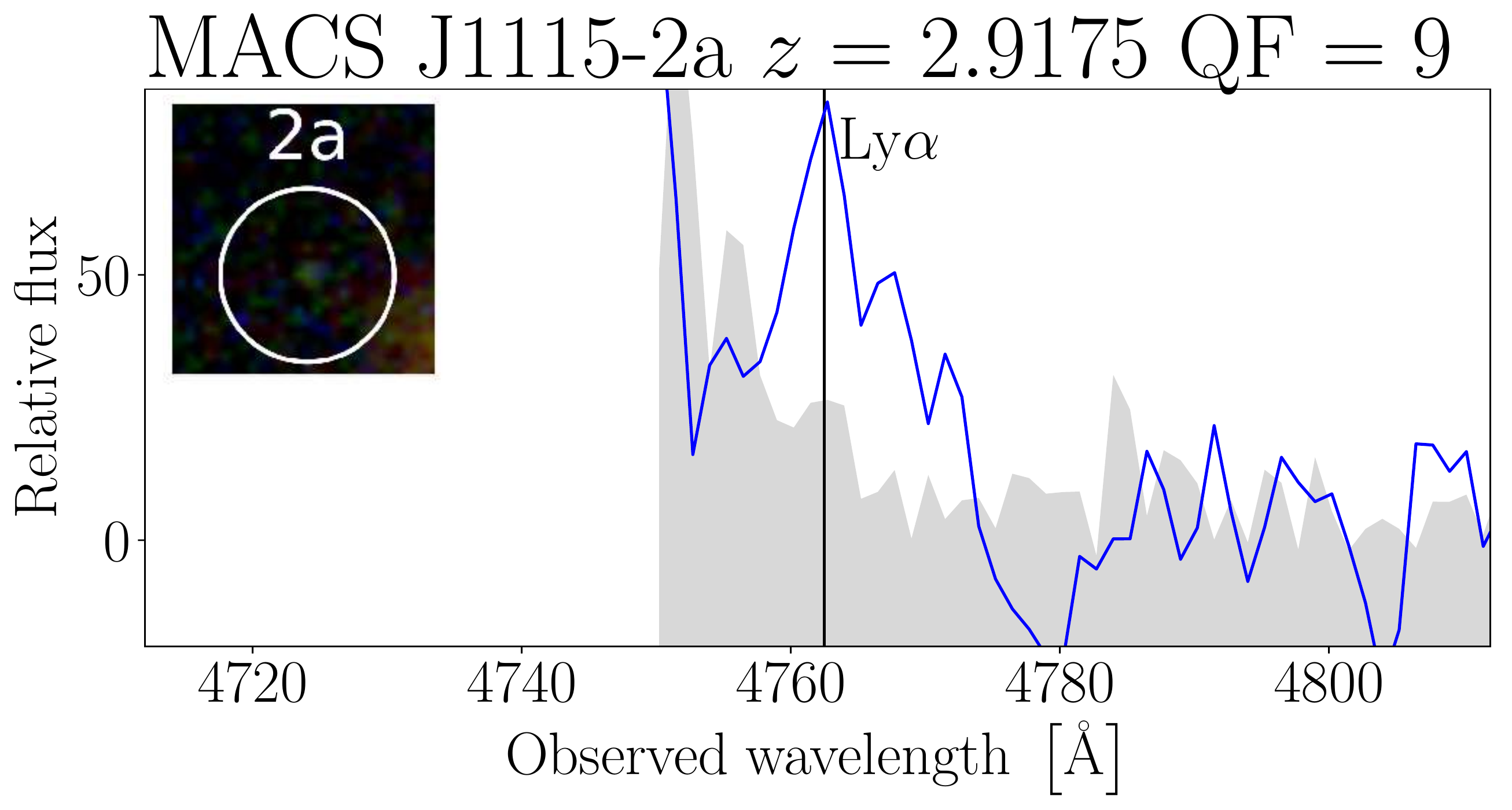}
   \includegraphics[width = 0.666\columnwidth]{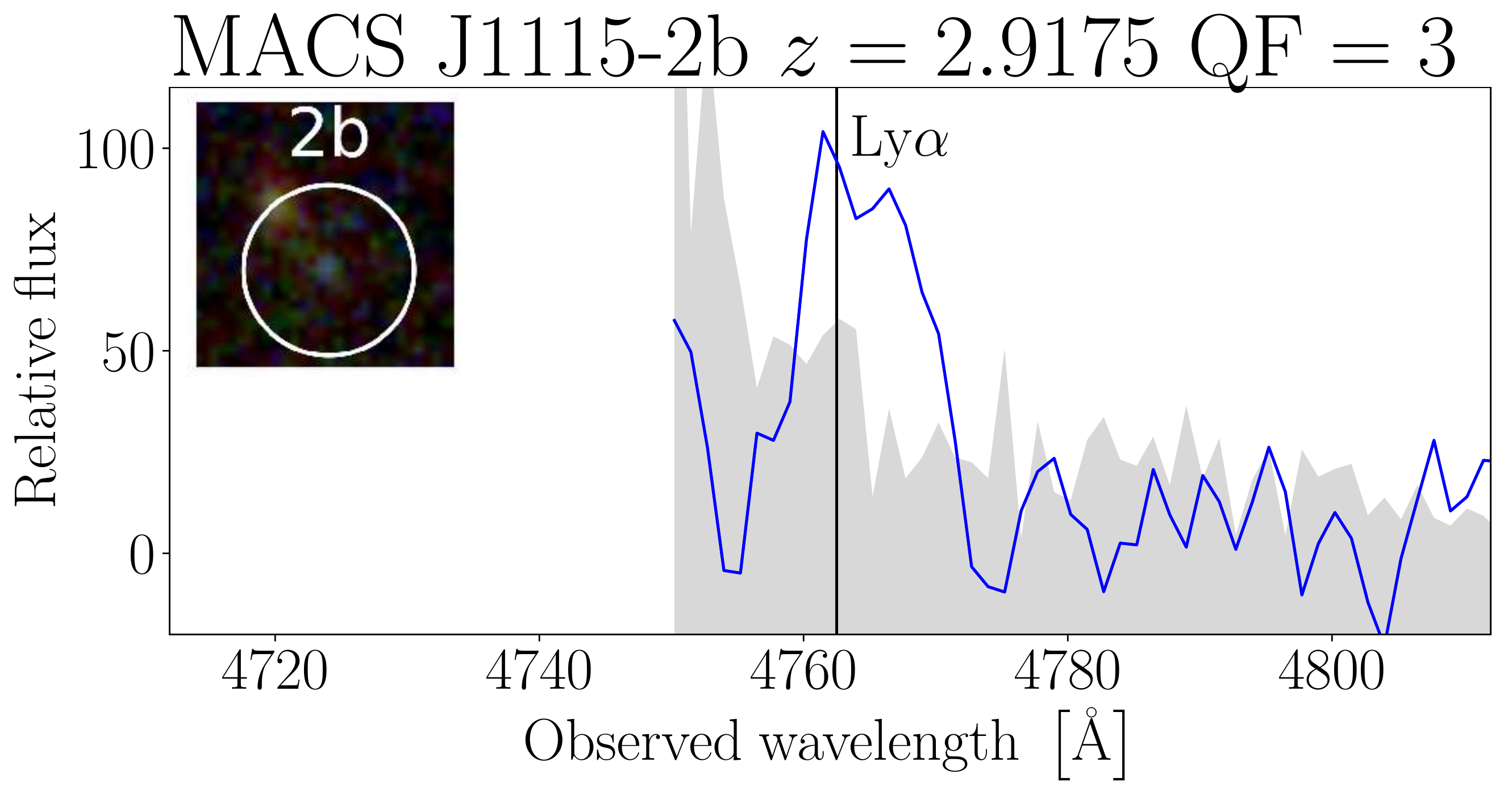}
   \includegraphics[width = 0.666\columnwidth]{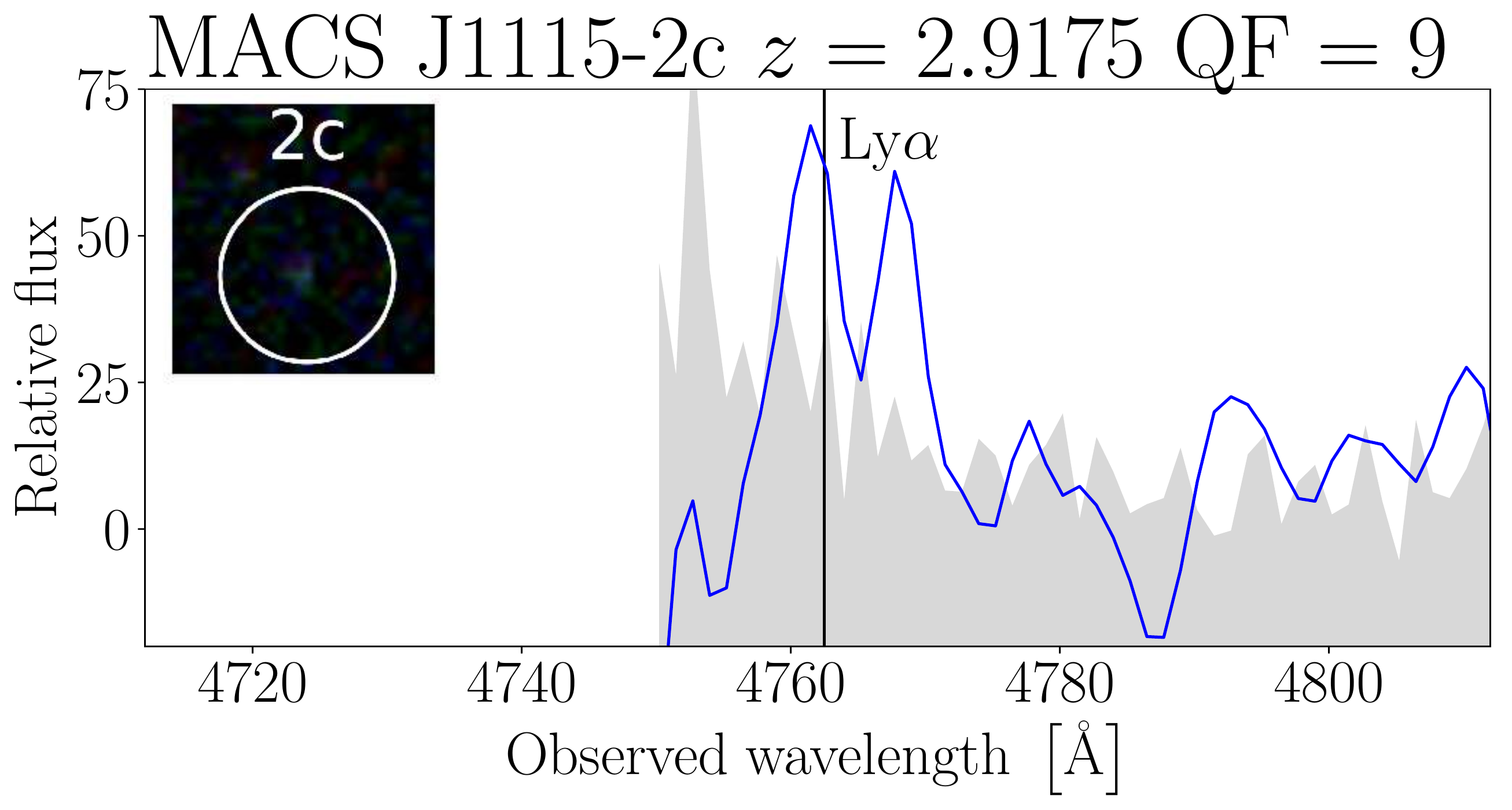}
   \includegraphics[width = 0.666\columnwidth]{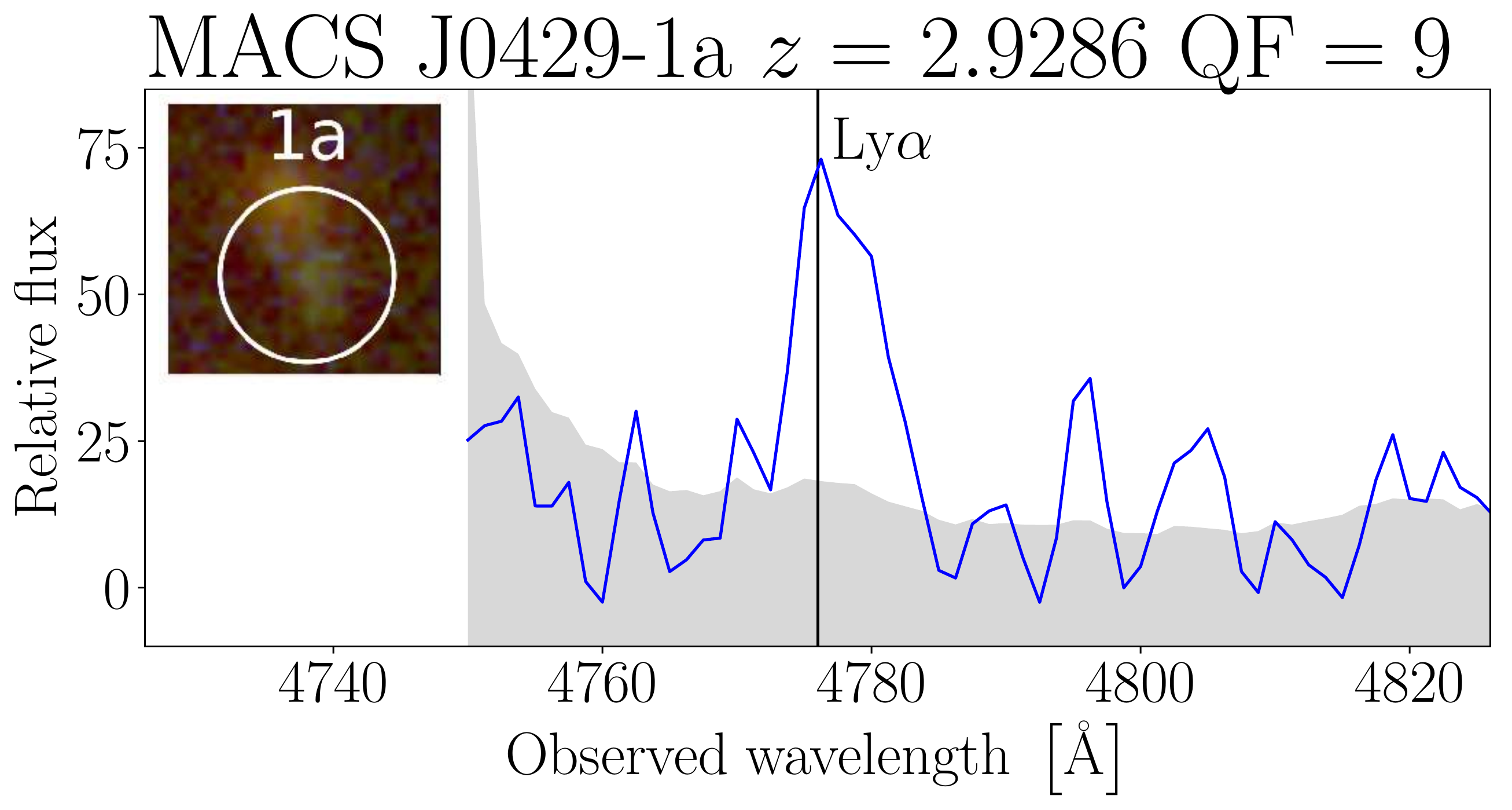}
   \includegraphics[width = 0.666\columnwidth]{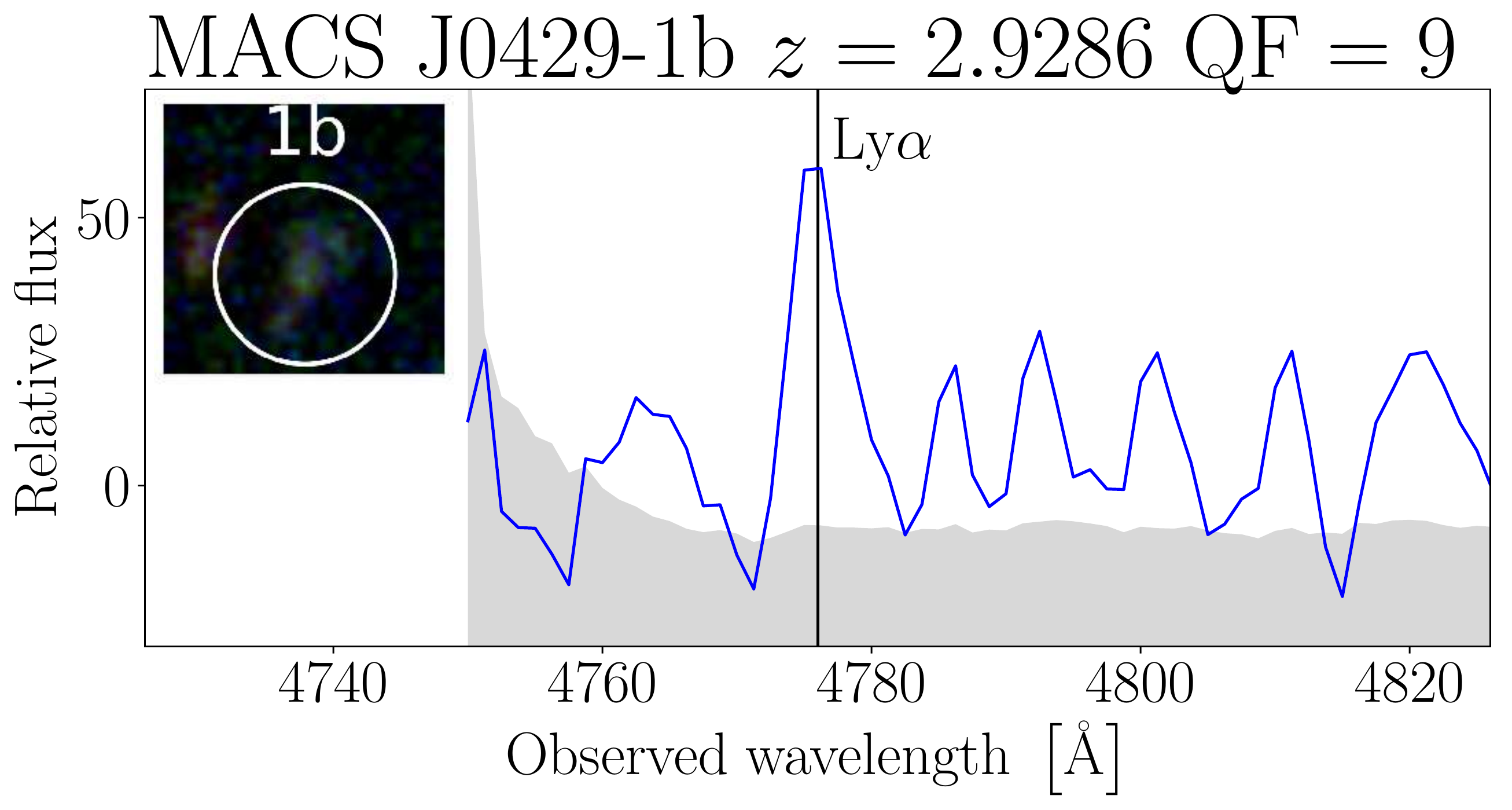}
   \includegraphics[width = 0.666\columnwidth]{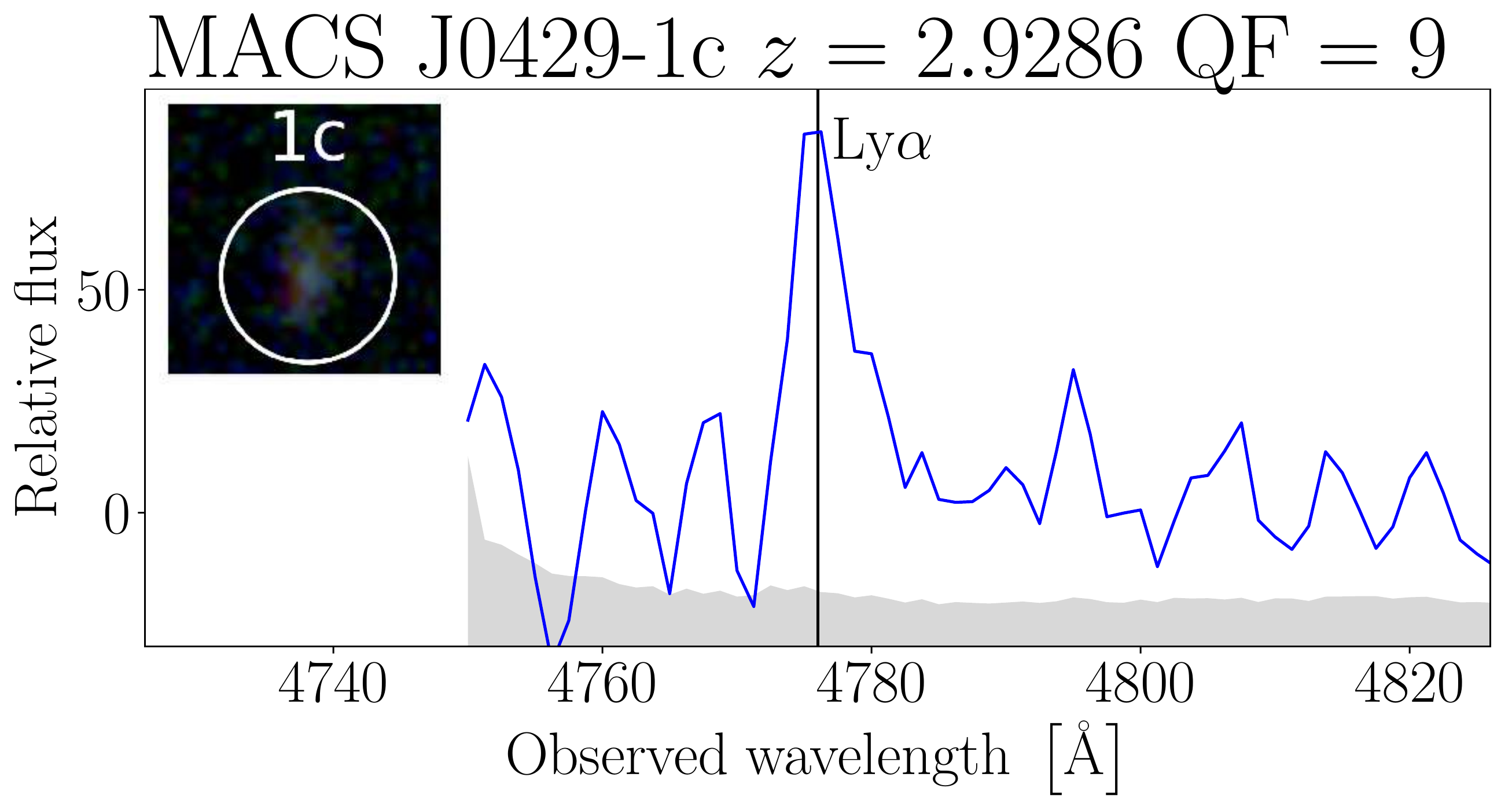}
   \includegraphics[width = 0.666\columnwidth]{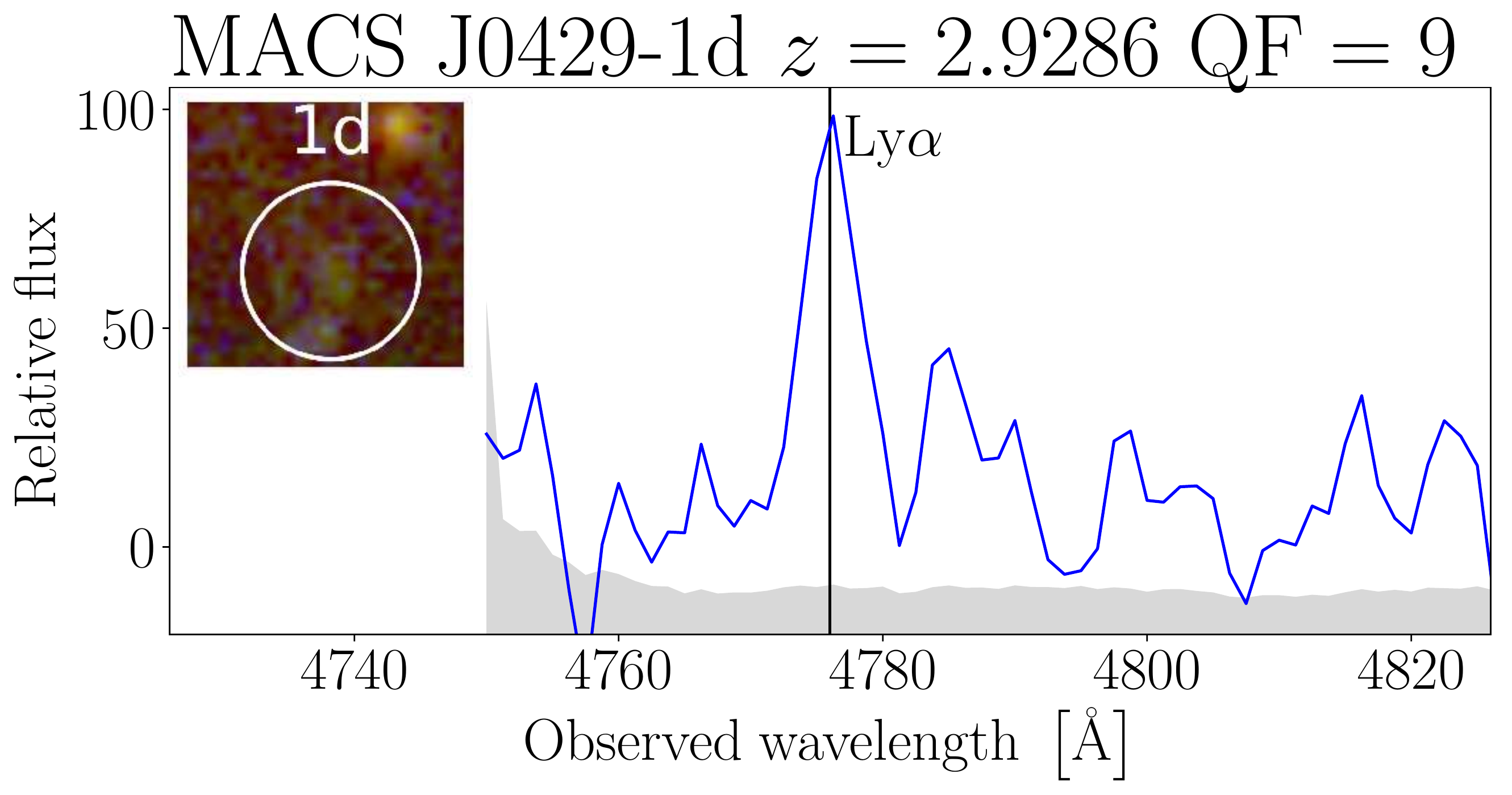}
   \includegraphics[width = 0.666\columnwidth]{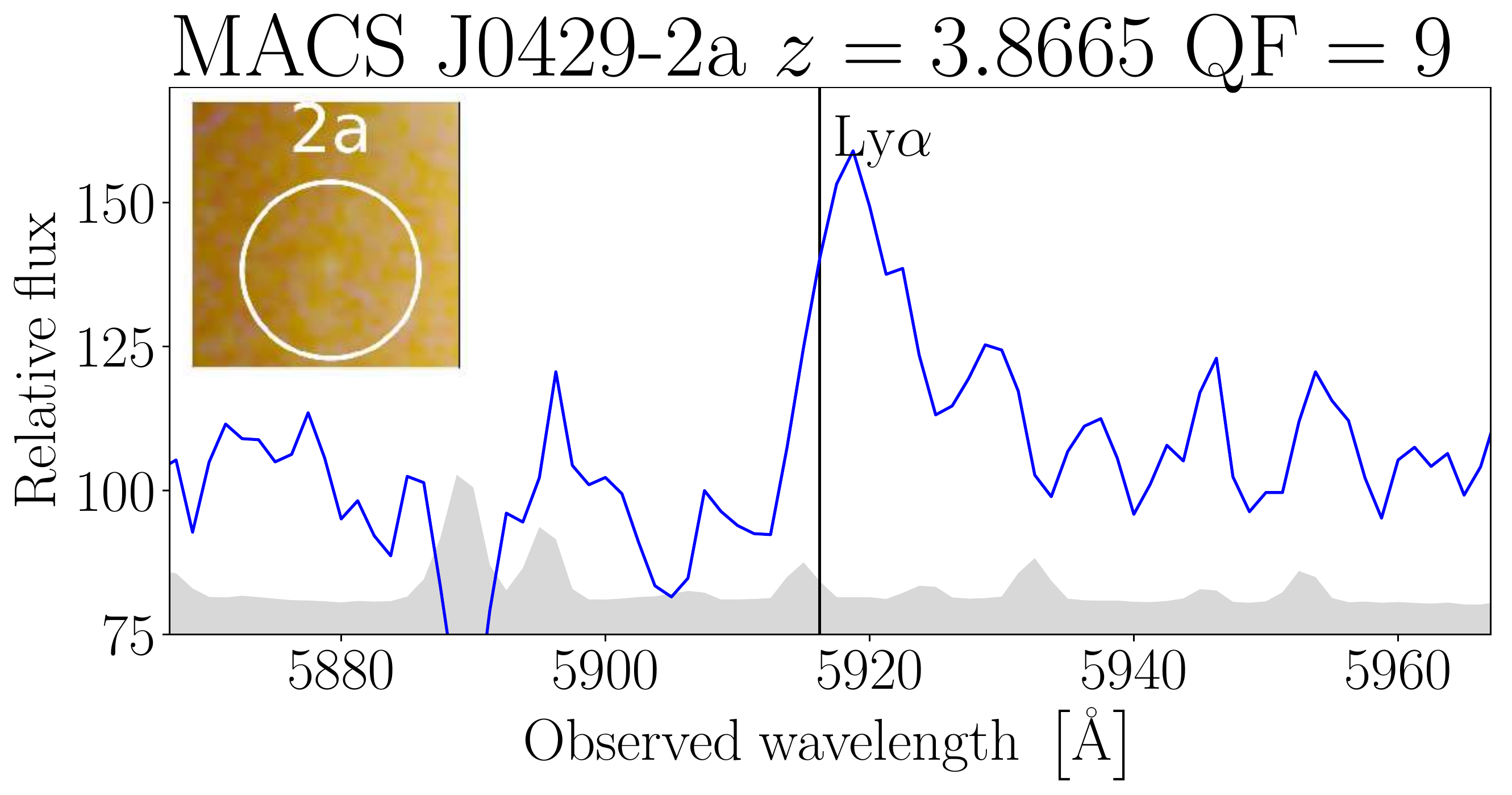}
   \includegraphics[width = 0.666\columnwidth]{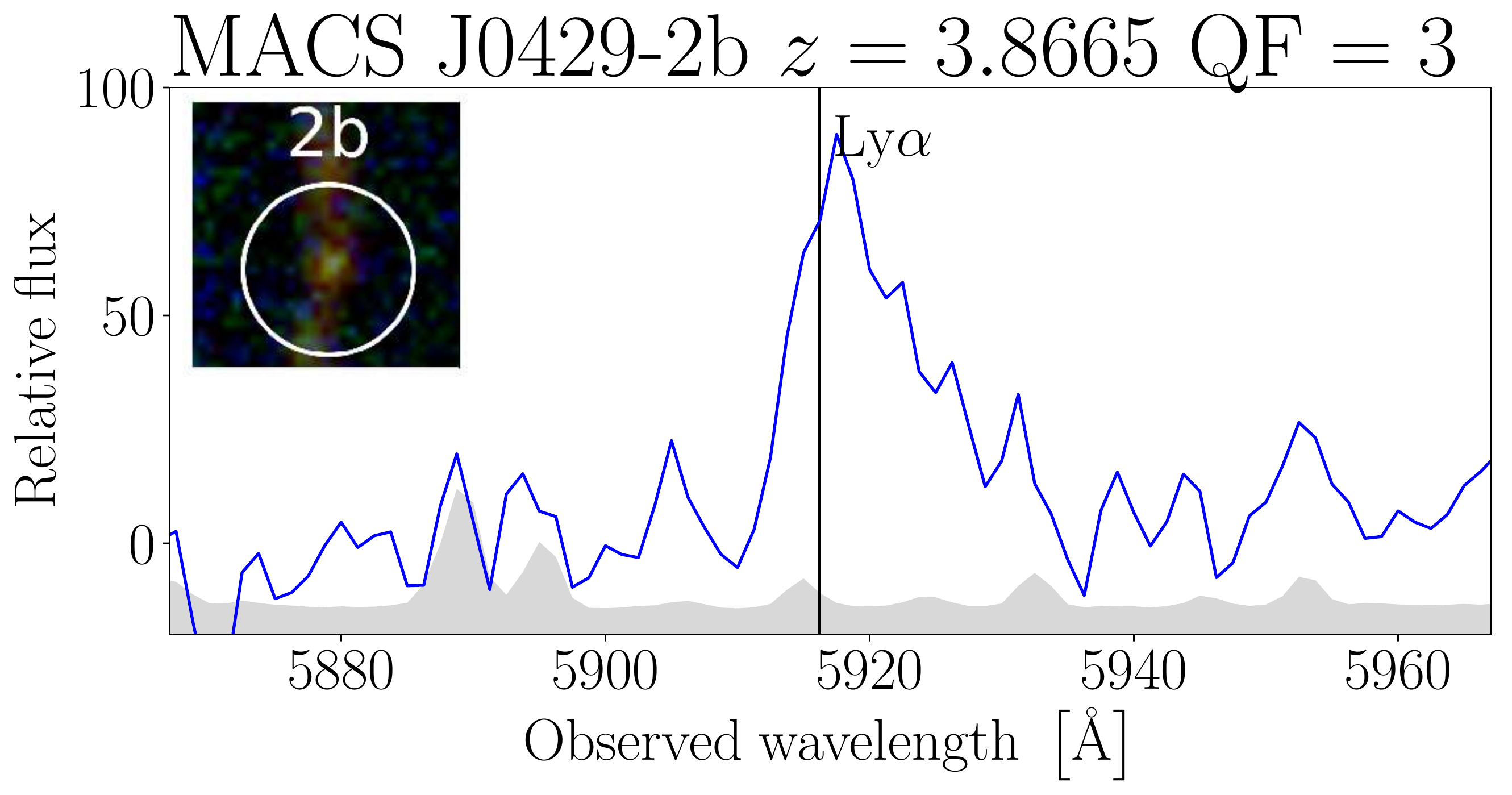}

  \caption{(Continued)}
  \label{fig:specs}
\end{figure*}

\begin{figure*}
\setcounter{figure}{\value{figure}-1}

   \includegraphics[width = 0.666\columnwidth]{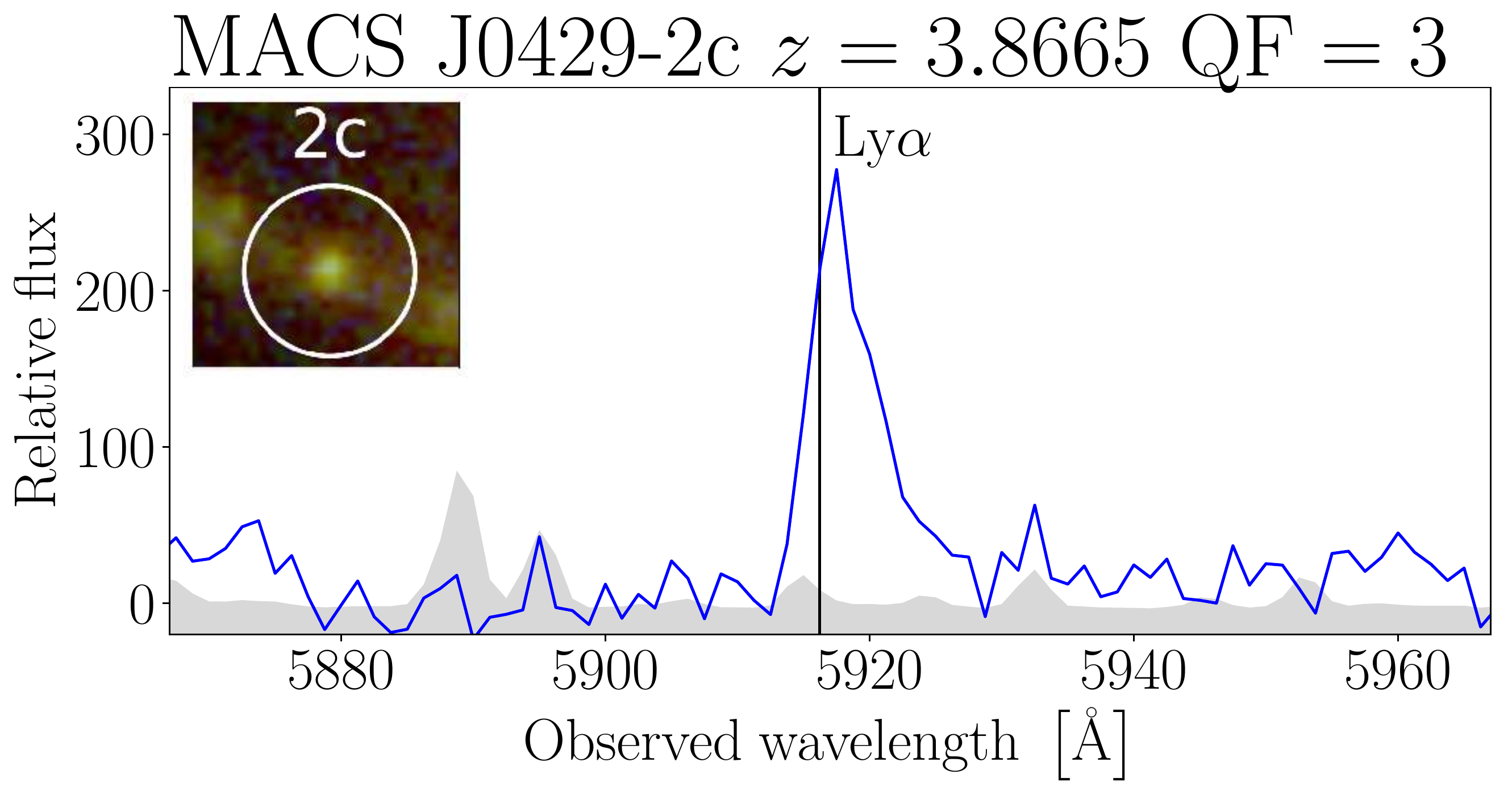}
   \includegraphics[width = 0.666\columnwidth]{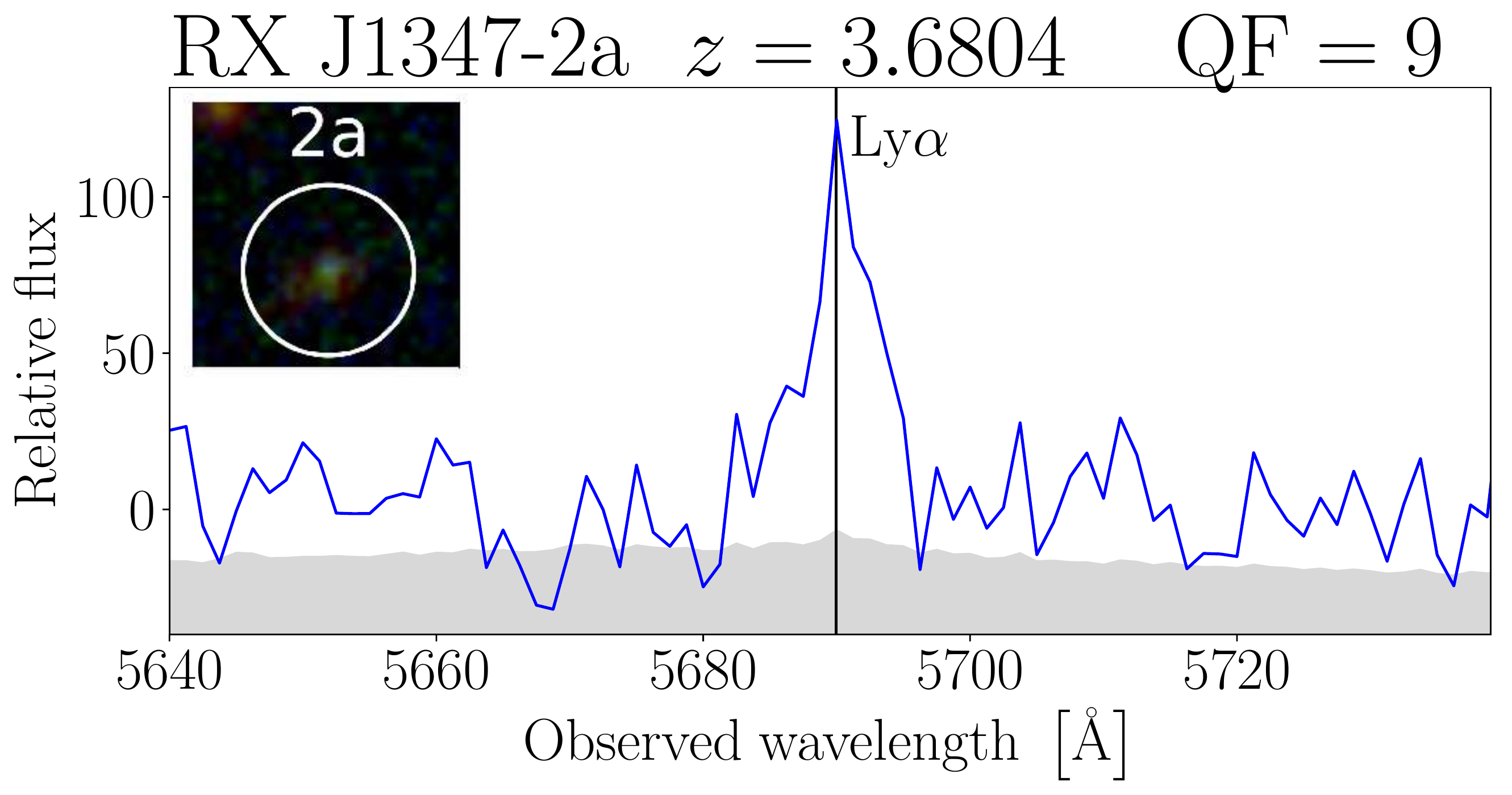}
   \includegraphics[width = 0.666\columnwidth]{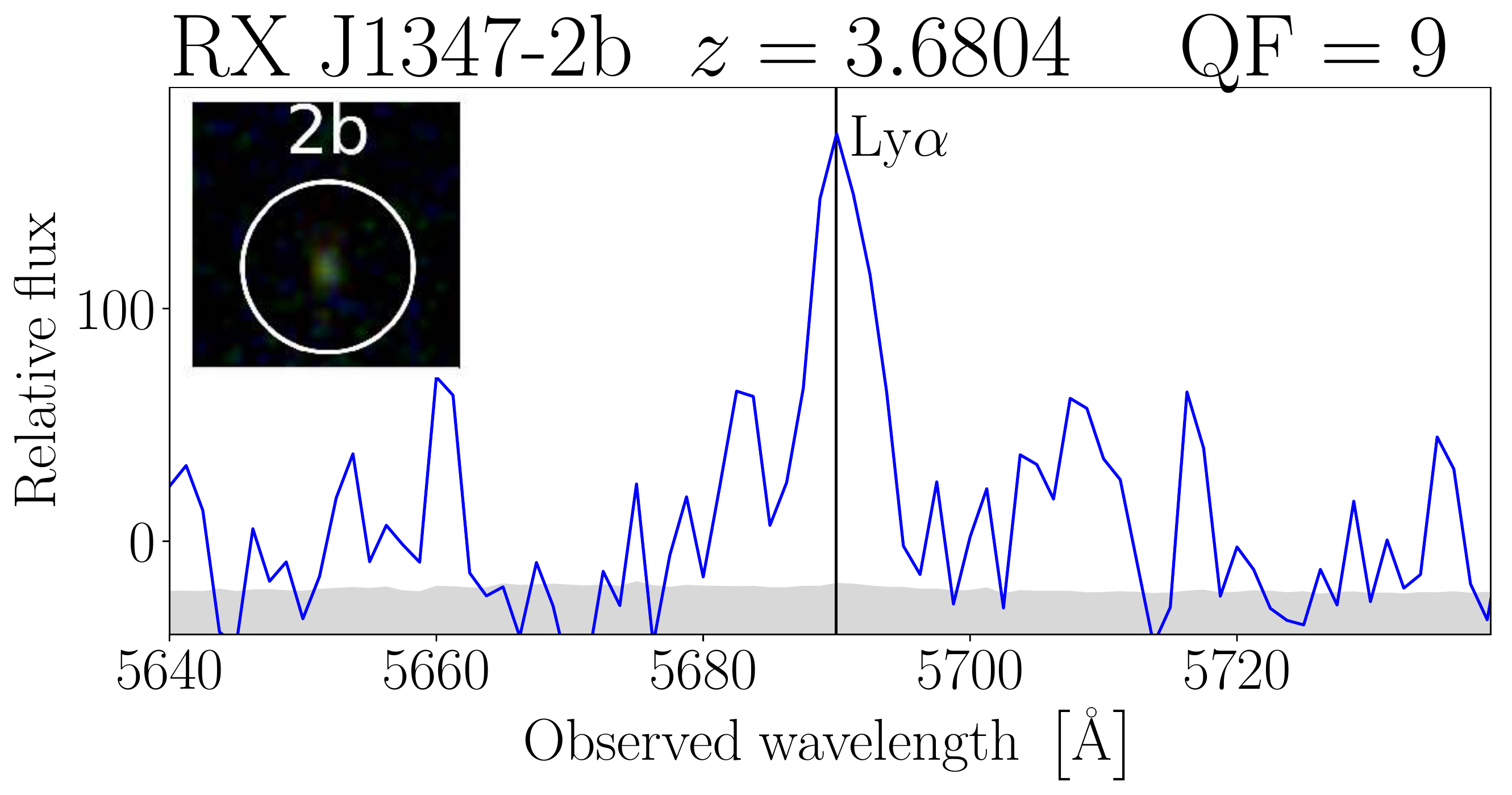}
   \includegraphics[width = 0.666\columnwidth]{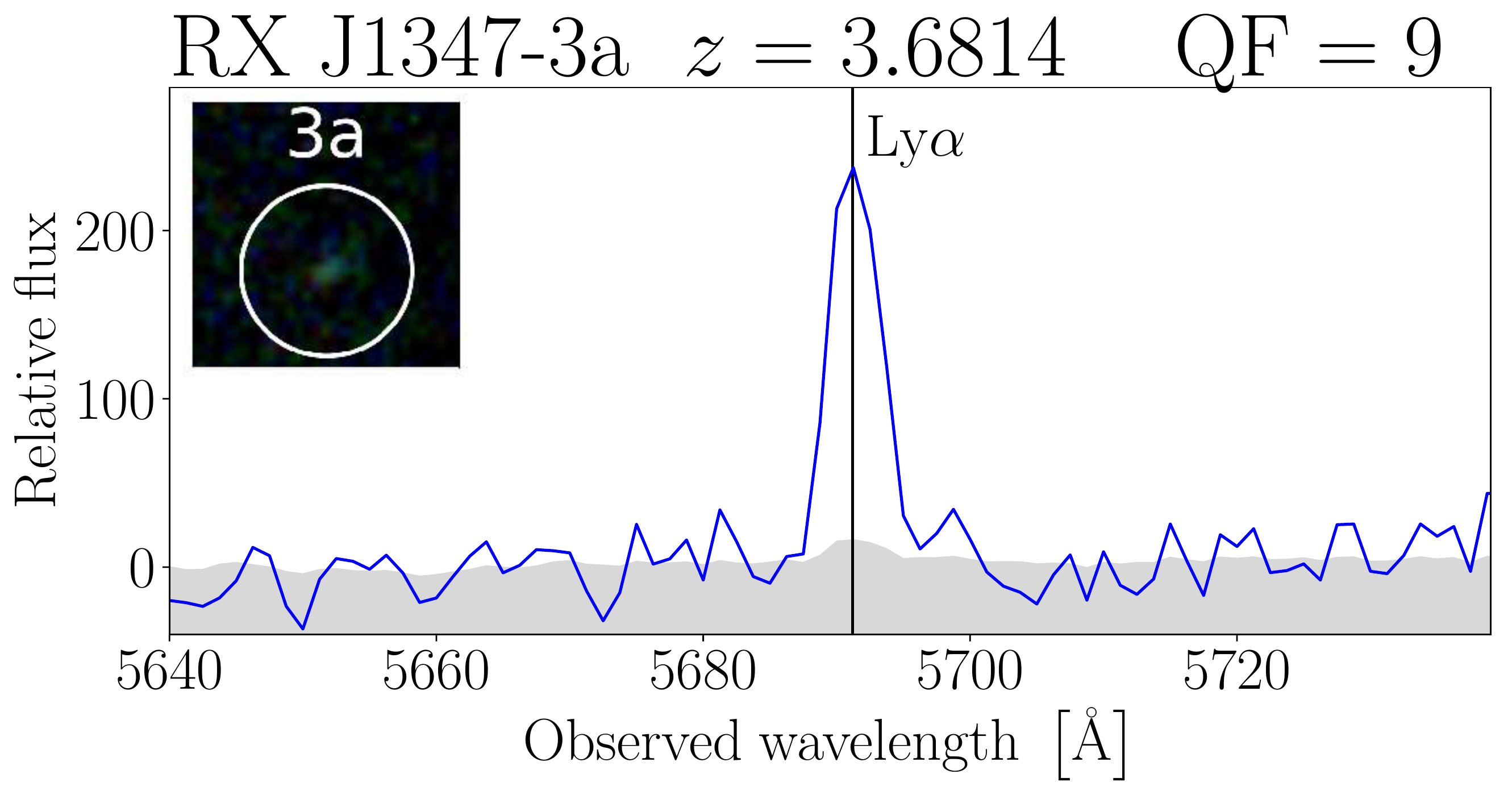}
   \includegraphics[width = 0.666\columnwidth]{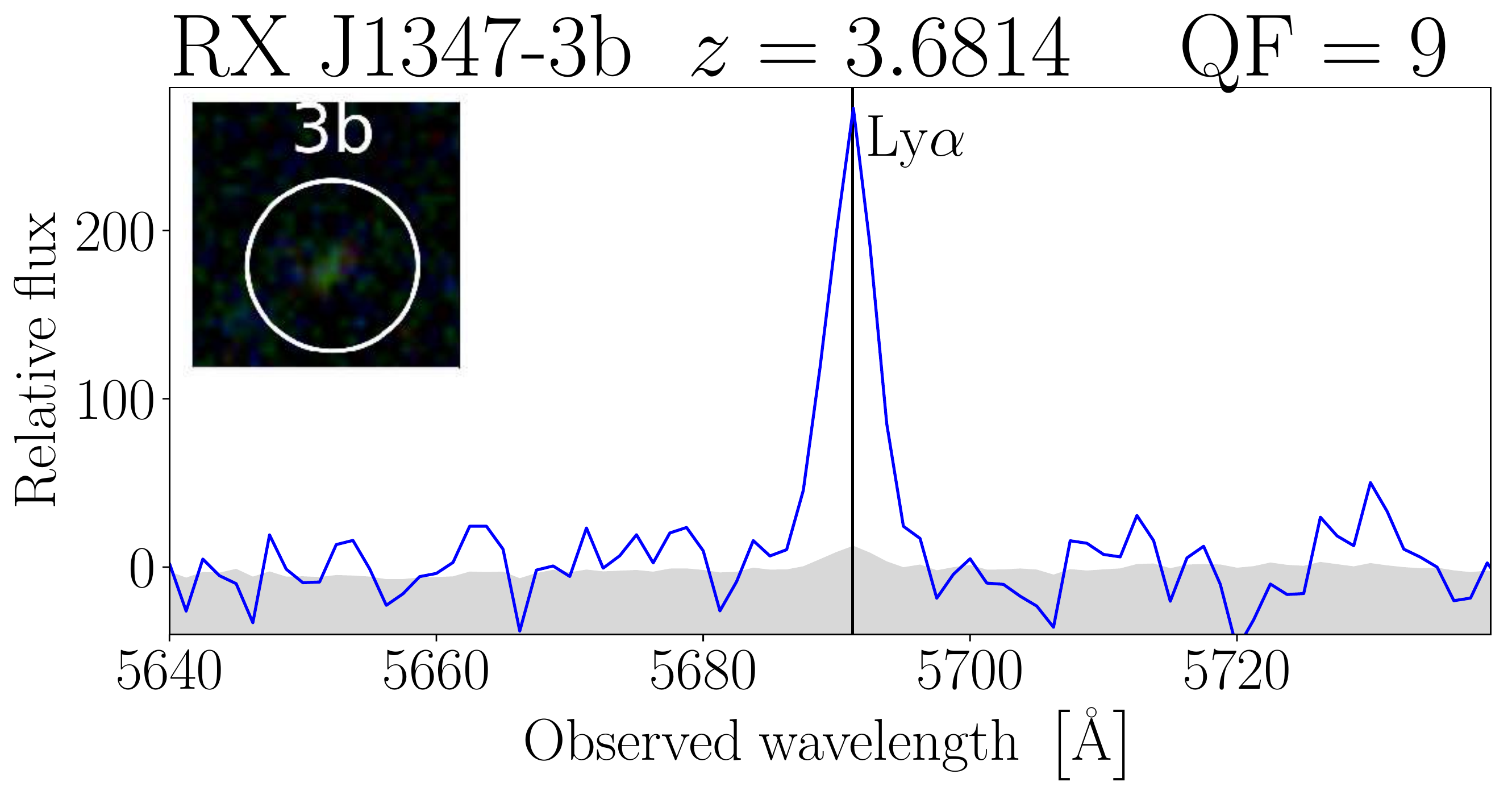}
   \includegraphics[width = 0.666\columnwidth]{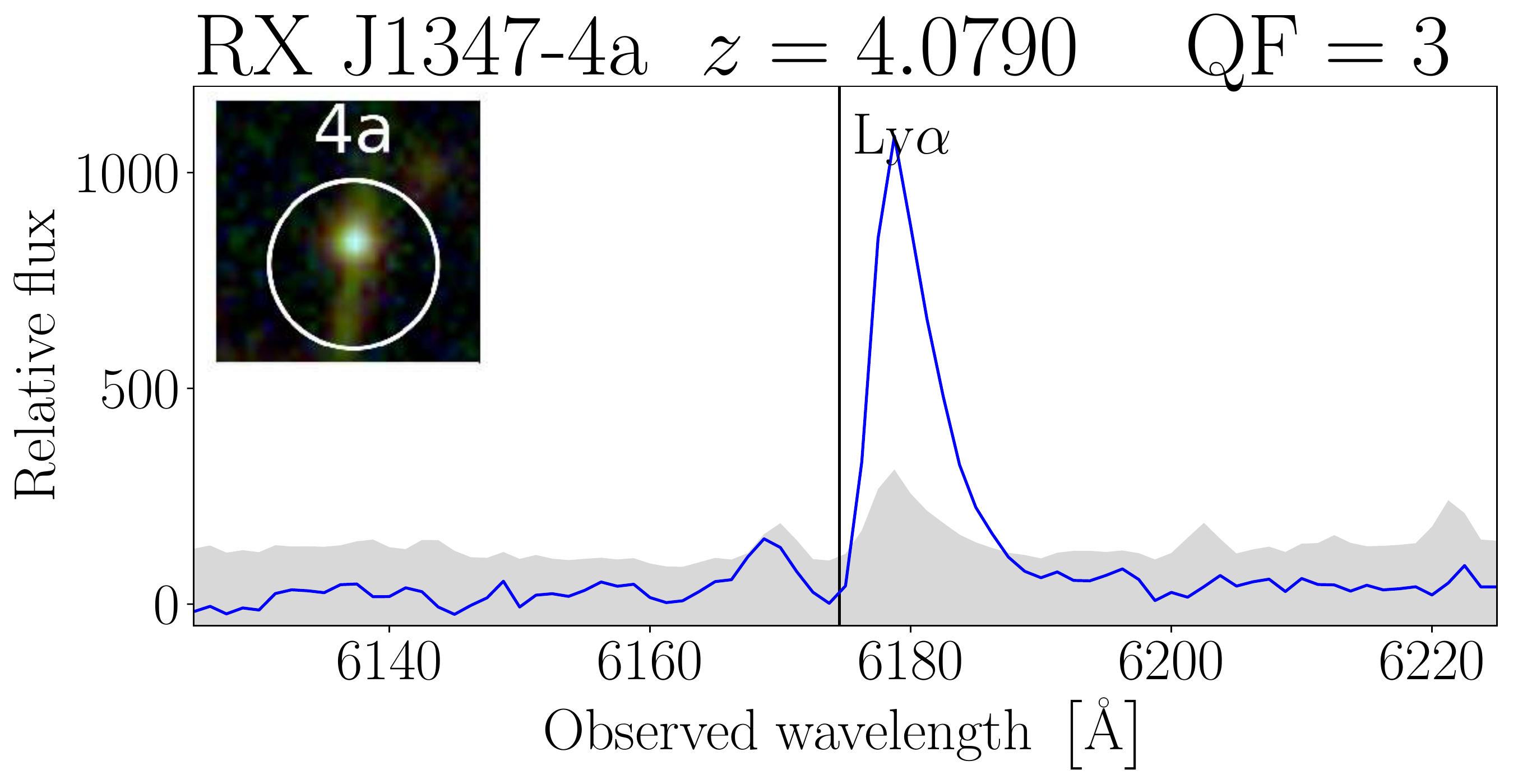}
   \includegraphics[width = 0.666\columnwidth]{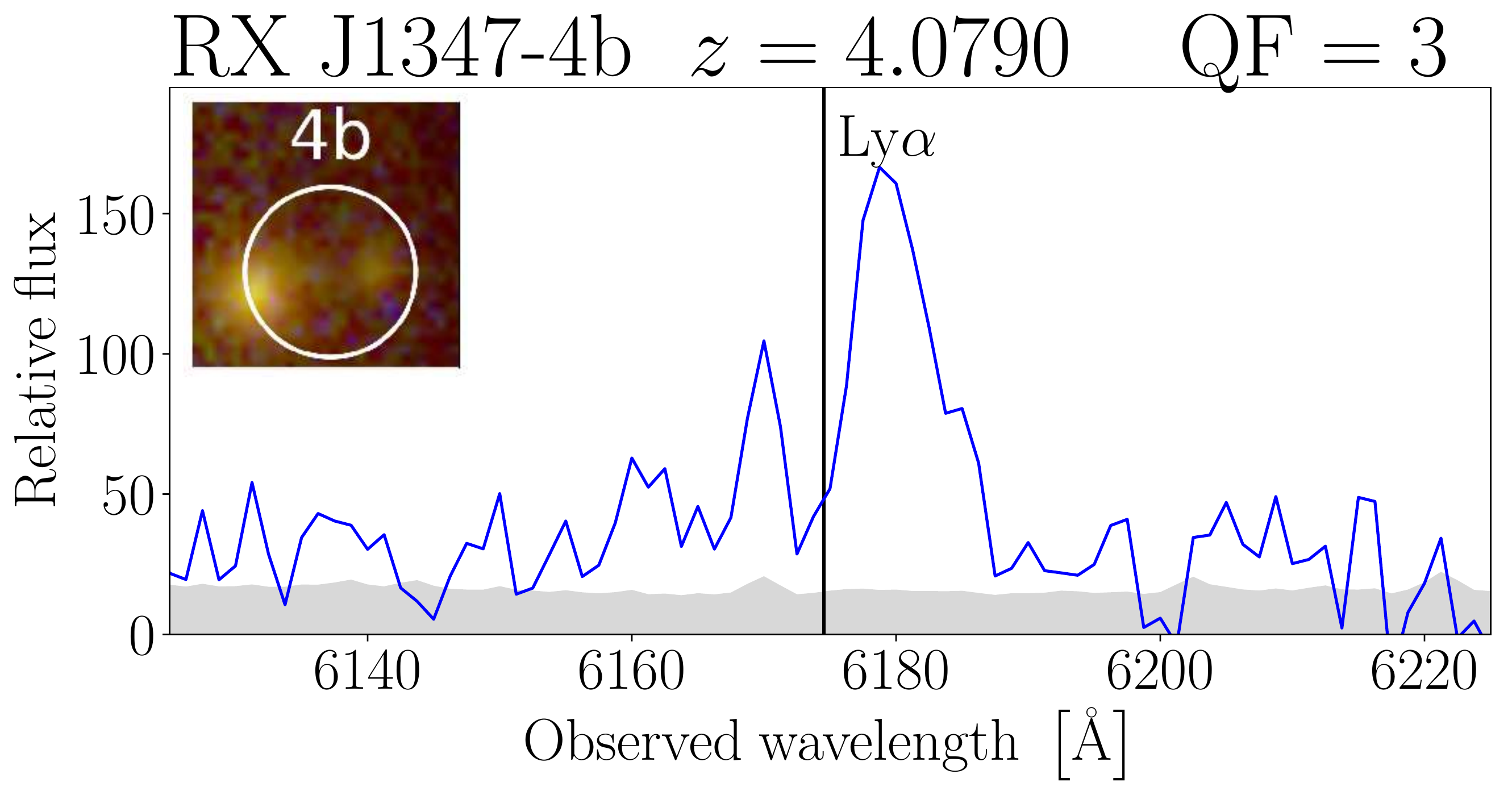}
   \includegraphics[width = 0.666\columnwidth]{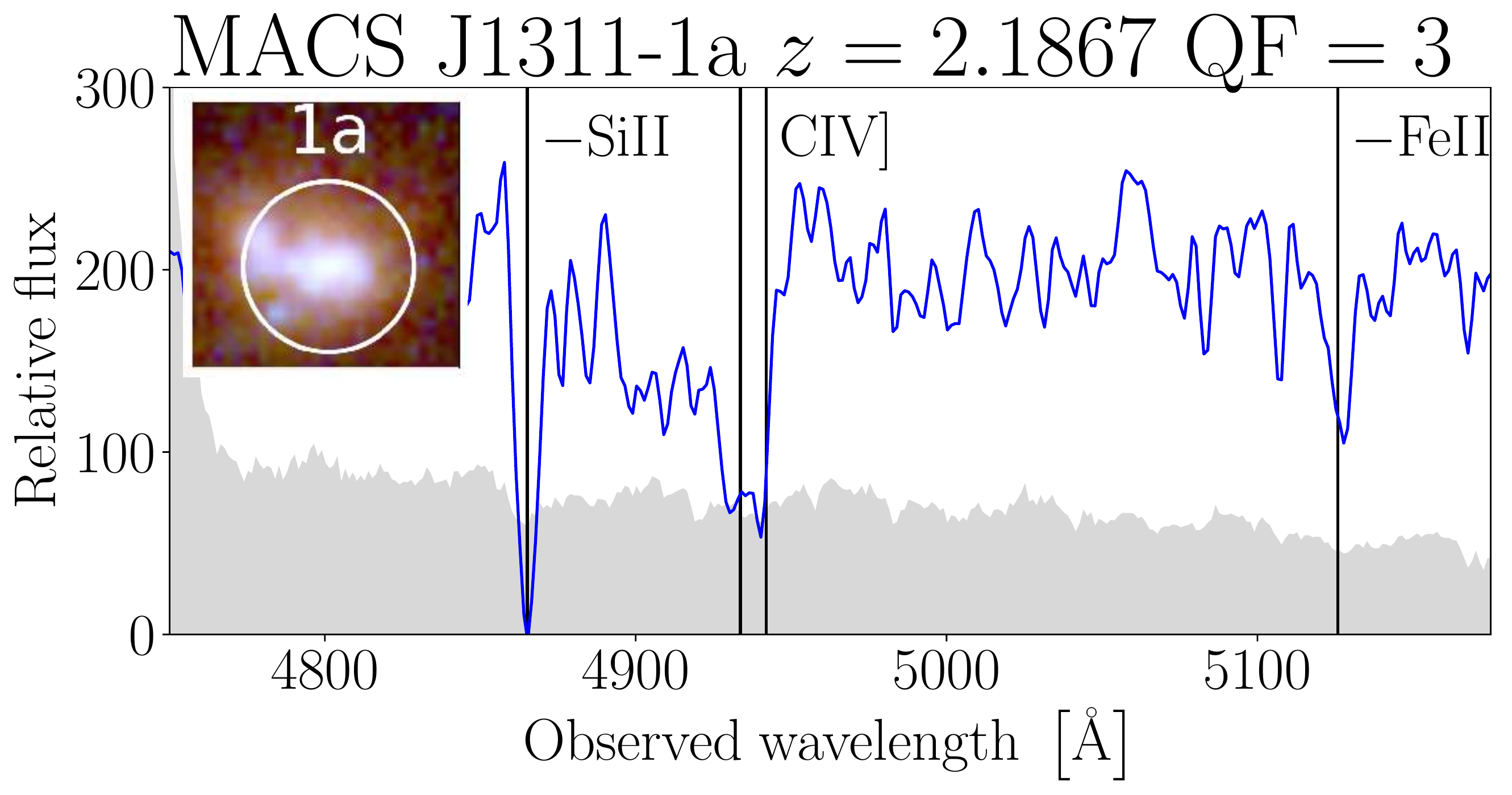}
   \includegraphics[width = 0.666\columnwidth]{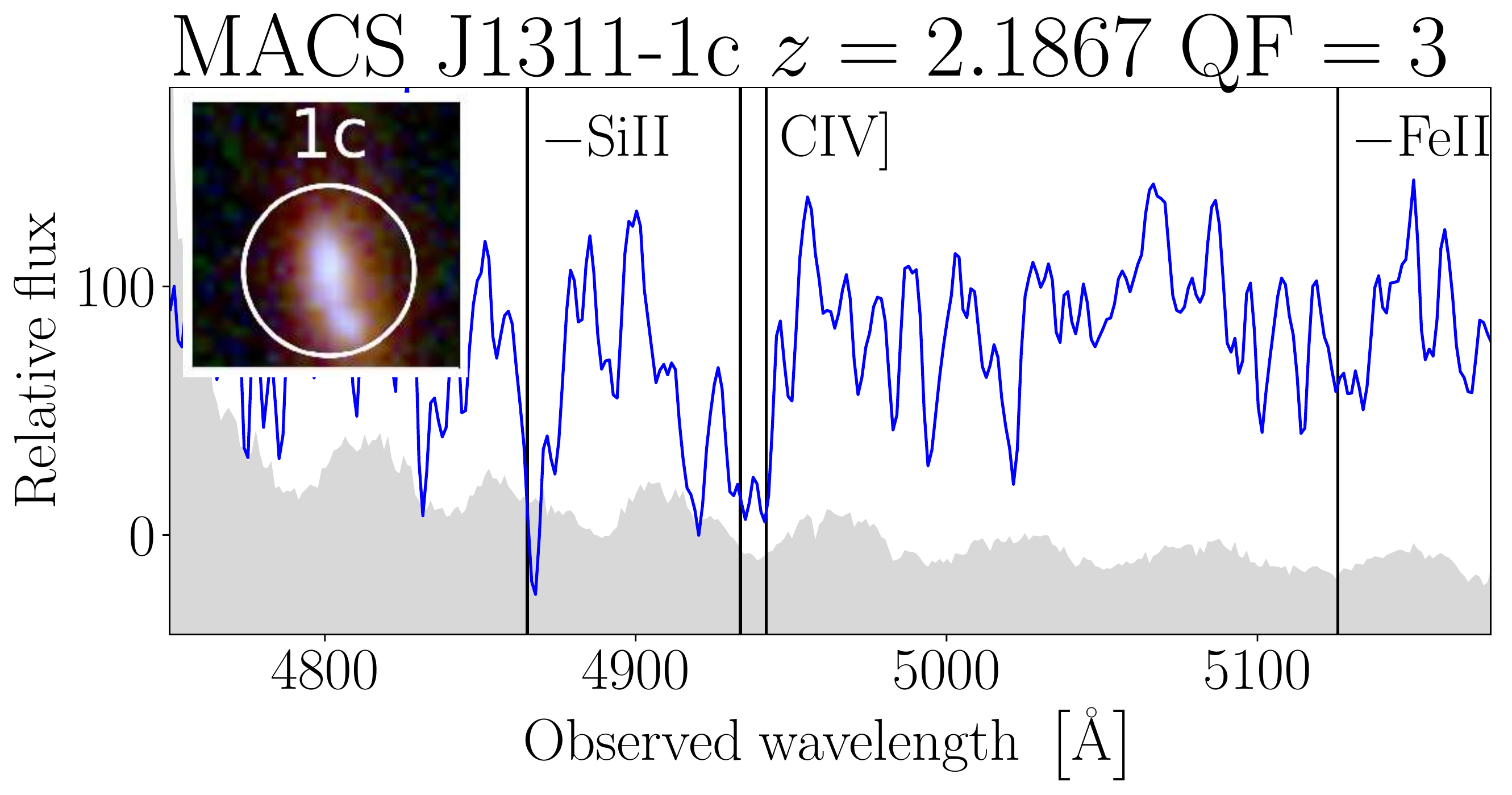}
   
  \caption{(Continued)}
  \label{fig:specs}
\end{figure*}

\end{document}